\PassOptionsToPackage{dvipsnames}{xcolor}
\documentclass[a4paper,11pt]{article}

\usepackage{float}
\usepackage{capt-of}
\usepackage{jheppub}
\usepackage{amsmath,amssymb,amsthm,mathtools}
\usepackage{bm}
\usepackage{booktabs,array,longtable}
\usepackage{enumitem}
\usepackage{microtype}
\usepackage{tikz}
\usetikzlibrary{calc,positioning,decorations.markings}
\hypersetup{
  colorlinks=true,
  linkcolor=MidnightBlue,
  citecolor=MidnightBlue,
  urlcolor=MidnightBlue,
  pdftitle={Hidden Region Finder: asymptotic expansions of Feynman integrals in Minkowski space}
}

\newcommand{\x}{\boldsymbol{x}}
\newcommand{\s}{\boldsymbol{s}}
\newcommand{\K}{\mathcal K}
\newcommand{\HR}{\mathrm{HR}}

\newcommand{\FSL}{\mathcal F_{\rm SL}}
\newcommand{\MSL}{\mathcal M_{\rm SL}}
\newcommand{\MObs}{\mathcal M_{\rm Obs}}
\newcommand{\FObs}{\mathcal F_{\rm Obs}}
\newcommand{\WSL}{W_{\rm SL}}
\newcommand{\WHR}{W_{\rm HR}}
\newcommand{\vHR}{\vec v_{\rm HR}}
\newcommand{\Rpos}{\mathbb R_{>0}}
\newcommand{\ideal}{\mathcal I}
\newcommand{\dd}{\mathrm d}
\newcommand{\SLcolour}[1]{{\color{red}#1}}
\newcommand{\HRcolour}[1]{{\color{blue}#1}}
\newcommand{\Othercolour}[1]{{\color{black}#1}}
\newcommand{\PosInv}[1]{{\color{ForestGreen}#1}}
\newcommand{\NegInv}[1]{{\color{BrickRed}#1}}

\tikzset{
 hrf internal/.style={thick},
 hrf contracted/.style={thick,densely dashed,Blue},
 hrf vertex/.style={circle,fill=black,draw=black,minimum size=3.2pt,
                    inner sep=0pt,outer sep=0pt},
 hrf edge label/.style={fill=white,inner sep=1pt,font=\scriptsize},
 hrf leg label/.style={font=\small},
 hrf momentum label/.style={fill=white,inner sep=.6pt,font=\scriptsize,text=blue}
}
\newcommand{\GlauberVertex}[3][blue]{%
 \begin{scope}[shift={#2},rotate=#3]
  \draw[fill=GreenYellow,draw=#1,line width=.55mm] (0,0) circle (6pt);
  \draw[->,>=stealth,draw=#1,line width=.75pt]
    (-4.2pt,0)--(4.2pt,0);
 \end{scope}%
}

\title{Hidden Region Finder: asymptotic expansions of Feynman integrals in Minkowski space}

\preprint{CERN-TH-2026-233}

\author{Einan Gardi}
\affiliation{Higgs Centre for Theoretical Physics, School of Physics and Astronomy,\\ The University of Edinburgh, Edinburgh EH9 3FD, United Kingdom}
\affiliation{CERN, Theoretical Physics Department, CH-1211 Geneva 23, Switzerland}

\emailAdd{Einan.Gardi@ed.ac.uk}

\abstract{The Method of Regions is a systematic way to derive asymptotic
expansions of Feynman integrals.  For Euclidean integrals there is a
well-established algorithm to determine the complete set of regions as
facets of a Newton polytope defined by the graph polynomials.  In many
physically relevant Minkowski limits, additional regions known as
\emph{hidden regions} are needed to obtain the correct asymptotic expansion.
By making a precise connection with pinch singularities in parameter space
and the associated Landau conditions, we identify the general mechanism
giving rise to these regions and devise an algorithm, the \emph{Hidden Region
Finder}, to determine them.  We show that hidden regions arise through a
delicate interplay between asymptotic scaling of the edge parameters,
which enhances certain monomials of the graph polynomial, and cancellations
amongst these monomials, which reduce their collective contribution to the
same order as other terms in the polynomial.
We explore massless four-, five- and six-point integrals in a variety of
wide-angle mass-shell, planar, collinear, double-collinear and Regge limits.  Across these
different expansions, hidden regions recur in the same small set of seed
topologies.  In all cases studied, they can be traced to wide-angle
configurations with multiple hard-scattering subdiagrams.  Distinct kinematic
limits of a given seed inherit a common singular locus, while each limit fixes
its own scaling vector and cancellation depth.  This exploratory study showcases the
potential of the proposed algorithm both to uncover missing contributions to
asymptotic expansions and to organise them through their underlying singular
geometry.}

\begin{document}
\maketitle
\flushbottom

\section{Introduction}

Identifying the leading regions of Feynman integrals is fundamental whenever
a scattering problem contains widely separated scales.  Their most immediate
use is to construct asymptotic expansions of individual Feynman integrals.
Such expansions have supplied essential ingredients in a wide range of
state-of-the-art computations
\cite{Pak:2009dg,Pak:2009bx,Anastasiou:2015ema,Jaskiewicz:2025dihiggs,
Guan:2024hlf,Gardi:2025ule,Gardi:2025massive}.
The same region analysis determines the momentum configurations that underlie
factorisation, evolution and resummation of large logarithms in scattering
amplitudes and cross sections.  In the classic analyses of Collins, Soper and
Sterman, for example, hard, collinear, soft and Glauber configurations enter the proof
of parton-distribution-function (PDF) factorisation, with the contour
deformation or cancellation of the Glauber contribution providing a central
subtlety \cite{Collins:1984kg,Collins:1985ue,Collins:1989gx,Becher:2024kmk}.
More generally,
factorisation into hard, jet and soft functions, followed by their evolution,
underlies perturbative resummations of threshold and transverse-momentum
logarithms \cite{Catani:1989ne,Contopanagos:1996nh,Catani:2000vq}.  Effective
field theories take this separation of momentum modes a step further by
encoding it directly in their degrees of freedom and Lagrangian.  In
heavy-quark effective theory and soft-collinear effective theory, the modes
relevant at a given scale are represented by explicit fields, while harder
fluctuations are integrated out
\cite{Grozin:2004yc,Bauer:2000yr,Bauer:2000ew}.  A systematic procedure for
identifying regions is therefore a starting point both for asymptotic
expansion at the level of individual integrals and for establishing the
factorisation and resummation properties of amplitudes and cross sections.

The Method of Regions (MoR) implements this separation of scales through an
integrand-level expansion.  Once an integration chart (a chosen local
coordinate system for the integration variables) and its overlap
prescription have been fixed, a region is characterised by the scaling law of
the integration variables as the limit is approached, which determines the
corresponding Taylor expansion of the integrand.  Remarkably, in dimensional
regularisation, the correct asymptotic expansion of the original integral is obtained by summing
the contributions of all regions, with each expanded integrand integrated
over the full original domain.  Extending all regional approximations in
this way makes them overlap.  The corresponding overlap terms
are obtained by applying the expansions associated with two or more regions
to the same integrand.  When all
ultraviolet and infrared singularities are appropriately regulated, the sum
of these overlap contributions, with the corresponding inclusion--exclusion
signs, normally reduces to scaleless integrals and therefore vanishes.
Jantzen established this relation in momentum space under explicit
conditions and showed that overlap contributions must instead be retained
when the chosen regularisation makes them non-scaleless
\cite{Jantzen:2011nz}; see also
refs.~\cite{Smirnov:2002pj,Semenova:2018cwy,Ma25}.  The construction assumes
that one is already given a set of regions together with non-overlapping
domains that cover the full momentum integration space and satisfy the
required admissibility conditions.  It therefore validates the MoR once a
complete admissible set has been supplied, but does not determine that set;
for a generic expansion, the corresponding separation must still be
constructed case by case.  Determining a complete region set for
non-Euclidean expansions is the central problem addressed in this paper.

The MoR has its origins in the expansion-by-subgraphs programme for Euclidean
large-momentum and large-mass limits
\cite{CtkGrshnTch82,Ctk83,GrshnLrnTch83,GrshnLrn87,SmthDVr88,Grshn89}.
Smirnov placed these expansions on a systematic footing and related their
terms to asymptotically irreducible subgraphs
\cite{Smirnov:1990rz,Smirnov:1994tg}.  The more general strategy of regions was
established by Beneke and Smirnov in their treatment of threshold expansion
\cite{Beneke:1997zp}, and was subsequently developed for other non-Euclidean
regimes, notably Sudakov limits
\cite{Smirnov:1998vk,Smirnov:1999bza}.  In such limits the hard
region is supplemented by soft and collinear modes and, where appropriate,
potential or Glauber modes.  For general non-Euclidean expansions, however,
no procedure is known that determines the complete set of regions: a
plausible list of momentum modes may be suggested by physical insight, but
omitting even one region gives an incorrect asymptotic expansion.

Parameter space provides an important route towards making this problem
algorithmic.\footnote{Early developments along this parametric route include the
alpha-parameter reformulation of the strategy of regions in
ref.~\cite{Smirnov:1999bza} and the sector-decomposition and Mellin--Barnes
algorithm of ref.~\cite{Pilipp:2008power}, which provides an independent
power expansion, supplies the ansatz for a differential-equation analysis,
and offers a numerical check on a regions analysis.}  A key advance
was the geometric formulation of Pak and Smirnov \cite{Pak:2010pt}: it maps
every monomial of a graph polynomial to its exponent vector and identifies
region vectors with inward normals to lower facets of the resulting Newton
polytope.  The Lee--Pomeransky
representation \cite{Lee:2013hzt}, in which the polynomial
\(\mathcal P=\mathcal U+\mathcal F\) is integrated over the positive orthant
\((x_e>0\ \forall e)\),
provides a particularly economical setting for this construction.  For
Euclidean integrals, and more generally when the relevant polynomial admits a
same-sign representation, the lower facets determine the regions directly
from the polynomial support.  The reason is that the asymptotics in parameter
space are then controlled by endpoints.  Under a generally non-uniform
scaling of the parameters towards a boundary, the leading monomials have a
common weight and their exponent vectors span a lower facet, whose inward
normal is the region vector.  We refer to such contributions as \emph{facet regions}.  In
this setting the geometric construction turns the endpoint analysis into a
convex-geometric problem.  It underlies implementations such as
\textsc{asy2}, \textsc{aspire} and the
expansion-by-regions module of \textsc{pySecDec}
\cite{Jantzen:2012mw,Ananthanarayan:2018tog,Heinrich:2021dbf}.

The limitation is that not every region is characterised by endpoint scaling
in the original parameters.  Unlike in the fixed-sign case, in Minkowski
kinematics \(\mathcal F\) generally contains terms of opposite sign, which can
cancel at finite positive values of the parameters.  Such cancellation
surfaces may pinch the integration contour and need not correspond to facets
of the original Newton polytope.  In momentum space, contour pinches are
intrinsic to the identification of contributing modes; once their analogues
are relevant in parameter space, the polynomial support alone is therefore
insufficient.  Potential and Glauber regions provided important early
examples of this difficulty.
For particular classes of such examples, the relevant cancellations can be
exposed before the Newton-polytope construction by suitable linear changes of
parameters, as implemented in \textsc{asy2} \cite{Jantzen:2012mw}.
This is a valuable but intrinsically restricted mechanism: it requires a
cancellation surface that can be resolved by the available linear
transformations.  It does not provide a general determination of regions in
Minkowski kinematics, where cancellation surfaces may be polynomial and may
only become relevant after a nontrivial asymptotic scaling.

The missing information is supplied by the singularity structure of the
integral.  The Landau equations give necessary conditions for a singularity:
propagator singularities must trap the integration contour, producing a
pinch surface \cite{Lnd59,Bjorken:1959fd,Nakanishi:1959jzx,
Collins:2020euz}.  Ref.~\cite{Gardi:2022khw} made the connection between such
pinches and \emph{facet} regions precise for the wide-angle on-shell
expansion: facets of the Newton polytope were related to infrared solutions
of the Landau equations and to hard, collinear and soft momentum
configurations.  This led to a graph-theoretical construction of the facet
regions, subsequently proved and developed further in
ref.~\cite{Ma:2023hrt}.

A Newton polytope retains the exponents of the monomials but not the signs and
kinematic information carried by their coefficients.  It therefore cannot
decide whether a cancellation locus is pinched or whether an asymptotic
scaling promotes it to a genuine region.  Supplying precisely this missing
information is the task required to complete the MoR beyond facet regions.

Ref.~\cite{Gardi:2024axt} subsequently showed that genuinely nonlinear
pinches, and hence non-facet regions, do occur in the same wide-angle on-shell
expansion.  In massless \(2\to2\) scattering, the first such example appears
for a special nonplanar three-loop topology known as the Crown.  A dissection
of parameter space at the singular locus
converts the interior pinch into endpoint singularities, after which ordinary
geometric sector-decomposition and MoR algorithms can be applied in the local
coordinates.  The resulting additional contributions, absent from the
facets of the original Newton polytope, were termed \emph{hidden regions}.
For the Crown, the hidden region describes Landshoff scattering in the
wide-angle on-shell expansion and a Glauber configuration in the Regge limit
\cite{Gardi:2024axt}.  The underlying fixed-angle momentum configuration
predates the HR terminology: related perturbative and independent-scattering
analyses include
Refs.~\cite{Halliday:1964fixedangle,Landshoff:1974wideangle,Botts:1989kf},
with Botts and Sterman providing the momentum-space power counting used in
Ref.~\cite{Gardi:2024axt}.  Identifying the wide-angle and Regge
configurations as HRs of the same seed was the first indication that a special
topology can generate hidden regions in more than one kinematic expansion.

Dissection provides a decisive certificate for a hidden region and a cover
by useful local coordinates for evaluating its contribution, but it presupposes knowledge
of the pinch locus.  The next step was taken in the five-point
spacelike-collinear problem of ref.~\cite{Chen:2026dnj}.\footnote{A
complementary study appearing concurrently analysed the spacelike-collinear
setting at amplitude level, deriving an all-order soft-collinear subgraph
factorisation for a Glauber-pinched gluon and relating the resulting
factorisation violation to Reggeisation~\cite{Barcaro:2026reggeization}.}  In that example the
leading graph polynomial separates into a factorised, individually
superleading sector and an obstruction.  The pinch conditions apply to the
former only after a non-uniform LP-parameter scaling has made the
obstruction subleading.  This example made explicit the essential interplay
between cancellation and scaling and motivated a general procedure for
discovering the region directly in the original parameters.

The purpose of the present paper is to formulate that procedure as the
\emph{Hidden Region Finder} (HRF), thereby addressing a central obstacle to
the practical use of the MoR in non-Euclidean limits.  The core construction
has two stages.
First, HRF extracts candidate polynomial cancellation factors and uses them
to decompose the relevant leading polynomial into a superleading
cancellation sector \(\mathcal F_{\rm SL}\) and an obstruction
\(\mathcal F_{\rm Obs}\).  It then determines a consistent scaling and
measures separately the first resolved weights of \(\mathcal F_{\rm SL}\)
and of the rest of the Lee--Pomeransky polynomial.  Their equality is
observed in all examples below rather than imposed by the algorithm.
Positivity of the LP parameters and a source-resolved dissection, culminating
in an exact local lower facet, certify the result.  Boundary
hidden regions are included by applying the same construction on contraction
strata.
Composite kinematic limits are handled by an outer step, termed
\emph{asymptotic-order alignment}, which first exposes monomials that occur at
different orders in the native expansion parameter but must participate in
the same cancellation.  The central HRF logic itself is unchanged.

We formulate HRF at the level of scalar Feynman integrals with unit
numerator.  This entails no loss of generality for identifying candidate
regions: their existence and location are fixed by the propagator
denominators and do not depend on the numerator.  The numerator can,
however, change the order at which a region contributes, or cause its
contribution to vanish at a given order.  This distinction is especially
important at
leading power in gauge theories, where numerator structure can produce
behaviour different from that of the corresponding scalar integral.
Accounting for such numerator-dependent effects in the various asymptotic
expansions lies beyond the scope of this paper.

We develop the construction for massless internal propagators.  In this case
both Symanzik polynomials are multiaffine, or equivalently linear in every
individual edge parameter.  Internal masses add
\(\mathcal U\sum_e m_e^2x_e\) to \(\mathcal F\), generating quadratic
dependence on individual parameters and new possible cancellation
structures.  Extending HRF to this case is non-trivial and lies outside the
scope of this paper.

The applications are exploratory rather than an exhaustive classification.
They are ordered to expose progressively the structures that a general
region-finding procedure must accommodate.  The four-point examples of
Sec.~\ref{sec:example-crown} begin with a direct interior cancellation with
no obstruction; their contraction descendants then motivate the search on
boundary strata and already show that the derivative harvest must retain
general polynomial cancellation factors, including factors with more than
two monomials.  The five-point Fish family of
Sec.~\ref{sec:example-five-point} has a non-vanishing obstruction already at
the level of the seed topology, and this obstruction fixes a non-uniform
parameter scaling, making the decomposition into \(\mathcal F_{\rm SL}\) and
\(\mathcal F_{\rm Obs}\) essential.  Further examples in that section test
the inheritance of this cancellation geometry under a local vertex correction
and introduce multi-generator cancellation ideals.  The six-point examples of
Sec.~\ref{sec:example-alignment} involve composite kinematic limits and show
why asymptotic orders must sometimes be aligned before the core construction
is applied and why more than one cancellation layer may be active.  At the
same time, the examples show both sides of the organising pattern: special graph
topologies recur across physically different expansions, while the existence
and form of a hidden region remain sensitive to the chosen limit and to the
permutation of external legs.  A separate momentum-space analysis reconstructs the
corresponding loop modes and distinguishes propagator virtualities, values at
the pinch and local integration widths.

The broader survey suggests a topological organisation of Landshoff hidden
regions by complete-bipartite incidence cores, while also showing that the
external-leg configuration and the kinematic limit remain essential.  It compares
several complete-bipartite cores and their four-, five- and six-point
external-leg configurations, distinguishing certified positive Landshoff regions from
algebraic stationary loci excluded by positivity.  These results motivate,
but do not establish, conjectures about minimal Landshoff seeds, inheritance
under label-preserving graph contraction and the transverse order of positive hidden-region
pinches.

Section~\ref{sec:principles} presents the parameter-space construction
independently of any particular graph.  Section~\ref{sec:applications}
develops it through a sequence of massless examples, each introducing an
additional feature of the algorithm.  Section~\ref{sec:momentum-reconstruction}
addresses the logically separate translation from certified parameter
regions to momentum modes, pole pinches and Glauber routings.
Section~\ref{sec:topological-organization} examines the complete-bipartite
organisation of Landshoff hidden regions and states the conjectures suggested
by the survey.  Appendix~\ref{app:squared-ideal} gives the local
ideal-theoretic relation between the Landau conditions and the squared
cancellation ideal.  A notation summary is provided in the final appendix.

\section{Hidden Regions Finder: Principles}
\label{sec:principles}

In this section we set out the principles underlying the Hidden Region
Finder.  After introducing the parameter-space setup and definitions in
Sec.~\ref{sec:notation}, we present the core construction in
Sec.~\ref{sec:ordinary}: the Landau conditions identify cancellation factors
whose common vanishing defines a singular locus, and an appropriate scaling
makes the associated cancellation sector uniformly superleading.
Section~\ref{sec:alignment} extends this construction to nontrivial starting
faces and boundary strata, making it applicable in a wide range of
circumstances.  Section~\ref{sec:search-certification} then develops the search-level
strategy needed to identify equivalent presentations and to certify either
the detection or the absence of hidden regions.  Finally,
Sec.~\ref{sec:workflow} summarises the complete algorithmic workflow.

\subsection{Parameter-space setup and definitions}
\label{sec:notation}

Consider a connected scalar graph \(\mathcal G\) with \(N\) internal edges
and \(L\) loops.  Edge \(e\) has propagator power \(\nu_e\), with total
propagator power \(\nu\equiv\sum_{e=1}^{N}\nu_e\).  We use dimensional
regularisation in \(D=4-2\epsilon\) space-time dimensions, where
\(\epsilon\) is the dimensional regulator.  We denote by \(\s\) a set of
real kinematic invariants formed from the external momenta.  Dependence on
the graph is normally suppressed.  When several graphs or inequivalent
regions are compared, a graph-dependent quantity is written as
\(X[\mathcal G;\alpha]\), where the optional second entry labels the
region or branch.  Thus graph identity is always an argument, whereas a
parenthesised superscript, as in \(\mathcal P^{(R)}[\mathcal G]\), denotes a
rescaled region expansion.
After choosing orientations and a basis of loop momenta
\(\boldsymbol k=(k_1,\ldots,k_L)\), the momentum through edge \(e\) can be
written as
\begin{equation}
 q_e^\mu(\boldsymbol k,\boldsymbol p)
 =\sum_{a=1}^{L}\eta_{ea}k_a^\mu+P_e^\mu(\boldsymbol p),
 \qquad \eta_{ea}\in\{0,\pm1\},
 \label{eq:edge-momentum-routing}
\end{equation}
where \(P_e\) is a linear combination of external momenta.  For the massless
internal propagators considered here, define
\(\mathcal D_e=q_e^2+i0\).  The momentum-space integral is
\begin{equation}
 I_{\mathcal G}(\s)
 =\int[\dd\boldsymbol k]\prod_{e=1}^{N}\mathcal D_e^{-\nu_e},
 \qquad
 [\dd\boldsymbol k]
 \equiv\prod_{a=1}^{L}\frac{\dd^D k_a}{i\pi^{D/2}}.
 \label{eq:momentum-integral}
\end{equation}
Introducing one proper-time parameter \(\widetilde x_e\) for each propagator while retaining the
loop integrations gives\footnotemark\ \cite{Weinzierl:2022eaz}
\footnotetext{For the Fourier-form Schwinger parametrisation with arbitrary
propagator powers, see Eqs.~(25) and (26) of
Ref.~\cite{Kim:2023parametrizations}; the original proper-time construction
is given in Eq.~(2.24) of Ref.~\cite{Schwinger:1951nm}.}
\begin{equation}
\begin{aligned}
 I_{\mathcal G}(\s)
 &=\frac{e^{-i\pi\nu/2}}{\prod_{e=1}^{N}\Gamma(\nu_e)}
 \int_0^\infty\!\left(\prod_{e=1}^{N}
       \dd\widetilde x_e\,\widetilde x_e^{\nu_e-1}\right)
 \int[\dd\boldsymbol k]\,
 \exp\!\left[i\mathcal S
 (\boldsymbol k,\boldsymbol p;\widetilde{\boldsymbol x})\right],\\
 \mathcal S(\boldsymbol k,\boldsymbol p;\widetilde{\boldsymbol x})
 &\equiv\sum_{e=1}^{N}\widetilde x_e\mathcal D_e
 =\sum_{e=1}^{N}\widetilde x_e
       \big[q_e^2(\boldsymbol k,\boldsymbol p)+i0\big].
\end{aligned}
 \label{eq:schwinger-representation}
\end{equation}
The sign of the exponent and the prefactor \(e^{-i\pi\nu/2}\) follow from
the Feynman causal prescription \(\mathcal D_e=q_e^2+i0\): for
\(\widetilde x_e>0\), the factor
\(e^{i\widetilde x_e\mathcal D_e}\) is damped.

For our purposes it is convenient to work in the Lee--Pomeransky (LP)
representation \cite{Lee:2013hzt}.  It is obtained from the Schwinger
representation by carrying out the Gaussian loop integrations followed by
an auxiliary-scale transformation, as described in
Ref.~\cite[App.~B]{Gardi:2022khw}:
\begin{equation}
\begin{aligned}
 I_{\mathcal G}(\s)
 &=\mathcal N_{\rm LP}
 (\mu_{\rm LP}^2)^{LD/2-\nu}
 \int_0^\infty
 \prod_{e=1}^{N}\frac{\dd x_e}{x_e}\,x_e^{\nu_e}\,
 \mathcal P_{\mu_{\rm LP}}(\x,\s)^{-D/2},\\
 \mathcal N_{\rm LP}
 &\equiv
 \frac{\Gamma(D/2)}{
  \Gamma\!\left((L+1)D/2-\nu\right)
 \prod_{e=1}^{N}\Gamma(\nu_e)},
 \qquad
 \mathcal P_{\mu_{\rm LP}}(\x,\s)
 =\mathcal U(\x)+\frac{\mathcal F(\x;\s)}{\mu_{\rm LP}^2}.
\end{aligned}
 \label{eq:LP-representation}
\end{equation}
Here \(\mu_{\rm LP}\) is an arbitrary reference mass and the LP parameters
\(x_e\) are dimensionless.  The power of \(\mu_{\rm LP}^2\) outside the
integral restores the mass dimension of the original momentum integral.
In both the Schwinger and LP representations the \(N\) edge parameters are
independent,
with each parameter integrated over \((0,\infty)\).

Before the division by \(\mu_{\rm LP}^2\) in
Eq.~\eqref{eq:LP-representation}, the first and second Symanzik graph
polynomials for massless internal propagators are
\cite{Weinzierl:2022eaz}
\begin{equation}
\begin{aligned}
 \mathcal U(\x)
 &=\sum_{T^1}\prod_{e\notin T^1}x_e,\\
 \mathcal F(\x;\s)
 &=\sum_{T^2}(-s_{T^2})\prod_{e\notin T^2}x_e-i0.
\end{aligned}
 \label{eq:graph-polynomials}
\end{equation}
Here \(T^1\) runs over spanning trees and \(T^2\) over spanning 2-forests of
\(\mathcal G\); \(s_{T^2}\) is the squared total external momentum entering
either connected component of \(T^2\).  Writing the coefficients as
\(-s_{T^2}\) makes manifest that every monomial
of the real part of \(\mathcal F\) is positive in Euclidean kinematics, where
all channel invariants \(s_{T^2}\) are spacelike.

For an \(L\)-loop graph, \(\mathcal U\) is homogeneous of degree \(L\), while
\(\mathcal F\) is homogeneous of degree \(L+1\) in the edge parameters.
Each edge occurs at most once in every complementary tree or 2-forest
product.  Hence both graph polynomials are linear in each individual
\(x_e\).  This multiaffinity is a standing assumption below.

In Eq.~\eqref{eq:graph-polynomials}, the causal Feynman \(i0\) prescription
for the propagators in
Eq.~\eqref{eq:schwinger-representation} translates into the overall \(-i0\)
prescription for the second Symanzik graph polynomial \(\mathcal F\), thereby
fixing the physical-sheet boundary value.  The term \(-i0\) is a limiting
prescription, not an additional constant monomial.  A necessary condition
for a singularity of the parametric integral is given by the Landau
conditions \cite{Lnd59,Collins:2020euz,Weinzierl:2022eaz}
\begin{equation}
 \mathcal F=0,
 \qquad
 x_e\frac{\partial\mathcal F}{\partial x_e}=0
 \quad(e=1,\ldots,N),
 \label{eq:parametric-Landau}
\end{equation}
where the parameters \(x_e\) are not all zero, and \(\mathcal F\) in these
algebraic equations denotes its real polynomial part.  By Euler's identity,
homogeneity of \(\mathcal F\) makes the first condition a consequence of the
logarithmic derivative conditions; we display it separately to make the
singular hypersurface explicit.  The factors \(x_e\) allow solutions on
boundary strata.  For an interior first-sheet pinch all
\(x_e>0\), and
Eq.~\eqref{eq:parametric-Landau} reduces to the stationary conditions
\(\partial\mathcal F/\partial x_e=0\) for every edge.  These equations give
the necessary condition for the positive integration contour to be pinched,
while the displayed \(-i0\) fixes the causal approach to the singular locus.

\paragraph{Geometric expansion by regions.}

Having introduced the parametric representation, we now turn to the
asymptotic expansion of the Feynman integral in a specified kinematic limit,
using the geometric parametric formulation of the Method of Regions
\cite{Pak:2010pt,Jantzen:2012mw,Ananthanarayan:2018tog,Heinrich:2021dbf}.
Let \(\K\) denote the physical domain in the real kinematic coordinates
\(\s\), specified by the appropriate inequalities among them, and let
\(\delta\to0^+\) parametrise the chosen asymptotic limit.\footnote{The natural
kinematic deformation need not itself be dimensionless.  If a chart uses a
dimensionful variable \(\delta_{\rm phys}\), its powers are understood as
powers of \(\delta_{\rm phys}/Q\) for a fixed nonzero reference scale \(Q\).
Retaining the chart's natural symbol when recording these powers does not
affect region vectors or weight comparisons.}  No universal sign
convention is imposed on the individual components of \(\s\); the defining
inequalities of \(\K\) fix the physical domain in each application.  This
allows the same variables to
be retained across different physical regions, making changes of sign under
crossing or analytic continuation explicit.

After expanding the kinematics, we write each monomial in the LP polynomial as
\begin{equation}
 m_i\equiv c_i(\s)\,\delta^{a_i}\x^{\boldsymbol r_i},
 \qquad
 \x^{\boldsymbol r_i}\equiv\prod_{e=1}^{N}x_e^{r_{i,e}}.
\label{eq:monomial-definition}
\end{equation}
To retain its origin in the two graph polynomials, we partition the monomial
index set as
\(\mathcal M\equiv
\mathcal M_{\mathcal U}\sqcup\mathcal M_{\mathcal F}\) and write
\begin{equation}
 \begin{aligned}
 \mathcal U(\x)
 &=\sum_{i\in\mathcal M_{\mathcal U}}m_i,
 &\qquad
 c_i(\s)&\equiv1,\quad a_i\equiv0
 \quad(i\in\mathcal M_{\mathcal U}),\\
 \frac{\mathcal F(\x;\delta,\s)}{\mu_{\rm LP}^2}
 &=\sum_{i\in\mathcal M_{\mathcal F}}m_i,\\
 \mathcal P_{\mu_{\rm LP}}(\x;\delta,\s)
 &\equiv\mathcal U(\x)
   +\frac{\mathcal F(\x;\delta,\s)}{\mu_{\rm LP}^2}
 =\sum_{i\in\mathcal M}m_i.
 \end{aligned}
\label{eq:Pmonomials}
\end{equation}
Unless stated otherwise, in the region analysis we choose \(\mu_{\rm LP}\)
fixed and nonzero as \(\delta\to0\).  We then express the kinematic
coefficients in units of \(\mu_{\rm LP}^2\), suppress the scale label, and use
the customary shorthand \(\mathcal P=\mathcal U+\mathcal F\).  Such a fixed
choice changes neither the exponent geometry nor the cancellation locus.  If
a \(\delta\)-dependent reference scale is useful, we state that choice
explicitly and include its scaling in the monomial weights.

In this and subsequent monomial decompositions, \(\mathcal F\) denotes its
real polynomial part; the \(-i0\) remains a boundary-value prescription and
is not included in \(\mathcal M_{\mathcal F}\).  Thus
\(\boldsymbol r_i=(r_{i,1},\ldots,r_{i,N})\) is the length-\(N\)
LP-parameter exponent vector of the monomial \(m_i\).  We define the
corresponding augmented exponent point by
\begin{equation}
 \vec r_i\equiv(\boldsymbol r_i;a_i)\in\mathbb R^{N+1}.
\end{equation}
The convex hull of these points is the Newton polytope of \(\mathcal P\); the
extremal exponent points are its vertices.  An LP-parameter
scaling is represented by an augmented vector whose last component is
normalised to one,
\begin{equation}
 \vec v_R=(\boldsymbol v_R;1),
 \qquad
 x_e\longmapsto \delta^{v_{R,e}}x_e.
\label{eq:regionvector}
\end{equation}
Under this scaling, the weight of a monomial is defined by
\begin{equation}
 w_R(m_i)\equiv \vec v_R\cdot\vec r_i
 =a_i+\boldsymbol v_R\cdot\boldsymbol r_i.
\label{eq:weight}
\end{equation}
For a given \(\vec v_R\), the monomials of minimum weight are the dominant
terms as \(\delta\to0^+\).  When their exponent points span a codimension-one
lower face of the Newton polytope, this face is a lower facet.  Its
inward-pointing normal is the augmented vector \(\vec v_R\), normalised as in
Eq.~\eqref{eq:regionvector}; it gives the corresponding facet-region scaling.

\paragraph{Sign cancellations and hidden regions.}

The geometric construction above depends only on the augmented exponent
points \(\vec r_i\), and not on the coefficients \(c_i(\s)\).  Once the
monomial support is fixed, it is therefore insensitive to the coefficient
signs and produces the same Newton polytope and facet regions in every
kinematic domain.  In particular, it cannot detect cancellations that depend
on relative signs between dominant monomials.

Sign cancellations missed by a direct Newton-polytope analysis were considered
in Ref.~\cite{Jantzen:2012mw}; their role in hidden regions and their resolution
by dissecting the positive integration domain were demonstrated in
Ref.~\cite{Gardi:2024axt}.  To determine whether such a cancellation is
possible, we must restore the coefficient information by specifying a physical
kinematic domain \(\K\).  Its defining inequalities fix the signs of the
relevant coefficients \(c_i(\s)\), away from any boundary locus explicitly
approached in the expansion.  Since
\(\delta^{a_i}\x^{\boldsymbol r_i}>0\) for \(\delta>0\) and \(\x>0\), the
sign of \(m_i\) is the sign of \(c_i(\s)\).  These signs may differ between
physical domains or crossed channels.

For the scaling vector under consideration, let \(\mathcal P_{\rm dom}\)
denote the sum of the monomials of minimum weight in Eq.~\eqref{eq:weight}.
If these dominant monomials cannot mutually cancel, as for a
same-sign polynomial, the facet normal completely specifies the conventional
region.  If instead \(\mathcal P_{\rm dom}\) contains terms of opposite sign,
it may vanish in the positive integration domain.  A hidden region can arise
when this zero forms a stationary first-sheet pinch.  The scaling vector
determines \(\mathcal P_{\rm dom}\), but not its cancellation locus; the
geometric data must therefore be supplemented by the cancellation locus and
its resolution.

Although all monomials enter the Newton polytope of \(\mathcal P\), only the
\(\mathcal F\)-monomials can participate in such a cancellation.  Suppose that
\(\mathcal P_{\rm dom}\) contains both kinds of monomial and decompose it as
\(\mathcal P_{\rm dom}=\mathcal U_{\rm dom}+\mathcal F_{\rm dom}\).  At an
interior first-sheet stationary pinch,
\begin{equation}
 \mathcal P_{\rm dom}=\mathcal U_{\rm dom}+\mathcal F_{\rm dom}=0,
 \qquad
 0=\sum_e x_e\frac{\partial\mathcal P_{\rm dom}}{\partial x_e}
 =L\mathcal U_{\rm dom}+(L+1)\mathcal F_{\rm dom},
 \label{eq:no-U-mixed-pinch}
\end{equation}
where the second equality follows from the respective LP-parameter degrees
\(L\) and \(L+1\).  These equations require
\(\mathcal U_{\rm dom}=\mathcal F_{\rm dom}=0\), but any nonempty sum
\(\mathcal U_{\rm dom}\) is strictly positive for \(\x>0\).  Thus
\(\mathcal U\)-monomials cannot form part of an interior first-sheet
pinch.\footnote{The same conclusion is immediate in the homogenised LP
representation
\(\widehat{\mathcal P}(\x,x_0)=\mathcal F(\x)+x_0\mathcal U(\x)\): stationarity
with respect to \(x_0\) requires \(\mathcal U=0\), which has no solution in the
positive interior for a connected graph \cite{Chen:2019mqc}.} They must
nevertheless be retained because they can affect which terms are dominant in
the complete LP polynomial.

For later use, we record how an LP-parameter region vector translates into
propagator virtualities.  In the Schwinger representation the relation is
immediate: \(\widetilde x_e\) multiplies \(\mathcal D_e\) in
Eq.~\eqref{eq:schwinger-representation}, so their regional scalings are
inverse.  The auxiliary-scale transformation relating the Schwinger and LP
representations ensures that \(x_e\) and \(\widetilde x_e\) have identical
regional scaling vectors \cite[App.~B]{Gardi:2022khw}.  Hence
\begin{equation}
 x_e\sim\delta^{v_{R,e}}
 \quad\Longleftrightarrow\quad
 Q^2\widetilde x_e\sim\delta^{v_{R,e}}
 \quad\Longrightarrow\quad
 \frac{\mathcal D_e}{Q^2}\sim\delta^{-v_{R,e}},
 \label{eq:schwinger-inverse-virtuality}
\end{equation}
where \(Q^2\) is a fixed hard scale.  Thus a parametrically large edge parameter
corresponds to a parametrically small propagator virtuality, and conversely.
Equation~\eqref{eq:schwinger-inverse-virtuality} determines the propagator
virtualities associated with a parameter-space region; recovering a
consistent set of loop-momentum modes requires the additional momentum-routing
analysis developed in Sec.~\ref{sec:momentum-reconstruction}.

\subsection{The core HRF construction}
\label{sec:ordinary}

The core construction takes as external input a specified starting polynomial
\(\mathcal F_\star\) and a specified subvector \(\x_{\rm active}\) of nonzero
LP parameters, with \(N_{\rm active}\) components.  All other LP parameters,
if any, have already been set to zero.  This selects the stratum on which the
core construction is applied: a sector of the closure of the positive LP
domain with a specified subset of parameters set to zero and all remaining
active parameters positive.  The choice of these inputs is not part of the
core construction.

The mechanism sought by HRF is the following.  A subset of the monomials of
\(\mathcal F_\star\) may sum to a polynomial \(\FSL\) that admits a
first-sheet Landau pinch in the positive LP domain, even though neither the
complete polynomial \(\mathcal F\) nor the starting polynomial
\(\mathcal F_\star\) does.  The remaining monomials obstruct this
cancellation.  A hidden-region scaling makes the individual monomials of
\(\FSL\) superleading relative to the obstruction, while their mutual
cancellation counters this individual enhancement, so that their sum
contributes at the physical leading order.  We therefore call \(\FSL\) the
superleading cancellation sector and the complementary part of
\(\mathcal F_\star\) the obstruction.  Concrete instances of this mechanism
are worked out in Sec.~\ref{sec:applications}.

The construction has two parts.  Part~I uses cancellation factors exposed by
derivatives of \(\mathcal F_\star\), and products of these factors, to
construct a candidate \(\FSL\) and validate its common positive Landau locus.
Part~II determines the candidate scaling from the complete LP polynomial and
imposes the required homogeneity and separation in weight.  After any
required order-alignment composition described in
Sec.~\ref{sec:alignment}, the scaleful lower-facet test supplies the final
certificate in Sec.~\ref{sec:scaleful}.

To state the ordinary core construction, expand
\begin{equation}
 \mathcal F(\x;\delta,\s)
 =
 \mathcal F_0(\x;\s)
 +\sum_{n>0}\delta^n\mathcal F_n(\x;\s).
\label{eq:Fexpansion}
\end{equation}
In the simplest interior case, \(\x_{\rm active}=\x\) and the cancellation
sector lies in \(\mathcal F_0\), so we take
\(\mathcal F_\star=\mathcal F_0\).  More generally, boundary hidden regions
are treated by first restricting the graph polynomial to a contraction
stratum, while cancellations involving monomials from different native orders
in \(\delta\) require preliminary asymptotic-order alignment.  These two
independent extensions are described in Sec.~\ref{sec:alignment}.

Writing the starting polynomial as
\(\mathcal F_\star=\sum_{i\in\mathcal M_\star}m_i\), with
\(\mathcal M_\star\subseteq\mathcal M_{\mathcal F}\), let
\(\MSL\subseteq\mathcal M_\star\) denote the subset of monomials selected for
a candidate cancellation sector, and define
\begin{equation}
 \FSL\equiv\sum_{i\in\MSL}m_i.
\label{eq:FSLsupport}
\end{equation}
Here \(\MSL\) denotes a subset of the monomial support of
\(\mathcal F_\star\); Eq.~\eqref{eq:FSLsupport} merely fixes the notation.
Section~\ref{sec:decomposition} explains how HRF proposes such a subset and
tests whether the corresponding \(\FSL\) defines an admissible cancellation
sector.  In the special case where \(\MSL=\mathcal M_\star\), one has
\(\FSL=\mathcal F_\star\) and the obstruction vanishes.

\subsubsection{Part I: constructing and validating the decomposition}
\label{sec:decomposition}

\paragraph{The Landau target.}

The purpose of the decomposition is to isolate a polynomial \(\FSL\) that
can admit a first-sheet pinch in the relative interior of the chosen active
stratum.  Applying the interior form of Eq.~\eqref{eq:parametric-Landau} to the
active LP variables gives the defining target
\begin{equation}
 \frac{\partial\FSL}{\partial x_e}=0
 \quad\text{for every active }x_e,
 \qquad
 \x_{\rm active}\in\Rpos^{N_{\rm active}},
 \qquad
 \s\in\K.
\label{eq:SLpinch}
\end{equation}
Here ``interior'' refers to the positive domain of the active parameters.
When \(\x_{\rm active}=\x\), this is an interior pinch of the full LP domain;
when some parameters have been fixed to zero, the same condition describes a
boundary pinch in the full domain.  Since \(\FSL\) is homogeneous in the
active LP parameters, Euler's identity
implies \(\FSL=0\) whenever the derivative equations hold.  Thus
Eq.~\eqref{eq:SLpinch} is the complete Landau condition on the chosen active
stratum; a separate equation \(\FSL=0\) would be redundant.

Formulating Landau loci directly through derivatives of an LP-type graph
polynomial also underlies the principal Landau determinant framework
\cite{Fevola:2023pld}.  That construction studies complex critical loci of
the Lee--Pomeransky polynomial \(\mathcal P=\mathcal U+\mathcal F\) and its
facial restrictions.  Here we instead require a first-sheet pinch in the
positive orthant of the selected asymptotic polynomial \(\FSL\).

Let \(C\) be a smooth component of this common pinch locus.  Smoothness allows
us to choose a nonredundant set of local defining polynomials
\(f_1,\ldots,f_r\), whose gradients are independent along \(C\), so that
\(C\) is locally given by \(f_1=\cdots=f_r=0\) and the \(f_i\) provide
coordinates transverse to \(C\).  Their radical local ideal in the active
LP-parameter space is \(\ideal_C=\langle f_1,\ldots,f_r\rangle\).\footnote{Locally,
\(\langle f_1,\ldots,f_r\rangle\) consists of combinations
\(\sum_i h_i f_i\), with coefficients \(h_i\) nonsingular near \(C\).  The
qualifier ``radical'' means that, if \(g^n\in\ideal_C\) for some \(n\geq1\),
then \(g\in\ideal_C\), so the ideal records the common zero locus without
algebraic multiplicity.}  The Landau equations imply
\begin{equation}
 \FSL\in\ideal_C^2.
 \label{eq:Landau-square-ideal}
\end{equation}
To see this, complete the \(f_i\) by coordinates along \(C\).  Homogeneity and
Eq.~\eqref{eq:SLpinch} imply that \(\FSL\) vanishes
on \(C\), eliminating the constant term in its local expansion, while the
derivative equations eliminate every term linear in the \(f_i\).
The expansion therefore begins with products \(f_if_j\).  The two-factor
structure used below is consequently a direct implication of the Landau
conditions, not an additional HRF assumption.
Appendix~\ref{app:squared-ideal} provides an intuitive derivation, followed by
the equivalent ideal-theoretic formulation
in terms of the conormal exact sequence.  The implication in
Eq.~\eqref{eq:Landau-square-ideal} does not require \(\FSL\) to be
multiaffine; rather, multiaffinity of the second Symanzik polynomial in the
massless case is what makes the cancellation factors accessible through
derivatives of \(\mathcal F_\star\).

\paragraph{Derivative harvesting.}

Because \(\FSL\) is not known in advance, HRF searches for traces of its
cancellation factors in derivatives of \(\mathcal F_\star\).  The harvested
objects are candidate non-monomial cancellation polynomials
\(f_j(\x,\s)\).  Each \(f_j\) must be capable of vanishing for \(\x>0\) and
\(\s\in\K\); at fixed \(\s\), a necessary condition is that its nonzero
LP-monomial coefficients are not all of the same sign.  In choosing a given
\(f_j\), any multiplicative factor that is nonzero throughout the selected
domain (such as a monomial in the active LP parameters or a kinematic factor
known to be nonzero throughout \(\K\)) is omitted, since this does not change
the zero locus of that \(f_j\), and hence does not change their common zero
locus.

The direct harvest has two complementary forms.  One may factorise the
complete derivatives.  When the kinematic dependence admits a useful
decomposition into independent structures, one may instead, or additionally,
collect each derivative in those structures and factorise the resulting
LP-parameter coefficient polynomials separately.  This coefficientwise
harvest is neither a universal step nor by itself sufficient.  In particular,
when an intended generator depends irreducibly on several kinematic variables,
such a decomposition may obscure rather than expose it.  Factors that remain
coupled in derivatives of an already selected \(\FSL\) may instead be recovered
from the saturated selected-sector gradient ideal, with branches supported
on factors required to be nonzero removed, in
Eq.~\eqref{eq:selected-gradient-saturation}.  The factors and kinematic
sectors exposed by the applicable direct harvest provide the input for
constructing the candidate generators below.

At this stage the derivatives of \(\mathcal F_\star\) are used only to identify
candidate factors, not imposed as equations.  When an obstruction is present,
the derivatives of \(\mathcal F_\star\) need not all vanish on the true HR
locus.  The derivatives of the subsequently selected \(\FSL\) do vanish there;
those derivatives of \(\mathcal F_\star\) that receive no contribution from
the obstruction therefore vanish as well, while the remaining ones may not.
Thus \(\mathcal F_\star\) itself need not have a stationary point in the
positive orthant, and generically does not.

\paragraph{Generators for massless graph polynomials and the obstruction
decomposition.}

Equation~\eqref{eq:Landau-square-ideal} identifies the square of the local
defining ideal,
\begin{equation}
 \ideal_C^2
 =\big\langle f_if_j\,\big|\,1\leq i\leq j\leq r\big\rangle.
 \label{eq:cancellation-square-ideal}
\end{equation}
The diagonal products make the full square ideal independent of the chosen
local defining polynomials.  Equation~\eqref{eq:cancellation-square-ideal} is
therefore the presentation-independent algebraic envelope implied by the Landau
conditions; it is not the list of generators admitted by a particular
massless graph polynomial.

In the original edge-variable basis, define the LP-parameter support of a
cancellation factor by
\begin{equation}
 \operatorname{supp}_{x}(f)
 \equiv\bigl\{e\,\big|\,\deg_{x_e}f>0\bigr\}.
 \label{eq:factor-x-support}
\end{equation}
For a massless graph, multiaffinity of the second Symanzik polynomial
restricts a candidate generator built from the cancellation factors to the
form
\begin{equation}
 g_k=\prod_{i\in S_k}f_i,
 \qquad |S_k|\geq2,
 \qquad
 \operatorname{supp}_{x}(f_i)
 \cap\operatorname{supp}_{x}(f_j)=\varnothing
 \quad (i\neq j,\ i,j\in S_k).
\label{eq:massless-generator}
\end{equation}
The factors in \(S_k\) must be distinct and have pairwise-disjoint
LP-parameter support: a repeated factor or overlapping dependence on any
\(x_e\) would produce degree two in that parameter and could not reproduce a
monomial of the massless graph polynomial.  The construction permits
\(|S_k|>2\).  The core HRF search used throughout this paper nevertheless
constructs pair products, \(g_k=f_if_j\): every generator required by the
examples contains precisely two cancellation factors.  Products of three or
more factors are also implemented and have been tested in exploratory
searches, but no instance has been found in which they are required.

Multiaffinity is only the first support test.  The expanded monomials of
\(g_k\) must also be embeddable in monomials already present in the starting
polynomial \(\mathcal F_\star\).  Writing \(\operatorname{Mon}(g_k)\) for the
set of monomials \(m\) in the expanded generator, and \(\boldsymbol r(m)\) for
the LP exponent vector of \(m\), this necessary condition is
\begin{equation}
 \text{for every }m\in\operatorname{Mon}(g_k),\quad
 \text{there exists }m_\ell\in\mathcal M_\star\text{ such that}
 \quad \boldsymbol r(m)\leq\boldsymbol r_\ell
 \quad\text{componentwise}.
 \label{eq:generator-support}
\end{equation}
This test concerns only the LP-parameter support.  Equivalently, after
disregarding kinematic coefficients, each generator monomial must divide the
LP-parameter part of an existing monomial of \(\mathcal F_\star\).  This
permits a multiplier to supply the remaining LP-parameter content, but it
excludes products which belong abstractly to \(\ideal_C^2\) and cannot occur
in the graph polynomial.

For a candidate generator set
\(G=\{g_1,\ldots,g_{n_{\rm gen}}\}\), with
\(n_{\rm gen}\equiv |G|\), define the candidate generator ideal
\begin{equation}
 \mathcal I_{\rm gen}(G)\equiv
 \langle g_1,\ldots,g_{n_{\rm gen}}\rangle
 \subseteq\ideal_C^2.
 \label{eq:generator-ideal}
\end{equation}
At this stage \(G\) and \(\mathcal I_{\rm gen}(G)\) remain tentative.  They
are retained only if the obstruction search finds a nonzero complement of
\(\mathcal F_\star\) that lies in \(\mathcal I_{\rm gen}(G)\); otherwise
\(G\) is discarded and another candidate generator set is tested.

To perform this search, HRF considers proper subsets of monomials
\(\MObs\subsetneq\mathcal M_\star\) to remove from \(\mathcal F_\star\).  For
each choice, with \(\MSL\) its complement, define
\begin{align}
 \MObs&\subsetneq\mathcal M_\star,
 &\MSL&\equiv\mathcal M_\star\setminus\MObs,
 \nonumber\\
 \FObs&\equiv\sum_{i\in\MObs}m_i,
 &\FSL&\equiv\sum_{i\in\MSL}m_i
       =\mathcal F_\star-\FObs.
 \label{eq:candidate-obstruction}
\end{align}
The obstruction support \(\MObs\) may be empty, but \(\MSL\) must be
nonempty.  The choice gives an algebraically admissible candidate
decomposition only if \(\FSL\) is nonzero and belongs to the ideal
\(\mathcal I_{\rm gen}(G)\) generated by \(G\) in Eq.~\eqref{eq:generator-ideal}:
\begin{equation}
 \mathcal F_\star=\FSL+\FObs,
 \qquad
 \FSL\neq0,
 \qquad
 \FSL\in\mathcal I_{\rm gen}(G).
\label{eq:obstruction}
\end{equation}
The ideal-membership condition is equivalently expressed as
\begin{equation}
 \FSL=\sum_{k=1}^{n_{\rm gen}}M_k(\x,\s)g_k.
\label{eq:idealpresentation}
\end{equation}
The multipliers supply the remaining monomial content.  By default,
Eq.~\eqref{eq:idealpresentation} is tested as exact membership in the
polynomial ideal \(\mathcal I_{\rm gen}(G)\), requiring the
polynomial-reduction remainder to vanish.  The support test in
Eq.~\eqref{eq:generator-support} is necessary,
whereas the exact identity in Eq.~\eqref{eq:idealpresentation} is what ensures
that the selected generators reproduce the actual terms of \(\FSL\).
Only this complete identity tests kinematic compatibility: no term-by-term or
generator-by-generator match is required.\footnote{In the five-point
near-planar example, the individual generator contributions contain the
rational stationary ratios \(\boldsymbol\rho=-K^{-1}\boldsymbol b\), whose
nonlinear kinematic dependence cancels only in the complete sum; see
Sec.~\ref{sec:five-point-near-planar-seed}, especially
Eq.~\eqref{eq:five-full-stationary-decomposition} and the discussion following
Eq.~\eqref{eq:five-gram-chain-rule}.}
Every such presentation belongs to the ambient square ideal and hence has the
quadratic vanishing required by Eq.~\eqref{eq:SLpinch}.  Conversely,
Appendix~\ref{app:squared-ideal} proves membership in \(\ideal_C^2\) at a
smooth pinch component; it does not select the graph-dependent set \(G\).

\paragraph{Validation of the common locus.}

A necessary preliminary condition for every distinct cancellation factor
\(f_i\) appearing in the generators \(G\) is that its nonzero terms include
both signs in the kinematic domain \(\K\).  This is not sufficient: the
harvested factors and the resulting decomposition of \(\mathcal F_\star\)
become physically admissible only if all these factors have a common positive
zero at one and the same kinematic point:
\begin{equation}
 \exists\,(\x_{\rm active},\s)
 \in\Rpos^{N_{\rm active}}\times\K:
 \qquad
 f_i(\x,\s)=0
 \quad\text{for every }i\in\bigcup_{k=1}^{n_{\rm gen}}S_k.
\label{eq:commonzero}
\end{equation}
On a smooth component \(C\) of this solution, after removing redundant
relations, the selected factors are taken to generate the radical local ideal
\(\ideal_C\) used in Eqs.~\eqref{eq:Landau-square-ideal} and
\eqref{eq:cancellation-square-ideal}.
Equivalently one solves Eq.~\eqref{eq:SLpinch} directly.  The common locus,
rather than any one factor in isolation, encodes the pinch.

The derivatives of an already selected \(\FSL\) need not display the
cancellation factors \(f_i\) as literal factors; they may contain only
coupled polynomial linear combinations of them.\footnote{An explicit example
is the five-point near-planar
Fish in Sec.~\ref{sec:five-point-near-planar-seed}, where saturation of the
selected gradient ideal recovers the local cancellation ideal
\(\langle f_A,f_B,f_C\rangle\), although literal factorisation of an
individual raw derivative does not expose these factors; see
Eqs.~\eqref{eq:five-near-planar-saturation-data} and
\eqref{eq:five-near-planar-saturated-ideal}.}  To recover the complete common
locus in this situation, package the stationarity equations for \(\FSL\) in
Eq.~\eqref{eq:SLpinch} into a gradient ideal and define
\(\mathcal I_\nabla^{\rm SL}\) and the localisation element \(S\) by
\begin{equation}
 \mathcal I_\nabla^{\rm SL}\equiv
 \left\langle\frac{\partial\FSL}{\partial x_e}\right\rangle_e,
 \qquad\text{and}\qquad
 S\equiv\prod_{e\in\mathrm{active}}x_e\prod_a h_a(\s),
 \label{eq:selected-gradient-ideal}
\end{equation}
where the \(h_a(\s)\) are kinematic factors known to be nonzero in the
domain under consideration.  When the coefficient matrix relating these
combinations to the \(f_i\) is invertible on the physical component,
multiplication by its adjugate shows that its determinant times each \(f_i\)
belongs to the gradient ideal once the kinematic constraints are imposed.
Here algebraic localisation means allowing division only by factors known to
be nonzero on the selected component.  Isolating the individual \(f_i\)
therefore requires localisation by this
nonvanishing determinant; the factors needed for this inversion are included
in \(S\).  The localisation is implemented by the saturation
\begin{equation}
 \bigl(\mathcal I_\nabla^{\rm SL}+\mathcal I_{\rm kin}\bigr):S^\infty,
 \label{eq:selected-gradient-saturation}
\end{equation}
with \(\mathcal I_{\rm kin}\) imposing the algebraic constraints of the
chosen expansion.  Equivalently, this saturation discards solution components
supported on \(S=0\).  It is an optional completion or validation of an
already selected cancellation sector, not a general pre-decomposition harvest of
\(\partial\mathcal F_\star\).

The cancellation factors identified in Part~I will later provide local
cancellation-resolving coordinates for the final dissection certificate.
No dissection is performed at this stage: Part~II first determines the
candidate core scaling from the complete layer structure.  If
asymptotic-order alignment is required, its composition with this core
scaling is defined in Sec.~\ref{sec:alignment}.  The dissection charts then
provide the final certificate as described in Sec.~\ref{sec:scaleful}.

\subsubsection{Part II: determining the candidate scaling}
\label{sec:layers}

\paragraph{Hidden-region conditions.}
\label{sec:definition}

Given the selected polynomial \(\FSL\) and its common positive pinch component
\(C\), validated in Part~I, Part~II determines a candidate scaling
\begin{equation}
 \vHR=(\boldsymbol v_{\HR};1).
\end{equation}
Using the monomial-weight definition in Eq.~\eqref{eq:weight}, we denote by
\(\WSL\) the common individual weight of the monomials in \(\MSL\), and by
\(\WHR\) the weight of the first nonvanishing layer of the complete LP
polynomial under this scaling, after all cancellations on \(C\) have been
taken into account.  These weights are determined together with
\(\boldsymbol v_{\HR}\).  A HR is characterised by these data when the
following four conditions hold:
\begin{enumerate}[label=(\alph*)]
\item \textbf{Homogeneity:} every monomial \(m_i\in\MSL\) has the same
      individual weight \(\WSL\);
\item \textbf{Common locus:} the validated \(\FSL\) obeys the simultaneous
      first-sheet equations \eqref{eq:SLpinch} in the positive LP orthant and
      in \(\K\);
\item \textbf{Hierarchy:} the cancellation removes the individually
      superleading behaviour, and the first nonvanishing layer of the
      complete LP polynomial has weight \(\WHR\), with
      \begin{equation}
        \WHR>\WSL;
      \label{eq:gap}
      \end{equation}
\item \textbf{Scalefulness:} in local cancellation-resolving coordinates, the
      transformed leading LP polynomial defines a scaleful lower-facet region.
\end{enumerate}
The gap \(\WHR-\WSL\) measures the depth of the cancellation.

\paragraph{The rescaled LP polynomial in a hidden region.}
\label{sec:weight-colours}

The structure to be produced by the HR scaling is most transparent at the
level of the rescaled complete LP polynomial.  For any region vector
\(\boldsymbol v_R\), define
\begin{equation}
 \mathcal P^{(R)}(\x;\delta,\s)
 \equiv
 \mathcal P(\delta^{\boldsymbol v_R}\x;\delta,\s),
\label{eq:rescaled-LP-polynomial}
\end{equation}
so every displayed power of \(\delta\) includes both the explicit kinematic
dependence and the scaling of the LP parameters \(x_e\).  We refer to the
monomials in the lower-weight cancelling sector as individually superleading
(red), the monomials at the first resolved nonzero order as resolved leading
(blue), and all other monomials as remaining (black).  In the generic
single-layer case, the hidden-region expansion and the corresponding facet
expansion for comparison are
\begin{subequations}
\label{eq:region-colour-schematics}
\begin{align}
 \mathcal P^{(\HR)}
 &\equiv\mathcal P(\delta^{\boldsymbol v_{\HR}}\x;\delta,\s)
 =\SLcolour{\underbrace{\delta^{\WSL}\FSL}_{\FSL^{(\HR)}}}
   +\delta^{\WHR}\HRcolour{\mathcal P_{\WHR}}
   +\Othercolour{\mathcal O(\delta^{>\WHR})},
 &&\text{hidden region}.
\label{eq:HR-colour-schematic}\\
 \mathcal P^{(R)}
 &=\delta^{W_R}\HRcolour{\mathcal P_{W_R}}
   +\Othercolour{\mathcal O(\delta^{>W_R})},
 &&\text{facet region}.
\label{eq:facet-colour-schematic}
\end{align}
\end{subequations}
In a hidden region the individually superleading monomials (red) have a lower
individual weight than the resolved leading monomials (blue), but their sum
vanishes on the positive pinch component \(C\).  By comparison, a facet
region has no individually superleading sector.  When condition~(a) holds,
the rescaled superleading polynomial is
\begin{equation}
 \FSL^{(\HR)}
 \equiv
 \left.\FSL\right|_{\x\mapsto\delta^{\boldsymbol v_{\HR}}\x}
 =\delta^{\WSL}\FSL.
\label{eq:rescaled-SL-polynomial}
\end{equation}
Thus \(\FSL^{(\HR)}\) includes the \(\delta\)-dependence induced by the
LP-parameter scaling.  It cancels on \(C\), while the first resolved nonzero
LP layer appears at \(\WHR>\WSL\).\footnote{In the five-point
spacelike-collinear example of
Sec.~\ref{sec:five-point-spacelike-collinear}, the monomials in the
individually superleading \(\FSL\) sector (red) of
Eq.~\eqref{eq:five-HR-polynomial} all have \(\WSL=-5\), while the resolved
leading obstruction and kinematically suppressed monomials (blue) in the same
equation, together with the \(\mathcal U\) layer in
Eq.~\eqref{eq:five-U-leading}, enter at \(\WHR=-4\); see
Eq.~\eqref{eq:five-HR-vector}.  The resulting unit gap and the obstruction
fix the non-uniform HR scaling.}

\paragraph{Determining the HR vector: homogeneity and hierarchy.}

In the generic case, \(\boldsymbol v_{\HR}\), \(\WSL\) and \(\WHR\) are
determined by the homogeneity and hierarchy conditions.  The first
requirement is homogeneity of the selected cancellation sector,
\begin{equation}
 w_{\HR}(m_i)=\WSL,
 \qquad \forall\,m_i\in\MSL.
\label{eq:SLhomogeneity}
\end{equation}
This condition is necessary because a cancellation can take place only among
monomials of the same individual weight.  If their weights
differ, the lowest-weight subset dominates by itself and the remaining
monomials cannot cancel it.  In the generic case, no additional occupied
layer with weight between \(\WSL\) and \(\WHR\) vanishes on the same locus.
The second (hierarchy) requirement is then
\begin{equation}
 w_{\HR}(m_i)\geq\WHR>\WSL,
 \qquad \forall\,m_i\in\mathcal M\setminus\MSL.
\label{eq:single-layer-hierarchy}
\end{equation}
This condition constrains only the monomials outside \(\MSL\).  Each must
have weight at least \(\WHR\).  Those saturating the inequality are the
dominant monomials at that weight outside the superleading sector \(\FSL\).
Because \(\mathcal M\) is the monomial support of the
complete LP polynomial, Eq.~\eqref{eq:single-layer-hierarchy} also includes
the required balance with \(\mathcal U\).  The strict gap is essential: if
the conditions are inconsistent, or if they admit only
\(\WHR=\WSL\), the decomposition does not define a HR and must be rejected.
A factorised cancellation sector and a positive pinch are not sufficient
without this hierarchy.

The weight definition \eqref{eq:weight} translates these conditions into a
finite linear problem.  For the selected monomials, and for the remaining
monomials \(m_j\) that saturate the hierarchy at weight \(\WHR\), one has
\begin{subequations}
\label{eq:active-weight-system}
 \begin{align}
  a_i+\boldsymbol v_{\HR}\!\cdot\boldsymbol r_i&=\WSL,
  &&m_i\in\MSL,
\label{eq:active-SL-weight}\\
  a_j+\boldsymbol v_{\HR}\!\cdot\boldsymbol r_j&=\WHR,
  &&m_j\in\mathcal M\setminus\MSL.
\label{eq:active-HR-weight}
 \end{align}
\end{subequations}
These active equalities form a linear system for the \(N+2\) unknowns
\((\boldsymbol v_{\HR},\WSL,\WHR)\); the unsaturated instances of
Eq.~\eqref{eq:single-layer-hierarchy} test the resulting solution as
inequalities.  If the active system has full rank, the normalisation of the
last component of \(\vHR\) to one gives a unique candidate vector.  If it is
rank deficient, Eqs.~\eqref{eq:SLhomogeneity} and
\eqref{eq:single-layer-hierarchy} define a family rather than a unique
vector, and the candidate is not certified until the dissection analysis
of Sec.~\ref{sec:scaleful} fixes or rejects it.  Thus these equations are the
practical determining conditions, but uniqueness is a property to be checked,
not assumed.

Before applying the hierarchy, all monomials suppressed at fixed LP
parameters must be restored.  If a monomial from \(\mathcal F_n\), \(n>0\),
is promoted to the SL level or below by the candidate scaling, the proposed
\(\FSL\) is incomplete and the analysis must be repeated with that monomial
included.  For each surviving candidate, it is useful to display the
occupied weight layers of the complete rescaled polynomial,
\begin{equation}
 \mathcal P^{(\HR)}
 \equiv \mathcal P(\delta^{\boldsymbol v_{\HR}}\x;\delta,\s)
 =
 \sum_k
 \delta^{W_k}\,\mathcal P_{W_k}(\x,\s),
 \qquad W_k<W_{k+1}.
\label{eq:weightlayers}
\end{equation}
The sum runs only over the weights which actually occur.  One must not insert
fictitious intermediate layers: for example, if a parametrisation produces
only even powers of the expansion parameter, the absence of an odd power is
not an additional cancellation.  In the generic case,
\(\mathcal P_{\WSL}=\FSL\), and the next resolved occupied weight is
\(\WHR\).

The natural expectation is that the cancellation suppresses the individually
superleading monomials in \(\FSL\) just enough for their sum to enter at the
resolved leading weight \(\WHR\).  Otherwise a sector selected because its
monomials are superleading would, after cancellation, become subleading to
the resolved leading LP polynomial.  The expected saturation occurs in every
certified example considered here, and we conjecture that it is a general
property of hidden regions satisfying the conditions above.  The
hierarchy conditions alone do not prove it, however.  Logically, the first
nonzero term in the expansion of \(\FSL\) in the cancellation-resolving
coordinates could lie above \(\WHR\), with the resolved
leading layer supplied entirely by the rest of the LP polynomial.  HRF
therefore keeps the transformed \(\FSL\), \(\FObs\), the remaining
\(\mathcal F\) layers and \(\mathcal U\) as separate sources in the
dissection, measures their first nonzero weights independently, and does not
impose the expected equality.

An analogous refinement is needed only if the restored polynomial contains
an intervening layer \(\mathcal P_{W_k}\), with
\(\WSL<W_k<\WHR\), whose sum itself vanishes on the same pinch component
\(C\).  Here \(W_k\) is the common weight of the monomials in
\(\mathcal P_{W_k}\) before this additional cancellation is resolved; we call
it the nominal weight of the layer.  Rather than rejecting the candidate by
Eq.~\eqref{eq:single-layer-hierarchy}, one first tests whether the complete
layer \(\mathcal P_{W_k}\) belongs to the local pinch ideal
\(\ideal_C=\langle f_1,\ldots,f_r\rangle\).  If no intervening layer does so,
the generic test above is complete.

If such a layer is present, choose local transverse coordinates \(y_j=f_j\)
and coordinates \(\boldsymbol u\) along \(C\), and expand only as far as its
first nonzero transverse term,
\begin{equation}
 \mathcal P_{W_k}(\boldsymbol u,\boldsymbol y)
 =\sum_{|\boldsymbol\beta|=d_k}
   \boldsymbol y^{\boldsymbol\beta}
   h_{k,\boldsymbol\beta}(\boldsymbol u)
   +\mathcal O(|\boldsymbol y|^{d_k+1}),
 \qquad
 y_j\sim\delta^{\tau_j},\quad \tau_j>0.
\label{eq:exceptional-layer-expansion}
\end{equation}
Each nonzero term at that first transverse order has resolved weight
\(W_k+\boldsymbol\beta\cdot\boldsymbol\tau\), which must satisfy
\begin{equation}
 W_k+\boldsymbol\beta\cdot\boldsymbol\tau
 \geq\WHR>\WSL,
\label{eq:finalhierarchy}
\end{equation}
Thus an intervening layer with \(W_k<\WHR\) is admissible only if its
vanishing on \(C\) suppresses every first nonzero transverse contribution to
weight \(\WHR\) or above.  In all examples requiring this refinement, at
least one contribution saturates Eq.~\eqref{eq:finalhierarchy}, so the layer
reaches precisely \(\WHR\), just as \(\FSL\) does.  HRF tests the more general
inequality and does not impose this saturation.  This ideal-layer test is a
fallback for an explicitly encountered intervening cancellation, not part
of the generic determination of \(\boldsymbol v_{\HR}\).  It becomes
particularly relevant when the full \(\delta\)-dependent polynomial is
restored after asymptotic-order alignment, as discussed in
Sec.~\ref{sec:alignment}.

In a chart centred on the full common pinch \(C\), every independent
transverse coordinate \(y_j=f_j\) tends to zero, so every \(\tau_j\) is
positive.  This does not require the hierarchy to determine each
\(\tau_j\) separately.  Pullback here means translating the local scaling
through the coordinate change to the original LP parameters \(x_e\).  A
continuous family represents only local ratio
freedom if its pullback to the original LP parameters is fixed.  If the
pullback varies, the HR scaling remains undetermined; the local lower-facet
analysis must fix it, otherwise the candidate is rejected.

The output of Part~II is therefore a candidate scaling in the variables
supplied to the core construction.  In the ordinary case this is already the
candidate total vector.  Under asymptotic-order alignment it becomes the core
vector that must first be composed with the alignment vector as described in
Sec.~\ref{sec:alignment}.  In either case final certification is deferred
until after the wrappers, in Sec.~\ref{sec:scaleful}.

\subsection{Wrappers around the core construction}
\label{sec:alignment}

Two extensions fix the inputs supplied to the core without changing its
internal logic.  Asymptotic-order alignment selects a nontrivial starting
face of the full \(\delta\)-dependent \(\mathcal F\) and thereby specifies
\(\mathcal F_\star\).  A boundary search specifies \(\x_{\rm active}\) by
setting the complementary subset of LP parameters to zero, thereby
restricting the graph polynomial to a contraction stratum; graphically, this
contracts the corresponding edges and merges their incident vertices.  Either
operation, or both together, is followed by the same decomposition and
scaling construction of Sec.~\ref{sec:ordinary}.

\subsubsection{Asymptotic-order alignment}

When the relevant cancellation structure is not visible in the naive
fixed-\(\x\) leading polynomial \(\mathcal F_0\), a preliminary LP-parameter
rescaling can bring monomials occurring at different native powers of
\(\delta\) to a common effective leading order before the cancellation
decomposition is applied.  This need for cross-order alignment is
characteristic of the composite limits encountered in the examples below,
although the construction does not assume that the limit is composite.  This
operation is called
\emph{asymptotic-order alignment}.

The polynomial relevant to this preliminary operation is the full
\(\delta\)-dependent \(\mathcal F\), rather than the complete LP polynomial
\(\mathcal P=\mathcal U+\mathcal F\).  As shown in
Eq.~\eqref{eq:no-U-mixed-pinch}, \(\mathcal U\)-monomials cannot participate in
a cancellation producing an interior first-sheet stationary pinch.  They
nevertheless remain part of the complete LP problem and are restored when the
candidate scaling is tested.

Let \(\boldsymbol\phi\in\mathbb Q^N\) be the preliminary alignment vector and
introduce new parameters \(\boldsymbol y\) by
\begin{equation}
 x_e=\delta^{\phi_e}y_e.
 \label{eq:first-alignment-map}
\end{equation}
The corresponding augmented alignment vector is
\(\vec\phi\equiv(\boldsymbol\phi;1)\).
For a monomial
\(m_i=c_i(\s)\delta^{a_i}\x^{\boldsymbol r_i}\), the rescaling changes its
effective expansion order from \(a_i\) to
\begin{equation}
 \omega_{\boldsymbol\phi}(m_i)
 \equiv
 \vec\phi\cdot\vec r_i
 =a_i+\boldsymbol\phi\cdot\boldsymbol r_i.
 \label{eq:alignment-weight}
\end{equation}
Monomials that belong to different fixed-\(\x\) orders in \(\delta\) can
thereby acquire a common effective weight.  The relevant value is the minimum
of \(\omega_{\boldsymbol\phi}\): the monomials attaining it form the exposed
face and provide the support from which the cancellation sector is selected.
Before alignment these monomials can occur at different native powers of
\(\delta\), so they need not all belong to \(\mathcal F_0\).  After alignment
they form a common leading polynomial, allowing derivative harvesting to
expose their joint cancellation structure.

Selecting the monomials of minimum \(\omega_{\boldsymbol\phi}\) turns
asymptotic-order alignment into a geometric problem analogous, but not
identical, to the conventional geometric Method of Regions.  Here the
relevant Newton polytope is the convex hull of the augmented exponent points
of the full \(\delta\)-dependent \(\mathcal F\), and the relevant objects are
its exposed lower faces, which need not be facets.  The conventional
construction instead uses the Newton polytope of the complete LP polynomial
\(\mathcal P\) and associates regions with its lower facets.

HRF enumerates these faces without presupposing the cancellation structure.
For asymptotic-order alignment, the relevant faces contain monomials with
different native powers of \(\delta\).  For each such face, one may choose any
rational \(\boldsymbol\phi\) for which \(\vec\phi\) lies in the relative
interior of its normal cone.  All such choices select the same lowest-weight
monomials: the alignment is specified by the exposed face, while \(\vec\phi\)
is a convenient normal that exposes it.

For an order-aligned search, let
\(\mathcal M_\star\subseteq\mathcal M_{\mathcal F}\) consist of these
lowest-weight monomials and define
\begin{equation}
 \mathcal F^{[\boldsymbol\phi]}
 \equiv
 \sum_{m_i\in\mathcal M_\star}m_i.
\label{eq:alignmentface}
\end{equation}
The general HRF starting object is therefore
\begin{equation}
 \mathcal F_\star=
 \begin{cases}
   \mathcal F_0, & \text{ordinary search},\\
   \mathcal F^{[\boldsymbol\phi]}, & \text{order-aligned search}.
 \end{cases}
\label{eq:Fstar}
\end{equation}
Alignment thus brings these monomials into the modified starting polynomial
\(\mathcal F_\star\), to which the same cancellation-factor and obstruction
analysis, followed by the pinch test in the positive orthant, is applied.

Asymptotic-order alignment is not itself the hidden-region scaling.  It is
the first of two coordinate transformations; HRF then determines the scaling
about the cancellation locus in the aligned coordinates.

\paragraph{The core vector and the composition law.}

Let \(\boldsymbol v_{\rm core}\in\mathbb Q^N\) denote the HRF scaling in the aligned
variables,
\begin{equation}
 y_e=\delta^{(v_{\rm core})_e}z_e.
 \label{eq:second-hrf-map}
\end{equation}
Combining Eqs.~\eqref{eq:first-alignment-map} and
\eqref{eq:second-hrf-map} gives
\begin{equation}
 x_e=\delta^{\phi_e+(v_{\rm core})_e}z_e,
 \qquad
 \boldsymbol v_{\HR}=\boldsymbol\phi+\boldsymbol v_{\rm core},
 \qquad
 \vHR=(\boldsymbol\phi+\boldsymbol v_{\rm core};1).
 \label{eq:alignment-composition}
\end{equation}
Correspondingly, the alignment and the subsequent HR rescaling assign the
same original monomial the successive weights
\begin{equation}
 \omega_{\boldsymbol\phi}(m_i)
 =a_i+\boldsymbol\phi\cdot\boldsymbol r_i,
 \qquad
 w_{\HR}(m_i)
 =a_i+(\boldsymbol\phi+\boldsymbol v_{\rm core})\cdot\boldsymbol r_i.
 \label{eq:successive-monomial-weights}
\end{equation}
The second is its absolute weight after the composed rescaling.

Only the \(N\) edge-parameter exponents are added.  The last component is the
normalisation of the single expansion parameter and is appended once after
the two transformations have been composed.  Thus one must never add
\(\vec\phi\) and \((\boldsymbol v_{\rm core};1)\), which would spuriously
produce a last component equal to two.

The notation of Sec.~\ref{sec:layers} is retained for the composed scaling.
Thus, in Eqs.~\eqref{eq:weightlayers} and \eqref{eq:HR-colour-schematic},
\begin{equation}
 \mathcal P^{(\HR)}
 =\mathcal P(\delta^{\boldsymbol\phi+\boldsymbol v_{\rm core}}\boldsymbol z;
             \delta,\s),
 \qquad
 \boldsymbol v_{\HR}=\boldsymbol\phi+\boldsymbol v_{\rm core}.
 \label{eq:composed-LP-rescaling}
\end{equation}
and \(\WSL\), \(\WHR\) and \(W_k\) denote absolute weights in the complete
native polynomial.  Likewise, \(\FSL^{(\HR)}\) denotes the selected
\(\FSL\subset\mathcal F^{[\boldsymbol\phi]}\) after the complete rescaling of
the original LP parameters,
\(\x=\delta^{\boldsymbol\phi+\boldsymbol v_{\rm core}}\boldsymbol z\).  The simplified expression
\(\FSL^{(\HR)}=\delta^{\WSL}\FSL\) in
Eq.~\eqref{eq:HR-colour-schematic} applies directly to the unaligned core
case.

Normally the aligned face alone determines the core vector.  Let
\(\boldsymbol v_{\rm core}^{(0)}\) denote the vector obtained at this stage.  Because
the aligned search initially retains only this face of \(\mathcal F\),
\(\boldsymbol v_{\rm core}^{(0)}\) remains provisional until the omitted layers of the
complete LP polynomial are restored.  If a restored layer
contributes at a lower weight and does not vanish on the same pinch locus,
the candidate is rejected.  In the exceptional case where the entire layer
also cancels there, it must instead be included in the simultaneous hierarchy
analysis of Eq.~\eqref{eq:finalhierarchy}, which may modify the provisional
vector.  In the generic case
\(\boldsymbol v_{\rm core}=\boldsymbol v_{\rm core}^{(0)}\); in either case
the resulting \(\boldsymbol v_{\rm core}\) is then combined with
\(\boldsymbol\phi\) in
Eq.~\eqref{eq:alignment-composition}.\footnote{This exceptional multi-layer
refinement occurs in the double spacelike-collinear twisted-hexagon example
of Sec.~\ref{sec:example-alignment}: the middle layer in
Eq.~\eqref{eq:dsc-two-layer-colours} vanishes to first order on the common
pinch and changes the face-only vector to the final result in
Eq.~\eqref{eq:dsc-vector-example}.}

\paragraph{Uniform-shift ambiguity.}

The second Symanzik polynomial is homogeneous of fixed parameter degree.
Consequently,
\begin{equation}
 \boldsymbol\phi\longmapsto
 \boldsymbol\phi+c(1,\ldots,1)
\label{eq:uniformshift}
\end{equation}
adds the same amount to the weight of every monomial of \(\mathcal F\) and
does not change the face \(\mathcal F^{[\boldsymbol\phi]}\).  The preliminary alignment
vector therefore has this universal equivalence in addition to the general
freedom to choose a representative within the normal cone.

The coordinate transformation itself shows how this ambiguity is removed.
For fixed original \(\x\), replacing
\(\boldsymbol\phi\) by \(\boldsymbol\phi+c\boldsymbol1\) replaces the aligned
coordinates by \(\boldsymbol y'=\delta^{-c}\boldsymbol y\).  The same second
transformation is therefore represented by
\begin{equation}
 \boldsymbol v_{\rm core}' = \boldsymbol v_{\rm core}-c\boldsymbol1,
 \qquad
 (\boldsymbol\phi+c\boldsymbol1)+\boldsymbol v_{\rm core}'
 =\boldsymbol\phi+\boldsymbol v_{\rm core}.
 \label{eq:shift-covariant-composition}
\end{equation}
Hence the physical total vector is independent of the chosen alignment
representative, provided the core vector is transformed consistently.

In practice this compensation is fixed by the complete LP polynomial.
The polynomials \(\mathcal U\) and \(\mathcal F\) have degrees \(L\) and
\(L+1\), respectively, so a uniform shift changes their relative weights.
Requiring the resolved cancellation layer to balance the appropriate
\(\mathcal U\) layer, while satisfying the hierarchy conditions of
Sec.~\ref{sec:layers}, fixes
the representative and the physical gap \(\WHR-\WSL\).  This gap is a
property of the region and is not an adjustable normalisation.

The consistent two-stage procedure is therefore:
\begin{enumerate}[label=(\roman*)]
 \item enumerate the relevant exposed faces of the augmented support and
 choose a convenient representative \(\boldsymbol\phi\) for each;
 \item run HRF in the aligned variables to obtain the cancellation ideal and
 a provisional \(\boldsymbol v_{\rm core}^{(0)}\);
 \item restore \(\mathcal U\) and every occupied layer of the original
 \(\delta\)-dependent \(\mathcal F\), test the generic monomial hierarchy,
 and invoke the ideal-layer refinement only if an intervening layer vanishes
 on the same pinch component;
 \item compose the edge exponents according to
 \(\boldsymbol v_{\HR}=\boldsymbol\phi+\boldsymbol v_{\rm core}\), append the final
 component one, then dissect the common pinch, certify the resolved local
 lower facet and scalefulness, and pull its normal back.
\end{enumerate}

\subsubsection{Boundary hidden regions}
\label{sec:boundary}

An interior HR has all \(x_e>0\).  A boundary HR occurs on a contraction
stratum
\begin{equation}
 x_{e_1}=\cdots=x_{e_k}=0.
\end{equation}
The graph polynomial is first restricted to that stratum, the vanished
variables are removed, and the complete analysis is repeated with every
remaining active parameter positive.  In graph language this is the
corresponding contraction minor; in Landau language it is a subleading
Landau singularity.  The contraction is performed on the externally labelled
graph, so the momentum label and orientation of every external half-edge are
retained in the contracted graph.

A complete search must include the relevant contraction strata.  Failure to
find an interior HR is not a proof that the graph has no HR.

If a contraction stratum also requires asymptotic-order alignment, the
restriction is made first and the alignment faces are then constructed from
the surviving support.  The resulting \(\mathcal F_\star\) is passed to the
core HRF construction.  Thus boundary restriction and order alignment are
independent wrappers which may be nested without modifying the core logic.

\subsection{Certification, equivalence and completeness}
\label{sec:search-certification}

The core construction, together with any required order-alignment
composition, first determines a candidate total vector in the original LP
parameters.  Final certification then requires a local dissection test of
that vector.  At search level, the HRF
construction may also produce several algebraic presentations of the same
physical hidden region, whereas establishing absence requires a search
independent of any preferred presentation of the cancellation factors.  We
formulate these certification, equivalence and completeness criteria here.

\subsubsection{Final certification by dissection}
\label{sec:scaleful}

Whether the region integral is scaleful is a property of the resolved
monomial support of the LP polynomial.  It does not depend on the
space-time dimension \(D\).  A mere count of variables appearing in selected
monomials is a useful diagnostic but is not a general certificate.
For \(N\) local variables, choose one exponent point \(\vec r_0\) as a
reference.  The resolved support has full affine rank when the differences
\(\vec r_i-\vec r_0\) contain \(N\) linearly independent vectors.  This is
the dimensionality required for a facet; a smaller rank leaves an additional
scaling freedom.

After the cancellations are resolved, HRF always performs the dissection of
Ref.~\cite{Gardi:2024axt}.  In that work both cancellation factors in each
generator were used as dissection coordinates.  The subsequent
spacelike-collinear Fish analysis of Ref.~\cite{Chen:2026dnj} instead used a
dissection based on one of the two factors, anticipating the more economical
construction adopted systematically here.\footnote{A complementary development
of the dissection idea is aimed at numerical
integration in the Minkowski regime~\cite{Jones:2024minkowski,Jones:2025jzc,Jones:2026contour}.
There the positive parameter domain is partitioned along the full
\(\mathcal F=0\) hypersurface, rather than only at its Landau-pinched locus,
and the resulting domains are mapped back to a standard integration domain.
Cylindrical algebraic decomposition provides a general way to construct this
partition.  Each resolved integrand then has a definite sign; for
\(\mathcal F<0\) the overall phase is extracted, leaving a non-negative
integrand.  This deliberately extensive dissection is advantageous for
numerical integration, whereas HRF uses the minimal dissection needed to
resolve and certify a candidate hidden region.}

For the pair-product generators
realised here, Eq.~\eqref{eq:idealpresentation} reads
\begin{equation}
 \FSL=\sum_{k=1}^{n_{\rm gen}}M_k(\x,\s)g_k
      =\sum_{k=1}^{n_{\rm gen}}M_k(\x,\s)f_{a_k}f_{b_k}.
 \label{eq:generator-factor-dissection}
\end{equation}
It is sufficient to select one factor from each generator, say
\(f_{a_k}\), as a local cancellation-resolving condition.  Then every
generator contribution contains an explicit selected factor,
\(M_k(\x,\s)f_{a_k}f_{b_k}\), while the complementary factor \(f_{b_k}\)
remains in the transformed LP polynomial.  The cancellation that made the
region hidden arose because derivatives cancelled at a stationary point
inside the positive domain.  Once an independent set of the selected factors
has been resolved into boundary coordinates, any surviving region must appear
as an ordinary lower facet of the transformed polynomial.  Introducing both
factors from every generator is therefore unnecessary and can introduce
dependent local coordinates.  Different choices of one factor from each
generator provide alternative dissection charts.

For each such choice compatible with the common positive solution, use the
selected factors as dissection coordinates \(y_k=f_{a_k}\), omitting repeated
or dependent entries, and complete them to a regular local coordinate system
\((y_k,u_\alpha)\).  Since only one factor from each generator has been
selected, the \(y_k\) need not form a complete set of coordinates normal to
the full physical pinch, and the remaining \(u_\alpha\) need not all be
tangent to it.  Divide a neighbourhood into the sign sectors compatible with
\(x_e>0\), retaining the Jacobian of every change of variables.  In each
sector the selected cancellation surface is a boundary or corner, and the
transformed LP polynomial is treated by the geometric Method of Regions.  Its
lower-facet normals determine the possible relative scalings of the selected
dissection coordinates; no separate scan over those scalings is required.
Radial--ratio charts\footnote{Here a chart is a local coordinate patch; in a
radial--ratio chart one variable measures the distance from the selected
cancellation surface while ratios specify the relative direction among the
selected dissection coordinates.} can be used to display the resulting
centred and endpoint descriptions explicitly.  The near-planar Fish example
gives an explicit construction of the signed local coordinates and their
division into sign sectors in
Eqs.~\eqref{eq:five-near-planar-signed-local-coordinates}
and~\eqref{eq:five-near-planar-sign-sectors}.

The lower-facet test is applied to the complete polynomial in each chart.  It
does not require
\(\nabla_{\x}\mathcal F_\star=0\): when an obstruction is present, this
gradient is generically nonzero on the cancellation locus.  The stationary
equations defining that locus apply instead to the selected polynomial
\(\FSL\).

Each local lower facet supplies a normal in the coordinates of its chart.  To
compare this normal with the candidate total vector obtained after Part~II
and any required order-alignment composition, substitute the corresponding
local scaling into the inverse coordinate map and read off the scaling of every original LP
parameter \(x_e\).  We call the resulting vector, normalised to have last
component one, the \emph{pullback} of the local normal.  The final certificate
is attached to the chart cover rather than to one preferred local facet: all
equivalent charts must have the same pullback, while their ratio coordinates
may distribute the fixed cancellation depth differently among the local
variables.

The local chart is not unique.  A centred chart keeps the relative approach
of the selected dissection coordinates explicit and may expose a
lower-dimensional face common to neighbouring facets.  An \emph{endpoint
chart} is the limiting allocation in which the full resolved suppression is
assigned to one selected dissection coordinate; the leading support of the
transformed complete LP polynomial must then define an exact lower facet with
a positive inward normal.

The boundary defined by the factors selected for a chart may be larger than
the physical cancellation locus, while the complementary factors remain in
the transformed polynomial.  The physical locus itself is chart independent
and remains the common zero of the complete set of defining factors.  The
facets or faces exposed by the charts live in cancellation-resolving
coordinates and need not be faces of the original Newton polytope.  The
centred and endpoint charts, and the agreement of their pulled-back normals,
are displayed explicitly for the spacelike-collinear Fish in
Sec.~\ref{sec:five-point-sc-charts}.

Before dissection, Part~II determines a candidate core vector.  After any
required order-alignment composition, this gives the candidate total vector
in the original LP parameters.  In every example considered below this vector is already
fixed before dissection; any remaining freedom concerns only ratios among
cancellation-resolving coordinates and leaves the pullback unchanged.
Agreement with every relevant pulled-back local normal therefore provides a
strong independent check.  Thus, in the examples studied here, dissection
neither modifies nor rejects an HR that passes the preceding algebraic
determination: it supplies the final geometric certificate and exposes its
local chart structure.  If the homogeneous weight equations were instead to
retain a residual rescaling of the total vector or leave a cancellation layer
unresolved, their output would remain underdetermined and would not yet
constitute a certified HR.  The original-coordinate support may likewise
display the anticipated facet, but this does not replace the final local
certificate.

In the examples below we display adapted local coordinates when they
illuminate a nontrivial feature of the cancellation geometry or its
certificate --- for example, freedom among local charts, a moving or boundary
pinch locus, or several independent transverse directions.  Throughout the
certified examples, dissection confirms the candidate total vector obtained
from the core hierarchy and any required order-alignment composition.  Where the local
analysis reveals no additional structure, we state this agreement without
reproducing the complete coordinate map and local facet analysis.

\subsubsection{Equivalence of presentations}
\label{sec:equivalence}

A physical HR is specified by a positive component of the common pinch locus
and the chart-independent scaling data obtained by resolving its normal
directions, in particular the resulting total vector in the original LP
parameters.  A lower facet belongs to a chosen dissection chart and is not by
itself an invariant label.  Facets in different charts, or a common face in a
centred chart and its bounding endpoint facets, represent the same HR when
they resolve the same positive component and their normals pull back to the
same original-coordinate scaling.  In particular, neither a change of local
generators of the cancellation ideal nor a redistribution of the fixed
cancellation depth among normal coordinates produces a new region.

Distinct HRs may instead arise from:
\begin{itemize}
\item different cancellation ideals, corresponding to different singular
      hypersurface systems; or
\item inequivalent resolved normal directions, giving different pulled-back
      total scaling vectors.
\end{itemize}
If a cancellation locus has several disconnected positive branches, each
must be tested separately; no such case occurs in the examples below.
The accepted constructions must therefore be classified only after identifying
local facets related by regular chart changes and comparing their pullbacks to
the original LP parameters.

\subsubsection{A scaling-first formulation of a hidden-region absence statement}
\label{sec:absence}

The constructive HRF strategy described above is cancellation-first: it uses
derivatives of \(\mathcal F_\star\) to identify candidate cancellation
factors, constructs the generators and \(\FSL\), and only then determines the
scaling.  A conceptually interesting alternative is to reverse this order and
begin with the supports exposed by possible scalings.  This scaling-first
route is presentation-independent at the level of candidate supports, but it
has not been implemented or tested.

The homogeneity condition \eqref{eq:SLhomogeneity}, together with strict
separation from the remaining monomials, implies that the exponent support
of \(\FSL\) is an exposed face of the relevant starting support.  This
observation suggests the following formally exhaustive scaling-first search:

\begin{enumerate}[label=\arabic*.]
\item enumerate the relevant exposed faces of \(\mathcal F_0\), or of the
      augmented support of the full \(\delta\)-dependent \(\mathcal F\) when
      asymptotic-order alignment is required;
\item for each face polynomial, solve the simultaneous positive Landau
      equations in the physical domain \(\K\);
\item test every positive solution against \(\mathcal U\), all restored
      \(\delta\)-dependent layers and the hierarchy conditions, then apply
      source-resolved dissection in every sign sector compatible with the
      original positive domain and enumerate the resulting exact local lower
      facets;
\item repeat the analysis on all required contraction strata.
\end{enumerate}

This construction should be regarded as a theoretical alternative, not as
the standard used for negative audits in this work.  In particular, the
four-loop No-Crown absence results of
Secs.~\ref{sec:no-crown-wide-angle} and~\ref{sec:crown-wide-angle-regge}
were obtained by exhaustive application of the derivative-first HRF search,
rather than by the scaling-first procedure above.  More generally, an
absence statement requires exhaustive coverage of the relevant possibilities
within whichever search strategy is used.  A non-exhaustive search may
discover HRs, but cannot establish their absence.

\subsection{Algorithm summary}
\label{sec:workflow}

The complete conceptual workflow is:
steps 1--3 configure the physical problem and its optional wrappers,
steps 4--9 apply the two parts of the core HRF construction,
step~10 supplies the final dissection certificate after any required
order-alignment composition,
and
steps 11--12 manage equivalence and completeness at search level.
\begin{enumerate}[label=\textbf{\arabic*.},leftmargin=*]
\item \textbf{Fix the kinematic chart.}
      Specify the expansion parameter~\(\delta\to0^+\), a common set of real kinematic coordinates~\(\s\), and the inequalities defining the physical domain~\(\K\).
      Introduce positive coordinates only locally if they simplify a
      particular positivity test.  Otherwise, retaining common signed
      coordinates explicitly reveals how cancellation loci appear or
      disappear under crossing or analytic continuation.
\item \textbf{Choose a stratum.}
      Start in the interior and repeat on the required contraction strata.
\item \textbf{Choose the starting face  \(\mathcal F_\star\).}
      Use \(\mathcal F_0\) in the ordinary case.  If the relevant cancellation
      may involve monomials at different native powers of \(\delta\), consider
      the order-aligned exposed faces \(\mathcal F^{[\boldsymbol\phi]}\) of the
      full \(\delta\)-dependent \(\mathcal F\).
\item \textbf{Generate candidate cancellation structures.}
      Form the derivatives of \(\mathcal F_\star\) and apply the mixed-sign
      test only as a necessary prefilter.  Harvest non-monomial cancellation
      factors directly by literal factorisation and, when compatible with the
      kinematic organisation, by collecting the derivatives into independent
      kinematic structures and factorising their LP-parameter coefficients.
      Do not require this coefficientwise decomposition or treat it as
      sufficient.  Do not solve the simultaneous derivative
      equations of \(\mathcal F_\star\), nor saturate their full ideal: they
      generically have no positive solution at this stage when an obstruction
      is present.
\item \textbf{Build the massless generator layer.}
      Form candidate generators as products of pairs of distinct harvested
      factors, \(g_k=f_if_j\), with \(i\neq j\).  Retain only pairs whose
      LP-parameter supports are disjoint and whose expanded products pass the
      LP-support test against \(\mathcal F_\star\).  Form the candidate ideal
      \(\mathcal I_{\rm gen}(G)\) from the surviving generators.
\item \textbf{Find the obstruction and construct the decomposition.}
      For each candidate ideal \(\mathcal I_{\rm gen}(G)\), search for subsets
      \(\MObs\subseteq\mathcal M_\star\) whose removal leaves a nonzero
      \(\FSL\in\mathcal I_{\rm gen}(G)\), with
      \(\MSL=\mathcal M_\star\setminus\MObs\).  Discard \(G\) if no such
      subset exists.
\item \textbf{Apply the first-sheet pinch test to \(\FSL\).}
      Only after \(\MSL\) has been isolated, solve
      \(\partial\FSL/\partial x_e=0\) simultaneously for all active
      parameters with \(\x>0\) and \(\s\in\K\).  The resulting common locus
      in the positive orthant defines the physical cancellation ideal; reject
      the candidate if this system has no solution.  When useful, complete these
      selected-sector equations, or explicitly selected derivative-sector
      relations, by the saturation \eqref{eq:selected-gradient-saturation}.

\item \textbf{Determine a candidate scaling.}
      Restore the complete \(\mathcal P=\mathcal U+\mathcal F\) and solve the
      homogeneity and single-layer hierarchy conditions
      \eqref{eq:SLhomogeneity} and \eqref{eq:single-layer-hierarchy} for
      \(\vHR=(\boldsymbol v_{\HR};1)\), \(\WSL\) and \(\WHR\).  Reject an
      inconsistent system or a solution with zero gap.  A full-rank system
      fixes a unique candidate vector; a rank-deficient system remains
      provisional until dissection.
\item \textbf{Inspect the scaled layers and resolve exceptional cases.}
      For each candidate vector, form the rescaled LP polynomial
      \(\mathcal P^{(\HR)}=\mathcal P(\delta^{\boldsymbol v_{\HR}}
      \boldsymbol x;\delta,\s)\) and decompose it into its occupied weight
      layers as in Eq.~\eqref{eq:weightlayers}.  If the first occupied
      layer above \(\WSL\) is nonzero on the common pinch component, the case
      is ordinary and the single-layer hierarchy is sufficient.  If one or
      more occupied layers above \(\WSL\) vanish there before the first
      resolved nonzero weight, determine the first nonzero order of each such
      layer in the local pinch ideal
      \(\ideal_C=\langle f_1,\ldots,f_r\rangle\) and impose the inequalities
      \eqref{eq:finalhierarchy}.  Apply this ideal-layer refinement to every
      intervening vanishing layer and reject an inconsistent system or a
      solution with zero gap.  This determines a candidate scaling, whose
      local lower-facet and scalefulness certificate is supplied by the
      dissection in the next step.
\item \textbf{Construct the dissection charts and certify the total vector.}
      For every candidate surviving either hierarchy, select one cancellation
      factor, say \(f_{a_k}\), from each pair-product generator
      \(g_k=f_{a_k}f_{b_k}\) as a cancellation-resolving condition, omit
      repeated or dependent selections, and complete the result to a regular
      local coordinate system.  Keep the complementary cancellation factors
      \(f_{b_k}\) in the transformed polynomial and keep \(\FSL\), \(\FObs\), the remaining
      \(\mathcal F\) layers and \(\mathcal U\) as separate sources.  Cover all
      sign sectors compatible with the original positive domain and enumerate
      the exact local lower facets of the transformed complete polynomial in
      each sector.  Measure the first nonzero local weight of each source.
      Test alternative admissible factor selections as needed;
      one exact positive certificate establishes existence, whereas rejection
      requires exhausting them.  Pull every certified local normal back to the
      original LP parameters.  The resulting total vector and positive gap
      provide the final HRF certificate; agreement with an
      original-coordinate determination is an independent check.
\item \textbf{Identify physical equivalence classes.}
      Merge generator presentations that describe the same pinch component
      in the positive orthant and local chart facets whose normals have the
      same original-coordinate pullback.
\item \textbf{State completeness honestly.}
      A positive HR requires all certification stages.  A no-HR conclusion
      requires exhaustive face and boundary coverage; incomplete searches
      remain unresolved.
\end{enumerate}

\section{Hidden Region Finder: Applications}
\label{sec:applications}

This section demonstrates the Hidden Region Finder through examples chosen to
isolate distinct algorithmic requirements.  The point is not merely to
catalogue hidden regions, but to make clear which stages of the search are
needed as the simplifying features of the most elementary case are removed.
We begin with the four-point three-loop Crown.  This is the simplest HR from
the viewpoint of discovery: the complete leading polynomial participates in a
simultaneous positive cancellation, so that \(\mathcal F_\star=\FSL\), no
obstruction is present and the HR has a uniform scaling.  Its nontrivial
discovery step is the separation of two generator sectors carrying the
independent kinematic coefficients \(s_{12}\) and \(s_{23}\); only after this
separation does factorisation expose the individual binomial cancellation
factors.  Its higher-loop descendants then illustrate hidden regions on
contraction strata and show that the corresponding factors can instead contain
three or more monomials.  The No-Crown analysis provides a converse test, while the Regge
limits test the correspondence across kinematic channels.  We next turn to
the five-point two-loop Fish family, based on the topology commonly called
the non-planar double box.  Its seed already contains a nonzero obstruction
and therefore requires a non-uniform parameter scaling.  A
vertex-corrected descendant shows how its cancellation locus survives a local
loop correction through a polynomial lift.  The near-planar wide-angle limit
provides the closest five-point parallel to the Crown and is the only example
here with a three-generator cancellation ideal.  The subsequent
rapidity-ordered and correlated MRK limits show the same unlabelled Fish seed
recurring in distinct kinematic expansions, while their different region
vectors demonstrate that the scaling is fixed by the details of the expansion,
not by the topology alone.  They separately expose the need for
asymptotic-order alignment and, in the MRK--planar case, multiple cancellation
layers.  Finally, the six-point twisted-box seed and its twisted-hexagon
realisation combine several of these ingredients.  The NMRK and
double-collinear regions inherit their cancellation geometry from a boundary
HR on the physical wide-angle self-crossing surface, while their different
total vectors again reflect the kinematic expansions used to reach that
surface.

\subsection{Four-point integrals: the Crown seed}
\label{sec:example-crown}

The four-point applications are organised by the three-loop Crown,
\(\mathcal G_{\rm C}\).  We first identify its interior wide-angle HR, then
follow the same seed onto
contraction strata of higher-loop graphs, complete the converse No-Crown
audit, and finally compare the resulting topology classification with the
three Regge channels.

\subsubsection{Wide-angle kinematics and the three-loop seed}

The Crown hidden region in the on-shell wide-angle expansion was first
identified and analysed in Ref.~\cite{Gardi:2024axt}.  From the viewpoint of
the finding problem, it is the simplest example considered here.  The full
leading polynomial is the cancellation sector, with no obstructing terms, and
the resulting HR scaling is uniform.  It is nevertheless not a
single-generator example: its derivatives mix two products of cancellation
factors, weighted by independent kinematic invariants.  We first show how the
derivative harvest separates these generator sectors and then certify the
region vector.  This provides the baseline against which the additional
search steps required by the later examples can be distinguished, as well as
the seed for the subsequent boundary and topology audits.

\begin{figure}[htbp]
\centering
\begin{tikzpicture}[scale=.52]
 \coordinate (v1) at (2,2);
 \coordinate (v2) at (2,8);
 \coordinate (v3) at (8,2);
 \coordinate (v4) at (8,8);
 \coordinate (v5) at (5,2.5);
 \coordinate (v6) at (5,7.5);

 \draw[hrf internal,draw=Green] (v6) to[bend right=10] (v1);
 \draw[hrf internal,draw=LimeGreen] (v6) to[bend right=10] (v2);
 \draw[hrf internal,draw=TealBlue] (v6) to[bend left=10] (v3);
 \draw[hrf internal,draw=OliveGreen] (v6) to[bend left=10] (v4);
 \foreach \a/\b/\bend/\jetcolour in {
   v5/v1/left/Green,
   v5/v2/left/LimeGreen,
   v5/v3/right/TealBlue,
   v5/v4/right/OliveGreen}{
   \draw[white,line width=4pt] (\a) to[bend \bend=10] (\b);
   \draw[hrf internal,draw=\jetcolour] (\a) to[bend \bend=10] (\b);
 }

 \node[hrf edge label] at (3.5,1.5) {$x_1$};
 \node[hrf edge label] at (2,3.5) {$x_2$};
 \node[hrf edge label] at (2,6.5) {$x_3$};
 \node[hrf edge label] at (3.5,8.5) {$x_4$};
 \node[hrf edge label] at (6.5,1.5) {$x_5$};
 \node[hrf edge label] at (8,3.5) {$x_6$};
 \node[hrf edge label] at (8,6.5) {$x_7$};
 \node[hrf edge label] at (6.5,8.5) {$x_8$};

 \node[hrf vertex,fill=Green,draw=Green] at (v1) {};
 \node[hrf vertex,fill=LimeGreen,draw=LimeGreen] at (v2) {};
 \node[hrf vertex,fill=TealBlue,draw=TealBlue] at (v3) {};
 \node[hrf vertex,fill=OliveGreen,draw=OliveGreen] at (v4) {};
 \node[hrf vertex,fill=Blue,draw=Blue] at (v5) {};
 \node[hrf vertex,fill=Blue,draw=Blue] at (v6) {};
 \node[font=\small\bfseries,text=Blue] at (5,3.25) {$H_1$};
 \node[font=\small\bfseries,text=Blue] at (5,6.75) {$H_2$};
 \draw[->,thick,Green] (.5,1)--(v1);
 \draw[->,thick,LimeGreen] (.5,9)--(v2);
 \draw[->,thick,TealBlue] (v3)--(9.5,1);
 \draw[->,thick,OliveGreen] (v4)--(9.5,9);
 \node[hrf leg label,text=Green] at (.5,.45) {$p_1$};
 \node[hrf leg label,text=LimeGreen] at (.5,9.55) {$p_2$};
 \node[hrf leg label,text=TealBlue] at (9.5,.45) {$p_3$};
 \node[hrf leg label,text=OliveGreen] at (9.5,9.55) {$p_4$};
\end{tikzpicture}
\caption{The three-loop Crown \(\mathcal G_{\rm C}\) with its LP-parameter
assignment and wide-angle momentum configuration.  The four shades identify
the four collinear directions: each external momentum and the corresponding
two-edge path have the same colour.  The blue four-point vertices \(H_1\) and
\(H_2\) are the two disconnected hard-scattering components.}
\label{fig:crown-graph}
\end{figure}
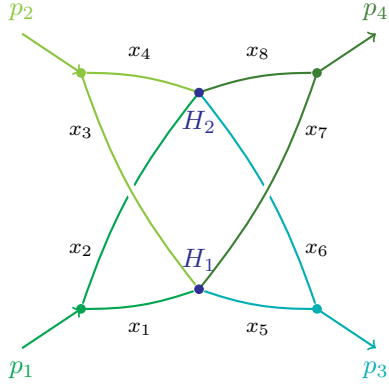

Take four external momenta approaching the physical \(2\to2\) domain, with
limiting invariants satisfying
\begin{equation}
 s_{12}>0,\qquad s_{23}<0,\qquad s_{12}>-s_{23},
 \label{eq:crown-domain}
\end{equation}
and regulate the on-shell limit by \(p_i^2=\delta P_i^2\), with fixed
nonzero \(P_i^2\) and \(\delta\to0^+\).  Momentum conservation gives the
exact off-shell relation
\(s_{12}+s_{23}+s_{13}=\sum_{i=1}^4p_i^2
=\delta\sum_{i=1}^4P_i^2\), which reduces to the usual massless relation
only in the limit.  A common normalisation of the \(P_i^2\) is immaterial
for the region vector.  The graph and its
LP-parameter assignment are shown in Fig.~\ref{fig:crown-graph}.

Each coloured two-edge path carries one of the four collinear directions
between the blue vertices \(H_1\) and \(H_2\).  These vertices represent two
separate hard scatterings connected by the four jet-like paths.  This is the
essential Landshoff feature of the Crown HR~\cite{Gardi:2024axt}.  It cannot
occur in a facet region of the wide-angle on-shell expansion, for which the
hard subgraph is necessarily connected \cite{Gardi:2022khw,Ma:2023hrt}.

Here and below, displayed LP polynomials use the weight-layer colour convention
of Sec.~\ref{sec:weight-colours}.  At fixed parameters the leading polynomial is
\begin{equation}
\begin{aligned}
 \mathcal F_0[\mathcal G_{\rm C}]
   &={\SLcolour{s_{12}g_1[\mathcal G_{\rm C}]}}
     {\SLcolour{-s_{23}g_2[\mathcal G_{\rm C}]}},\\
 g_1[\mathcal G_{\rm C}]&=f_1f_2,
 &g_2[\mathcal G_{\rm C}]&=f_3f_4,\\
 f_1&=x_4x_5-x_3x_6,
 &f_2&=x_2x_7-x_1x_8,\\
 f_3&=x_2x_3-x_1x_4,
 &f_4&=x_6x_7-x_5x_8.
\end{aligned}
 \label{eq:crown-F0}
\end{equation}
Here the four \(f_i\) provide the first explicit example of the cancellation
factors introduced in Sec.~\ref{sec:decomposition}.  The full derivatives do
not factorise directly into them because they mix the two generator sectors.
For example,
\begin{align*}
 \frac{\partial\mathcal F_0[\mathcal G_{\rm C}]}{\partial x_1}
   &=-s_{12}x_8f_1+s_{23}x_4f_4,
 &
 \frac{\partial\mathcal F_0[\mathcal G_{\rm C}]}{\partial x_5}
   &= s_{12}x_4f_2+s_{23}x_8f_3.
\end{align*}
Separating the coefficients of the independent kinematic structures and
omitting the monomial prefactors, which are nonzero in the positive orthant,
therefore harvests all four~\(f_i\).  Pairing factors with disjoint
LP-parameter support gives the two generators
\(g_1[\mathcal G_{\rm C}]=f_1f_2\) and
\(g_2[\mathcal G_{\rm C}]=f_3f_4\).
The corresponding candidate generator ideal is
\(\mathcal I_{\rm gen}(\mathcal G_{\rm C})
=\langle g_1[\mathcal G_{\rm C}],g_2[\mathcal G_{\rm C}]\rangle\).  The
positive pinch test then selects the component of its zero set on which all
four \(f_i\) vanish simultaneously.
Once the two generator sectors multiplying \(s_{12}\) and \(-s_{23}\) have
been isolated, the candidate common cancellation locus
\(f_1=f_2=f_3=f_4=0\) is visible directly in the full leading polynomial,
without first removing an obstruction.

To display the structure of this locus in its simplest form, we now introduce
a ratio parametrisation.  This is an analytic description of the result, not
the parametrisation used in the HRF search.  Regard the graph as four two-edge
paths between the two four-point vertices
\(H_1\) and \(H_2\).  Factoring one positive overall LP-parameter scale from
each path leaves the four ratios
\begin{equation}
 r_A=\frac{x_1}{x_2},\qquad r_B=\frac{x_3}{x_4},\qquad
 r_C=\frac{x_5}{x_6},\qquad r_D=\frac{x_7}{x_8}.
 \label{eq:crown-path-ratios}
\end{equation}
We collect them in the column vector
\(\boldsymbol r\equiv(r_A,r_B,r_C,r_D)^T\).  Then
\begin{align}
 \mathcal F_0[\mathcal G_{\rm C}]
 &=x_2x_4x_6x_8\,\Psi_C(\boldsymbol r),\nonumber\\
 \Psi_C(\boldsymbol r)
 &=s_{12}(r_C-r_B)(r_D-r_A)
   -s_{23}(r_B-r_A)(r_D-r_C)
   =\frac12\boldsymbol r^T K_C\boldsymbol r,
 \label{eq:crown-ratio-quadratic}\\[-1mm]
 K_C&=\begin{pmatrix}
 0&s_{12}&-s_{12}-s_{23}&s_{23}\\
 s_{12}&0&s_{23}&-s_{12}-s_{23}\\
 -s_{12}-s_{23}&s_{23}&0&s_{12}\\
 s_{23}&-s_{12}-s_{23}&s_{12}&0
 \end{pmatrix}.
 \nonumber
\end{align}
The ratio Landau equations are \(K_C\boldsymbol r=0\).  Every row of \(K_C\)
sums to zero, so \((1,1,1,1)^T\) is a null vector and
\(\det K_C=0\).  On the other hand, in the
domain~\eqref{eq:crown-domain} a principal three-by-three minor is nonzero:
\begin{equation}
 \det (K_C)_{1\ldots3,1\ldots3}
 =-2s_{12}s_{23}(s_{12}+s_{23})\ne0.
 \label{eq:crown-ratio-rank}
\end{equation}
Consequently \(K_C\) has rank three and its complete stationary locus is
\begin{equation}
 \boldsymbol r=\rho(1,1,1,1),\qquad \rho>0,
 \qquad \Psi_C(\boldsymbol r)=0.
 \label{eq:crown-ratio-stationarity}
\end{equation}
Here \((1,1,1,1)^T\) specifies the null direction and \(\rho\) is the single
positive scalar coordinate along it, namely the common value of the four path
ratios.  Physically, \(\rho\) sets the common division of the momentum from
each jet between the hard scatterings \(H_1\) and \(H_2\).  The last equality
follows directly from
\(\Psi_C=\boldsymbol r^TK_C\boldsymbol r/2\).  Thus a positive stationary
family exists at every generic point of the physical wide-angle domain; no
additional kinematic condition is required.  In the analytic treatment of
Ref.~\cite{Gardi:2024axt}, this ratio change is the rescaling step that
precedes the local dissection of the Crown pinch.  It turns
each binomial factor in~\eqref{eq:crown-F0} into a positive path scale times
a linear difference of two ratio points.  Their common zero is
\(r_A=r_B=r_C=r_D\), which is fixed by only three relative distances, for
example \(r_B-r_A\), \(r_C-r_A\) and \(r_D-r_A\).  The four displayed
binomial constraints therefore define a codimension-three common locus.
This establishes the structure of the cancellation locus, but the
decomposition follows separately from Eq.~\eqref{eq:crown-F0}:
\(\mathcal F_0[\mathcal G_{\rm C}]
=s_{12}g_1[\mathcal G_{\rm C}]-s_{23}g_2[\mathcal G_{\rm C}]\) belongs entirely to
\(\mathcal I_{\rm gen}(\mathcal G_{\rm C})\), with no remaining monomials to
form an obstruction.  Therefore
\begin{equation}
 \mathcal F_\star[\mathcal G_{\rm C}]
 =\mathcal F_0[\mathcal G_{\rm C}]
 =\FSL[\mathcal G_{\rm C}],
 \qquad \FObs[\mathcal G_{\rm C}]=0.
 \label{eq:crown-no-obstruction}
\end{equation}
At fixed \(s_{12}\) and \(s_{23}\), the external-virtuality expansion is
exactly
\(\mathcal F[\mathcal G_{\rm C}]
=\mathcal F_0[\mathcal G_{\rm C}]
+\delta\mathcal F_1[\mathcal G_{\rm C}]\), where
\begin{equation*}
 \mathcal F_1[\mathcal G_{\rm C}]
 =\left.\sum_{i=1}^4 P_i^2
   \frac{\partial\mathcal F[\mathcal G_{\rm C}]}{\partial p_i^2}
  \right|_{p_1^2=\cdots=p_4^2=0}.
\end{equation*}
Thus \(\mathcal F_1\) is linear in the four fixed virtuality coefficients,
with the corresponding two-forest polynomials as coefficients.  To determine
the HR vector, the monomials of
\(\FSL[\mathcal G_{\rm C}]=\mathcal F_0[\mathcal G_{\rm C}]\) are assigned a
common weight according to Eq.~\eqref{eq:SLhomogeneity}, while the
\(\delta\mathcal F_1[\mathcal G_{\rm C}]\) and
\(\mathcal U[\mathcal G_{\rm C}]\) monomials at resolved-leading weight
saturate the hierarchy in Eq.~\eqref{eq:single-layer-hierarchy}.  The resulting
linear system forces the edge part of the region vector to take the form
\(\boldsymbol v_{\HR}=v(1,\ldots,1)\).  Since
\(\mathcal F_1\) has degree four and \(\mathcal U\) degree three, their
equality at resolved-leading weight gives \(1+4v=3v\), and hence \(v=-1\).
Using the augmented-vector convention of Sec.~\ref{sec:layers},
\(\vHR=(\boldsymbol v_{\HR};1)\), the resulting vector and weights are
\begin{equation}
\begin{aligned}
 \vHR[\mathcal G_{\rm C}]&=(-1,-1,-1,-1,-1,-1,-1,-1;1),\\
 (\WSL,\WHR)&=(-4,-3).
\end{aligned}
 \label{eq:crown-vector}
\end{equation}
Consequently, applying the definition in
Eq.~\eqref{eq:rescaled-LP-polynomial}, the complete LP polynomial after the HR
rescaling is
\begin{equation}
\begin{aligned}
 \mathcal P^{(\HR)}[\mathcal G_{\rm C}](\x;\delta,\s)
 &\equiv
 \mathcal P[\mathcal G_{\rm C}](\delta^{-1}\x;\delta,\s)\\
 &=\delta^{-4}\SLcolour{\mathcal F_0[\mathcal G_{\rm C}]}
  +\delta^{-3}\HRcolour{\bigl(\mathcal F_1[\mathcal G_{\rm C}]
                              +\mathcal U[\mathcal G_{\rm C}]\bigr)}.
\end{aligned}
 \label{eq:crown-colours}
\end{equation}
Because the second Symanzik polynomial is linear in the external virtualities
and every \(p_i^2\) is proportional to \(\delta\), this expansion terminates
at \(\mathcal F_1\).  Thus, in the occupied-layer terminology of
Eq.~\eqref{eq:weightlayers}, the complete rescaled Crown polynomial has
exactly two nonempty layers.  The weight-\(-4\) layer
\(\mathcal P_{-4}[\mathcal G_{\rm C}]=\mathcal F_0[\mathcal G_{\rm C}]\)
is the superleading cancellation sector, while the weight-\(-3\) layer
\(\mathcal P_{-3}[\mathcal G_{\rm C}]
=\mathcal F_1[\mathcal G_{\rm C}]+\mathcal U[\mathcal G_{\rm C}]\) is the
resolved-leading layer.  There is no remaining layer in this example.
Neither \(\mathcal F_0[\mathcal G_{\rm C}]\), the factors \(f_i\), nor the
generator ideal depends on the \(P_i^2\).  For the fixed native scaling
\(p_i^2\sim\delta\), any proper subset of the external legs may therefore be
kept strictly on shell, provided at least one nonzero virtuality remains.  In
the Crown this changes \(\mathcal F_1[\mathcal G_{\rm C}]\), but not the
cancellation locus or the vector in Eq.~\eqref{eq:crown-vector}.  The nonzero
virtualities remain useful as regulators of the ordinary infrared
singularities.

The role of the virtuality layer in Part~II, despite its absence from the
Part-I cancellation locus, becomes clearer upon generalising its native
order.  Suppose the nonzero virtualities scale as
\(p_i^2=\delta^{a_i}P_i^2\), and let
\(a_{\min}=\min_i a_i\).  The first virtuality layer contains the terms from
the legs with \(a_i=a_{\min}\).  For the Crown, the two-forest support supplied
by any nonempty set of such legs is sufficient to complete the hierarchy
system; their number does not affect its solution.  The resolved-leading
balance is \(a_{\min}+4v=3v\), and therefore
\begin{equation*}
\begin{aligned}
 \boldsymbol v_{\HR}&=-a_{\min}(1,\ldots,1),
 &\vHR&=(\boldsymbol v_{\HR};1)
       =(-a_{\min},\ldots,-a_{\min};1),\\
 (\WSL,\WHR)&=(-4a_{\min},-3a_{\min}).
\end{aligned}
\end{equation*}
Terms with \(a_i>a_{\min}\) occur at higher weights and do not alter this
augmented vector.  Equivalently, using \(\lambda=\delta^{a_{\min}}\) as the
expansion parameter gives the normalised augmented vector
\((-1,\ldots,-1;1)\).  A trial branch that omits the first virtuality layer is
instead rank deficient: after dissection its resolved support contains no
monomial of positive native \(\delta\)-degree and spans only a lower-dimensional
face, not a scaleful lower facet.  Restoring the \(a_{\min}\) layer fixes the
scaling and completes the facet.  This conclusion uses the support of the
Crown virtuality polynomials and is not asserted here as a
topology-independent rule.

The Crown HR corresponds in momentum space to the Landshoff
independent-scattering configuration.  The central collinear directions can
be read directly from the on-shell Landau conditions.  In the
Coleman--Norton picture of a physical-region pinch, the pinched propagators
describe on-shell classical propagation~\cite{Coleman:1965cn,Collins:2020euz}.
For each massless two-edge path of the Crown this picture becomes degenerate:
the two propagator momenta are null at the pinch and differ by the null
external momentum attached to that path.  Their scalar product with that
external momentum therefore vanishes, which on the physical real branch
places both momenta on its collinear ray.  These four rays are indicated by
the four shades in Fig.~\ref{fig:crown-graph}.

To obtain a momentum-space measure one must additionally determine the local
widths about these rays and their correlations.  The analysis of
Ref.~\cite{Gardi:2024axt}\footnote{The momentum-space analysis there builds
on the earlier independent-scattering treatment of Botts and
Sterman~\cite{Botts:1989kf}.} adopted the corresponding jet-like fluctuation
envelopes and then used momentum conservation and the pinch constraints to
fix the correlated longitudinal support.  Its momentum-space and dissected
parameter-space counts give the same leading degree, providing a strong
consistency check of the mode realisation.

This completes the baseline HRF case: after separating two kinematic
coefficient sectors, the derivative harvest exposes the two generators and
their simultaneous positive zero; the selected cancellation sector retains
all of \(\mathcal F_0[\mathcal G_{\rm C}]\), and the hierarchy admits the uniform scaling
certified by the dissection.  No obstruction removal or preliminary
asymptotic-order alignment is needed.  The later examples are designed to
show what must be added when one or more of these simplifications do not
occur.

\subsubsection{Boundary descendants: SuperCrown and HyperCrown}
\label{sec:example-crown-boundaries}

Ref.~\cite{Gardi:2024axt} had already observed that the Crown organises the
three- and four-loop graphs that can support wide-angle HRs.  At three loops,
the complete class identified there comprises the Crown itself and nine graphs
obtained by opening either or both of its four-point hard vertices.  At four
loops, the mixed-sign screen retained 1097 topologies: 1081 contain the Crown
as a contraction minor, whereas 16 do not.
Because this sign pattern is only
a necessary prefilter, the census did not by itself certify an HR in every
Crown-containing graph.  Ref.~\cite{Gardi:2024axt} therefore conjectured that,
through four loops in massless \(2\to2\) scattering, HRs occur precisely for
the topologies in which the Crown structure can be exposed by contractions.

Contractions are performed within the externally labelled parent graph, as in
Sec.~\ref{sec:boundary}.  For the Crown, however, the external labels impose no
additional restriction: its symmetry relates the external-leg permutations,
and the corresponding crossed channels all support the HR.  In the Crown
family below, a Crown contraction minor may therefore be identified at the
level of the unlabelled topology.  The stronger externally labelled condition
becomes essential for less symmetric seeds, such as the Fish, for which the
incident seed vertex and momentum assignment must also be preserved.

Here we sharpen both sides of this picture with exact HRF statements.  On the
positive side, the examples below certify boundary HRs in representative
four- and five-loop Crown descendants.  Conversely,
Sec.~\ref{sec:no-crown-wide-angle} gives complete absence certificates for
the 16 exceptional four-loop topologies.  In the positive examples below,
setting selected additional LP parameters to zero contracts edges without
reducing the loop order of the parent graph: it exposes the disconnected
Crown hard structure while the remaining propagators accommodate the
additional loop momenta.

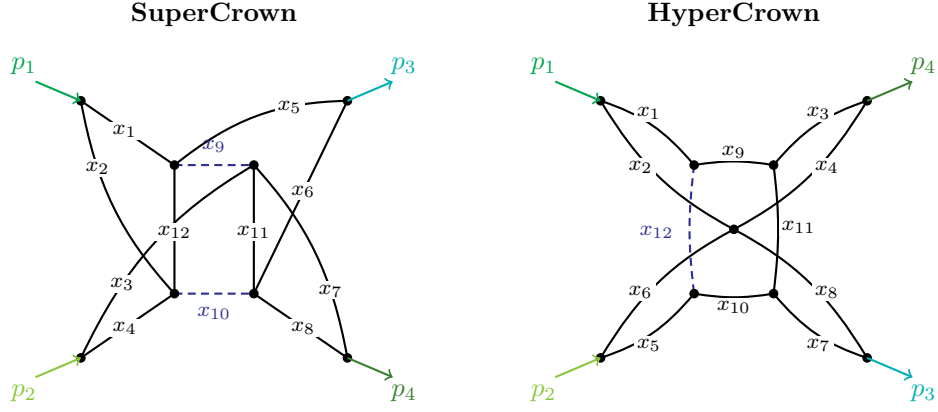
\begin{figure}[htbp]
\centering
\begin{tikzpicture}[xscale=.75,yscale=.85]
 \coordinate (v1) at (-2.35,2.00);
 \coordinate (v2) at (-2.35,-2.00);
 \coordinate (v3) at (2.35,2.00);
 \coordinate (v4) at (2.35,-2.00);
 \coordinate (v5) at (-0.70,1.00);
 \coordinate (v6) at (-0.70,-1.00);
 \coordinate (v7) at (0.70,1.00);
 \coordinate (v8) at (0.70,-1.00);
 \node[font=\small\bfseries] at (0,3.35) {SuperCrown};

 \draw[hrf internal] (v1)--node[hrf edge label,pos=.47] {$x_1$}(v5);
 \draw[hrf internal] (v1) to[bend right=17]
   node[hrf edge label,pos=.30] {$x_2$}(v6);
 \draw[hrf internal] (v2) to[bend left=17]
   node[hrf edge label,pos=.30] {$x_3$}(v7);
 \draw[hrf internal] (v2)--node[hrf edge label,pos=.47] {$x_4$}(v6);
 \draw[hrf internal] (v3) to[bend right=17]
   node[hrf edge label,pos=.30] {$x_5$}(v5);
 \draw[hrf internal] (v3)--node[hrf edge label,pos=.47] {$x_6$}(v8);
 \draw[hrf internal] (v4) to[bend right=17]
   node[hrf edge label,pos=.30] {$x_7$}(v7);
 \draw[hrf internal] (v4)--node[hrf edge label,pos=.47] {$x_8$}(v8);
 \draw[hrf contracted] (v5)--node[font=\scriptsize,text=Blue,above=1pt] {$x_9$}(v7);
 \draw[hrf contracted] (v6)--node[font=\scriptsize,text=Blue,below=1pt] {$x_{10}$}(v8);
 \draw[hrf internal] (v7)--node[hrf edge label,pos=.52] {$x_{11}$}(v8);
 \draw[hrf internal] (v5)--node[hrf edge label,pos=.52] {$x_{12}$}(v6);

 \foreach \v in {1,...,8} \node[hrf vertex] at (v\v) {};
 \draw[->,thick,Green] (-3.15,2.30)--(v1);
 \draw[->,thick,LimeGreen] (-3.15,-2.30)--(v2);
 \draw[->,thick,TealBlue] (v3)--(3.15,2.30);
 \draw[->,thick,OliveGreen] (v4)--(3.15,-2.30);
 \node[hrf leg label,text=Green] at (-3.35,2.55) {$p_1$};
 \node[hrf leg label,text=LimeGreen] at (-3.35,-2.55) {$p_2$};
 \node[hrf leg label,text=TealBlue] at (3.35,2.55) {$p_3$};
 \node[hrf leg label,text=OliveGreen] at (3.35,-2.55) {$p_4$};
\end{tikzpicture}
\hspace{10mm}
\begin{tikzpicture}[xscale=.75,yscale=.85]
 \coordinate (v1) at (-2.35,2.00);
 \coordinate (v2) at (-2.35,-2.00);
 \coordinate (v4) at (2.35,2.00);
 \coordinate (v3) at (2.35,-2.00);
 \coordinate (v5) at (-0.70,1.00);
 \coordinate (v9) at (0.70,1.00);
 \coordinate (v8) at (0.70,-1.00);
 \coordinate (v7) at (-0.70,-1.00);
 \coordinate (v6) at (0,0);

 \node[font=\small\bfseries] at (0,3.35) {HyperCrown};

 \draw[hrf internal]
   (v1) to[bend left=15]
   node[hrf edge label,pos=.47,above] {$x_1$} (v5);
 \draw[hrf internal]
   (v1) to[bend right=15]
   node[hrf edge label,pos=.35,below] {$x_2$} (v6);

 \draw[hrf internal]
   (v4) to[bend right=15]
   node[hrf edge label,pos=.47,above] {$x_3$} (v9);
 \draw[hrf internal]
   (v4) to[bend left=15]
   node[hrf edge label,pos=.35,below] {$x_4$} (v6);

 \draw[hrf internal]
   (v2) to[bend right=15]
   node[hrf edge label,pos=.47,below] {$x_5$} (v7);
 \draw[hrf internal]
   (v2) to[bend left=15]
   node[hrf edge label,pos=.35,above] {$x_6$} (v6);

 \draw[hrf internal]
   (v3) to[bend left=15]
   node[hrf edge label,pos=.47,below] {$x_7$} (v8);
 \draw[hrf internal]
   (v3) to[bend right=15]
   node[hrf edge label,pos=.35,above] {$x_8$} (v6);

 \draw[hrf internal]
   (v9) to[bend right=8]
   node[hrf edge label,pos=.50,above] {$x_9$} (v5);
 \draw[hrf internal]
   (v7) to[bend right=8]
   node[hrf edge label,pos=.50,below] {$x_{10}$} (v8);
 \draw[hrf internal]
   (v8) to[bend right=8]
   node[hrf edge label,pos=.52,right] {$x_{11}$} (v9);
 \draw[hrf contracted]
   (v5) to[bend right=8]
   node[font=\scriptsize,text=Blue,pos=.52,left=2pt] {$x_{12}$} (v7);

 \foreach \v in {1,...,9}
   \node[hrf vertex] at (v\v) {};

 \draw[->,thick,Green]
   (-3.15,2.30)--(v1);
 \draw[->,thick,LimeGreen]
   (-3.15,-2.30)--(v2);
 \draw[->,thick,TealBlue]
   (v3)--(3.15,-2.30);
 \draw[->,thick,OliveGreen]
   (v4)--(3.15,2.30);

 \node[hrf leg label,text=Green]
   at (-3.35,2.55) {$p_1$};
 \node[hrf leg label,text=LimeGreen]
   at (-3.35,-2.55) {$p_2$};
 \node[hrf leg label,text=TealBlue]
   at (3.35,-2.55) {$p_3$};
 \node[hrf leg label,text=OliveGreen]
   at (3.35,2.55) {$p_4$};
\end{tikzpicture}
\caption{Higher-loop Crown descendants: the five-loop SuperCrown
\(\mathcal G_{\rm SC}\) (left) and the four-loop HyperCrown
\(\mathcal G_{\rm HC}\) (right).  Blue dashed edges are set to zero on the
representative contraction strata used below: \(x_9=x_{10}=0\) for the
SuperCrown and \(x_{12}=0\) for the HyperCrown.}
\label{fig:crown-descendant-graphs}
\end{figure}

We denote the SuperCrown and HyperCrown by \(\mathcal G_{\rm SC}\) and
\(\mathcal G_{\rm HC}\), respectively.  The two graphs and their parameter
assignments are shown in Fig.~\ref{fig:crown-descendant-graphs}.

The boundary is essential to this inheritance mechanism, rather than merely
a convenient simplification of the parent graph.  The Crown cancellation
geometry is tied to four jet-like flows meeting at two disconnected hard
components.  Contracting the indicated edges merges their incident vertices and
exposes precisely this hard structure.  The propagators that remain beyond
the Crown minor must then carry momenta compatible with
the inherited flow,
without joining the two hard components into a single connected hard
subdiagram.  This momentum-space picture motivates the boundary strata shown
in Fig.~\ref{fig:crown-descendant-graphs}, which we test below.  Their HR status is
established first in parameter space, where this
compatibility is reflected in the way the extra active parameters enter the
cancellation and scaling equations.

The inheritance has two distinct algebraic signatures.  Additional active
parameters can occur as common positive monomial factors of the superleading
polynomial, thereby altering its weight but not its cancellation locus.  They
can instead enter longer cancellation factors that remain multihomogeneous in
the relevant parameter groups and reduce, upon further contraction, to
monomial multiples of the Crown factors.  The parameter-space analysis below
exhibits both patterns; their momentum-space interpretation is given
subsequently.

\paragraph{SuperCrown: a codimension-two inherited Crown.}

Applying the boundary wrapper of Sec.~\ref{sec:boundary} at
\(x_9=x_{10}=0\), followed by the Part-I decomposition test of
Sec.~\ref{sec:decomposition}, HRF finds
\begin{equation}
 \mathcal F_{\rm SL}[\mathcal G_{\rm SC};\,x_9=x_{10}=0]
 =x_{11}x_{12}\,\mathcal F_{\rm SL}[\mathcal G_{\rm C}](x_1,\ldots,x_8),
 \qquad
 \mathcal F_{\rm Obs}[\mathcal G_{\rm SC};\,x_9=x_{10}=0]=0.
 \label{eq:SC-decomposition}
\end{equation}
Thus the surviving superleading polynomial is nothing but the Crown polynomial, multiplied by the two additional active edge parameters. 
For the cancellation locus, the common positive monomial factor
\(x_{11}x_{12}\) is immaterial.  The two generator polynomials are therefore
precisely \(g_1[\mathcal G_{\rm C}]\) and \(g_2[\mathcal G_{\rm C}]\) defined~in~Eq.~\eqref{eq:crown-F0}, with the same simultaneous positive component as
the Crown.  The Part-II hierarchy determines, in the active order
\((x_1,\ldots,x_8,x_{11},x_{12})\),
\begin{equation}
\begin{aligned}
 \vHR[\mathcal G_{\rm SC};\,x_9=x_{10}=0]
 &=(-1,-1,-1,-1,-1,-1,-1,-1,-1,-1;1),\\
 (\WSL,\WHR)&=(-6,-5).
\end{aligned}
 \label{eq:SC-vector}
\end{equation}
The subsequent dissection supplies the final lower-facet certificate.
The weights include the monomial factor \(x_{11}x_{12}\).  HRF finds a single
HR on this stratum.  This is the simplest form of parametric inheritance: the
boundary exposes the Crown cancellation ideal, while the parameters
\(x_{11}\) and \(x_{12}\) factor out as a positive monomial.  They change the
overall weights but not the positive cancellation locus.

\paragraph{HyperCrown: two boundary-HR types.}

The central square of the HyperCrown has dihedral symmetry \(D_4\).
External, edge and kinematic
labels are permuted together.  Consequently one direct positive witness
establishes its entire orbit, including crossing-related representatives.
Applying the boundary wrapper and core HRF construction to one representative
of each \(D_4\) orbit gives the following wide-angle results through
codimension two:
\begin{table}[H]
\centering
\small
\begin{tabular}{@{}
 >{\raggedright\arraybackslash}p{0.10\textwidth}
 >{\raggedright\arraybackslash}p{0.18\textwidth}
 >{\raggedright\arraybackslash}p{0.31\textwidth}
 >{\raggedright\arraybackslash}p{0.32\textwidth}@{}}
\toprule
Type & Representative boundary
& Certified \(\vec v_{\rm HR}=(\boldsymbol v_{\rm HR};1)\)
& Full orbit and conclusion\\
\midrule
I & \(x_{12}=0\)
& \(v_{11}=-2\), \(v_e=-1\) otherwise;
  \((\WSL,\WHR)=(-6,-5)\)
& four single-edge boundaries \(x_e=0\), \(e=9,\ldots,12\);
  one HR on each\\
\addlinespace[2mm]
II & \(x_9=x_{11}=0\)
& ten active entries \(v_e=-1\);
  \((\WSL,\WHR)=(-5,-4)\)
& \(\{x_9,x_{11}\}\), \(\{x_{10},x_{11}\}\),
  \(\{x_{10},x_{12}\}\), \(\{x_9,x_{12}\}\); one HR on each boundary\\
\addlinespace[2mm]
Excluded & \(x_9=x_{10}=0\)
& no admissible HR scaling
& \(\{x_9,x_{10}\}\), \(\{x_{11},x_{12}\}\); no HR on either boundary\\
\bottomrule
\end{tabular}
\caption{Certified HyperCrown boundary-HR orbits through codimension two.}
\label{tab:hypercrown-boundary-orbits}
\end{table}
Here Types~I and II label boundary-stratum orbits;
Table~\ref{tab:hypercrown-boundary-orbits} specifies one
representative of each orbit.  Graphically, each displayed boundary condition
\(x_e=0\) contracts edge \(e\), whose parameter is therefore absent from the
active variables in the Landau-locus search.

The two types have different layer structures because their certified vectors
are different.  For Type~I, \(v_{11}=-2\), whereas every other active parameter
has weight \(-1\).  Since \(\mathcal F_0[\mathcal G_{\rm HC}]\) is homogeneous
of degree five, its terms proportional to \(x_{11}\) have weight \(-6\), while
the surviving terms independent of \(x_{11}\) have weight \(-5\).  In the
notation of Sec.~\ref{sec:weight-colours}, the Type-I rescaled LP polynomial is
\begin{equation}
\begin{aligned}
 \mathcal P^{(\HR)}[\mathcal G_{\rm HC};{\rm I}](\x;\delta,\s)
 &\equiv
 \left.\mathcal P[\mathcal G_{\rm HC}]
   (\delta^{\boldsymbol v_{\HR}[\mathcal G_{\rm HC};{\rm I}]}\x;
    \delta,\s)\right|_{x_{12}=0}
 \\[-1mm]
 &=\SLcolour{\delta^{-6}x_{11}Q_{\rm I}}
  +\delta^{-5}\HRcolour{\mathcal P_{-5}[\mathcal G_{\rm HC};{\rm I}]}
  +\Othercolour{\mathcal O(\delta^{>-5})}.
\end{aligned}
 \label{eq:HC-type-I-decomposition}
\end{equation}
Here \(\mathcal F_{\rm SL}[\mathcal G_{\rm HC};{\rm I}]=x_{11}Q_{\rm I}\),
where the cancellation polynomial is
\begin{align}
 Q_{\rm I}&=s_{12}G_1+s_{23}G_2,\nonumber\\
 G_1&=(x_2x_3-x_1x_4+x_2x_9)
       (x_6x_7-x_5x_8+x_6x_{10}),\nonumber\\
 G_2&=(x_2x_5-x_1x_6)
       (x_4x_7-x_3x_8-x_8x_9+x_4x_{10}).
 \label{eq:HC-type-I-generators}
\end{align}
Thus multi-term polynomial cancellation factors occur already in this Crown
descendant: three of the four primitive factors in
Eq.~\eqref{eq:HC-type-I-generators} contain three or four monomials.  This is
the mixed algebraic form of Crown inheritance.  The parameter \(x_{11}\)
factors from \(\mathcal F_{\rm SL}=x_{11}Q_{\rm I}\), whereas \(x_9\) and
\(x_{10}\) enter the primitive factors themselves.  Setting
\(x_9=x_{10}=0\) reduces them to Crown-form binomials; with these parameters
active, multihomogeneity in the corresponding parameter groups gives
factor-length profiles \(3\mathbin{\times}3\) and
\(2\mathbin{\times}4\).  Applying the coefficientwise derivative harvest
illustrated for the Crown, followed by disjoint-support pairing, gives
\(G_1\) and \(G_2\).  The accepted ideal \(\langle G_1,G_2\rangle\) has a
component in the positive orthant.  This component, together with the
derivative conditions, is the Landshoff cancellation locus.

The blue resolved-leading polynomial
\(\mathcal P_{-5}[\mathcal G_{\rm HC};{\rm I}]\) includes every contribution
of the complete LP polynomial at weight \(-5\).  In particular, it contains
the leading monomials of the restricted \(\mathcal U[\mathcal G_{\rm HC}]\),
as well as the contribution from \(\mathcal F_0[\mathcal G_{\rm HC}]\), which
is the nonzero polynomial
\[
 \mathcal F_{\rm Obs}[\mathcal G_{\rm HC};{\rm I}]
 \equiv
 \Bigl(\mathcal F_0[\mathcal G_{\rm HC}]
 \big|_{x_{12}=0}\Bigr)_{-5}\ne0.
\]
This is the first example in which HRF retains a nontrivial obstruction after
extracting the superleading cancellation sector.  Its role begins in the
decomposition stage of the core algorithm: the pinch conditions must be
applied to \(x_{11}Q_{\rm I}\), rather than to the complete restricted
\(\mathcal F_0[\mathcal G_{\rm HC}]\), whose obstruction need not be
stationary on the selected positive component.  In the subsequent scaling
stage, placing this obstruction and the leading \(\mathcal U\) monomials at
the same resolved-leading weight fixes the extra unit of scaling on the
opposite square edge, \(v_{11}=-2\).

For Type~II, by contrast, all ten active parameters have weight \(-1\).
Consequently every surviving term of
\(\mathcal F_0[\mathcal G_{\rm HC}]\) has weight \(-5\): the complete
restricted \(\mathcal F_0\) is the superleading sector and
\(\mathcal F_{\rm Obs}[\mathcal G_{\rm HC};{\rm II}]=0\).  The HRF search
returns the following generator presentation for this superleading sector:
\begin{equation}
\begin{aligned}
 \mathcal F_{\rm SL}[\mathcal G_{\rm HC};{\rm II}]
={}&s_{12}(x_2x_3-x_1x_4)
 \Bigl[x_{12}(x_6x_7-x_5x_8-x_8x_{10})
       +x_{10}(x_6x_7-x_5x_8)\Bigr]\\
 &+s_{23}(x_4x_7-x_3x_8)
 \Bigl[x_{12}(x_2x_5-x_1x_6+x_2x_{10})
       +x_{10}(x_2x_5-x_1x_6)\Bigr]\\
={}&\left.\mathcal F_0[\mathcal G_{\rm HC}]
\right|_{x_9=x_{11}=0}.
\end{aligned}
\label{eq:HC-type-II-generators}
\end{equation}
Here the Crown inheritance is realised entirely through the second algebraic
pattern.  Setting \(x_{10}=0\) reduces the two five-term partners to
\(x_{12}\) times their respective Crown binomials, whereas setting
\(x_{12}=0\) gives the same binomials multiplied by \(x_{10}\).  With both
parameters active, their multihomogeneous lift gives the complete five-term
polynomials displayed above.

This factorisation also makes the reciprocal character of the derivative harvest
particularly transparent.  The variables in each outer binomial are disjoint
from those in its five-term partner.  Thus, for example, the \(s_{12}\) and
\(s_{23}\) coefficients of \(\partial\mathcal F_{\rm SL}/\partial x_3\)
are respectively \(x_2\) times the first five-term polynomial and \(-x_8\)
times the second.  After monomial factors are removed, differentiation in the
variables of either factor therefore discovers its partner, as expected from
the stationary conditions.  Differentiation within the five-term factors also
produces further candidates, such as
\(x_2x_5-x_1x_6+x_2x_{10}\).  The derivative harvest is consequently larger
than the factor set entering the final decomposition.  The subsequent
positive-locus, disjoint-support pairing, decomposition and exact-scaling
conditions are essential: here they select the two complete channel products
displayed above and give zero obstruction.

With the uniform scaling of the ten active parameters, the complete rescaled
LP polynomial is therefore
\begin{equation}
\begin{aligned}
 \mathcal P^{(\HR)}[\mathcal G_{\rm HC};{\rm II}](\x;\delta,\s)
 &\equiv
 \left.\mathcal P[\mathcal G_{\rm HC}]
   (\delta^{\boldsymbol v_{\HR}[\mathcal G_{\rm HC};{\rm II}]}\x;
    \delta,\s)\right|_{x_9=x_{11}=0}
 \\[-1mm]
 &=\SLcolour{\delta^{-5}\mathcal F_{\rm SL}[\mathcal G_{\rm HC};{\rm II}]}
  +\delta^{-4}\HRcolour{\mathcal P_{-4}[\mathcal G_{\rm HC};{\rm II}]}
  +\Othercolour{\mathcal O(\delta^{>-4})}.
\end{aligned}
 \label{eq:HC-type-II-decomposition}
\end{equation}
The blue polynomial \(\mathcal P_{-4}[\mathcal G_{\rm HC};{\rm II}]\)
is the resolved-leading layer and contains the leading \(\mathcal U\) and
off-shell contributions.

The opposite-pair boundaries do not support an HR.  On
\(x_9=x_{10}=0\), factorisation of the relevant derivatives and
kinematic-channel polynomials produces an \(s_{12}\)-supported Crown-like
generator but no \(s_{23}\)-supported companion.  The resulting
single-generator form admits no scaling satisfying the
homogeneous-weight, strict-separation and active-variable coverage conditions.
The \(D_4\)-related boundary \(x_{10}=x_{11}=0\) is excluded in the same way.
These are algebraic absence statements for the two boundary strata.  The
HyperCrown interior has not been investigated here.

For comparison across the Crown family, we define the
\emph{generator-factor profile} of a pair generator
\(g_k=f_{i_k}f_{j_k}\) to be \(m\mathbin{\times}n\) when the two primitive
factors \(f_{i_k}\) and \(f_{j_k}\) contain \(m\) and \(n\) monomials,
respectively; we order the pair so that \(m\le n\).
Table~\ref{tab:crown-inheritance-patterns} lists the profiles of the two
generators in each topology, separated by a comma, together with the
corresponding parametric inheritance pattern.
\begin{table}[H]
\centering
\small
\begin{tabular}{@{}llll@{}}
\toprule
Case & Generator-factor profiles & Parametric inheritance pattern
& \(\mathcal F_{\rm Obs}\)\\
\midrule
Crown & \(2\mathbin{\times}2,\ 2\mathbin{\times}2\)
& seed binomial pairing & zero\\
SuperCrown & inherited \(2\mathbin{\times}2,\ 2\mathbin{\times}2\)
& common positive monomial factor & zero\\
HyperCrown I & \(3\mathbin{\times}3,\ 2\mathbin{\times}4\)
& monomial factor and polynomial lift & nonzero\\
HyperCrown II & \(2\mathbin{\times}5,\ 2\mathbin{\times}5\)
& polynomial lift & zero\\
\bottomrule
\end{tabular}
\caption{Generator structure and parametric realisation of Crown inheritance
in the two-generator examples.}
\label{tab:crown-inheritance-patterns}
\end{table}

The family therefore exposes three indispensable features of HRF.  First, the
search must descend to contraction strata: an interior-only analysis misses
regions present in the parent integral because the contraction can expose the
disconnected hard subdiagram required by the inherited Crown mechanism.
Second, the cancellation-factor
harvest must admit general polynomials: this requirement appears already in
the HyperCrown boundary regions, before the five-point examples below.
Third, the single-edge HyperCrown boundary \(x_{12}=0\) displays the dual role
of a nonzero obstruction.  During decomposition it is essential for isolating
the polynomial to which the pinch conditions apply; during scaling
determination its balance with the leading \(\mathcal U\) monomials fixes the
non-uniform component \(v_{11}=-2\).  Both roles of the obstruction recur
beyond the Crown family.

\begin{figure}[H]
\centering
\begin{tikzpicture}[xscale=.58,yscale=.68]
 \coordinate (v1) at (-2.35,2.00);
 \coordinate (v2) at (-2.35,-2.00);
 \coordinate (v3) at (2.35,2.00);
 \coordinate (v4) at (2.35,-2.00);
 \coordinate (v5) at (-0.70,1.00);
 \coordinate (v6) at (-0.70,-1.00);
 \coordinate (v7) at (0.70,1.00);
 \coordinate (v8) at (0.70,-1.00);
 \node[font=\scriptsize\bfseries] at (0,3.55) {SuperCrown};

 \draw[hrf internal,draw=Green] (v1)--
   node[hrf edge label,pos=.47] {$x_1$}(v5);
 \draw[hrf internal,draw=Green] (v1) to[bend right=17]
   node[hrf edge label,pos=.30] {$x_2$}(v6);
 \draw[hrf internal,draw=LimeGreen] (v2) to[bend left=17]
   node[hrf edge label,pos=.30] {$x_3$}(v7);
 \draw[hrf internal,draw=LimeGreen] (v2)--
   node[hrf edge label,pos=.47] {$x_4$}(v6);
 \draw[hrf internal,draw=TealBlue] (v3) to[bend right=17]
   node[hrf edge label,pos=.30] {$x_5$}(v5);
 \draw[hrf internal,draw=TealBlue] (v3)--
   node[hrf edge label,pos=.47] {$x_6$}(v8);
 \draw[hrf internal,draw=OliveGreen] (v4) to[bend right=17]
   node[hrf edge label,pos=.30] {$x_7$}(v7);
 \draw[hrf internal,draw=OliveGreen] (v4)--
   node[hrf edge label,pos=.47] {$x_8$}(v8);
 \draw[hrf contracted] (v5)--
   node[font=\scriptsize,text=Blue,pos=.67,below=-10pt] {$x_9$}(v7);
 \draw[hrf contracted] (v6)--
   node[font=\scriptsize,text=Blue,below=1pt] {$x_{10}$}(v8);
 \draw[very thick,draw=red] (v7)--
   node[hrf edge label,pos=.52,right,text=red] {$x_{11}$}(v8);
 \draw[very thick,draw=red] (v5)--
   node[hrf edge label,pos=.52,left,text=red] {$x_{12}$}(v6);

 \foreach \v in {1,...,4} \node[hrf vertex] at (v\v) {};
 \foreach \v in {5,...,8}
   \node[hrf vertex,fill=Blue,draw=Blue] at (v\v) {};
 \node[font=\scriptsize\bfseries,text=Blue] at (1.18,1.38) {$H_1$};
 \node[font=\scriptsize\bfseries,text=Blue] at (-.18,-.62) {$H_2$};

 \draw[->,thick,Green] (-3.15,2.30)--(v1);
 \draw[->,thick,LimeGreen] (-3.15,-2.30)--(v2);
 \draw[->,thick,TealBlue] (v3)--(3.15,2.30);
 \draw[->,thick,OliveGreen] (v4)--(3.15,-2.30);
 \node[hrf leg label,text=Green] at (-3.35,2.55) {$p_1$};
 \node[hrf leg label,text=LimeGreen] at (-3.35,-2.55) {$p_2$};
 \node[hrf leg label,text=TealBlue] at (3.35,2.55) {$p_3$};
 \node[hrf leg label,text=OliveGreen] at (3.35,-2.55) {$p_4$};
\end{tikzpicture}
\hspace{4mm}
\begin{tikzpicture}[xscale=.58,yscale=.68]
 \coordinate (v1) at (-2.35,2.00);
 \coordinate (v2) at (-2.35,-2.00);
 \coordinate (v4) at (2.35,2.00);
 \coordinate (v3) at (2.35,-2.00);
 \coordinate (v5) at (-0.70,1.00);
 \coordinate (v9) at (0.70,1.00);
 \coordinate (v8) at (0.70,-1.00);
 \coordinate (v7) at (-0.70,-1.00);
 \coordinate (v6) at (0,0);
 \node[font=\scriptsize\bfseries] at (0,3.55) {HyperCrown I: soft};

 \draw[hrf internal,draw=Green] (v1) to[bend left=15]
   node[hrf edge label,pos=.47,above] {$x_1$} (v5);
 \draw[hrf internal,draw=Green] (v1) to[bend right=15]
   node[hrf edge label,pos=.35,below] {$x_2$} (v6);
 \draw[hrf internal,draw=OliveGreen] (v4) to[bend right=15]
   node[hrf edge label,pos=.47,above] {$x_3$} (v9);
 \draw[hrf internal,draw=OliveGreen] (v4) to[bend left=15]
   node[hrf edge label,pos=.35,below] {$x_4$} (v6);
 \draw[hrf internal,draw=LimeGreen] (v2) to[bend right=15]
   node[hrf edge label,pos=.47,below] {$x_5$} (v7);
 \draw[hrf internal,draw=LimeGreen] (v2) to[bend left=15]
   node[hrf edge label,pos=.35,above] {$x_6$} (v6);
 \draw[hrf internal,draw=TealBlue] (v3) to[bend left=15]
   node[hrf edge label,pos=.47,below] {$x_7$} (v8);
 \draw[hrf internal,draw=TealBlue] (v3) to[bend right=15]
   node[hrf edge label,pos=.35,above] {$x_8$} (v6);
 \draw[hrf internal,draw=OliveGreen] (v9) to[bend right=8]
   node[hrf edge label,pos=.50,above] {$x_9$} (v5);
 \draw[hrf internal,draw=TealBlue] (v7) to[bend right=8]
   node[hrf edge label,pos=.50,below] {$x_{10}$} (v8);
 \draw[very thick,draw=red] (v8) to[bend right=8]
   node[hrf edge label,pos=.52,right,text=red] {$x_{11}$} (v9);
 \draw[hrf contracted] (v5) to[bend right=8]
   node[font=\scriptsize,text=Blue,pos=.18,right=.5pt] {$x_{12}$} (v7);

 \foreach \v in {1,2,3,4,8,9} \node[hrf vertex] at (v\v) {};
 \foreach \v in {5,6,7}
   \node[hrf vertex,fill=Blue,draw=Blue] at (v\v) {};
 \node[font=\scriptsize\bfseries,text=Blue] at (-1.32,-.08) {$H_1$};
 \node[font=\scriptsize\bfseries,text=Blue] at (0,-.48) {$H_2$};

 \draw[->,thick,Green] (-3.15,2.30)--(v1);
 \draw[->,thick,LimeGreen] (-3.15,-2.30)--(v2);
 \draw[->,thick,TealBlue] (v3)--(3.15,-2.30);
 \draw[->,thick,OliveGreen] (v4)--(3.15,2.30);
 \node[hrf leg label,text=Green] at (-3.35,2.55) {$p_1$};
 \node[hrf leg label,text=LimeGreen] at (-3.35,-2.55) {$p_2$};
 \node[hrf leg label,text=TealBlue] at (3.35,-2.55) {$p_3$};
 \node[hrf leg label,text=OliveGreen] at (3.35,2.55) {$p_4$};
\end{tikzpicture}
\hspace{4mm}
\begin{tikzpicture}[xscale=.58,yscale=.68]
 \coordinate (v1) at (-2.35,2.00);
 \coordinate (v2) at (-2.35,-2.00);
 \coordinate (v4) at (2.35,2.00);
 \coordinate (v3) at (2.35,-2.00);
 \coordinate (v5) at (-0.70,1.00);
 \coordinate (v9) at (0.70,1.00);
 \coordinate (v8) at (0.70,-1.00);
 \coordinate (v7) at (-0.70,-1.00);
 \coordinate (v6) at (0,0);
 \node[font=\scriptsize\bfseries] at (0,3.55) {HyperCrown II: collinear};

 \draw[hrf internal,draw=Green] (v1) to[bend left=15]
   node[hrf edge label,pos=.47,above] {$x_1$} (v5);
 \draw[hrf internal,draw=Green] (v1) to[bend right=15]
   node[hrf edge label,pos=.35,below] {$x_2$} (v6);
 \draw[hrf internal,draw=OliveGreen] (v4) to[bend right=15]
   node[hrf edge label,pos=.47,above] {$x_3$} (v9);
 \draw[hrf internal,draw=OliveGreen] (v4) to[bend left=15]
   node[hrf edge label,pos=.35,below] {$x_4$} (v6);
 \draw[hrf internal,draw=LimeGreen] (v2) to[bend right=15]
   node[hrf edge label,pos=.47,below] {$x_5$} (v7);
 \draw[hrf internal,draw=LimeGreen] (v2) to[bend left=15]
   node[hrf edge label,pos=.35,above] {$x_6$} (v6);
 \draw[hrf internal,draw=TealBlue] (v3) to[bend left=15]
   node[hrf edge label,pos=.47,below] {$x_7$} (v8);
 \draw[hrf internal,draw=TealBlue] (v3) to[bend right=15]
   node[hrf edge label,pos=.35,above] {$x_8$} (v6);
 \draw[hrf contracted] (v9) to[bend right=8]
   node[font=\scriptsize,text=Blue,pos=0.50,above=-1.95pt] {$x_9$} (v5);
 \draw[hrf internal,draw=LimeGreen] (v7) to[bend right=8]
   node[hrf edge label,pos=.50,below] {$x_{10}$} (v8);
 \draw[hrf contracted] (v8) to[bend right=8]
   node[font=\scriptsize,text=Blue,pos=.80,left=-3.15pt] {$x_{11}$} (v9);
 \draw[hrf internal,draw=LimeGreen] (v5) to[bend right=8]
   node[hrf edge label,pos=.52,left] {$x_{12}$} (v7);

 \foreach \v in {1,2,3,4,7} \node[hrf vertex] at (v\v) {};
 \foreach \v in {5,6,8,9}
   \node[hrf vertex,fill=Blue,draw=Blue] at (v\v) {};
 \node[font=\scriptsize\bfseries,text=Blue] at (1.32,.10) {$H_1$};
 \node[font=\scriptsize\bfseries,text=Blue] at (0,-.48) {$H_2$};

 \draw[->,thick,Green] (-3.15,2.30)--(v1);
 \draw[->,thick,LimeGreen] (-3.15,-2.30)--(v2);
 \draw[->,thick,TealBlue] (v3)--(3.15,-2.30);
 \draw[->,thick,OliveGreen] (v4)--(3.15,2.30);
 \node[hrf leg label,text=Green] at (-3.35,2.55) {$p_1$};
 \node[hrf leg label,text=LimeGreen] at (-3.35,-2.55) {$p_2$};
 \node[hrf leg label,text=TealBlue] at (3.35,-2.55) {$p_3$};
 \node[hrf leg label,text=OliveGreen] at (3.35,2.55) {$p_4$};
\end{tikzpicture}
\caption{Edge-wise momentum-mode realisations of the Crown descendants.
The four green shades denote the external-direction collinear modes, as in
Fig.~\ref{fig:crown-graph}; blue dashed edges are contracted into the blue
hard components \(H_1\) and \(H_2\), and red lines are soft.}
\label{fig:crown-descendant-momentum-modes}
\end{figure}
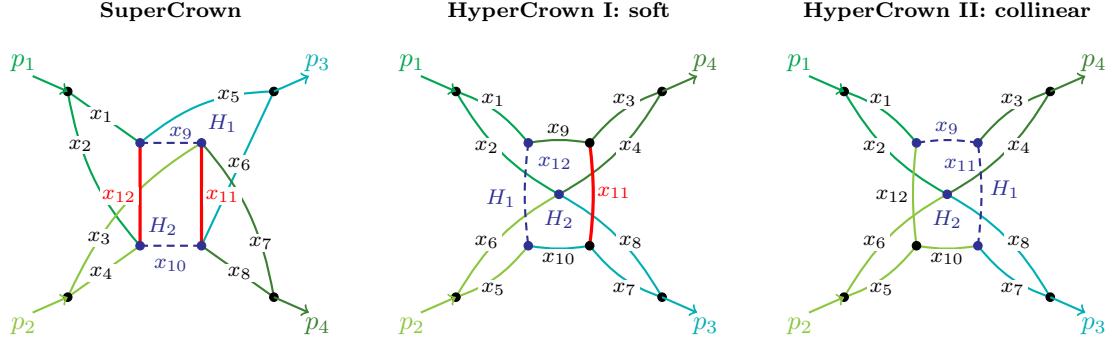

Having completed the parameter-space certification, we now record a compatible
momentum-space realisation.
The construction combines momentum conservation at each vertex with the
inverse relation between Schwinger-parameter and propagator-virtuality
scalings, derived in Ref.~\cite[App.~B]{Gardi:2022khw} and reviewed in
Eq.~\eqref{eq:schwinger-inverse-virtuality}.
Figure~\ref{fig:crown-descendant-momentum-modes} displays the resulting
edge-wise momentum modes.

At the loop-momentum level, choose a basis adapted to the embedded three-loop
Crown flow of Fig.~\ref{fig:crown-graph}, whose leading momentum modes remain
unchanged.  On the SuperCrown stratum, the two additional independent loop
momenta can be chosen soft and routed through \(x_{11}\) and \(x_{12}\).  For
HyperCrown Type~I, the single additional loop momentum can likewise be chosen
soft and routed through \(x_{11}\) after \(x_{12}\) is contracted.  For
Type~II, after contracting \(x_9\) and \(x_{11}\), the additional loop momentum
can be chosen \(p_2\)-collinear and circulates through the two parallel lines
\(x_{10}\) and \(x_{12}\).  In all three cases, the added loops preserve,
rather than join, the two hard components: their soft or collinear momentum
scalings remain compatible with the inherited Crown flow.  This compatibility
is universal and explains why any topology containing a compatible Crown
contraction minor must contain an HR on the corresponding
contraction stratum.

With these loop-momentum assignments established,
Fig.~\ref{fig:crown-descendant-momentum-modes} gives the physical
interpretation of the parametric inheritance patterns summarised in
Table~\ref{tab:crown-inheritance-patterns}.
For the SuperCrown, \(x_{11}\) and \(x_{12}\) are the red soft lines; their
appearance as a common positive monomial factor is therefore the parametric
signature of the factor \(x_{11}x_{12}\) in
Eq.~\eqref{eq:SC-decomposition}.  HyperCrown Type~I combines the two patterns:
the red soft line \(x_{11}\) factors out, while the green \(p_4\)- and
\(p_3\)-collinear lines \(x_9\) and \(x_{10}\) enter the longer
multihomogeneous cancellation factors in
Eq.~\eqref{eq:HC-type-I-generators}.  In Type~II, \(x_{10}\) and \(x_{12}\)
are both \(p_2\)-collinear, accounting for the purely polynomial lift in
Eq.~\eqref{eq:HC-type-II-generators}.

\subsubsection{The four-loop No-Crown converse}
\label{sec:no-crown-wide-angle}

The converse wide-angle audit revisits the four-loop graph sample isolated in
Ref.~\cite{Gardi:2024axt} and sharpens the picture using the HRF absence
test.  Sixteen
mixed-sign topologies without a Crown contraction minor have
no interior or boundary HR.  The search is complete on the stated graph sample,
so these are
no-HR certificates rather than failures to find a candidate.  Together with
the certified positive HyperCrown cases above, these results provide exact
evidence on both sides of the Crown-minor conjecture.  They do not, however,
constitute a topology-by-topology HR certificate for all 1081
Crown-containing four-loop graphs.

\subsubsection{The Crown family in Regge limits: an audited correspondence}
\label{sec:crown-wide-angle-regge}

The relation between the wide-angle Landshoff region and the Regge Glauber
region was first observed in Ref.~\cite{Gardi:2024axt}.  The comparison below
turns this observation into a topology-by-topology audit through four loops.
The audited Crown-containing set includes the graph obtained by opening both
four-point hard vertices of the Crown in the \(t\) channel.  This is
Mandelstam's graph \(G_{tt}\), the simplest of a class of diagrams introduced
as examples whose Regge asymptotic behaviour is governed by Regge cuts
\cite{Mandelstam:1963cuts2}.
From this point on, we work strictly on shell.  Momentum conservation gives
\(s_{13}=-s_{12}-s_{23}\), and the three channel charts used in the audit are
\begin{align}
 T_{23}:&\quad s_{23}=-\delta s_{12}, &&s_{12}>0,\nonumber\\
 T_{12}:&\quad s_{12}=-\delta s_{23}, &&s_{23}<0,\nonumber\\
 T_{13}:&\quad s_{23}=-s_{12}+\delta s_{12}, &&s_{12}>0,
 \label{eq:crown-regge-charts}
\end{align}
with \(\delta\to0^+\).  In the ratio coordinates of
Eq.~\eqref{eq:crown-path-ratios}, the leading Crown polynomial in each chart
is, up to a nonzero kinematic and positive monomial factor,
\begin{align}
 \Psi_C^{T_{23}}&\propto(r_C-r_B)(r_D-r_A),\nonumber\\
 \Psi_C^{T_{12}}&\propto(r_B-r_A)(r_D-r_C),\nonumber\\
 \Psi_C^{T_{13}}&\propto(r_C-r_A)(r_D-r_B).
 \label{eq:crown-regge-ratio-factors}
\end{align}
The corresponding positive cancellation loci are therefore
\begin{align}
 T_{23}:&\quad r_B=r_C,\quad r_A=r_D,\nonumber\\
 T_{12}:&\quad r_A=r_B,\quad r_C=r_D,\nonumber\\
 T_{13}:&\quad r_A=r_C,\quad r_B=r_D.
 \label{eq:crown-regge-loci}
\end{align}
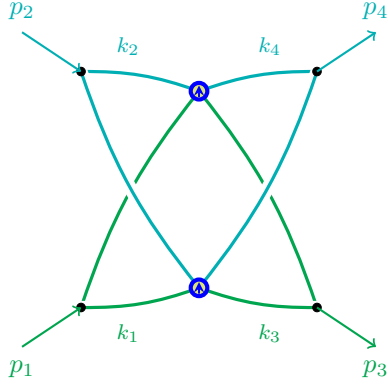
\begin{figure}[htbp]
\centering
\begin{tikzpicture}[scale=.52]
 \coordinate (v1) at (2,2);
 \coordinate (v2) at (2,8);
 \coordinate (v3) at (8,2);
 \coordinate (v4) at (8,8);
 \coordinate (v5) at (5,2.5);
 \coordinate (v6) at (5,7.5);

 \draw[very thick,Green] (v6) to[bend right=10] (v1);
 \draw[very thick,TealBlue] (v6) to[bend right=10] (v2);
 \draw[very thick,Green] (v6) to[bend left=10] (v3);
 \draw[very thick,TealBlue] (v6) to[bend left=10] (v4);
 \foreach \a/\b/\bend/\colour in {
   v5/v1/left/Green,v5/v2/left/TealBlue,
   v5/v3/right/Green,v5/v4/right/TealBlue}{
   \draw[white,line width=5pt] (\a) to[bend \bend=10] (\b);
   \draw[very thick,\colour] (\a) to[bend \bend=10] (\b);
 }

 \foreach \v in {1,...,4} \node[hrf vertex] at (v\v) {};
 \draw[->,thick,Green] (.5,1)--(v1);
 \draw[->,thick,TealBlue] (.5,9)--(v2);
 \draw[->,thick,Green] (v3)--(9.5,1);
 \draw[->,thick,TealBlue] (v4)--(9.5,9);
 \node[hrf leg label,text=Green] at (.5,.45) {$p_1$};
 \node[hrf leg label,text=TealBlue] at (.5,9.55) {$p_2$};
 \node[hrf leg label,text=Green] at (9.5,.45) {$p_3$};
 \node[hrf leg label,text=TealBlue] at (9.5,9.55) {$p_4$};

 \node[hrf momentum label,text=Green] at (3.2,1.35) {$k_1$};
 \node[hrf momentum label,text=TealBlue] at (3.2,8.65) {$k_2$};
 \node[hrf momentum label,text=Green] at (6.8,1.35) {$k_3$};
 \node[hrf momentum label,text=TealBlue] at (6.8,8.65) {$k_4$};

 \GlauberVertex[blue]{(v5)}{90}
 \GlauberVertex[blue]{(v6)}{90}
\end{tikzpicture}
\caption{Momentum modes of the Crown HR in the \(T_{23}\) Regge channel.
All external momenta are exactly lightlike.  Of the eight propagator momenta,
those shown in green are approximately collinear to both \(p_1\) and \(p_3\),
while those in teal are approximately collinear to both \(p_2\) and \(p_4\);
the Glauber momentum is instead the loop combination
\(\ell_G=k_1-k_3\sim Q(\delta,\delta,\delta^{1/2})\), with
\(\delta=-s_{23}/s_{12}\) in this chart.  This is the independent Glauber
loop of the HR.  The markers at the hard vertices identify its local
\(t\)-channel segments.  In the present embedding both local arrows point
upward; placing the vertices side by side and opening them into propagators
makes the closed-loop orientation explicit, with the Glauber momentum flowing
upward on one side and downward on the other.  This is the same momentum
configuration as Fig.~11(b) of Ref.~\cite{Gardi:2024axt}, where the direction
of the Glauber exchange was identified; the different vertex marker used here
is purely notational.  The drawing convention used throughout the paper is as
follows.  A GreenYellow disc marks a vertex through which a Glauber loop
flows; its arrow follows the local \(t\)-channel flow, while its rim
colour records whether the momentum in the complementary local channel is
hard, soft or collinear to the indicated jet direction.  Here both arrows
follow the upward local \(t\)-channel flow and the blue rims record hard
complementary channels.  Reversing every arrow gives the equivalent opposite
loop-momentum convention.  The other Regge channels follow by crossing.}
\label{fig:crown-regge-glauber}
\end{figure}
This exposes an important change in codimension.  At generic wide angle the
Crown locus identifies all four ratio points and hence has three independent
relative-distance equations.  In each Regge chart the leading polynomial
reduces to one product of two difference factors, leaving only two independent
equations.  Each Regge locus is therefore larger and contains the wide-angle diagonal
\(r_A=r_B=r_C=r_D\).

Nevertheless, the reduction from three equations to two does not produce
new seed topologies in the completed low-loop survey.  The positive side is
inherited by every audited graph that exposes a Crown
contraction minor: the Crown itself has an HR in all three channels, and its
higher-loop descendants carry the corresponding region on the contraction
stratum.  On the negative
side, the same sixteen four-loop No-Crown topologies were checked in
\(T_{23}\), \(T_{12}\) and \(T_{13}\).  All 48 interiors and all 343356
nonempty contraction strata with between one and \(E-2\) edges contracted were
resolved, with no HR.  Deeper contractions leave at most one active parameter,
so no nontrivial cancellation locus remains.

Thus, after the channel relabellings implied by crossing, the set of
topologies with a Regge HR agrees exactly with the set having a wide-angle
Landshoff HR at three and four loops.  Equivalently, in this complete audited
sample both are precisely the topologies containing the corresponding
Crown contraction minor.  We call this the
wide-angle-to-Regge conjecture when
extrapolated beyond four loops.  Its validity is not a trivial consequence
of the Landau equations: algebraically the Regge pinch requires fewer
conditions and could have admitted additional topologies, but none occurs in
the completed survey.

So far we have focussed on $2\to2$ scattering, where the three-loop Crown
emerged as the unique seed topology for HRs in both the wide-angle and Regge
limits. We now turn to massless $2\to3$ scattering, where a similar phenomenon
occurs already at two loops: again, a special seed topology supports HRs in a
variety of kinematic limits.

\subsection{Five-point integrals: the Fish seed}
\label{sec:example-five-point}

The examples below are organised around a single unlabelled two-loop
six-propagator topology commonly called the non-planar double box, which we
refer to as the Fish seed, but involve different permutations of the external
legs and kinematic expansions. The seed was first identified
in the spacelike-collinear limit studied in the recent
Letter~\cite{Chen:2026dnj}, where it was shown to be the unique two-loop
topology supporting an HR. The second setting is an exactly on-shell
near-planar expansion of five-particle wide-angle scattering. The cancellation
ideals and total vectors are different in these two settings, but in both cases
the same sparse topology supports an HR.  Imposing standard fixed-transverse
MRK and near-planarity simultaneously then yields a continuous one-parameter
family of HR scalings that retains the three-generator wide-angle locus.  We
subsequently consider a central-soft rapidity-ordered limit, which reaches the
normalised planar configuration by a more degenerate route and requires order
alignment before the core HRF construction.

\subsubsection{The spacelike-collinear limit}
\label{sec:five-point-spacelike-collinear}

The spacelike-collinear application of HRF was presented in the recent
Letter~\cite{Chen:2026dnj}.  Here we apply HRF in parameter space to identify
the seed cancellation locus and certify its scaling.  These data then
provide the baseline for tracking inheritance under graph enlargement.  We
take \(p_1\) and \(p_2\) incoming and
\(p_3,p_4,p_5\) outgoing, with \(p_3\parallel p_2\) in the collinear limit.
For this analysis we use the kinematic parametrisation introduced in
Ref.~\cite{Henn:2024qjq} and also employed in the recent
Letter~\cite{Chen:2026dnj}.  It originates from
a momentum-twistor construction.  In our notation the spacelike-collinear
chart is
\begin{equation}
 \begin{gathered}
 s_{12}=sz,\qquad s_{23}=-4\delta^2,\qquad
 s_{34}=-s\chi(z-1),\\
 s_{45}=s,\qquad s_{15}=s\chi+c\delta,\qquad \delta\to0^+,
 \end{gathered}
 \label{eq:five-kinematics}
\end{equation}
where \(\delta\) tends to zero through positive real values, and
\begin{equation}
 s>0,\qquad z>1,\qquad -1<\chi<0,
 \label{eq:five-domain}
\end{equation}
with fixed nonzero \(c\).  In particular,
\(s_{23}=-4\delta^2<0\) away from the strict limit, so
Eq.~\eqref{eq:five-kinematics} parametrises the spacelike-collinear approach.
This form makes the dependence on the transverse
collinear variable \(\delta\) especially transparent: the deformation away
from the strict collinear surface enters linearly through \(c\delta\), while
the vanishing collinear invariant starts at \(s_{23}=-4\delta^2\).  Thus the
Mandelstam invariants are polynomial, and at most quadratic, in the expansion
variable.  The effect of crossing to the timelike channel on the cancellation
factors is described below.

\begin{figure}[H]
\centering
\begin{tikzpicture}[scale=.48]
 \coordinate (v1) at (2,3);
 \coordinate (v2) at (2,7);
 \coordinate (v3) at (5,8);
 \coordinate (v4) at (8,5);
 \coordinate (v5) at (5,2);

 \draw[very thick,Green] (v3) to[bend right=20] (v1);
 \draw[very thick,LimeGreen] (v3) to[bend right=20] (v2);
 \draw[white,double=white,double distance=3pt,thick]
   (v5) to[bend left=20] (v2);
 \draw[very thick,LimeGreen] (v5) to[bend left=20] (v2);
 \draw[very thick,Green] (v5) to[bend left=20] (v1);
 \draw[very thick,PineGreen] (v3) to[bend left=20] (v4);
 \draw[very thick,PineGreen] (v5) to[bend right=20] (v4);

 \foreach \v in {1,...,5} \node[hrf vertex] at (v\v) {};
 \draw[->,thick,Green] (0,2)--(v1);
 \draw[->,thick,OliveGreen] (0,8)--(v2);
 \draw[->,thick,OliveGreen] (v3)--(7,9);
 \draw[->,thick,LimeGreen] (v4)--(10,5);
 \draw[->,thick,TealBlue] (v5)--(7,1);

 \node[hrf leg label,text=Green] at (.5,1.5) {$p_1$};
 \node[hrf leg label,text=OliveGreen] at (.5,8.5) {$p_2$};
 \node[hrf leg label,text=OliveGreen] at (6.2,9.1) {$p_3$};
 \node[hrf leg label,text=LimeGreen] at (9.5,5.6) {$p_4$};
 \node[hrf leg label,text=TealBlue] at (6.2,.8) {$p_5$};

 \node[hrf edge label] at (3.50,5.75) {$x_1$};
 \node[hrf edge label] at (3.05,1.95) {$x_2$};
 \node[hrf edge label] at (3.05,8.05) {$x_3$};
 \node[hrf edge label] at (3.55,4.15) {$x_4$};
 \node[hrf edge label] at (6.65,7.45) {$x_5$};
 \node[hrf edge label] at (6.65,2.55) {$x_6$};

\end{tikzpicture}
\caption{The two-loop non-planar double box, referred to here as the Fish
seed topology, with a representative
spacelike-collinear permutation of the external legs and the LP parameters used in
Eqs.~\eqref{eq:five-facet-polynomial}--\eqref{eq:five-HR-polynomial}.
The same unlabelled six-edge topology underlies the other Fish-seed limits
considered below; each kinematic limit specifies its permutation of external legs
separately.}
\label{fig:five-point-seed}
\end{figure}
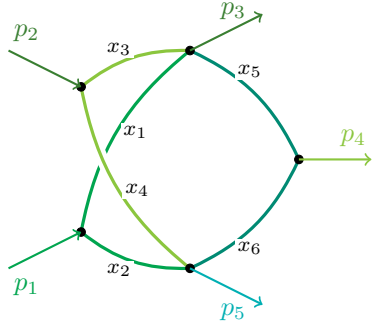

The two-loop seed, one representative permutation of the external legs and its
LP-parameter assignment are shown in Fig.~\ref{fig:five-point-seed}.  Its
\(p_4\leftrightarrow p_5\) image is an equally relevant permutation in the
spacelike-collinear limit.  Both permutations, together with the additional
graphs whose contractions expose the seed with the required external
labelling, were included in the amplitude-level integral analysis of
Ref.~\cite{Chen:2026dnj}.  To maintain
direct correspondence with that analysis, throughout this subsection we
retain the one-based LP-parameter labelling \(x_1,\ldots,x_6\) used in the
Letter.

\paragraph{The Fish HRF construction.}

We first apply the two parts of the HRF construction: Part~I identifies the
hidden-region decomposition, and Part~II determines its scaling.  At fixed LP
parameters, the Part~I decomposition of the strict leading polynomial is
\begin{equation}
 \begin{aligned}
 \mathcal F_0={}&
 \underbrace{s\bigl(x_2x_5+\chi x_1x_6\bigr)
       \bigl[(z-1)x_3-x_4\bigr]}_{s f_1f_2}
 +\underbrace{sz(\chi+1)x_2x_3x_6}_{\FObs},\\
 f_1={}&x_2x_5+\chi x_1x_6,
 \qquad f_2=(z-1)x_3-x_4.
 \end{aligned}
 \label{eq:five-F0-obstruction}
\end{equation}
In the HR decomposition, the second term is identified as \(\FObs\).  It is
itself part of \(\mathcal F_0\), and plays its defining role because the HR
scaling suppresses it relative to the monomials in ${\cal F}_{\text{SL}}=s f_1f_2$, which are individually superleading.

The derivative harvest supplies \(f_1\) and \(f_2\), which HRF pairs into the
single generator
\begin{equation}
 g_1=f_1f_2,
 \qquad G=\{g_1\},
 \qquad \mathcal I_{\rm gen}(G)=\langle g_1\rangle,
 \qquad \FSL=s\,g_1\in\mathcal I_{\rm gen}(G).
 \label{eq:five-generator}
\end{equation}
Part~II then determines the hidden-region vector
\begin{equation}
 \vHR=(-2,-1,-2,-2,-2,-1;1),
 \qquad (\WSL,\WHR)=(-5,-4).
 \label{eq:five-HR-vector}
\end{equation}
The dissection below provides its final certificate.
\paragraph{Comparison between the HR and the ordinary facet region.}

As already discussed in Ref.~\cite{Chen:2026dnj}, it is insightful to compare
the layer structure of the HR with that of the ordinary facet region.
For comparison, the ordinary facet region has scaling vector
\begin{equation}
 \vec v_C=(-2,0,-2,-2,-2,0;1).
 \label{eq:five-facet-vector}
\end{equation}
Its momentum-space interpretation as a collinear region is displayed in
Table~\ref{tab:five-sc-edge-flow} below.
To compare their layer structures directly, we explicitly rescale each graph
polynomial as in Eq.~\eqref{eq:rescaled-LP-polynomial}, writing
\(\mathcal F^{(R)}=\mathcal F(\delta^{\boldsymbol v_R}\x;\delta,\s)\)
and analogously for \(\mathcal U^{(R)}\).  For the ordinary facet this gives
\begin{equation}
 \begin{aligned}
 \mathcal F^{(C)}={}&
 \delta^{-4}\HRcolour{\Bigl[
  s\bigl(x_2x_5+\chi x_1x_6\bigr)
   \bigl[(z-1)x_3-x_4\bigr]+4x_1x_4x_5\Bigr]}\\
 &+\Othercolour{\Bigl[-c\delta^{-3}x_1x_4x_6
  +\delta^{-2}\bigl(sz(\chi+1)x_2x_3x_6-4x_2x_3x_5\bigr)
  +c\delta^{-1}x_2x_3x_6\Bigr]}.
 \end{aligned}
 \label{eq:five-facet-polynomial}
\end{equation}
There is no superleading sector.  By contrast, the HR scaling gives
\begin{equation}
 \begin{aligned}
 \mathcal F^{(\HR)}={}&
 \delta^{-5}\SLcolour{\underbrace{
  s\bigl(x_2x_5+\chi x_1x_6\bigr)
   \bigl[(z-1)x_3-x_4\bigr]}_{\FSL}}\\
 &+\delta^{-4}\HRcolour{\Bigl[
   \underbrace{sz(\chi+1)x_2x_3x_6}_{\FObs}
   -cx_1x_4x_6+4x_1x_4x_5\Bigr]}
 +\delta^{-3}\Othercolour{\Bigl[
   cx_2x_3x_6-4x_2x_3x_5\Bigr]}.
 \end{aligned}
 \label{eq:five-HR-polynomial}
\end{equation}
The complete LP comparison also requires the first Symanzik polynomial, which
supplies the same leading layer in the two regions,
\begin{equation}
 \begin{aligned}
 \mathcal U^{(C)}&=\delta^{-4}\HRcolour{\mathcal U_{-4}}
                  +\Othercolour{\mathcal O(\delta^{-2})},
 &\qquad
 \mathcal U^{(\HR)}&=\delta^{-4}\HRcolour{\mathcal U_{-4}}
                  +\Othercolour{\mathcal O(\delta^{-3})},\\
 \mathcal U_{-4}&=x_1x_3+x_1x_4+x_1x_5+x_3x_5+x_4x_5.
 \end{aligned}
 \label{eq:five-U-leading}
\end{equation}
In the facet region, the blue terms of
Eqs.~\eqref{eq:five-facet-polynomial} and \eqref{eq:five-U-leading}
together form the ordinary lower-facet polynomial at weight \(-4\).  The
monomial \(sz(\chi+1)x_2x_3x_6\) appears only at black weight \(-2\), while
the first \(c\)-dependent term is black at weight \(-3\); hence the leading
power in the asymptotic expansion is independent of \(c\).  This is in line
with the fact that this region appears in both the timelike and spacelike
collinear limits, while the former, which admits strict collinear
factorisation, cannot feature \(c\) dependence~\cite{Chen:2026dnj}.  In the HR,
the monomials constituting the factorised part of \(\mathcal F_0\) are
individually superleading at weight \(-5\).  The remaining monomial
\(sz(\chi+1)x_2x_3x_6\), identified as \(\FObs\) in the HR decomposition, is
suppressed by one power relative to them: it has weight \(-4\) and joins the
other terms in the blue physical layer.  The obstruction is also essential to
the HRF hierarchy conditions selecting the non-uniform vector in
Eq.~\eqref{eq:five-HR-vector}.  The \(c\)-dependent
term \(-cx_1x_4x_6\) is promoted to that same layer.  The HR therefore depends
on \(c\) at leading power and exhibits factorisation-violating
dependence.

The channel dependence of this positive cancellation locus is already visible
in these two factors.  To obtain the corresponding physical timelike-collinear
limit while retaining a \(2\to3\) process, cross particle~2 to the final state
and particle~4 to the initial state, so that the process becomes
\(1+4\to2+3+5\).  On the strict collinear surface the physical ranges in the
same variables are
\begin{equation}
 s<0,\qquad 0<z<1,\qquad \chi>0.
 \label{eq:five-timelike-domain}
\end{equation}
For positive LP parameters these ranges give \(f_1>0\) and \(f_2<0\), so
neither factor can vanish.  Thus the positive common zero in the spacelike
domain is lost under the simultaneous crossing, in agreement with the absence
of this HR in the timelike-collinear limit~\cite{Chen:2026dnj}.

\paragraph{Dissection charts and the invariant region.}
\label{sec:five-point-sc-charts}
This example also displays explicitly why a local facet is not an invariant
label for the HR.  Under the original scaling
Eq.~\eqref{eq:five-HR-vector},
\begin{equation}
 f_1=\delta^{-3}\widehat f_1,
 \qquad f_2=\delta^{-2}\widehat f_2,
 \qquad \FSL=\delta^{-5}s\,\widehat f_1\widehat f_2.
 \label{eq:five-sc-normalised-factors}
\end{equation}
The gap \(\WHR-\WSL=1\) fixes only the total transverse suppression.
A chart centred on the simultaneous locus may distribute it as
\begin{equation}
 \widehat f_1\sim\delta^\alpha,
 \qquad \widehat f_2\sim\delta^{1-\alpha},
 \qquad 0<\alpha<1.
 \label{eq:five-sc-dissection-family}
\end{equation}
Using \(y_i=f_i\) and eliminating \(x_2,x_4\), respectively, the local
weights are
\begin{equation}
 \bigl(v_{x_1},v_{x_3},v_{x_5},v_{x_6},v_{y_1},v_{y_2};1\bigr)
 =\bigl(-2,-2,-2,-1,-3+\alpha,-1-\alpha;1\bigr),
 \label{eq:five-sc-centred-chart}
\end{equation}
with Jacobian \(1/x_5\).  Every \(0<\alpha<1\) approaches the full locus
\(f_1=f_2=0\), and every such scaling pulls back to the same original vector
\eqref{eq:five-HR-vector}.  Throughout this open interval, the leading support
is the same and has affine dimension five, hence codimension two in the
seven-dimensional transformed Newton polytope.  It is the common lower face
of the two endpoint facets, rather than a facet by itself.

The two endpoint facets provide alternative dissections of the same HR.  At the
\(\alpha\to0\) endpoint one may set
\begin{equation}
 \begin{aligned}
  y_2&=f_2, & x_4&=(z-1)x_3-y_2,\\
  \boldsymbol\xi_{(2)}&=(x_1,x_2,x_3,x_5,x_6,y_2),&
  \vec v_{(2)}&=(-2,-1,-2,-2,-1,-1;1),
 \end{aligned}
 \label{eq:five-sc-f2-chart}
\end{equation}
whereas at \(\alpha\to1\) one may use
\begin{equation}
 \begin{aligned}
  y_1&=f_1, & x_2&=\frac{y_1-\chi x_1x_6}{x_5},\\
  \boldsymbol\xi_{(1)}&=(x_1,x_3,x_4,x_5,x_6,y_1),&
  \vec v_{(1)}&=(-2,-2,-2,-2,-1,-2;1).
 \end{aligned}
 \label{eq:five-sc-f1-chart}
\end{equation}
In both cases the transformed leading support has affine dimension six and
defines an exact lower facet.  The first chart has unit Jacobian.  In the
second, the coordinate measure has weight \(-11\) and the Jacobian \(1/x_5\)
has weight \(+2\); both therefore give the measure weight \(-9\).  Pulling
either facet normal back gives Eq.~\eqref{eq:five-HR-vector}.

Thus either Eq.~\eqref{eq:five-sc-f2-chart} or
Eq.~\eqref{eq:five-sc-f1-chart} may be used as an integration chart for the
same invariant cancellation component \(f_1=f_2=0\); they are not distinct
physical pinches or contributions to be combined.  The calculation of
Ref.~\cite{Chen:2026dnj} used the \(y_2=f_2\) dissection in
Eq.~\eqref{eq:five-sc-f2-chart} alone.

It is useful for comparison with later examples to separate the overall
approach to the common zero from the relative approach to its two factors.
Here the natural kinematic deformation \(\delta\) is dimensionful, so let
\(Q=\sqrt{s}\) and \(\lambda=\delta/Q\).  The LP parameters, and hence the
cancellation factors \(f_1\) and \(f_2\), are already dimensionless in the
convention of Eq.~\eqref{eq:LP-representation}.  Their natural regional
weights are removed by defining the dimensionless combinations
\[
 t_1=\lambda^3 f_1,
 \qquad
 t_2=\lambda^2 f_2.
\]
Within a fixed sign sector,
a centred radial--ratio parameterisation is
\begin{equation}
 |t_1|=\rho e^\eta,\qquad |t_2|=\rho e^{-\eta},\qquad
 \rho\sim\lambda^{1/2}.
 \label{eq:five-sc-radial-ratio-chart}
\end{equation}
Here \(\rho\) measures the approach to the product locus, while \(\eta\)
records how its suppression is distributed between the two factors.
The scaling
\(\eta=(\alpha-\tfrac12)\log\lambda+O(1)\) represents the freedom in
Eq.~\eqref{eq:five-sc-dissection-family}; fixed \(\eta\) gives the symmetric
centred chart.  This form will be used only for comparison below; the
calculation of Ref.~\cite{Chen:2026dnj} used the endpoint chart
Eq.~\eqref{eq:five-sc-f2-chart}.

\paragraph{Parameter-space power counting.}

For unit propagator powers, the scalar LP representation contains
\(\mathcal P^{-D/2}\).  In the original LP variables, the six edge
components of \(\vec v_{\rm HR}\) in Eq.~\eqref{eq:five-HR-vector} sum to
\(\sum_{e=1}^{6}v_{{\rm HR},e}=-10\).  Equations~\eqref{eq:five-sc-centred-chart}--
\eqref{eq:five-sc-f1-chart} show in three equivalent ways that resolving the
total cancellation depth contributes one additional power.  The parameter
measure is therefore
\begin{equation}
 \delta^{-10}\,\delta^{+1}=\delta^{-9}.
 \label{eq:five-sc-parameter-measure}
\end{equation}
Since the resolved LP polynomial has weight \(-4\),
\(\mathcal P^{-D/2}\sim\delta^{2D}\), and hence
\begin{equation}
 I_{\rm SC}^{\rm scalar}\sim
 \delta^{-9+2D}=\delta^{-1-4\epsilon},
 \qquad D=4-2\epsilon.
 \label{eq:five-sc-parameter-power}
\end{equation}
The power enhancement is a property of the unit-numerator scalar integral.
For the gauge-theory numerator relevant to the amplitude, one additional
power of \(\delta\) gives the leading-power behaviour
\(\delta^{-4\epsilon}\).

\paragraph{Momentum-space interpretation.}
To connect the parameter-space result with its momentum-space realisation,
and to contrast it with the ordinary facet region, we first fix conventions
common to both regions.  Whenever a component triple is displayed, the
subscript on \((+,-,\perp)_{(n_+,n_-)}\) names the null reference directions
that define its plus and minus components, in that order.  Here we use the
global frame \((+,-,\perp)_{(p_2,p_1)}\) defined by the two incoming momenta,
\(p_1=(0,p_1^-,\boldsymbol 0_\perp)\) and
\(p_2=(p_2^+,0,\boldsymbol 0_\perp)\),
so that the plus direction is aligned with \(p_2\), and hence with the
collinear momentum \(p_3\), while the minus direction is aligned with
\(p_1\).
We orient \(q_1,q_2\) away from the \(p_1\) vertex, \(q_3,q_4\) away
from the \(p_2\) vertex, \(q_5\) from the \(p_3\) vertex towards the
\(p_4\) vertex, and \(q_6\) from the \(p_5\) vertex towards the \(p_4\)
vertex.  These orientation conventions are common to the ordinary facet
region and the HR; their virtuality and component scalings distinguish the
two cases.  The momentum \(q_e\) flows through the edge carrying \(x_e\).
With the hard scale \(Q=\sqrt{s}\) held fixed, we write
\[
 \frac{q_e^2}{Q^2}\sim
 \left(\frac{\delta}{Q}\right)^{-v_e},
 \qquad
 \frac{1}{Q}\bigl(q_e^+,q_e^-,|q_{e\perp}|\bigr)_{(p_2,p_1)}
 \sim
 \left(\left(\frac{\delta}{Q}\right)^{a_e},
       \left(\frac{\delta}{Q}\right)^{b_e},
       \left(\frac{\delta}{Q}\right)^{c_e}\right).
\]
The first relation follows directly from the parameter-space vector through
the inverse Schwinger scaling relation in
Eq.~\eqref{eq:schwinger-inverse-virtuality}.  Determining the separate
component exponents \((a_e,b_e,c_e)\) requires additional momentum-space
input, notably momentum conservation and the pinch conditions.  Below we
abbreviate the component scaling as
\(q_e\sim(\delta^{a_e},\delta^{b_e},\delta^{c_e})_{(p_2,p_1)}\); this shorthand records
only powers of \(\delta\), with the fixed factors of \(Q\) understood.

The ordinary facet region admits a homogeneous collinear loop basis and no
transverse-dominated loop rerouting.  Its component and virtuality scalings
are recorded in the first two columns of Table~\ref{tab:five-sc-edge-flow}.

For the HR, \(q_1\) and \(q_5\) are separately soft at the pinch.  With the orientations
fixed above, conservation at the vertex carrying \(p_3\) reads
\(q_1+q_3=p_3+q_5\).  A compatible mode-adapted routing is obtained by
choosing the independent loop momenta
\begin{equation}
 \ell_s=q_5\sim(\delta,\delta,\delta),
 \qquad
 \ell_G=q_1-q_5
 =p_3-q_3\sim(\delta,\delta^2,\delta),
 \label{eq:five-sc-glauber-rerouting}
\end{equation}
at the pinch.  The extra suppression of \(\ell_G^-\) is a consequence of the pinch
conditions: they correlate the leading \(O(\delta)\) minus components of
\(q_1\) and \(q_5\) so that these cancel in the difference.  A generic
difference of two soft momenta would remain soft; here
\(\ell_G^+\ell_G^-\sim\delta^3\) is subleading to
\(\ell_{G\perp}^2\sim\delta^2\), making the Glauber loop transverse dominated.
A complete oriented routing is then
\[
 \begin{aligned}
 q_1&=\ell_G+\ell_s, &
 q_2&=p_1-\ell_G-\ell_s, &
 q_3&=p_3-\ell_G,\\
 q_4&=p_2-p_3+\ell_G, &
 q_5&=\ell_s, &
 q_6&=p_4-\ell_s.
 \end{aligned}
\]
This routing gives the compatible HR component scalings in the last two columns
of Table~\ref{tab:five-sc-edge-flow}; the corresponding virtualities follow
from the inverse Schwinger relation.
Here and below, a momentum valuation is the leading exponent of \(\delta\) in
the indicated momentum component or virtuality.

\begin{table}[H]
\centering
\small
\begin{tabular}{@{}c cc cc@{}}
\toprule
& \multicolumn{2}{c}{facet \(\vec v_C\)}
& \multicolumn{2}{c}{hidden \(\vHR\)}\\
\cmidrule(lr){2-3}\cmidrule(l){4-5}
edge & \((a_e,b_e,c_e)_{(p_2,p_1)}\) & \(-v_e\)
     & \((a_e,b_e,c_e)_{(p_2,p_1)}\) & \(-v_e\)\\
\midrule
\(q_1\) & \((0,2,1)\) & 2 & \((1,1,1)\) & 2\\
\(q_2\) & \((0,0,1)\) & 0 & \((1,0,1)\) & 1\\
\(q_3\) & \((0,2,1)\) & 2 & \((0,2,1)\) & 2\\
\(q_4\) & \((0,2,1)\) & 2 & \((0,2,1)\) & 2\\
\(q_5\) & \((0,2,1)\) & 2 & \((1,1,1)\) & 2\\
\(q_6\) & \((0,0,0)\) & 0 & \((0,0,0)\) & 1\\
\bottomrule
\end{tabular}
\caption{Edge-momentum valuations of a compatible routing for the five-point
spacelike-collinear facet and HR.  The component exponents refer to the
\((+,-,\perp)_{(p_2,p_1)}\) lightcone frame defined in the text; local loop-integration
widths are determined below.}
\label{tab:five-sc-edge-flow}
\end{table}

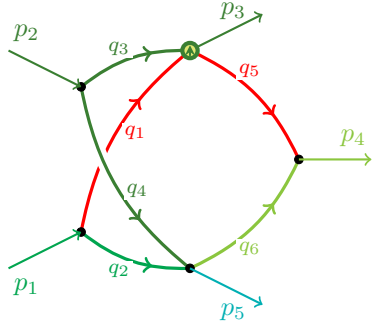
\begin{figure}[H]
\centering
\begin{tikzpicture}[scale=.48]
 \tikzset{flow arrow/.style={postaction={decorate},decoration={markings,
   mark=at position .68 with {\arrow{>}}}}}
 \coordinate (v1) at (2,3);
 \coordinate (v2) at (2,7);
 \coordinate (v3) at (5,8);
 \coordinate (v4) at (8,5);
 \coordinate (v5) at (5,2);

 \draw[very thick,red,flow arrow] (v1) to[bend left=20] (v3);
 \draw[very thick,OliveGreen,flow arrow] (v2) to[bend left=20] (v3);
 \draw[white,line width=5pt] (v5) to[bend left=20] (v2);
 \draw[very thick,OliveGreen,flow arrow] (v2) to[bend right=20] (v5);
 \draw[very thick,Green,flow arrow] (v1) to[bend right=20] (v5);
 \draw[very thick,red,flow arrow] (v3) to[bend left=20] (v4);
 \draw[very thick,LimeGreen,flow arrow] (v5) to[bend right=20] (v4);

 \foreach \v in {1,...,5} \node[hrf vertex] at (v\v) {};
 \draw[->,thick,Green] (0,2)--(v1);
 \draw[->,thick,OliveGreen] (0,8)--(v2);
 \draw[->,thick,OliveGreen] (v3)--(7,9);
 \draw[->,thick,LimeGreen] (v4)--(10,5);
 \draw[->,thick,TealBlue] (v5)--(7,1);

 \node[hrf leg label,text=Green] at (.5,1.5) {$p_1$};
 \node[hrf leg label,text=OliveGreen] at (.5,8.5) {$p_2$};
 \node[hrf leg label,text=OliveGreen] at (6.2,9.1) {$p_3$};
 \node[hrf leg label,text=LimeGreen] at (9.5,5.6) {$p_4$};
 \node[hrf leg label,text=TealBlue] at (6.2,.8) {$p_5$};

 \node[hrf edge label,text=red] at (3.50,5.75) {$q_1$};
 \node[hrf edge label,text=Green] at (3.05,1.95) {$q_2$};
 \node[hrf edge label,text=OliveGreen] at (3.05,8.05) {$q_3$};
 \node[hrf edge label,text=OliveGreen] at (3.55,4.15) {$q_4$};
 \node[hrf edge label,text=red] at (6.65,7.45) {$q_5$};
 \node[hrf edge label,text=LimeGreen] at (6.65,2.55) {$q_6$};

 \GlauberVertex[OliveGreen]{(v3)}{90}
\end{tikzpicture}
\caption{Momentum-space interpretation of the spacelike-collinear Fish HR.
Arrowheads define the edge-momentum orientations used in the text.
Red denotes the soft edges \(q_1\) and \(q_5\), olive the edges collinear to
the \(p_2,p_3\) direction, and green and lime the remaining jet directions.
The oriented GreenYellow disc at the vertex carrying \(p_3\) marks the
Glauber loop \(\ell_G=q_1-q_5\) at the collinear-to-soft vertex.  It follows
the convention defined in Fig.~\ref{fig:crown-regge-glauber}; here its olive
rim records the collinear complementary local channel.}
\label{fig:five-sc-mode}
\end{figure}

The virtuality exponent \(-v_e\) is fixed by the parameter-space region in
both columns of Table~\ref{tab:five-sc-edge-flow}.  If the leading terms in
\(q_e^2=2q_e^+q_e^- -q_{e\perp}^2\) do not cancel, the displayed component
exponents reproduce it as
\(-v_e=\min(a_e+b_e,2c_e)\).  This holds for every facet entry in the table
and for \(q_1,\ldots,q_5\) in the HR column, but is not a defining property
of facet regions: leading component cancellations may occur in either type
of region, and whether a collinear direction is manifest also depends on the
chosen lightcone frame.  The HR entry for \(q_6\) illustrates the latter
point.  Indeed, the explicit
routing gives \(q_6=p_4-\ell_s\).  Since \(p_4\) is wide-angle relative to
the two reference directions, all three components displayed in the table
are of order \(Q\), yielding \((a_6,b_6,c_6)=(0,0,0)\).  Nevertheless its
leading momentum is the lightlike vector \(p_4\), and hence
\[
 q_6^2=(p_4-\ell_s)^2=-2p_4\mathbin{\cdot}\ell_s+\ell_s^2
 \sim Q\delta,
\]
so that \(-v_6=1\).  In a local lightcone basis with \(p_4\) as the plus
reference direction and \(\bar p_4\) a conjugate null direction, the
abbreviated component scaling is instead
\(q_6\sim(1,\delta,\delta)_{(p_4,\bar p_4)}\), making
its \(p_4\)-collinear character manifest.  This is the scaling shown by the
lime edge in Fig.~\ref{fig:five-sc-mode}.  The HR therefore has one soft and
one Glauber loop, and the reconstruction is mode complete.

For the momentum-space power count we need the local widths, not only the
routing valuations in Table~\ref{tab:five-sc-edge-flow}.  In the loop basis of
Eq.~\eqref{eq:five-sc-glauber-rerouting}, the opposite-side poles of the
\((q_3,q_4)\) pair restrict \(\Delta\ell_G^-\) to \(O(\delta^2)\).  The
\((q_1,q_2)\) pair restricts the corresponding plus transfer to
\(O(\delta)\), while the \((q_5,q_6)\) and \((q_1,q_2)\) pairs fix the two
soft longitudinal widths at \(O(\delta)\).  Finally, preserving the
virtualities in Table~\ref{tab:five-sc-edge-flow} fixes both transverse
widths at \(O(\delta)\), by the homogeneity criterion of
Sec.~\ref{sec:pole-pinches-local-widths}.  Thus this mode-adapted basis has
\begin{equation}
 \ell_s\big|_{\rm pinch}\sim\Delta\ell_s
 \sim(\delta,\delta,\delta),
 \qquad
 \ell_G\big|_{\rm pinch}\sim\Delta\ell_G
 \sim(\delta,\delta^2,\delta),
 \label{eq:five-sc-modes}
\end{equation}
where the entries denote \((q^+,q^-,|q_\perp|)\).  Thus the compatible routing
and the independently determined fluctuation widths have the same scaling in
this suitable independent loop basis.  The two measures scale as
\(\delta^D\) and \(\delta^{D+1}\), respectively.  Using the virtuality
valuations \(-v_e\) already listed in Table~\ref{tab:five-sc-edge-flow},
the six edge virtualities obey
\begin{equation}
 \left(\frac{q_1^2}{Q^2},\ldots,\frac{q_6^2}{Q^2}\right)
 \sim(\delta^2,\delta,\delta^2,\delta^2,\delta^2,\delta),
 \qquad -\boldsymbol v_{\HR}=(2,1,2,2,2,1).
 \label{eq:five-sc-virtualities}
\end{equation}
The six virtuality exponents therefore sum to ten.  Hence
\begin{equation}
 I_{\rm SC}^{\rm scalar}\sim
 \delta^{D+(D+1)-10}=\delta^{-1-4\epsilon},
 \label{eq:five-sc-momentum-power}
\end{equation}
in agreement with the parameter-space count in
Eq.~\eqref{eq:five-sc-parameter-power}.  In the original edge routing the
Glauber momentum is the difference of two soft edge momenta, which is why
inspection of individual propagator modes alone does not reveal it.

\subsubsection{A vertex correction: polynomial lifting of the seed locus}

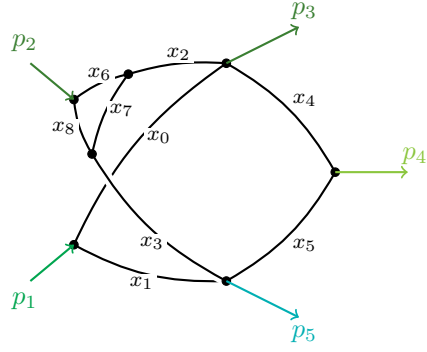
\begin{figure}[htbp]
\centering
\begin{tikzpicture}[scale=.48]

 \coordinate (v1) at (0.8,3);
 \coordinate (v2) at (0.8,7);
 \coordinate (v3) at (5,8);
 \coordinate (v4) at (8,5);
 \coordinate (v5) at (5,2);
 \coordinate (v7) at (2.3,7.7);
 \coordinate (v6) at (1.3,5.5);

 \draw[hrf internal]
   (v1) to[bend left=15]
   node[hrf edge label,pos=.53,right=1pt] {$x_0$} (v3);
 \draw[hrf internal]
   (v1) to[bend right=15]
   node[hrf edge label,pos=.46,below] {$x_1$} (v5);
 \draw[hrf internal]
   (v7) to[bend left=10]
   node[hrf edge label,pos=.52,above] {$x_2$} (v3);

 \draw[white,line width=4pt]
   (v6) to[bend right=15] (v5);
 \draw[hrf internal]
   (v6) to[bend right=15]
   node[hrf edge label,pos=.52,below=1pt] {$x_3$} (v5);

 \draw[hrf internal]
   (v3) to[bend left=15]
   node[hrf edge label,pos=.50,above right] {$x_4$} (v4);
 \draw[hrf internal]
   (v4) to[bend left=15]
   node[hrf edge label,pos=.50,below right] {$x_5$} (v5);

 \draw[hrf internal]
   (v2) to[bend left=12]
   node[hrf edge label,pos=.50,above] {$x_6$} (v7);
 \draw[hrf internal]
   (v6) to[bend left=12]
   node[hrf edge label,pos=.52,right] {$x_7$} (v7);
 \draw[hrf internal]
   (v2) to[bend right=12]
   node[hrf edge label,pos=.50,left] {$x_8$} (v6);

 \foreach \v in {1,...,7} \node[hrf vertex] at (v\v) {};
 \draw[->,thick,Green] (-.40,2)--(v1);
 \draw[->,thick,OliveGreen] (-.40,8)--(v2);
 \draw[->,thick,OliveGreen] (v3)--(7,9);
 \draw[->,thick,LimeGreen] (v4)--(10,5);
 \draw[->,thick,TealBlue] (v5)--(7,1);
 \node[hrf leg label,text=Green]
  at (-.55,1.45) {$p_1$};
\node[hrf leg label,text=OliveGreen]
  at (-.55,8.55) {$p_2$};
 \node[hrf leg label,text=OliveGreen] at (7.15,9.45) {$p_3$};
 \node[hrf leg label,text=LimeGreen] at (10.20,5.45) {$p_4$};
 \node[hrf leg label,text=TealBlue] at (7.15,.55) {$p_5$};

\end{tikzpicture}
\caption{The three-loop vertex-corrected five-point graph, drawn as the
two-loop seed of Fig.~\ref{fig:five-point-seed} with a local vertex correction
adjacent to \(p_2\).  In particular, \(x_7\) lies outside the other loops.
The labels \(x_0,\ldots,x_8\) are those used in
Eqs.~\eqref{eq:five-vertex-factors}--\eqref{eq:five-vertex-HR-data}.}
\label{fig:five-point-vertex}
\end{figure}

The three-loop vertex-corrected graph in the same spacelike-collinear
kinematics is shown in Fig.~\ref{fig:five-point-vertex}.

Here the seed cancellation geometry is inherited in a polynomially lifted
form.  The polynomial-factor harvest returns two cancellation factors.  The
first is the seed factor \(f_1\) defined in
Eq.~\eqref{eq:five-F0-obstruction}, expressed here in the edge labelling of
Fig.~\ref{fig:five-point-vertex}; the second is a polynomial lift of the seed
factor \(f_2\), which we denote by \(\widetilde f_2\).  Collecting all terms
proportional to \(z-1\), including the \(x_6x_7\) term, gives
\begin{align}
 f_1={}&x_1x_4+\chi x_0x_5,
 \nonumber\\
 \widetilde f_2={}&(z-1)A-B,
 &A={}&x_2(x_6+x_7+x_8)+x_6x_7,
 \nonumber\\
 &&B={}&x_3(x_6+x_7+x_8)+x_7x_8,
 \label{eq:five-vertex-factors}
\end{align}
Both \(A\) and \(B\) are positive in the LP domain.  Thus \(\widetilde f_2\) has the
same \((z-1)A-B\) sign structure as the seed factor
\(f_2=(z-1)x_3-x_4\), while \(f_1\) retains the two-term seed structure on
the four edges unaffected by the vertex correction.  HRF pairs these factors
into the single generator \(\widetilde g_1=f_1\widetilde f_2\).  The ensuing Part-I
decomposition and Part-II hierarchy solution are
\begin{equation}
\begin{aligned}
 \mathcal F_\star={}&
 \underbrace{\SLcolour{-s\,\widetilde g_1}}_{\FSL}
 +\underbrace{\HRcolour{-s(1+\chi)z\,x_1x_5
 (x_2x_6+x_2x_7+x_6x_7+x_2x_8)}}_{\FObs},\\
 \vHR={}&(-2,-1,-2,-2,-2,-1,-2,-2,-2;1),
 \qquad (\WSL,\WHR)=(-7,-6).
\end{aligned}
 \label{eq:five-vertex-HR-data}
\end{equation}
As in the seed, the unit gap fixes the total suppression of the product but
need not assign it uniquely to \(f_1\) and \(\widetilde f_2\).  A centred normal chart
and its endpoint facets are equivalent only after their normals are pulled
back to the vector in Eq.~\eqref{eq:five-vertex-HR-data}; the discussion in
Sec.~\ref{sec:five-point-sc-charts} applies unchanged to these general
polynomial factors.
The simultaneous crossing described above acts in precisely the same way on
the lifted factors: in the timelike domain of
Eq.~\eqref{eq:five-timelike-domain}, \(\widetilde f_2<0\) and \(f_1>0\) for positive LP
parameters.  Hence the vertex-corrected graph has no positive cancellation
locus there.  Its channel dependence is therefore inherited from the Fish
seed rather than qualitatively new.  The additional loop provides a nontrivial
check that HRF retains this cancellation geometry when a seed factor is lifted
to a general polynomial and occurs with polynomial multipliers in \(\FSL\).

\paragraph{Momentum-flow inheritance.}\par
\begin{figure}[H]
\centering
\begin{tikzpicture}[scale=.48]
 \coordinate (v1) at (0.8,3);
 \coordinate (v2) at (0.8,7);
 \coordinate (v3) at (5,8);
 \coordinate (v4) at (8,5);
 \coordinate (v5) at (5,2);
 \coordinate (v7) at (2.3,7.7);
 \coordinate (v6) at (1.3,5.5);

 \draw[very thick,red]
   (v1) to[bend left=15]
   node[hrf edge label,pos=.53,right=1pt,text=red] {$x_0$} (v3);
 \draw[very thick,Green]
   (v1) to[bend right=15]
   node[hrf edge label,pos=.46,below,text=Green] {$x_1$} (v5);
 \draw[very thick,OliveGreen]
   (v7) to[bend left=10]
   node[hrf edge label,pos=.52,above,text=OliveGreen] {$x_2$} (v3);
 \draw[white,line width=5pt] (v6) to[bend right=15] (v5);
 \draw[very thick,OliveGreen]
   (v6) to[bend right=15]
   node[hrf edge label,pos=.52,below=1pt,text=OliveGreen] {$x_3$} (v5);
 \draw[very thick,red]
   (v3) to[bend left=15]
   node[hrf edge label,pos=.50,above right,text=red] {$x_4$} (v4);
 \draw[very thick,LimeGreen]
   (v4) to[bend left=15]
   node[hrf edge label,pos=.50,below right,text=LimeGreen] {$x_5$} (v5);
 \draw[very thick,OliveGreen]
   (v2) to[bend left=12]
   node[hrf edge label,pos=.50,above,text=OliveGreen] {$x_6$} (v7);
 \draw[very thick,OliveGreen]
   (v6) to[bend left=12]
   node[hrf edge label,pos=.52,right,text=OliveGreen] {$x_7$} (v7);
 \draw[very thick,OliveGreen]
   (v2) to[bend right=12]
   node[hrf edge label,pos=.50,left,text=OliveGreen] {$x_8$} (v6);

 \foreach \v in {1,...,7} \node[hrf vertex] at (v\v) {};
 \draw[->,thick,Green] (-.40,2)--(v1);
 \draw[->,thick,OliveGreen] (-.40,8)--(v2);
 \draw[->,thick,OliveGreen] (v3)--(7,9);
 \draw[->,thick,LimeGreen] (v4)--(10,5);
 \draw[->,thick,TealBlue] (v5)--(7,1);
 \node[hrf leg label,text=Green] at (-.55,1.45) {$p_1$};
 \node[hrf leg label,text=OliveGreen] at (-.55,8.55) {$p_2$};
 \node[hrf leg label,text=OliveGreen] at (7.15,9.45) {$p_3$};
 \node[hrf leg label,text=LimeGreen] at (10.20,5.45) {$p_4$};
 \node[hrf leg label,text=TealBlue] at (7.15,.55) {$p_5$};

 \GlauberVertex[OliveGreen]{(v3)}{90}
\end{tikzpicture}
\caption{Momentum-space interpretation of the vertex-corrected
spacelike-collinear HR.  Red denotes the two soft edges, olive the enlarged
\(p_2,p_3\)-collinear subgraph, and green and lime the remaining jet
directions.  At the \(p_3\) vertex the olive-rimmed marker shows the inherited
Glauber loop \(\ell_G=q_0-q_4\) at the collinear-to-soft vertex and follows
the convention defined in Fig.~\ref{fig:crown-regge-glauber}.}
\label{fig:five-vertex-mode}
\end{figure}
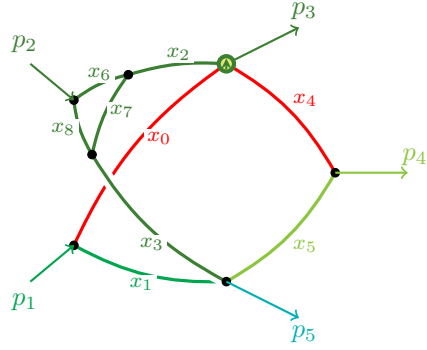

\smallskip
The vertex-corrected graph has an externally labelled Fish contraction minor
and preserves the HR flow of Fig.~\ref{fig:five-sc-mode}: the two soft edges
and their Glauber difference remain in place.  The local
correction replaces the two \(p_2\)-collinear seed edges by the larger
homogeneous collinear subgraph shown in
Fig.~\ref{fig:five-vertex-mode}.  The additional loop is therefore
collinear and introduces no second Glauber loop.  This is also the
momentum-space counterpart of the polynomial lift: the longer factor \(\widetilde f_2\)
reflects the enlarged collinear subgraph, whereas \(f_1\) retains the
two-term structure of the seed.

\subsubsection{Massless five-point kinematics and the near-planar expansion}
\label{sec:five-point-near-planar}

Let all five external momenta be exactly massless and keep all adjacent
Mandelstam invariants at wide angle, with the physical signs
\begin{equation}
 s_{12},s_{34},s_{45}>0,\qquad s_{23},s_{15}<0.
 \label{eq:five-wide-angle-signs}
\end{equation}
Define
\begin{equation}
 \Gamma_5=\det(2p_i\!\cdot p_j)_{i,j=1}^{4}=\epsilon_5^2,
 \qquad \epsilon_5=4i\,\varepsilon(p_1,p_2,p_3,p_4).
 \label{eq:five-gram-definition}
\end{equation}
Geometrically, \(\Gamma_5\) is the signed squared four-volume of the
parallelotope spanned by \(p_1,\ldots,p_4\), up to the conventional factor
\(2^4\); equivalently it is \(2^4(4!)^2\) times the signed squared volume of
the corresponding four-simplex.  Its vanishing is therefore the condition
that the external momenta become coplanar.  For real momenta on the physical
non-coplanar sheet, \(\epsilon_5\) is purely imaginary and \(\Gamma_5<0\).

To define the approach to this boundary, use an exact parametrisation of
general wide-angle five-point kinematics.  These variables were introduced
in the MRK analysis of
Ref.~\cite{Abreu:2024xoh}.\footnote{In the
five-point MRK parametrisation of Ref.~\cite{Caron-Huot:2020vlo}, the
transverse variables denoted there by \(z,\bar z\) coincide, with the present
momentum labels, with \(w,\bar w\).  Related multi-Regge-inspired
parametrisations beyond five points are discussed in
Ref.~\cite{Byrne:2025phh}.}  In the global
\((+,-,\perp)_{(p_2,p_1)}\) lightcone frame, with $p_2$ defining the plus
direction and $p_1$ the minus direction, write the complex transverse momentum as
\(\mathbf p=p^x+ip^y\) and its conjugate as
\(\bar{\mathbf p}=p^x-ip^y\), and define
\begin{equation}
 w=-\frac{\mathbf p_3}{\mathbf p_4},\qquad
 \bar w=-\frac{\bar{\mathbf p}_3}{\bar{\mathbf p}_4},\qquad
 X_{34}=\frac{p_3^+}{p_4^+}=\frac{s_{13}}{s_{14}},\qquad
 X_{45}=\frac{p_4^+}{p_5^+}=\frac{s_{14}}{s_{15}}.
 \label{eq:five-exact-kinematic-chart}
\end{equation}
The final equalities express the longitudinal variables directly as
Lorentz-invariant ratios of Mandelstam invariants.
For physical real momenta away from coplanarity, \(\bar w=w^*\).
For comparing the two expansions below it is useful to replace the inverse
transverse ratios by
\begin{equation}
 \zeta=-\frac1w=\frac{\mathbf p_4}{\mathbf p_3},\qquad
 \bar\zeta=-\frac1{\bar w}
 =\frac{\bar{\mathbf p}_4}{\bar{\mathbf p}_3}.
 \label{eq:five-zeta-chart}
\end{equation}
These four dimensionless ratios and the transverse scale
\(|\mathbf p_4|^2\) form an exact chart, up to an irrelevant longitudinal
boost, and therefore determine all five independent Mandelstam invariants.
Transverse momentum
conservation gives
\(\mathbf p_3=-w\mathbf p_4\) and
\(\mathbf p_5=-(1-w)\mathbf p_4\), together with their barred counterparts;
on-shellness then fixes the minus components from the plus components.  In
this chart the parity-odd invariant factorises exactly as
\begin{equation}
 \epsilon_5=s_{12}|\mathbf p_4|^2(w-\bar w)
 =s_{12}|\mathbf p_4|^2
  \frac{\zeta-\bar\zeta}{\zeta\bar\zeta}.
 \label{eq:five-exact-parity-odd-chart}
\end{equation}
Equations~\eqref{eq:five-exact-kinematic-chart} and
\eqref{eq:five-zeta-chart} describe generic five-point kinematics before
any expansion is taken.  We shall select three distinct paths from this
same generic domain.  The near-planar path makes the angular factor
\(\zeta-\bar\zeta\) vanish at fixed \(|\mathbf p_4|^2\).  The central-soft
rapidity-ordered path defined later instead sends
\(\zeta,\bar\zeta\to0\) with the phase ratio \(\zeta/\bar\zeta\) held at a
generic value, while holding
\[
 Q_\perp^2\equiv |\mathbf p_3|^2
 =\frac{|\mathbf p_4|^2}{\zeta\bar\zeta}
\]
fixed.  Equation~\eqref{eq:five-exact-parity-odd-chart} then becomes
\(\epsilon_5=s_{12}Q_\perp^2(\zeta-\bar\zeta)\to0\).  This path therefore
provides a second approach to the normalised planar surface, through
transverse softening of \(p_4\) rather than angular alignment at fixed
\(|\mathbf p_4|^2\).  It is rapidity ordered but differs from standard MRK
because the transverse ratio
\(|\mathbf p_4|^2/|\mathbf p_3|^2=\zeta\bar\zeta\) also tends to zero.  The
third path combines standard fixed-transverse MRK with \(w-\bar w\to0\);
unlike the central-soft path, neither transverse scale is sent to zero.

The near-planar wide-angle expansion is defined by
\begin{align}
 w&=-\frac1\xi+\frac{i}{2}\lambda,&
 \bar w&=-\frac1\xi-\frac{i}{2}\lambda,\nonumber\\
 \zeta&=\frac{\xi}{1-i\xi\lambda/2},&
 \bar\zeta&=\frac{\xi}{1+i\xi\lambda/2},\nonumber\\
 &\lambda\to0^+,&
 &\xi,\ X_{34},\ X_{45},\ |\mathbf p_4|^2\ \text{fixed},
 \label{eq:five-near-planar-chart}
\end{align}
where \(\xi>0\).  Thus exact planarity is simply
\(\zeta=\bar\zeta=\xi\), or equivalently
\(w=\bar w=-1/\xi\).  The minus sign makes the first-sheet branch with
positive LP parameters manifest in the formulae below.
There is no rapidity hierarchy or soft external momentum.  Thus
\begin{equation}
 \epsilon_5=i c_\epsilon\lambda,\qquad
 \Gamma_5=-c_\epsilon^2\lambda^2,\qquad
 c_\epsilon=s_{12}|\mathbf p_4|^2>0.
 \label{eq:five-near-planar-limit}
\end{equation}
This kinematic construction is independent of the loop order and of the
integral topology considered below.

\subsubsection{The Fish seed in near-planar scattering}
\label{sec:five-point-near-planar-seed}

\medskip
\noindent\textbf{Symmetric coordinates for the Fish topology.}\par\smallskip

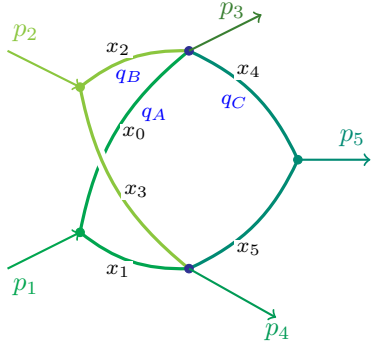
\begin{figure}[H]
\centering
\begin{tikzpicture}[scale=.48]
 \coordinate (v1) at (2,3);
 \coordinate (v2) at (2,7);
 \coordinate (v3) at (5,8);
 \coordinate (v4) at (8,5);
 \coordinate (v5) at (5,2);

 \draw[very thick,Green] (v3) to[bend right=20] (v1);
 \draw[very thick,LimeGreen] (v3) to[bend right=20] (v2);
 \draw[white,double=white,double distance=3pt,thick]
   (v5) to[bend left=20] (v2);
 \draw[very thick,LimeGreen] (v5) to[bend left=20] (v2);
 \draw[very thick,Green] (v5) to[bend left=20] (v1);
 \draw[very thick,PineGreen] (v3) to[bend left=20] (v4);
 \draw[very thick,PineGreen] (v5) to[bend right=20] (v4);

 \node[hrf momentum label] at (4.05,6.25) {$q_A$};
 \node[hrf momentum label] at (3.35,7.30) {$q_B$};
 \node[hrf momentum label] at (6.25,6.60) {$q_C$};

 \node[hrf vertex,fill=Green,draw=Green] at (v1) {};
 \node[hrf vertex,fill=LimeGreen,draw=LimeGreen] at (v2) {};
 \node[hrf vertex,fill=Blue,draw=Blue] at (v3) {};
 \node[hrf vertex,fill=PineGreen,draw=PineGreen] at (v4) {};
 \node[hrf vertex,fill=Blue,draw=Blue] at (v5) {};
 \draw[->,thick,Green] (0,2)--(v1);
 \draw[->,thick,LimeGreen] (0,8)--(v2);
 \draw[->,thick,OliveGreen] (v3)--(7,9);
 \draw[->,thick,ForestGreen] (v5)--(7.4,.65);
 \draw[->,thick,PineGreen] (v4)--(10,5);

 \node[hrf leg label,text=Green] at (.5,1.5) {$p_1$};
 \node[hrf leg label,text=LimeGreen] at (.5,8.5) {$p_2$};
 \node[hrf leg label,text=OliveGreen] at (6.2,9.1) {$p_3$};
 \node[hrf leg label,text=ForestGreen] at (7.45,.25) {$p_4$};
 \node[hrf leg label,text=PineGreen] at (9.5,5.6) {$p_5$};

 \node[hrf edge label] at (3.50,5.75) {$x_0$};
 \node[hrf edge label] at (3.05,1.95) {$x_1$};
 \node[hrf edge label] at (3.05,8.05) {$x_2$};
 \node[hrf edge label] at (3.55,4.15) {$x_3$};
 \node[hrf edge label] at (6.65,7.45) {$x_4$};
 \node[hrf edge label] at (6.65,2.55) {$x_5$};

\end{tikzpicture}
\caption{A representative permutation of the external legs and LP-parameter assignment for
the near-planar wide-angle analysis.  Relative to
Fig.~\ref{fig:five-point-seed}, \(p_4\) and \(p_5\) are interchanged.  The
five shades of green distinguish the five wide-angle directions.
The pairs of propagators forming paths \(A,B,C\) have respectively the
colours of \(p_1,p_2,p_5\), making their collinear assignments explicit.
The vertices carrying \(p_3\) and \(p_4\) are blue to identify the two hard
vertices, while the other three vertices match their path colours.
The blue labels associate the path momenta \(q_A,q_B,q_C\) used below with
the \(x_0,x_2,x_4\) edges, respectively.  All three are oriented outward
from the vertex carrying \(p_3\), so momentum conservation there gives
\(p_3+q_A+q_B+q_C=0\).
}
\label{fig:five-point-near-planar-attachment}
\end{figure}

The same six-edge topology analysed in
Sec.~\ref{sec:five-point-spacelike-collinear} has a second certified HR in the
near-planar expansion defined above.  For convenience we use the
\(p_4\leftrightarrow p_5\)-interchanged labelling relative to the
representative in Fig.~\ref{fig:five-point-seed},
\begin{equation}
 (p_1,p_2,p_3,p_5,p_4)\longrightarrow(1,2,3,4,5),
 \label{eq:five-near-planar-attachment}
\end{equation}
as shown in Fig.~\ref{fig:five-point-near-planar-attachment}.  This
interchange is immaterial both here and in the spacelike-collinear limit; it
becomes consequential only in the later soft and rapidity-ordered limits
that distinguish \(p_4\) from \(p_5\).

For the remainder of the Fish-seed analysis we use the zero-based edge labels
\((x_0,\ldots,x_5)\), corresponding respectively to
\((x_1,\ldots,x_6)\) in Fig.~\ref{fig:five-point-seed}.  As indicated by the
matching propagator colours in
Fig.~\ref{fig:five-point-near-planar-attachment}, group them as the three
two-edge paths \((x_0,x_1)\),
\((x_2,x_3)\) and \((x_4,x_5)\) between the two trivalent hard vertices, and put
\begin{equation}
 r_A=\frac{x_0}{x_1},\qquad r_B=\frac{x_2}{x_3},\qquad
 r_C=\frac{x_4}{x_5},\qquad
 \mathcal F=x_1x_3x_5\,\Phi(r_A,r_B,r_C).
 \label{eq:five-path-ratios}
\end{equation}
Here \(x_1,x_3,x_5>0\) are coordinates for the magnitudes on the three
two-edge paths, while \(r_A,r_B,r_C\) specify the ratios within them; they
are together six independent coordinates on the positive LP-parameter
orthant.  Write
\begin{align}
 \Phi(\boldsymbol r)&=q_{AB}r_Ar_B+q_{BC}r_Br_C
 +q_{CA}r_Cr_A+b_A r_A+b_B r_B+b_C r_C
 =\frac12\boldsymbol r^T K\boldsymbol r+\boldsymbol b^T\boldsymbol r,
 \nonumber\\
 K&=\begin{pmatrix}
 0&q_{AB}&q_{CA}\\ q_{AB}&0&q_{BC}\\ q_{CA}&q_{BC}&0
 \end{pmatrix},
 \qquad \boldsymbol b=(b_A,b_B,b_C)^T.
 \label{eq:five-ratio-quadratic}
\end{align}
For the permutation in Eq.~\eqref{eq:five-near-planar-attachment},
the entries are Mandelstam invariants, which may be expressed in terms of
the adjacent ones using five-point momentum conservation:
\begin{align}
 q_{AB}&=s_{35}=s_{12}-s_{34}-s_{45},
 &b_A&=s_{14}=-s_{15}+s_{23}-s_{45},\nonumber\\
 q_{BC}&=s_{13}=-s_{12}-s_{23}+s_{45},
 &b_B&=s_{24}=s_{15}-s_{23}-s_{34},\nonumber\\
 q_{CA}&=s_{23},
 &b_C&=s_{45}.
 \label{eq:five-ratio-coefficients}
\end{align}
\medskip
\noindent\textbf{Landau analysis and the origin of the expansion.}\par\smallskip

We first use the Landau equations at generic five-point
kinematics to determine which kinematic limit can support the pinch.  Once
that limit has been identified, we will fix the corresponding expansion
and return to the standard HRF workflow.

Direct differentiation with respect to the ratios gives
\begin{equation}
 \nabla_{\boldsymbol r}\Phi=K\boldsymbol r+\boldsymbol b.
 \label{eq:five-ratio-gradient}
\end{equation}
The determinant of the coefficient matrix is
\begin{equation}
 \det K=2q_{AB}q_{BC}q_{CA}.
 \label{eq:five-ratio-determinant}
\end{equation}
Hence, whenever \(q_{AB}q_{BC}q_{CA}\ne0\), ratio stationarity at
\(\boldsymbol r=\boldsymbol\rho\) determines a unique ratio vector and its
stationary value:
\begin{equation}
 \begin{aligned}
 K\boldsymbol\rho&=-\boldsymbol b,\qquad
 \boldsymbol\rho=(\rho_A,\rho_B,\rho_C)^T=-K^{-1}\boldsymbol b,\\
 \Phi(\boldsymbol\rho)&=\frac12\boldsymbol b^T\boldsymbol\rho
 =-\frac12\boldsymbol b^TK^{-1}\boldsymbol b.
 \end{aligned}
 \label{eq:five-ratio-stationarity}
\end{equation}
Already at this generic kinematic point, define the three polynomials normal
to the ratio-stationary locus,
\begin{equation}
 f_A=x_0-\rho_Ax_1,\qquad f_B=x_2-\rho_Bx_3,
 \qquad f_C=x_4-\rho_Cx_5.
 \label{eq:five-path-normals}
\end{equation}
Here the components \(\rho_I\) retain their full kinematic dependence; no
kinematic limit has yet been taken.

Ratio stationarity is necessary but is not yet the complete Landau
condition.  To see the missing equation, use the independent path-magnitude
coordinates \(u_A=x_1\), \(u_B=x_3\), \(u_C=x_5\).  At fixed ratios,
\begin{equation}
 \mathcal F=u_Au_Bu_C\Phi,
 \qquad
 \frac{\partial\mathcal F}{\partial u_A}
 =u_Bu_C\Phi,
 \quad\text{and cyclically}.
 \label{eq:five-path-magnitude-equations}
\end{equation}
Since the path magnitudes are positive, stationarity in these directions
also requires
\begin{equation}
 \Phi(\boldsymbol\rho)=0.
 \label{eq:five-ratio-vanishing}
\end{equation}
In the original LP parameters this condition follows automatically
from homogeneity once all derivative equations are imposed.  In the ratio
chart, however, \(\Phi\) contains both quadratic and linear terms and is not
homogeneous, so Eq.~\eqref{eq:five-ratio-vanishing} does not follow from
Eq.~\eqref{eq:five-ratio-stationarity}.  To relate the stationary value in
Eq.~\eqref{eq:five-ratio-stationarity} to the external kinematics, assemble
the coefficients in \(K\) and \(\boldsymbol b\) into the symmetric matrix
\[
 \mathcal B=\begin{pmatrix}K&\boldsymbol b\\
                  \boldsymbol b^T&0\end{pmatrix}.
\]
Using the Mandelstam identifications in
Eq.~\eqref{eq:five-ratio-coefficients}, five-point momentum conservation
gives
\(\det\mathcal B=\det(2p_i\!\cdot p_j)_{i,j=1}^{4}=\Gamma_5\).
The relation of this determinant to the stationary value follows from
Schur's formula~\cite{Zhang:2005SchurComplement},
\[
 \det\begin{pmatrix}A&B\\ C&D\end{pmatrix}
 =\det A\,\det\!\left(D-CA^{-1}B\right).
\]
Taking \(A=K\), \(B=\boldsymbol b\),
\(C=\boldsymbol b^T\), and \(D=0\), and using
Eq.~\eqref{eq:five-ratio-stationarity}, gives
\[
 \det\mathcal B
 =-\det K\,\boldsymbol b^TK^{-1}\boldsymbol b
 =2\det K\,\Phi(\boldsymbol\rho)
 =4q_{AB}q_{BC}q_{CA}\,\Phi(\boldsymbol\rho).
\]
Consequently,
\begin{equation}
 \Phi(\boldsymbol\rho)=
 \frac{\Gamma_5}{4q_{AB}q_{BC}q_{CA}}.
 \label{eq:five-stationary-gram}
\end{equation}
Combining the ratio-stationarity conditions in
Eq.~\eqref{eq:five-ratio-stationarity} with the remaining path-magnitude
condition in Eq.~\eqref{eq:five-ratio-vanishing}, the full Landau system for
wide-angle kinematics (here \(\det K\ne0\)) fixes \(\boldsymbol\rho\)
uniquely and requires \(\Gamma_5=0\).  Thus vanishing \(\Gamma_5\) is the
only possible kinematic solution in this domain.  A positive Landau pinch
additionally requires \(\rho_A,\rho_B,\rho_C>0\); the corresponding chamber
is identified below.

\medskip
\noindent\textbf{The wide-angle near-planar Landau structure.}\par\smallskip

This structure contrasts directly with the Crown.  The Crown matrix
\(K_C\) in Eq.~\eqref{eq:crown-ratio-quadratic} has rank three and the
kinematics-independent zero mode \((1,1,1,1)\).  Its stationary system
therefore leaves one common ratio free, while its stationary value vanishes
throughout the generic wide-angle domain.  For the Fish, by contrast,
\(K\) is generically invertible and no common-ratio zero mode remains.

To expose the cancellation factors, complete the quadratic form in
Eq.~\eqref{eq:five-ratio-quadratic} about its stationary point:
\[
 \Phi(\boldsymbol r)=\Phi(\boldsymbol\rho)
 +q_{AB}(r_A-\rho_A)(r_B-\rho_B)
 +q_{BC}(r_B-\rho_B)(r_C-\rho_C)
 +q_{CA}(r_C-\rho_C)(r_A-\rho_A),
\]
where the term linear in \(\boldsymbol r-\boldsymbol\rho\) vanishes by
Eq.~\eqref{eq:five-ratio-stationarity}.  Using
\(r_A-\rho_A=f_A/x_1\), \(r_B-\rho_B=f_B/x_3\),
\(r_C-\rho_C=f_C/x_5\), and Eq.~\eqref{eq:five-stationary-gram}, this gives
the exact identity
\begin{align}
 \mathcal F={}&\frac{\Gamma_5}{4q_{AB}q_{BC}q_{CA}}x_1x_3x_5
 +q_{AB}x_5f_Af_B+q_{BC}x_1f_Bf_C+q_{CA}x_3f_Cf_A.
 \label{eq:five-full-stationary-decomposition}
\end{align}
This is an identity for the full massless \(\mathcal F\) at generic
five-point kinematics, not merely its restriction to the Gram surface.

We now express this generic Landau solution in the original \(x_e\)
variables.  The purpose is twofold: to show how the derivatives available to
HRF recover the cancellation factors \(f_A,f_B,f_C\), and to identify the
remaining condition they impose on the external kinematics.  HRF starts
from the usual Mandelstam representation of \(\mathcal F\), rather than the
completed form in Eq.~\eqref{eq:five-full-stationary-decomposition}.  Direct
differentiation therefore gives, for example,
\begin{equation}
 D_0\equiv\frac{\partial\mathcal F}{\partial x_0}
 =q_{AB}x_2x_5+q_{CA}x_3x_4+b_Ax_3x_5.
 \label{eq:five-raw-D0}
\end{equation}
Literal collection in \(q_{AB}\), \(q_{CA}\) and \(b_A\) exposes only
monomials.  The first component of \(K\boldsymbol\rho=-\boldsymbol b\),
\begin{equation}
 b_A=-q_{AB}\rho_B-q_{CA}\rho_C,
\end{equation}
instead reorganises the same derivative as
\begin{equation}
 D_0=q_{AB}x_5\bigl(x_2-\rho_Bx_3\bigr)
     +q_{CA}x_3\bigl(x_4-\rho_Cx_5\bigr)
 =q_{AB}x_5f_B+q_{CA}x_3f_C.
 \label{eq:five-adapted-D0}
\end{equation}
Equation~\eqref{eq:five-adapted-D0} shows that the \(f_I\) emerge from a
nontrivial stationary reorganisation of the raw derivatives.  Since the
final term of Eq.~\eqref{eq:five-full-stationary-decomposition} is independent
of \(x_0,x_2,x_4\), the three stationary-adapted derivative identities are
\begin{align}
 \frac{\partial\mathcal F}{\partial x_0}
 &=q_{AB}x_5f_B+q_{CA}x_3f_C,\nonumber\\
 \frac{\partial\mathcal F}{\partial x_2}
 &=q_{AB}x_5f_A+q_{BC}x_1f_C,\nonumber\\
 \frac{\partial\mathcal F}{\partial x_4}
 &=q_{BC}x_1f_B+q_{CA}x_3f_A.
 \label{eq:five-near-planar-gradient-system}
\end{align}
Let
\[
 D_u=\operatorname{diag}(u_A,u_B,u_C)
 =\operatorname{diag}(x_1,x_3,x_5),
 \qquad U=u_Au_Bu_C=x_1x_3x_5.
\]
The coefficient matrix multiplying \((f_A,f_B,f_C)^T\) on the right-hand
side of the derivative identities in
Eq.~\eqref{eq:five-near-planar-gradient-system} is the original-variable
representation of the ratio-space matrix \(K\):
\begin{align}
 M_f&=U D_u^{-1}KD_u^{-1}
 =\begin{pmatrix}
 0&q_{AB}x_5&q_{CA}x_3\\
 q_{AB}x_5&0&q_{BC}x_1\\
 q_{CA}x_3&q_{BC}x_1&0
 \end{pmatrix},\nonumber\\
 \det M_f&=U\det K
 =2q_{AB}q_{BC}q_{CA}x_1x_3x_5.
 \label{eq:five-near-planar-normal-matrix}
\end{align}
With \(\boldsymbol f=(f_A,f_B,f_C)^T\), imposing stationarity on the three
derivatives in Eq.~\eqref{eq:five-near-planar-gradient-system} gives the
linear system
\[
 M_f\boldsymbol f=0.
\]
The relation in Eq.~\eqref{eq:five-near-planar-normal-matrix} incorporates
both \(f_I=u_I(r_I-\rho_I)\) and the derivative Jacobian.  In the interior
positive orthant \(D_u\) is invertible, and at wide angle \(K\) is
invertible; hence \(M_f\) is invertible and the stationary system has the
unique solution \(\boldsymbol f=0\), or \(f_A=f_B=f_C=0\).  Thus \(M_f\)
encodes the original-coordinate derivative
system, whereas the bordered matrix \(\mathcal B\) above relates the
stationary value \(\Phi(\boldsymbol\rho)\) to \(\Gamma_5\).

The remaining three derivatives then reduce to
\begin{equation}
 \left.\left(
 \frac{\partial\mathcal F}{\partial x_1},
 \frac{\partial\mathcal F}{\partial x_3},
 \frac{\partial\mathcal F}{\partial x_5}
 \right)\right|_{f_A=f_B=f_C=0}
 =\frac{\Gamma_5}{4q_{AB}q_{BC}q_{CA}}
 (x_3x_5,x_1x_5,x_1x_3).
 \label{eq:five-remaining-gradient-system}
\end{equation}
Likewise, Eq.~\eqref{eq:five-full-stationary-decomposition} restricted to
this locus equals
\(\Gamma_5x_1x_3x_5/(4q_{AB}q_{BC}q_{CA})\).  The original-variable Landau
system therefore reproduces the ratio-space result in two stages: the first
three derivatives identify the local defining ideal
\(\langle f_A,f_B,f_C\rangle\), while the remaining equations require
\(\Gamma_5=0\).  This separation explains why the near-planar surface must
be selected before the HRF expansion begins.

Positivity of the resulting \(\rho_I\) is precisely the usual interior
first-sheet condition \(x_e>0\), because the \(\rho_I\) are the stationary
values of the positive ratios in Eq.~\eqref{eq:five-path-ratios}.  This
condition can be stated directly in Mandelstam variables.  Solving
Eq.~\eqref{eq:five-ratio-stationarity} gives
\begin{align}
 \rho_A&=\frac{q_{BC}b_A-q_{CA}b_B-q_{AB}b_C}
                   {2q_{AB}q_{CA}},\nonumber\\
 \rho_B&=\frac{-q_{BC}b_A+q_{CA}b_B-q_{AB}b_C}
                   {2q_{AB}q_{BC}},\nonumber\\
 \rho_C&=\frac{-q_{BC}b_A-q_{CA}b_B+q_{AB}b_C}
                   {2q_{BC}q_{CA}},
 \label{eq:five-positive-ratios-mandelstam}
\end{align}
where every \(q_{IJ}\) and \(b_I\) is the Mandelstam combination in
Eq.~\eqref{eq:five-ratio-coefficients}.  Where these three \(q_{IJ}\) are
nonzero, the nondegenerate positive planar Fish chamber is therefore
\begin{equation}
 \mathfrak C_{\rm Fish}^{+}
 =\left\{\boldsymbol s\in\mathcal K:\
 \Gamma_5=0,\quad
 \rho_A(\boldsymbol s)>0,\quad
 \rho_B(\boldsymbol s)>0,\quad
 \rho_C(\boldsymbol s)>0\right\}.
 \label{eq:five-positive-planar-chamber}
\end{equation}
The Gram equation supplies planarity, while the three inequalities select
the first-sheet component of the singular locus.  This is the generic
kinematic statement; it does not assign an asymptotic scaling to any
particular approach to the chamber or to its closure.

The spacelike-collinear example above gives the simplest boundary test.  Both
the wide-angle and \(p_2\parallel p_3\) collinear problems are invariant under
the permutation \(p_4\leftrightarrow p_5\), which does not affect the
relevant two-edge path.
Translating the collinear factor \(f_2\) in
Eq.~\eqref{eq:five-F0-obstruction} to the zero-based edge labels used here
gives
\begin{equation}
 f_2^{\rm coll}=(z-1)x_2-x_3.
 \qquad
 f_2^{\rm coll}=0
 \quad\Longrightarrow\quad
 \frac{x_3}{x_2}=z-1>0,
 \qquad
 r_B=\frac{x_2}{x_3}=\frac{1}{z-1}>0.
 \label{eq:five-collinear-positive-chamber-test}
\end{equation}
Thus the spacelike range \(z>1\) approaches the positive closure of
Eq.~\eqref{eq:five-positive-planar-chamber}.  In the timelike-collinear
range \(0<z<1\), the same ratio is negative, so planarity alone does not
preserve the pinch.  At this boundary \(q_{CA}=s_{23}\to0\), the matrix
\(K\) becomes singular and the three-ratio locus degenerates to the
two-factor collinear locus.  The positive chamber, rather than the Gram
surface alone, is therefore the appropriate general organising object.

\medskip
\noindent\textbf{Returning to HRF: the near-planar leading polynomial.}\par\smallskip

The generic Landau analysis has now selected \(\Gamma_5=0\).  We therefore
fix the near-planar expansion before invoking HRF.  This is the point at which
the standard HRF workflow begins: the expansion is specified and its leading
polynomial can be constructed.  Specialising the generic construction, the
normals become especially transparent in the exact chart
\eqref{eq:five-exact-kinematic-chart}.  Taking \(\lambda\to0\) in
Eq.~\eqref{eq:five-near-planar-chart} gives the exact-planarity relation
\(\zeta=\bar\zeta=\xi\).  To write the resulting solution of
\(K\boldsymbol\rho=-\boldsymbol b\) compactly, we introduce the kinematic
ratio
\begin{equation}
 R\equiv\frac{\xi+X_{45}(\xi+1)}
       {1+X_{34}X_{45}(\xi+1)}.
 \label{eq:five-near-planar-ratio-variables}
\end{equation}
The solution of \(K\boldsymbol\rho=-\boldsymbol b\) is then
\begin{equation}
 \rho_A=X_{34}\xi R,\qquad
 \rho_B=R,\qquad \rho_C=\xi,
 \label{eq:five-near-planar-rhos}
\end{equation}
and therefore
\begin{equation}
 f_A=x_0-X_{34}\xi R x_1,\qquad
 f_B=x_2-Rx_3,\qquad
 f_C=x_4-\xi x_5.
 \label{eq:five-near-planar-factors}
\end{equation}
On the physical positive-Landau branch
\(X_{34},X_{45},\xi>0\), so \(R>0\) and all three stationary ratios are
positive.  Beyond the near-planar expansion, no further kinematic restrictions
are imposed (in particular, no MRK hierarchy).

To define the near-planar expansion, let \(\boldsymbol{\kappa}\) denote four
coordinates tangent to the Gram surface and use \(\Gamma_5\) as a local
normal coordinate.  Then
\begin{equation}
 \mathcal F(\Gamma_5,\boldsymbol{\kappa})
 =\mathcal F_0(\boldsymbol{\kappa})
  +\Gamma_5\mathcal F_{\Gamma_5}(\boldsymbol{\kappa})
  +O(\Gamma_5^2).
 \label{eq:five-gram-expansion}
\end{equation}
There is no conflict between the apparently nonlinear kinematic dependence
in Eq.~\eqref{eq:five-full-stationary-decomposition} and the linearity of
\(\mathcal F\) in Mandelstam invariants.  The Gram determinant \(\Gamma_5\)
is itself a nonlinear polynomial in those invariants.  Replacing one
Mandelstam variable by \(\Gamma_5\) is therefore a nonlinear change of
kinematic coordinates, and at fixed \(\boldsymbol\kappa\)
\begin{equation}
 \mathcal F_{\Gamma_5}
 =\left.\frac{\partial\mathcal F}{\partial\Gamma_5}
  \right|_{\boldsymbol\kappa}
 =\sum_a\frac{\partial\mathcal F}{\partial s_a}
  \left.\frac{\partial s_a}{\partial\Gamma_5}
  \right|_{\boldsymbol\kappa}.
 \label{eq:five-gram-chain-rule}
\end{equation}
The Jacobian factors on the right need not be linear functions of the
remaining kinematics.  Equivalently, the separate terms in
Eq.~\eqref{eq:five-full-stationary-decomposition} contain the rational
ratios \(\boldsymbol\rho=-K^{-1}\boldsymbol b\); their nonlinear dependence
cancels in the sum, restoring the original Mandelstam-linear polynomial.

The complete coefficient \(\mathcal F_{\Gamma_5}\) away from the stationary
locus depends on the choice of the tangent coordinates
\(\boldsymbol\kappa\).  Its restriction to \(f_A=f_B=f_C=0\), however, is
fixed by Eq.~\eqref{eq:five-full-stationary-decomposition}:
\begin{equation}
 \left.\mathcal F_{\Gamma_5}\right|_{f_A=f_B=f_C=0}
 =\frac{x_1x_3x_5}{4q_{AB}q_{BC}q_{CA}}.
 \label{eq:five-gram-coefficient-on-locus}
\end{equation}
This stationary contribution is not an
``obstruction'' in the technical HRF sense: it is absent from the leading
polynomial by the definition of the chosen kinematic expansion, rather than
being made subleading by the HR scaling.  Since
\(\Gamma_5=O(\lambda^2)\), this contribution first appears two powers beyond
\(\mathcal F_0\).  Its precise weight relative to the cancelled leading
sector and to \(\mathcal U\) will follow only after the region vector has
been determined below.  The leading polynomial itself has the symmetric
factorised form
\begin{equation}
 \mathcal F_0\equiv\left.\mathcal F\right|_{\Gamma_5=0}
 =q_{AB}x_5f_Af_B+q_{BC}x_1f_Bf_C+q_{CA}x_3f_Cf_A.
 \label{eq:five-coplanar-factorisation}
\end{equation}
Thus the HR generators are the factorised products
\(f_Af_B,f_Bf_C,f_Cf_A\), not the individual normal equations.  This is a
three-generator ideal and is more general than the pair-sector search used
in the earlier examples.

\newpage
\medskip
\noindent\textbf{HRF factor harvesting, generators and scaling.}\par\smallskip

With the expansion fixed and \(\mathcal F_0\) displayed, we now apply the
core HRF construction to recover the cancellation locus from its derivatives
and determine the region scaling.  HRF starts from \(\mathcal F_0\); in this
example its decomposition finds no obstruction, so that
\(\mathcal F_\star=\mathcal F_0=\FSL\) and the complete leading polynomial is
the cancellation sector.  The selected-sector saturation of
Eq.~\eqref{eq:selected-gradient-saturation} is therefore applicable directly
to \(\mathcal F_0\).

In this example the abstract saturation in
Eq.~\eqref{eq:selected-gradient-saturation} becomes completely explicit.
Let \(\mathcal R\) denote the polynomial ring in the LP parameters and
the chosen kinematic variables, and define
\begin{equation}
 \mathcal I_\nabla^{(0)}
 =\left\langle\frac{\partial\mathcal F_0}{\partial x_e}
  \right\rangle_{e=0}^{5},\qquad
 \mathcal I_{\rm kin}=\langle\Gamma_5\rangle,\qquad
 S_{\rm np}=\left(\prod_{e=0}^{5}x_e\right)
 q_{AB}q_{BC}q_{CA}.
 \label{eq:five-near-planar-saturation-data}
\end{equation}
Here and below, the subscript ``np'' denotes the near-planar sector.
The selected-sector ideal used by HRF is therefore
\begin{equation}
 \mathcal J_{\rm np}
 =\bigl(\mathcal I_\nabla^{(0)}+\mathcal I_{\rm kin}\bigr)
 :S_{\rm np}^{\infty}.
 \label{eq:five-near-planar-saturated-ideal}
\end{equation}
The algebraic recovery of the cancellation-factor vector
\(\boldsymbol f=(f_A,f_B,f_C)^T\) follows directly from
Eq.~\eqref{eq:five-near-planar-normal-matrix}.  Let \(\boldsymbol D\) denote
the vector of the first three derivatives in
Eq.~\eqref{eq:five-near-planar-gradient-system}.  Then
\(\boldsymbol D=M_f\boldsymbol f\), and the adjugate identity gives
\[
 \det(M_f)\boldsymbol f
 =\operatorname{adj}(M_f)\boldsymbol D.
\]
Every component on the right belongs to \(\mathcal I_\nabla^{(0)}\).
Moreover,
\(\det M_f=2q_{AB}q_{BC}q_{CA}x_1x_3x_5\) is, up to a nonzero constant,
a factor of \(S_{\rm np}\).  Saturation therefore makes \(\det M_f\)
invertible, so the adjugate identity implies
\(f_A,f_B,f_C\in\mathcal J_{\rm np}\).  This is a genuine recovery step:
the \(q_{IJ}\) entering the matrix elements of \(M_f\) in
Eq.~\eqref{eq:five-near-planar-normal-matrix} are known kinematic
coefficients, whose nonvanishing specifies the generic wide-angle sector,
whereas \(\boldsymbol f\) is recovered from the derivatives rather than
supplied to HRF.  The Gr\"obner-basis elimination described in the footnote
below implements the polynomial combinations and permitted divisions made
explicit by the adjugate identity.

In physical terms, this saturation says that the active LP parameters
and the kinematic factors \(q_{AB},q_{BC},q_{CA}\) are nonzero, so division
by their products is allowed.  It performs all such divisions systematically
and removes solution components supported entirely on the excluded surface
\(S_{\rm np}=0\).\footnote{A standard computational implementation introduces
an auxiliary variable \(\tau\) and defines
\(\widetilde{\mathcal J}_{\rm np}
 =\langle\partial\mathcal F_0/\partial x_e\ (e=0,\ldots,5),
 \Gamma_5,1-\tau S_{\rm np}\rangle\subset\mathcal R[\tau]\).
A Gr\"obner basis in an elimination order for \(\tau\) then gives
\(\mathcal J_{\rm np}=\widetilde{\mathcal J}_{\rm np}\cap\mathcal R\).
The added equation sets \(\tau=S_{\rm np}^{-1}\), so elimination retains
precisely the polynomial consequences obtainable after division by powers of
\(S_{\rm np}\).  This is the Rabinowitsch trick; see, for example,
Ref.~\cite{Cox:2015iva}.}
Conversely, on \(\Gamma_5=0\),
all derivatives of \(\mathcal F_0\) vanish on that locus.  Thus the
LP-parameter content of \(\mathcal J_{\rm np}\) is precisely the
local defining ideal \(\langle f_A,f_B,f_C\rangle\).

To connect this ideal calculation back to the HRF harvesting step, return to
the three derivative identities in
Eq.~\eqref{eq:five-near-planar-gradient-system}.  Each identity contains two
terms.  Across the three identities these six terms fall into the three
kinematically independent sectors
\(q_{AB}=s_{35}\), \(q_{BC}=s_{13}\) and \(q_{CA}=s_{23}\), using
Eq.~\eqref{eq:five-ratio-coefficients}.  Collecting the first identity by
\(q_{AB}\) and \(q_{CA}\), for example, exposes \(x_5f_B\) and \(x_3f_C\);
the other two identities similarly expose all three local defining polynomials.

For the Crown, direct invariant-sector harvesting already produces this
adapted form.  Here the \(f_I\) contain the kinematic stationary ratios
\(\boldsymbol\rho\), so literal factorisation or collection of an individual
raw derivative does not reveal them.  The selected-sector saturation in
Eq.~\eqref{eq:five-near-planar-saturated-ideal} supplies precisely this
algebraic reorganisation and recovers the three normals without assuming
them in advance.  HRF then combines the harvested normals into candidate
factorised generators; their occurrence in \(\mathcal F_0\) identifies the
generator ideal
\(\langle f_Af_B,f_Bf_C,f_Cf_A\rangle\).

For the expansion defined in Eqs.~\eqref{eq:five-near-planar-chart} and
\eqref{eq:five-near-planar-limit}, the original-coordinate vector determined
by HRF and its weights are
\begin{equation}
 \vHR=(-2,-2,-2,-2,-2,-2;1),
 \qquad (\WSL,\WHR)=(-6,-4).
 \label{eq:five-near-planar-vector}
\end{equation}
The comparison announced above is now fixed.  Every cubic monomial in
\(\mathcal F_0\) has raw weight \(-6\).  The Gram-normal factor
\(\Gamma_5=O(\lambda^2)\) raises the weight of
\(\Gamma_5\mathcal F_{\Gamma_5}\) to \(-4\), while the quadratic polynomial
\(\mathcal U\) also has weight \(-4\).  Cancellation on the singular locus
promotes \(\mathcal F_0\) from \(\WSL=-6\) to the same resolved weight
\(\WHR=-4\).  Thus the cancelled leading sector, the first Gram-normal
correction and \(\mathcal U\) all enter the resolved LP polynomial at
\(\WHR\).

\medskip
\noindent\textbf{Parameter- and momentum-space power counting.}\par\smallskip

The HRF construction has now produced the generators and determined the
scaling vector in the original coordinates.  We first count directly in the
original LP-parameter domain.  This count does not itself require dissection;
the final HR certificate is supplied below.  The vector in
Eq.~\eqref{eq:five-near-planar-vector} gives
\(x_e\sim\lambda^{-2}\) for all six parameters.  HRF also computes the
cancellation depth as the gap
\begin{equation}
 \Delta W=\WHR-\WSL=(-4)-(-6)=2.
 \label{eq:five-near-planar-cancellation-depth}
\end{equation}
Every term of \(\FSL\) in
Eq.~\eqref{eq:five-full-stationary-decomposition} has the form
\(x_e f_I f_J\).  If \(d_I\) denotes the reduction of the natural normal
width by powers of \(\lambda\), equal resolved weights require
\(d_A+d_B=d_B+d_C=d_C+d_A=\Delta W=2\), and hence
\(d_A=d_B=d_C=1\).  The cancellation domain therefore restricts the three
independent normal combinations to
\begin{equation}
 |x_0-\rho_Ax_1|=O(\lambda^{-1}),\qquad
 |x_2-\rho_Bx_3|=O(\lambda^{-1}),\qquad
 |x_4-\rho_Cx_5|=O(\lambda^{-1}).
 \label{eq:five-near-planar-restricted-domain}
\end{equation}
Unlike the single-product Fish decomposition, the three pair products form a
full-rank system for the normal depths, so no continuous allocation freedom
remains.
Without these restrictions the six measures would give
\(\lambda^{-12}\).  Because of the cancellation, each normal has width
\(O(\lambda^{-1})\) rather than the unrestricted width
\(O(\lambda^{-2})\), restoring one power of \(\lambda\).  Since the
resolved LP polynomial has weight \(-4\), the
original-variable parameter-space count is therefore
\begin{equation}
 I_{\rm par}^{\rm scalar}
 \sim
 \underbrace{\lambda^{-12}}_{\text{unrestricted}}
 \;\underbrace{\lambda^{3}}_{\text{restricted widths}}
 \;\underbrace{\bigl(\lambda^{-4}\bigr)^{-D/2}}_{\mathcal P^{-D/2}}
 =\lambda^{2D-9}=\lambda^{-1-4\epsilon},
 \qquad D=4-2\epsilon.
 \label{eq:five-near-planar-parameter-power}
\end{equation}

The dissection supplying that certificate independently reproduces this
count, although it is not used here to evaluate the integral.  Introduce the
signed local coordinates in a first step,
\begin{equation}
 \begin{array}{l@{\hspace{3em}}l@{\hspace{3em}}l}
 u_A=x_1,&y_A=f_A,&x_0=\rho_Au_A+y_A,\\
 u_B=x_3,&y_B=f_B,&x_2=\rho_Bu_B+y_B,\\
 u_C=x_5,&y_C=f_C,&x_4=\rho_Cu_C+y_C.
 \end{array}
 \label{eq:five-near-planar-signed-local-coordinates}
\end{equation}
Thus the six new coordinates are
\((u_A,u_B,u_C;y_A,y_B,y_C)\): three tangential path magnitudes and three
signed normal displacements from the cancellation locus.  At fixed
kinematics the transformation from the six \(x_e\) to these six coordinates
has unit absolute Jacobian.  Only in a second step is the domain divided into
sign sectors,
\begin{equation}
 y_I=\sigma_I t_I,\qquad t_I\geq0,\qquad \sigma_I=\pm1.
 \label{eq:five-near-planar-sign-sectors}
\end{equation}
The cancellation locus is now the boundary \(t_A=t_B=t_C=0\).  In every
sector the corresponding ordinary lower facet has
\(u_I\sim\lambda^{-2}\) and \(t_I\sim\lambda^{-1}\), whose pullback is
Eq.~\eqref{eq:five-near-planar-vector}.  The dissected parameter-space count
is consequently
\begin{equation}
 I_{\rm dis}^{\rm scalar}
 \sim
 \underbrace{(\lambda^{-2})^3(\lambda^{-1})^3}
             _{\prod_I\dd u_I\,\dd t_I}
 \bigl(\lambda^{-4}\bigr)^{-D/2}
 =\lambda^{2D-9}=\lambda^{-1-4\epsilon},
 \label{eq:five-near-planar-dissection-power}
\end{equation}
in agreement with the direct count in
Eq.~\eqref{eq:five-near-planar-parameter-power}.

The paired sign sectors also expose a symmetry of the subleading HR
expansion.  Reversing the orientation transverse to the planar configuration
sends \(\epsilon_5\to-\epsilon_5\), equivalently \(\lambda\to-\lambda\),
while leaving all scalar products and
\(\Gamma_5\propto\epsilon_5^2\) unchanged.  In the local coordinates this
transformation pairs sectors through
\((y_A,y_B,y_C)\to(-y_A,-y_B,-y_C)\).  The local expansion is invariant
under the simultaneous reversal of \(\lambda\) and the three normal
coordinates, so contributions at odd relative powers of \(\lambda\) cancel
after the paired sectors are summed.  For scalar integrals and parity-even
numerators, the HR contribution therefore has the all-order structure
\begin{equation}
 I_{\rm HR}\sim\lambda^{-1-4\epsilon}
 \left[c_0(\epsilon)+c_1(\epsilon)\lambda^2
 +c_2(\epsilon)\lambda^4+\cdots\right].
 \label{eq:five-near-planar-even-expansion}
\end{equation}
Parity-odd numerators are not constrained in this way.

Since Eq.~\eqref{eq:five-near-planar-chart} gives
\(w-\bar w=i\lambda\) before any MRK hierarchy is imposed,
Eq.~\eqref{eq:five-near-planar-even-expansion} also describes the local
two-sided behaviour about the planar locus at general five-point
kinematics.  This two-sided structure is therefore inherited by its MRK
specialisation, in which $w,\bar w$ become the customary variables
$z,\bar z$.  Earlier MRK studies encountered the same non-trivial
continuation of parity-odd non-planar integrals across $z=\bar z$, together
with related cancellations in the ${\cal N}=4$ sYM hard function and in
four-dimensional QCD finite remainders; the complete two-loop Lipatov
vertex has no physical singularity on this locus
\cite{Caron-Huot:2020vlo,Buccioni:2024gzo,Abreu:2024xoh}.

Finally, we give a momentum-space consistency count for a physical
realisation of this HR.  Throughout this momentum-space count, let
\[
 P_i\equiv p_i\big|_{\lambda=0}
\]
denote the complete external four-momenta at the exactly planar point.  The
parameter-space vector fixes the six propagator
virtualities through Eq.~\eqref{eq:schwinger-inverse-virtuality}, but a
compatible assignment of momentum components requires additional physical
input.  Let us use the routing shown in
Fig.~\ref{fig:five-point-near-planar-attachment}.  The two independent loop
momenta are \(q_A\) and \(q_B\), while conservation at the upper common hard
vertex fixes
\begin{equation}
 q_C=-p_3-q_A-q_B.
 \label{eq:five-near-planar-loop-routing}
\end{equation}
The matching edge colours in
Fig.~\ref{fig:five-point-near-planar-attachment} display the three path
assignments: the two edges on path \(A\) carry \(q_A\) and \(q_A+p_1\), those
on path \(B\) carry \(q_B\) and \(q_B+p_2\), and those on path \(C\) carry
\(q_C\) and \(q_C-p_5\).  The parameter-space vector makes both propagators
on every path null at \(\lambda=0\).  For example,
\((q_A^\star)^2=(q_A^\star+P_1)^2=P_1^2=0\) implies
\(P_1\mathbin\cdot q_A^\star=0\), and the analogous relations hold on paths
\(B\) and \(C\).  On the physical real branch, each pair of null momenta
with vanishing scalar product is collinear.  Hence the central path momenta
lie on the \(P_1\), \(P_2\) and \(P_5\) rays, respectively; this conclusion
follows from the virtualities and the on-shell Landau branch rather than
being an independent mode assumption.

Expanding each two-propagator pair about its central ray supplies the usual
collinear fluctuation envelope.  For each path \(I=A,B,C\), let
\(\bar P_I\) denote a conjugate null reference direction and use the local
basis \((+,-,\perp)_{(P_I,\bar P_I)}\), with \(P_I=P_1,P_2,P_5\), respectively.
The component widths are
\begin{equation}
 \bigl(\Delta q_I^+,\Delta q_I^-,|\Delta q_I^\perp|\bigr)_{(P_I,\bar P_I)}
 \sim (1,\lambda^2,\lambda)_{(P_I,\bar P_I)},
 \qquad I=A,B,C,
 \label{eq:five-near-planar-collinear-modes}
\end{equation}
up to interchanging the two lightcone components according to orientation.
The companion edge on the same path has the same envelope.  Consequently
each marginal independent collinear loop measure scales as
\begin{equation}
 \dd^Dq_I
 \sim\dd q_I^+\,\dd q_I^-\,\dd^{D-2}q_I^\perp
 \sim\lambda^D,
\qquad I=A,B.
 \label{eq:five-near-planar-collinear-measures}
\end{equation}
These two marginal envelopes are not yet the complete joint support: the
dependent path \(C\) correlates them.  Before imposing that correlation, the allowed ranges of the
longitudinal fraction, the conjugate lightcone component and each transverse
component are \(O(1)\), \(O(\lambda^2)\) and \(O(\lambda)\), respectively.
The additional restriction on the two independent longitudinal fractions
can be formulated covariantly.  For the two independent loop momenta write
\begin{equation}
 q_A=\alpha_AP_1+k_A,
 \qquad
 q_B=\alpha_BP_2+k_B,
 \label{eq:five-near-planar-independent-loop-decomposition}
\end{equation}
where the large components are absorbed into \(\alpha_A,\alpha_B\), while in the
respective collinear bases the residual momenta have conjugate lightcone
components of order \(\lambda^2\) and transverse components of order
\(\lambda\).  The third path momentum is not independent: as defined in
Eq.~\eqref{eq:five-near-planar-loop-routing},
\(q_C=-p_3-q_A-q_B\).  At the exact pinch the residual momenta vanish and
the three path momenta obey
\[
 q_A^\star=\alpha_A^\star P_1,\qquad
 q_B^\star=\alpha_B^\star P_2,\qquad
 q_C^\star=\alpha_C^\star P_5.
\]
Consequently momentum conservation at \(\lambda=0\) imposes the constraint
\begin{equation}
 P_5=-\frac{P_3
 +\alpha_A^\star P_1+\alpha_B^\star P_2}
 {\alpha_C^\star}.
 \label{eq:five-near-planar-pinch-conservation}
\end{equation}

To express the allowed departure of \(q_C\) from this ray invariantly, choose
a null vector \(\bar P_5\) along the conjugate lightcone direction, with
\(P_5\mathbin\cdot\bar P_5\ne0\), and decompose
\begin{equation}
 q_C=\alpha_CP_5+k_{C\perp}+\beta_C\bar P_5,
 \qquad
 k_{C\perp}\mathbin\cdot P_5
 =k_{C\perp}\mathbin\cdot\bar P_5=0.
 \label{eq:five-near-planar-p5-decomposition}
\end{equation}
The collinear envelope of the \(p_5\) path means
\(k_{C\perp}=O(Q\lambda)\) and
\(\beta_C=O(\lambda^2)\), where \(\bar P_5\) is taken to be hard.
Two dimensionless Lorentz scalars that vanish on the \(P_5\) ray
are
\begin{equation}
\begin{aligned}
 \Delta_5(q_C)&=
 \frac{P_1\mathbin\cdot q_C}
      {P_1\mathbin\cdot P_5}
 -\frac{P_2\mathbin\cdot q_C}
      {P_2\mathbin\cdot P_5},\\[1mm]
 \Sigma_5(q_C)&=
 \frac{
 (P_1\mathbin\cdot P_2)
 (P_5\mathbin\cdot q_C)}
 {(P_1\mathbin\cdot P_5)
  (P_2\mathbin\cdot P_5)}.
\end{aligned}
\label{eq:five-near-planar-invariant-mismatches}
\end{equation}
For \(q_C=\alpha_CP_5\), the two ratios in \(\Delta_5(q_C)\) both
equal \(\alpha_C\), so they cancel in the difference and
\(\Delta_5(P_5)=0\).  Likewise \(\Sigma_5(P_5)=0\) because \(P_5^2=0\).

All denominator factors are fixed by
Eq.~\eqref{eq:five-near-planar-chart} at \(\lambda=0\), and are hard and
nonzero at generic wide angle.
Substitution of Eq.~\eqref{eq:five-near-planar-p5-decomposition} gives
\begin{align}
 \Delta_5(q_C)
 &=\frac{P_1\mathbin\cdot k_{C\perp}}
        {P_1\mathbin\cdot P_5}
   -\frac{P_2\mathbin\cdot k_{C\perp}}
        {P_2\mathbin\cdot P_5}
   +\beta_C\left(
      \frac{P_1\mathbin\cdot\bar P_5}
           {P_1\mathbin\cdot P_5}
     -\frac{P_2\mathbin\cdot\bar P_5}
           {P_2\mathbin\cdot P_5}\right)
 =O(\lambda),\nonumber\\
 \Sigma_5(q_C)
 &=\beta_C\,
   \frac{
   (P_1\mathbin\cdot P_2)
   (P_5\mathbin\cdot\bar P_5)}
   {(P_1\mathbin\cdot P_5)
    (P_2\mathbin\cdot P_5)}
 =O(\lambda^2).
 \label{eq:five-near-planar-invariant-widths}
\end{align}
Thus these invariant scalings express the local collinear envelope of the
dependent path.  Their effect on the joint support of the two independent
loop variables follows from the routing constraint.

Set \(\delta\alpha_I=\alpha_I-\alpha_I^\star\) for \(I=A,B\).  Inserting the
decompositions in
Eq.~\eqref{eq:five-near-planar-independent-loop-decomposition} into the
routing equation~\eqref{eq:five-near-planar-loop-routing}, and subtracting
its \(\lambda=0\) form in
Eq.~\eqref{eq:five-near-planar-pinch-conservation}, gives
\begin{equation}
 q_C-\alpha_C^\star P_5
 =-\delta\alpha_AP_1-\delta\alpha_BP_2+\ell_{\rm res},
 \qquad
 \ell_{\rm res}=-\bigl(p_3-P_3\bigr)-k_A-k_B.
 \label{eq:five-near-planar-residual-routing}
\end{equation}
Applying the two invariant projections gives
\begin{equation}
 \begin{pmatrix}
  \Delta_5(q_C)\\[1mm]
  \Sigma_5(q_C)
 \end{pmatrix}
 -
 \begin{pmatrix}
  \Delta_5(\ell_{\rm res})\\[1mm]
  \Sigma_5(\ell_{\rm res})
 \end{pmatrix}
 =
 (P_1\mathbin\cdot P_2)
 \begin{pmatrix}
  \dfrac{1}{P_2\mathbin\cdot P_5}
   &-\dfrac{1}{P_1\mathbin\cdot P_5}\\[3mm]
  -\dfrac{1}{P_2\mathbin\cdot P_5}
   &-\dfrac{1}{P_1\mathbin\cdot P_5}
 \end{pmatrix}
 \begin{pmatrix}
  \delta\alpha_A\\[1mm]
  \delta\alpha_B
 \end{pmatrix}.
 \label{eq:five-near-planar-longitudinal-fractions}
\end{equation}
The two vectors in Eq.~\eqref{eq:five-near-planar-longitudinal-fractions}
have opposite hierarchies.  The \(P_5\)-collinear path pair constrains
the complete dependent momentum \(q_C\), and
Eq.~\eqref{eq:five-near-planar-invariant-widths} gives
\(\Delta_5(q_C)=O(\lambda)\) and
\(\Sigma_5(q_C)=O(\lambda^2)\).  For the residual vector, the contractions
with \(P_1\) and \(P_2\) entering \(\Delta_5\) see the conjugate lightcone
components of \(k_A,k_B\), and the corresponding components of
\(p_3-P_3\), all of which are \(O(\lambda^2)\).  By contrast,
\(P_5\mathbin\cdot\ell_{\rm res}\) entering \(\Sigma_5\) generically sees
the \(O(\lambda)\) in-plane transverse components of \(k_A,k_B\).  Hence
\[
 \begin{pmatrix}
  \Delta_5(q_C)\\[1mm]\Sigma_5(q_C)
 \end{pmatrix}
 \sim
 \begin{pmatrix}\lambda\\[1mm]\lambda^2\end{pmatrix},
 \qquad
 \begin{pmatrix}
  \Delta_5(\ell_{\rm res})\\[1mm]\Sigma_5(\ell_{\rm res})
 \end{pmatrix}
 \sim
 \begin{pmatrix}\lambda^2\\[1mm]\lambda\end{pmatrix}.
\]
Thus the upper row of
Eq.~\eqref{eq:five-near-planar-longitudinal-fractions} is dominated by
\(\Delta_5(q_C)\).  In the lower row the \(O(\lambda)\) contribution from
\(\Sigma_5(\ell_{\rm res})\) must instead be cancelled by the longitudinal
fractions, leaving the required
\(\Sigma_5(q_C)=O(\lambda^2)\).  At fixed \(k_A,k_B\), this cancellation
fixes the centre of the allowed \((\delta\alpha_A,\delta\alpha_B)\) domain.
The determinant of the coefficient matrix is
\[
 -\frac{2(P_1\mathbin\cdot P_2)^2}
 {(P_1\mathbin\cdot P_5)
  (P_2\mathbin\cdot P_5)}\ne0
\]
at generic wide angle.  Since a translation does not change the measure, the
conditional volume follows from the two collinear ranges and this
nonsingular constant Jacobian:
\[
 \dd(\delta\alpha_A)\,\dd(\delta\alpha_B)
 =\frac{
 |(P_1\mathbin\cdot P_5)
  (P_2\mathbin\cdot P_5)|}
 {2|P_1\mathbin\cdot P_2|^2}
 \,\dd\Delta_5(q_C)\,\dd\Sigma_5(q_C)
 \sim\lambda\,\lambda^2=\lambda^3.
\]
The remaining transverse components are supplied by the residual momenta
already counted in the two collinear measures.  As required by the
parameter-space vector, all six propagator virtualities are then
\(q_e^2=O(\lambda^2)\).  Therefore
\begin{equation}
 I_{\rm mom}^{\rm scalar}
 \sim
 \underbrace{(\lambda^D)^2}_{\text{collinear measures}}
 \;\underbrace{\lambda^3}_{\text{restricted support}}
 \;\underbrace{(\lambda^{-2})^6}_{\text{propagators}}
 =\lambda^{2D-9}=\lambda^{-1-4\epsilon}.
 \label{eq:five-near-planar-momentum-power}
\end{equation}
The agreement of Eqs.~\eqref{eq:five-near-planar-parameter-power},
\eqref{eq:five-near-planar-dissection-power} and
\eqref{eq:five-near-planar-momentum-power} checks the cancellation depth
first in the original LP-parameter domain, then in local dissected coordinates,
and finally as restricted loop-momentum support.  In the
\((q_A,q_B)\) basis, the separate collinear envelopes have the same component
powers as their average momenta, but the additional factor \(\lambda^3\) is a
correlation in their joint support.  It cannot be recovered by comparing the
average-momentum scaling and marginal width of each loop separately.

The algebraic comparison with the Crown also has a direct momentum-space
interpretation.  At the vertex joining the two edges of each collinear path,
the attached external momentum is shared between the two propagator momenta.
This relation follows from the momentum-space Landau
equations before the Gaussian loop integrations are performed.  Using the
independent loop momenta \(q_A\) and \(q_B\) introduced above
Eq.~\eqref{eq:five-near-planar-loop-routing}, differentiate the Schwinger
exponent in Eq.~\eqref{eq:schwinger-representation} with respect to each.
The difference of the two equations cancels the common contribution from
the dependent momentum \(q_C\) and gives
\begin{equation}
 0=\frac12\left(
 \frac{\partial\mathcal S}{\partial q_A}
 -\frac{\partial\mathcal S}{\partial q_B}\right)
 =\widetilde x_0q_A+\widetilde x_1(q_A+p_1)
  -\widetilde x_2q_B-\widetilde x_3(q_B+p_2).
 \label{eq:five-near-planar-qA-Landau}
\end{equation}
At the pinch, Eq.~\eqref{eq:five-near-planar-qA-Landau} becomes
\[
 0=\bigl[(\widetilde x_0+\widetilde x_1)\alpha_A^\star
          +\widetilde x_1\bigr]P_1
   -\bigl[(\widetilde x_2+\widetilde x_3)\alpha_B^\star
          +\widetilde x_3\bigr]P_2.
\]
Since \(P_1\) and \(P_2\) are linearly independent at generic wide angle,
their coefficients vanish separately.  In particular,
\[
 \alpha_A^\star=-\frac{\widetilde x_1}
                         {\widetilde x_0+\widetilde x_1},
 \qquad
 q_A^\star=-\frac{\widetilde x_1}
                    {\widetilde x_0+\widetilde x_1}P_1,
 \qquad
 q_A^\star+P_1=\frac{\widetilde x_0}
                       {\widetilde x_0+\widetilde x_1}P_1.
\]
Thus the proper-time ratio
\(\widetilde r_A=\widetilde x_0/\widetilde x_1\) fixes the two
longitudinal shares.
The second coefficient above fixes the sharing on path \(B\); substituting
these results into either unsubtracted Landau equation fixes the analogous
sharing on path \(C\).

Equation~\eqref{eq:five-near-planar-qA-Landau} thereby supplies a
momentum-space interpretation of the three collinear-path sharings.
Physically, each external lightlike momentum splits at the vertex between
the two propagators of its path, and the two propagator momenta remain in the
same collinear sector.  This is the Coleman--Norton picture of the positive
pinch: the active massless propagators describe on-shell classical
trajectories, and the two null momenta on each path differ by the attached
null external momentum, which places them on the same collinear
ray~\cite{Coleman:1965cn,Collins:2020euz}.  Accommodating the three distinct
collinear sectors
simultaneously produces the momentum-space constraints on their longitudinal
shares.  The comparison with HRF is structural: the momentum-space Landau
equations fix these three collinear-splitting fractions, while the LP analysis
independently fixes the three ratios through
\(K\boldsymbol\rho=-\boldsymbol b\) in
Eq.~\eqref{eq:five-ratio-stationarity}.

In the Crown there are four such collinear two-edge paths: each external
lightlike particle splits into two propagator momenta in its own collinear
sector, and no external momentum is inserted at either common hard vertex.
Momentum conservation among the four collinear sectors leaves one common
sharing parameter free, which is the momentum-space counterpart of the zero
mode of \(K_C\).  In the Fish there are only three collinear two-edge paths,
carrying the \(p_1\)-, \(p_2\)- and \(p_5\)-collinear sectors, while \(p_3\)
and \(p_4\) enter at their two common hard vertices.  These hard insertions
supply the inhomogeneous terms \(\boldsymbol b\); since \(K\) is invertible,
the balance equations fix the unique ratio vector
\(\boldsymbol\rho=-K^{-1}\boldsymbol b\).  The remaining path-magnitude
Landau equation is \(\Phi(\boldsymbol\rho)=0\), which by
Eq.~\eqref{eq:five-stationary-gram} is equivalent to \(\Gamma_5=0\); it
ensures that these uniquely fixed collinear splittings are compatible with
the full five-point kinematics.
The contrast is therefore explicit: the Crown condition
\(r_A=r_B=r_C=r_D\) supplies only three independent relative constraints and
leaves their common value free, whereas the Fish fixes all three ratios
\(\rho_A,\rho_B,\rho_C\).  In momentum space this complete fixing restricts
the longitudinal integration domain; its measure is the \(\lambda^3\) factor
labelled ``restricted support'' in Eq.~\eqref{eq:five-near-planar-momentum-power}.

\subsubsection{Simultaneous fixed-transverse MRK and near-planarity}
\label{sec:five-point-mrk-planar}

An interesting limit, natural to consider here, combines the approach to
planarity with MRK.  The general kinematic chart in
Eq.~\eqref{eq:five-exact-kinematic-chart} provides a straightforward
parametrisation of this limit.  To separate its two ingredients, introduce
two independent positive dimensionless parameters.  The parameter \(x\)
controls the longitudinal hierarchy, and hence the rapidity separations,
through \(X_{34},X_{45}\propto x^{-1}\), while \(\lambda\) controls the
departure from planarity through the imaginary parts of \(w\) and
\(\bar w\):
\begin{align}
 X_{34}&=\frac{\widehat X_{34}}{x},&
 X_{45}&=\frac{\widehat X_{45}}{x},
 \nonumber\\
 w&=-\frac1\xi+\frac{i}{2}\lambda,&
 \bar w&=-\frac1\xi-\frac{i}{2}\lambda.
 \label{eq:five-mrk-planar-independent-limits}
\end{align}
with \(\widehat X_{34},\widehat X_{45},\xi>0\) and
\(|\mathbf p_4|^2>0\) fixed.  Thus \(x\to0\) produces the
fixed-transverse MRK hierarchy, whereas \(\lambda\to0\) approaches
planarity.  We correlate the two limits into a one-parameter family by
setting
\begin{equation}
 \lambda=\delta^a,\qquad x=\delta^b,\qquad a>0,\quad b\geq0.
 \label{eq:five-mrk-planar-correlated-rates}
\end{equation}
For \(b>0\), both longitudinal ratios grow while all transverse scales remain
fixed, giving standard fixed-transverse MRK.  In this hierarchy \(p_2,p_3\)
carry large \(p^+\), while \(p_1,p_5\) carry large \(p^-\).  These rapidity
groupings do not identify either pair as a single collinear mode.  At
\(b=0\) the longitudinal hierarchy and
the corresponding grouping are absent, and the kinematics returns to the
near-planar wide-angle case analysed in
Sec.~\ref{sec:five-point-near-planar-seed}.

The transverse approach to planarity is controlled independently by
\(\lambda\).  To display it directly in momentum variables, the
\(w,\bar w\) relations in
Eq.~\eqref{eq:five-mrk-planar-independent-limits} may be rewritten as
\begin{equation}
 |\mathbf p_3|^2=|\mathbf p_4|^2
 \left(\frac1{\xi^2}+\frac{\lambda^2}4\right),
 \qquad
 2\mathbf p_3\!\cdot\!\mathbf p_4=\frac{2|\mathbf p_4|^2}{\xi}.
 \label{eq:five-mrk-planar-transverse-chart}
\end{equation}
It follows that
\begin{equation}
 \frac{\Gamma_5}{s_{12}^2}
 =-|\mathbf p_4|^4\lambda^2
 =-|\mathbf p_4|^4\delta^{2a}.
 \label{eq:five-mrk-planar-normalized-gram}
\end{equation}
The Gram invariant \(\Gamma_5\) encodes the departure from planarity.  In the
normalisation of Eq.~\eqref{eq:five-mrk-planar-normalized-gram}, and since
\(|\mathbf p_4|^2\) is fixed in the present limit, this departure is governed
directly by \(\lambda=\delta^a\).

\begin{figure}[!b]
\centering
\begin{tikzpicture}[scale=.39]
\begin{scope}[xshift=0cm]
 \coordinate (v1) at (2,3);
 \coordinate (v2) at (2,7);
 \coordinate (v3) at (5,8);
 \coordinate (v4) at (8,5);
 \coordinate (v5) at (5,2);
 \draw[very thick] (v3) to[bend right=20] (v1);
 \draw[very thick] (v5) to[bend left=20] (v1);
 \draw[very thick] (v3) to[bend right=20] (v2);
 \draw[very thick] (v5) to[bend left=20] (v2);
 \draw[very thick] (v3) to[bend left=20] (v4);
 \draw[very thick] (v5) to[bend right=20] (v4);
 \draw[->,thick,Green] (0,2)--(v1);
 \draw[->,thick,olive] (0,8)--(v2);
 \draw[->,thick,olive!55!black] (v3)--(7,9);
 \draw[->,thick,blue] (v5)--(7.4,.65);
 \draw[->,thick,ForestGreen] (v4)--(10,5);
 \foreach \v in {1,2,4} \node[hrf vertex] at (v\v) {};
 \foreach \v in {3,5}
   \node[circle,fill=blue,draw=blue,minimum size=5.2pt,
         inner sep=0pt,outer sep=0pt] at (v\v) {};
 \node[hrf leg label,text=Green] at (.45,1.55) {$p_1$};
 \node[hrf leg label,text=olive] at (.45,8.45) {$p_2$};
 \node[hrf leg label,text=olive!55!black] at (6.2,9.1) {$p_3$};
 \node[hrf leg label,text=blue] at (7.45,.25) {$p_4$};
 \node[hrf leg label,text=ForestGreen] at (9.5,5.55) {$p_5$};
 \node[hrf edge label] at (3.50,5.75) {$x_0$};
 \node[hrf edge label] at (3.05,1.95) {$x_1$};
 \node[hrf edge label] at (3.05,8.05) {$x_2$};
 \node[hrf edge label] at (3.55,4.15) {$x_3$};
 \node[hrf edge label] at (6.65,7.45) {$x_4$};
 \node[hrf edge label] at (6.65,2.55) {$x_5$};
 \node[font=\scriptsize,align=center] at (5,-1.05)
   {(a) reference assignment: HR};
\end{scope}

\begin{scope}[xshift=11cm]
 \coordinate (v1) at (2,3);
 \coordinate (v2) at (2,7);
 \coordinate (v3) at (5,8);
 \coordinate (v4) at (8,5);
 \coordinate (v5) at (5,2);
 \draw[very thick] (v3) to[bend right=20] (v1);
 \draw[very thick] (v5) to[bend left=20] (v1);
 \draw[very thick] (v3) to[bend right=20] (v2);
 \draw[very thick] (v5) to[bend left=20] (v2);
 \draw[very thick] (v3) to[bend left=20] (v4);
 \draw[very thick] (v5) to[bend right=20] (v4);
 \draw[->,thick,Green] (0,2)--(v1);
 \draw[->,thick,olive] (0,8)--(v2);
 \draw[->,thick,blue] (v3)--(7,9);
 \draw[->,thick,olive!55!black] (v5)--(7.4,.65);
 \draw[->,thick,ForestGreen] (v4)--(10,5);
 \foreach \v in {1,2,4} \node[hrf vertex] at (v\v) {};
 \foreach \v in {3,5}
   \node[circle,fill=blue,draw=blue,minimum size=5.2pt,
         inner sep=0pt,outer sep=0pt] at (v\v) {};
 \node[hrf leg label,text=Green] at (.45,1.55) {$p_1$};
 \node[hrf leg label,text=olive] at (.45,8.45) {$p_2$};
 \node[hrf leg label,text=blue] at (6.2,9.1) {$p_4$};
 \node[hrf leg label,text=olive!55!black] at (7.45,.25) {$p_3$};
 \node[hrf leg label,text=ForestGreen] at (9.5,5.55) {$p_5$};
 \node[font=\scriptsize,align=center] at (5,-1.05)
   {(b) hard-vertex reflection: HR};
\end{scope}

\begin{scope}[xshift=22cm]
 \coordinate (v1) at (2,3);
 \coordinate (v2) at (2,7);
 \coordinate (v3) at (5,8);
 \coordinate (v4) at (8,5);
 \coordinate (v5) at (5,2);
 \draw[very thick] (v3) to[bend right=20] (v1);
 \draw[very thick] (v5) to[bend left=20] (v1);
 \draw[very thick] (v3) to[bend right=20] (v2);
 \draw[very thick] (v5) to[bend left=20] (v2);
 \draw[very thick] (v3) to[bend left=20] (v4);
 \draw[very thick] (v5) to[bend right=20] (v4);
 \draw[->,thick,Green] (0,2)--(v1);
 \draw[->,thick,olive] (0,8)--(v2);
 \draw[->,thick,blue] (v3)--(7,9);
 \draw[->,thick,ForestGreen] (v5)--(7.4,.65);
 \draw[->,thick,olive!55!black] (v4)--(10,5);
 \foreach \v in {1,2,4} \node[hrf vertex] at (v\v) {};
 \foreach \v in {3,5}
   \node[circle,fill=blue,draw=blue,minimum size=5.2pt,
         inner sep=0pt,outer sep=0pt] at (v\v) {};
 \node[hrf leg label,text=Green] at (.45,1.55) {$p_1$};
 \node[hrf leg label,text=olive] at (.45,8.45) {$p_2$};
 \node[hrf leg label,text=blue] at (6.2,9.1) {$p_4$};
 \node[hrf leg label,text=ForestGreen] at (7.45,.25) {$p_5$};
 \node[hrf leg label,text=olive!55!black] at (9.5,5.55) {$p_3$};
 \node[font=\scriptsize,align=center] at (5,-1.05)
   {(c) $p_3\leftrightarrow p_5$: no HR};
\end{scope}
\end{tikzpicture}
\caption{External-leg permutations relevant to the simultaneous MRK and
near-planar analysis.  The internal Fish topology and the positions of its
three trivalent vertices and two blue four-valent vertices are fixed in all
panels.  The LP labels \(x_0,\ldots,x_5\) are displayed in panel~(a) and
apply to the corresponding edges in all three panels.  Panels~(a) and~(b)
are the only two assignments among the twelve inequivalent permutations for
which HRF finds an HR; they place \(p_1,p_2,p_5\) at the
trivalent vertices and \(p_3,p_4\) at the four-valent vertices.  Panel~(c)
interchanges \(p_3\) and \(p_5\) relative to panel~(b), leaving \(p_4\) at a
four-valent vertex, but has no HR for \(b>0\).  The external-leg colours
retain a distinct shade for each resolved direction.  Related green shades
group \(p_1,p_5\) at the large-\(p^-\) end of the rapidity ordering, and
related olive shades group \(p_2,p_3\) at the large-\(p^+\) end; this grouping
does not identify either pair as a single collinear mode.  The mid-rapidity
outgoing line \(p_4\) is blue.}
\label{fig:five-mrk-planar-attachments}
\end{figure}
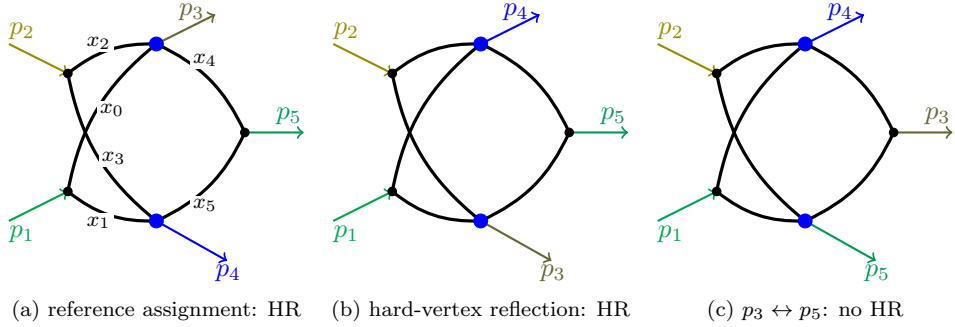

We now apply this composite limit to the Fish seed.  For the fixed incoming-leg
convention, the attachment audit finds an HR precisely for the two permutations
in which \(p_1,p_2,p_5\) are attached to the three trivalent path vertices and
\(p_3,p_4\) to the two four-valent vertices.  These two HR-supporting
permutations,
which differ only by exchanging \(p_3\) and \(p_4\) between the four-valent
vertices, are shown in Fig.~\ref{fig:five-mrk-planar-attachments}(a,b).
Panel~(c) displays the distinct \(p_3\leftrightarrow p_5\) interchange used
below to illustrate why putting \(p_4\) at a four-valent vertex is not by
itself sufficient.

To see how the MRK hierarchy changes the wide-angle near-planar stationary
locus, recall for the representative permutation in
Fig.~\ref{fig:five-mrk-planar-attachments}(a) the exact stationary identity
in Eq.~\eqref{eq:five-full-stationary-decomposition}:
\begin{align}
 \mathcal F={}&
 \underbrace{\frac{\Gamma_5}{4q_{AB}q_{BC}q_{CA}}\,x_1x_3x_5}
 _{\text{Gram-normal term}}
 \; + \;q_{AB}\,x_5\,g_{AB}
 \; + \;q_{BC}\,x_1\,g_{BC}
 \; + \;q_{CA}\,x_3\,g_{CA},
 \nonumber\\[-1mm]
 &g_{AB}=f_A\,f_B,\qquad
  g_{BC}=f_B\,f_C,\qquad
  g_{CA}=f_C\,f_A.
 \label{eq:five-mrk-planar-generators}
\end{align}
Here
\(f_A=x_0-\rho_Ax_1\), \(f_B=x_2-\rho_Bx_3\) and
\(f_C=x_4-\rho_Cx_5\).  The preceding subsection showed how the saturated
derivative ideal recovers these factors from the leading polynomial and
pairs them into the generator ideal
\(\langle f_Af_B,f_Bf_C,f_Cf_A\rangle\); here we retain that result without
repeating the construction.

Equation~\eqref{eq:five-mrk-planar-generators} is not the representation
supplied to HRF.  After inserting the composite scaling, HRF starts from the
unaligned leading polynomial \(\mathcal F_0\) in its ordinary Mandelstam
form.  The Gram-normal term is subleading in \(\delta\) and is therefore
absent from \(\mathcal F_0\).  We use the exact identity above only to track
how the previously established stationary structure and generator sectors
degenerate in the MRK hierarchy.  As established in
Eqs.~\eqref{eq:five-path-ratios} and~\eqref{eq:five-ratio-stationarity},
\(\boldsymbol\rho=(\rho_A,\rho_B,\rho_C)^T=-K^{-1}\boldsymbol b\) gives the
stationary values of the three path ratios
\((r_A,r_B,r_C)\).  Hence the generator locus lies in the positive LP
domain precisely when all three \(\rho_I\) are positive.

For this diagram, the composite limit defined by
Eqs.~\eqref{eq:five-mrk-planar-independent-limits} and
\eqref{eq:five-mrk-planar-correlated-rates} gives
\begin{equation}
 (\rho_A,\rho_B,\rho_C)
 \sim\left(\xi,\frac{x}{\widehat X_{34}},\xi\right).
 \label{eq:five-mrk-planar-ratios}
\end{equation}
For every nonzero \(x\) these ratios are positive, so the candidate locus
remains in the physical domain and passes the HRF positivity test.  The
trajectory approaches the boundary of the positive planar chamber through
\(\rho_B\to0\) as \(x\to0\).  Exchanging
\(p_3\) and \(p_4\) between the two four-valent vertices gives the
hard-vertex-reflected configuration in
Fig.~\ref{fig:five-mrk-planar-attachments}(b), which passes the same test;
both candidates are certified below.
The corresponding edge-momentum configuration for the representative
permutation is displayed below in Fig.~\ref{fig:five-mrk-planar-modes}.

The situation changes for Fig.~\ref{fig:five-mrk-planar-attachments}(c),
obtained from panel~(b) by interchanging \(p_3\) and \(p_5\) while keeping
\(p_1,p_2,p_4\) fixed.  At \(b=0\) this configuration still supports the
near-planar wide-angle HR of Sec.~\ref{sec:five-point-near-planar-seed}.
For \(b>0\), however, \(p_3\) and \(p_5\) approach the \(p_2\)- and
\(p_1\)-directed ends of the rapidity ordering, respectively, so their
interchange alone is not a symmetry of the MRK kinematics.  Its stationary
ratios behave as
\begin{equation}
 (\rho_A,\rho_B,\rho_C)_{p_3\leftrightarrow p_5}
 \sim\left(
 \widehat X_{45}\frac{(1+\xi)^2}{\xi^2x},
 -\frac{1+\xi}{\xi},
 -\frac{1+\xi}{\xi}\right).
 \label{eq:five-mrk-planar-p3-p5-negative-ratios}
\end{equation}
Two ratios are negative throughout the positive physical parameter domain,
so the stationary locus leaves \(\mathfrak C_{\rm Fish}^{+}\) and the HR is
lost.  A complete beam reflection would also interchange
\(p_1\leftrightarrow p_2\); it is distinct from the single outgoing-leg
interchange in panel~(c).

For either permutation with positive stationary ratios, solving the
homogeneity and hierarchy conditions of
Eq.~\eqref{eq:active-weight-system} for the unaligned expansion of the
complete \(\mathcal P=\mathcal U+\mathcal F\) determines, for each fixed
pair of rates \(a,b\), the candidate total vector
\begin{equation}
 \vHR^{\rm MRK+pl}=
 (-2a,-2a,b-2a,-2a,-2a,-2a;1).
 \label{eq:five-mrk-planar-vector}
\end{equation}
At \(b=0\), this reduces to the uniform near-planar wide-angle vector in
Eq.~\eqref{eq:five-near-planar-vector}, with \(\lambda=\delta^a\).
For the hard-vertex-reflected HR-supporting permutation, the corresponding
vector follows from the graph automorphism
\(x_0\leftrightarrow x_1\), \(x_2\leftrightarrow x_3\) and
\(x_4\leftrightarrow x_5\); its layer weights are unchanged.

The hierarchy solution can then be read directly from the stationary form in
Eq.~\eqref{eq:five-mrk-planar-generators}.  For \(b>0\), the relevant source
terms of the rescaled polynomial are
\begin{align}
 \mathcal F^{(\rm HR)}={}&
 \underbrace{\bigl(q_{AB}x_5g_{AB}+q_{BC}x_1g_{BC}\bigr)}_{
             \substack{W_{\rm SL}=-6a-b\\ \in\mathcal I_C^2}}
 +\underbrace{q_{CA}x_3g_{CA}}_{
             \substack{W_k=-6a\\ \in\mathcal I_C^2}}
 \nonumber\\[-1mm]
 &+\underbrace{
 \frac{\Gamma_5}{4q_{AB}q_{BC}q_{CA}}x_1x_3x_5}_{
 \substack{W_{\rm HR}=-4a\\ \notin\mathcal I_C}},
 \qquad
 \mathcal I_C=\langle f_A,f_B,f_C\rangle.
 \label{eq:five-mrk-planar-F-layers}
\end{align}
The first term defines \(W_{\rm SL}=-6a-b\).  The second has the intervening
nominal weight \(W_k=-6a\), strictly between \(W_{\rm SL}\) and
\(W_{\rm HR}=-4a\), yet its complete coefficient also vanishes on the same
pinch component.\footnote{Additional occupied layers need not be
intervening.  In the spacelike-collinear Fish, for example, the further
layers satisfy \(W_k>W_{\rm HR}\); none lies between \(W_{\rm SL}\) and
\(W_{\rm HR}\).}  The weights of the individual monomials in the first line
do not change.  Rather, both underbraced contributions vanish on the common
pinch component \(C=\{f_A=f_B=f_C=0\}\).  Resolving the cancellation in the
transverse normal directions suppresses their first nonzero contributions,
relative to their nominal layers, by \(\delta^{2a+b}\) and \(\delta^{2a}\),
respectively.  After cancellation, both therefore contribute effectively at
\(\delta^{-4a}\), alongside the nonvanishing Gram-normal term in the second line of
Eq.~\eqref{eq:five-mrk-planar-F-layers} and the leading term of
\(\mathcal U^{(\rm HR)}\), which also has weight \(-4a\).  This is the first
explicit application of the intervening ideal-layer test in
Eq.~\eqref{eq:finalhierarchy}.  At \(b=0\) the
first two weights in Eq.~\eqref{eq:five-mrk-planar-F-layers} coincide, so no
intervening layer remains.

What is inherited from the wide-angle Fish is the cancellation locus, not
its scaling: the relative MRK and Gram-normal rates determine both the total
vector and this layer structure.  In projective LP-parameter space the locus
itself moves towards a boundary, with only the middle path ratio tending to
zero.  Dissection is now applied as the final certification step.

This moving locus is most transparent in local coordinates
\begin{align}
 x_0&=\rho_Au_A+y_A,& x_1&=u_A,\nonumber\\
 x_2&=\rho_Bu_B+y_B,& x_3&=u_B,\nonumber\\
 x_4&=\rho_Cu_C+y_C,& x_5&=u_C.
 \label{eq:five-mrk-planar-local-coordinates}
\end{align}
Here \((u_A,u_B,u_C)\) are tangential path-magnitude coordinates along the
cancellation locus, whereas \((y_A,y_B,y_C)=(f_A,f_B,f_C)\) are signed
normal displacements from it, in the same notation as
Eq.~\eqref{eq:five-near-planar-signed-local-coordinates}.
The local dissection and lower-facet test give the weights
\begin{equation}
 (u_A,u_B,u_C,y_A,y_B,y_C)\sim
 \left(\delta^{-2a},\delta^{-2a},\delta^{-2a},
       \delta^{-a},\delta^{-a+2b},\delta^{-a}\right).
 \label{eq:five-mrk-planar-local-vector}
\end{equation}
Their pullback reproduces Eq.~\eqref{eq:five-mrk-planar-vector}, independently
checking the scaling already determined by the hierarchy equations.  The
exact leading local Lee--Pomeransky polynomial contains seven monomials of
affine rank six in the six local variables.  It therefore defines a full
lower facet and upgrades the candidate to a certified scaleful HR.

Having certified the region, we now determine its power counting in parameter
space.  For unit propagator powers, the local-coordinate weights in
Eq.~\eqref{eq:five-mrk-planar-local-vector} give the measure weight
\(-9a+2b\).  The physical layer in
Eq.~\eqref{eq:five-mrk-planar-F-layers} has weight \(-4a\), so
\(\mathcal P^{-D/2}\) contributes \(2aD\).  The resulting count is
\begin{equation}
 I_{\rm MRK+pl}^{\rm scalar}
 \sim\delta^{-9a+2b+2aD}.
 \label{eq:five-mrk-planar-parameter-power}
\end{equation}
Normalising the equal-rate point by a redefinition of \(\delta\) gives
\begin{equation}
 \begin{aligned}
 \vHR^{\rm MRK+pl}&=(-2,-2,-1,-2,-2,-2;1),\qquad
 W_{\rm SL}=-7,\qquad W_{\rm HR}=-4,\\
 \text{intermediate layer:}\quad W&=-6,\qquad
 I^{\rm scalar}\sim\delta^{1-4\epsilon}.
 \end{aligned}
 \label{eq:five-mrk-planar-equal-rate}
\end{equation}

Taken together, the hierarchy solution, local certification and power count
define a one-parameter family of certified HR scalings, since only the ratio
\(b/a\) is invariant under a redefinition of \(\delta\).  At \(b=0\) the
uniform near-planar wide-angle result is recovered and the intermediate layer
merges with the superleading layer; for \(b>0\) the MRK hierarchy makes the
vector non-uniform and separates these layers.  The boundary \(a=0\) removes
the approach to the Gram surface, so generic fixed-transverse MRK alone is not
certified by this construction.

\paragraph{Momentum-space interpretation.}
Because \(b=0\) recovers the near-planar wide-angle HR, the natural question
is how the MRK hierarchy deforms the average edge-momentum configuration and
the local integration widths within the same three-path organisation.  We keep
\(a,b\) general in the external kinematics, the average configuration and the
independent-loop power count below.  For the representative HR-supporting
permutation choose
\begin{equation}
 q_0=r,\quad q_1=p_1-r,\quad q_2=p_3-\ell,\quad
 q_3=\ell+p_2-p_3,\quad q_4=r-\ell,\quad
 q_5=p_5-r+\ell .
 \label{eq:five-mrk-planar-mode-routing}
\end{equation}
The variables \((r,\ell)\) provide a compact exact routing, but do not isolate
the asymptotic modes in the most transparent loop basis.  For the width
analysis we instead use
\(q_A\equiv q_1=p_1-r\) and \(q_C\equiv q_4=r-\ell\).  This affine change has
unit absolute Jacobian and aligns the two independent integrations with the
pinched propagator pairs used in the power count.
Choose the otherwise irrelevant longitudinal boost so that the nonzero
components of \(p_4\) are \(O(1)\).  Since the transverse scales remain
fixed, and writing each triple in the global frame as
\((p^+,p^-,|\boldsymbol p_\perp|)_{(p_2,p_1)}\), the external momenta then have the
schematic global-frame scalings
\begin{equation}
 \begin{aligned}
 p_1&\sim(0,\delta^{-b},\boldsymbol0_\perp),&
 p_2&\sim(\delta^{-b},0,\boldsymbol0_\perp),\\
 p_3&\sim(\delta^{-b},\delta^b,1),&
 p_4&\sim(1,1,1),&
 p_5&\sim(\delta^b,\delta^{-b},1),
 \end{aligned}
 \qquad (+,-,\perp)_{(p_2,p_1)},
 \label{eq:five-mrk-planar-external-modes}
\end{equation}
up to signs and nonzero \(O(1)\) coefficients.  Thus \(p_2,p_3\) share the
large-\(p^+\) end of the rapidity ordering, while \(p_1,p_5\) share the
large-\(p^-\) end.  Their transverse momenta remain fixed while the
corresponding large lightcone components grow as \(\delta^{-b}\), so the
normalised opening angles within both pairs scale as \(\delta^b\).
Equivalently,
\(-s_{23}/s_{12}\sim-s_{15}/s_{12}\sim\delta^{2b}\), although
\(s_{23}\) and \(s_{15}\) themselves remain of fixed order.  The two pairs
therefore occupy corresponding ends of the MRK rapidity ordering.  They must
not be identified as single collinear modes: their relative transverse
momenta remain of fixed order even while their normalised opening angles
shrink against the growing beam scale.

We first determine the average edge momenta \(q_e^\star\), in the
mixed-representation sense defined in
Sec.~\ref{sec:momentum-reconstruction}.  Writing
their global-frame component scaling as
\(\bigl((q_e^\star)^+,(q_e^\star)^-,|q_{e\perp}^\star|\bigr)
\sim(\delta^{a_e},\delta^{b_e},\delta^{c_e})_{(p_2,p_1)}\), the
mixed-representation stationary equations give
Table~\ref{tab:five-mrk-planar-edge-flow} for general \(a>0\) and \(b\geq0\);
Fig.~\ref{fig:five-mrk-planar-modes} displays the corresponding momentum
configuration.
\begin{figure}[!t]
\centering
\begingroup
\small
\begin{tabular}{@{}c c c c l@{}}
\toprule
edge & average \((a_e,b_e,c_e)_{(p_2,p_1)}\) & saddle \((q_e^\star)^2\)
 & region-scale \(q_e^2\) & interpretation\\
\midrule
\(q_0\) & \((2a+b,-b,a)\) & \(\delta^{2a}\) & \(\delta^{2a}\) & \(p_1\)-directed, path \(A\)\\
\(q_1\) & \((2a+b,-b,a)\) & \(\delta^{2a}\) & \(\delta^{2a}\) & \(p_1\)-directed, path \(A\)\\
\(q_2\) & \((-b,2a+b,a)\) & \(\delta^{2a}\) & \(\delta^{2a-b}\) & \(p_2\)-directed; deeper saddle value\\
\(q_3\) & \((0,2a+b,a)\) & \(\delta^{2a}\) & \(\delta^{2a}\) & vanishing fraction on path \(B\)\\
\(q_4\) & \((b,-b,0)\) & \(\delta^{2a}\) & \(\delta^{2a}\) & \(p_5\)-directed, fixed \(p_\perp\)\\
\(q_5\) & \((b,-b,0)\) & \(\delta^{2a}\) & \(\delta^{2a}\) & \(p_5\)-directed, fixed \(p_\perp\)\\
\bottomrule
\end{tabular}
\captionof{table}{Average edge-momentum modes, saddle virtualities and region-scale
virtualities for the simultaneous MRK and near-planar HR at general rates
\(a>0\) and \(b\geq0\).}
\label{tab:five-mrk-planar-edge-flow}
\endgroup
\vspace{1em}
\begin{tikzpicture}[scale=.52]
 \coordinate (v1) at (2,3);
 \coordinate (v2) at (2,7);
 \coordinate (v3) at (5,8);
 \coordinate (v4) at (8,5);
 \coordinate (v5) at (5,2);

 \draw[very thick,Green] (v3) to[bend right=20] (v1);
 \draw[very thick,Green] (v5) to[bend left=20] (v1);
 \draw[very thick,olive] (v3) to[bend right=20] (v2);
 \draw[very thick,olive] (v5) to[bend left=20] (v2);
 \draw[very thick,ForestGreen!55!black] (v3) to[bend left=20] (v4);
 \draw[very thick,ForestGreen!55!black] (v5) to[bend right=20] (v4);

 \draw[->,thick,Green] (0,2)--(v1);
 \draw[->,thick,olive] (0,8)--(v2);
 \draw[->,thick,olive!55!black] (v3)--(7,9);
 \draw[->,thick,blue] (v5)--(7.4,.65);
 \draw[->,thick,ForestGreen!55!black] (v4)--(10,5);

 \foreach \v in {1,2,4} \node[hrf vertex] at (v\v) {};
 \foreach \v in {3,5}
   \node[circle,fill=blue,draw=blue,minimum size=5.2pt,
         inner sep=0pt,outer sep=0pt] at (v\v) {};

 \node[hrf leg label,text=Green] at (.45,1.55) {$p_1$};
 \node[hrf leg label,text=olive] at (.45,8.45) {$p_2$};
 \node[hrf leg label,text=olive!55!black] at (6.25,9.15) {$p_3$};
 \node[hrf leg label,text=blue] at (7.45,.25) {$p_4$};
 \node[hrf leg label,text=ForestGreen!55!black] at (9.5,5.55) {$p_5$};

 \node[font=\scriptsize,anchor=west] at (3.15,5.68) {$q_0$};
 \node[font=\scriptsize,anchor=north] at (3.10,2.35) {$q_1$};
 \node[font=\scriptsize,anchor=south] at (3.12,7.58) {$q_2$};
 \node[font=\scriptsize,anchor=west] at (3.15,4.10) {$q_3$};
 \node[font=\scriptsize,anchor=south] at (6.88,7.25) {$q_4$};
 \node[font=\scriptsize,anchor=north] at (6.55,3.68) {$q_5$};
\end{tikzpicture}
\caption{Average edge-momentum configuration at the positive
Landau saddle of the MRK--planar HR for the representative permutation in
Fig.~\ref{fig:five-mrk-planar-attachments}(a).  The edge labels correspond to
Table~\ref{tab:five-mrk-planar-edge-flow}; green, olive and dark green show
paths \(A,B,C\), whose average flows approach the \(p_1,p_2,p_5\) directions.
For \(b>0\), \(q_3\) carries a fraction \(O(\delta^b)\) relative to \(q_2\);
at \(b=0\) they are of the same order.  This reconstruction has no independent
Glauber loop.}
\label{fig:five-mrk-planar-modes}
\end{figure}
The average-component and saddle-virtuality columns follow from the exact
mixed-representation stationary solution, whereas the region-scale column follows from the
LP-parameter vector.  Neither virtuality column is an integration width.  At
the stationary point, the leading \(O(\delta^b)\) minus and \(O(1)\) in-plane
transverse components cancel in \(q_2^\star=p_3-\ell^\star\).  Thus
\(q_2^\star\sim(\delta^{-b},\delta^{2a+b},\delta^a)_{(p_2,p_1)}\) and
\((q_2^\star)^2=O(\delta^{2a})\), although the region vector assigns
\(q_2^2=O(\delta^{2a-b})\) across the local integration region.
The routing variables in Eq.~\eqref{eq:five-mrk-planar-mode-routing} themselves
have average scalings
\begin{equation}
 r^\star\sim(\delta^{2a+b},\delta^{-b},\delta^a)_{(p_2,p_1)},\qquad
 \ell^\star\sim(\delta^b,\delta^b,1)_{(p_2,p_1)}.
 \label{eq:five-mrk-planar-routing-averages}
\end{equation}
Here \(r^\star=q_0^\star\), while
\(\ell^\star=r^\star-q_4^\star=p_3-q_2^\star\).  The
\(O(\delta^{-b})\) minus components cancel in the first difference and the
\(O(\delta^{-b})\) plus components cancel in the second.  For determining
integration widths we therefore use the mode-adapted variables \((q_A,q_C)\)
defined above.

With the average momentum directions established, we next determine the
integration widths and use them for the momentum-space power count.  We begin
with the LP-parameter vector in
Eq.~\eqref{eq:five-mrk-planar-vector}, which fixes the region-scale
propagator virtualities according to \(q_e^2\sim\delta^{-v_e}\); for example,
\begin{equation}
 q_1^2\sim\delta^{-v_1}=\delta^{2a},\qquad
 q_2^2\sim\delta^{-v_2}=\delta^{2a-b}.
 \label{eq:five-mrk-planar-q1-q2-virtualities}
\end{equation}
\begin{samepage}
Momentum conservation is built into the routing in
Eq.~\eqref{eq:five-mrk-planar-mode-routing}.  For general \(a>0\) and \(b>0\),
use the independent loop momenta \(q_A=q_1\) on path \(A\) and \(q_C=q_4\)
on path \(C\), introduced above.  The \(q_0,q_1\) and \(q_4,q_5\) pole pairs
pinch, respectively, the small lightcone components of \(q_A\) and \(q_C\).
In the corresponding mode-adapted frames, the separation of each pole pair
constrains the integration width of that component to
\(\delta^{2a+b}\).  Variations along the separate \(p_1\)- and
\(p_5\)-collinear rays scale as
\(\delta^{-b}\), while the region virtualities fix the transverse widths at
\(\delta^a\).  For these independent-loop widths we use a frame adapted to
each mode.  Let \(\bar p_i\) denote a conjugate null reference direction.
We may choose \(\bar p_1\parallel p_2\), so \((\bar p_1,p_1)\) agrees with the
global frame up to normalisation.  Since \(p_5\) has nonzero transverse
momentum in the global frame, \((\bar p_5,p_5)\) is instead a genuinely
different local basis.  The marginal widths are therefore
\begin{equation}
 \Delta q_A\sim
 (\delta^{2a+b},\delta^{-b},\delta^a)_{(\bar p_1,p_1)},\qquad
 \Delta q_C\sim
 (\delta^{2a+b},\delta^{-b},\delta^a)_{(\bar p_5,p_5)}.
 \label{eq:five-mrk-planar-independent-modes}
\end{equation}
\end{samepage}
Equation~\eqref{eq:five-mrk-planar-independent-modes} shows that the average
\(q_A=q_1\) triple in Table~\ref{tab:five-mrk-planar-edge-flow} agrees
component by component with its marginal width for general \(a,b\).  The
average \(q_C=q_4\) triple
\((2a+b,-b,a)_{(\bar p_5,p_5)}\) in its mode-adapted frame corresponds to the
global entry \((b,-b,0)_{(p_2,p_1)}\) in the table and likewise agrees with
its marginal width.  Thus the deeper saddle virtuality of the dependent edge
\(q_2\) does not contradict the average--width correspondence for the chosen
independent loop basis.

We can now perform the momentum-space power count.  Each balanced on-shell marginal
width contributes \(\delta^{aD}\).  The wide-angle correlation factor is
\(\lambda^3=\delta^{3a}\): it is the conditional longitudinal volume obtained
from the two invariant ranges and the nonsingular Jacobian displayed below
Eq.~\eqref{eq:five-near-planar-longitudinal-fractions}.  The MRK hierarchy
supplies one further restriction.  The path-\(B\) component of the stationary
equation~\eqref{eq:five-near-planar-qA-Landau} gives
\[
 1+\alpha_B^\star
 =\frac{\widetilde x_2}{\widetilde x_2+\widetilde x_3}
 =O(\rho_B)=O(\delta^b),
\]
where the Schwinger and LP ratios have the same regional scaling by
Eq.~\eqref{eq:schwinger-inverse-virtuality}.  Thus the two edges on the middle
path carry highly unequal longitudinal shares: in the routing of
Eq.~\eqref{eq:five-mrk-planar-mode-routing},
\(q_3=p_2-q_2\) is smaller than \(q_2\) by \(O(\delta^b)\).  Producing
\(q_3^+=O(1)\) from \(p_2^+,q_2^+=O(\delta^{-b})\) requires their leading
components to cancel to relative accuracy \(\delta^b\).  Since \(q_2\) is a
dependent combination of the two mode-adapted loop variables, this narrows
their joint longitudinal support, rather than either marginal width, by one
factor \(\delta^b\).  The resulting correlation factor is therefore
\(\delta^{3a+b}\).  The region vector fixes
\(-\sum_e v_e=12a-b\), so the six propagators contribute in total
\(\delta^{-(12a-b)}\).  Momentum space therefore gives
\begin{equation}
 I_{\rm MRK+pl}^{\rm scalar}
 \sim \delta^{aD}\,\delta^{aD}\,
       \delta^{3a+b}\,\delta^{-(12a-b)}
 =\delta^{-9a+2b+2aD},
 \label{eq:five-mrk-planar-momentum-power}
\end{equation}
in exact agreement with Eq.~\eqref{eq:five-mrk-planar-parameter-power}.
Thus the extra restriction in the count is a correlation of the collinear
loop variables.

\newpage

The qualitative path-flow asymmetry illustrated in
Table~\ref{tab:five-mrk-planar-edge-flow} and
Fig.~\ref{fig:five-mrk-planar-modes} persists for \(b>0\) and explains the
external-leg condition.
In the reference orientation, \(q_2\) is the only internal edge carrying the
leading \(p^+\) flow; \(q_3\) has a vanishing longitudinal fraction, while
paths \(A\) and \(C\) are directed towards the large-\(p^-\) end.  Momentum
conservation at the hard vertices therefore requires \(q_2\) to meet the
other large-\(p^+\) momentum, \(p_3\), at the upper four-valent vertex,
leaving the fixed-transverse momentum \(p_4\) at the lower one.
Interchanging \(q_2\) and \(q_3\) alone does not preserve this matching.  The
second HR-supporting assignment is instead the complete hard-vertex reflection: it
exchanges \(q_2\leftrightarrow q_3\) together with
\(p_3\leftrightarrow p_4\), \(q_0\leftrightarrow q_1\), and
\(q_4\leftrightarrow q_5\).

This example separates three logically distinct operations in HRF.  The
decomposition of \(\mathcal F\) identifies the cancellation factors and
generator ideal; positivity of the stationary ratios selects the external-leg
permutations for which the candidate locus lies in the positive LP domain;
and the local hierarchy together with the lower-facet test of the complete
\(\mathcal P=\mathcal U+\mathcal F\) determines the region vector and
certifies scalefulness.  Thus the three-generator cancellation ideal found for
the wide-angle near-planar Fish is inherited under the MRK deformation, while
the MRK hierarchy moves its stationary locus towards a boundary or, for a
different external-leg permutation, outside the positive domain.

\subsubsection{The central-soft rapidity-ordered limit}
\label{sec:five-point-central-soft-status}

This limit provides an alternative way to approach the normalised planar
configuration.  Rather than suppressing the parity-odd invariant at fixed
transverse scales, it combines strong rapidity ordering with one central
momentum becoming transversely soft.  It thereby separates features tied to
the planar singular locus from those that depend on the particular
asymptotic path used to reach it.

\paragraph{The soft limit.}

Starting again from the generic chart
\eqref{eq:five-exact-kinematic-chart}--\eqref{eq:five-zeta-chart}, introduce
two independent dimensionless parameters.  The parameter \(x\) controls the
two large rapidity gaps, while \(\tau\) measures the transverse softness of
the mid-rapidity momentum relative to the fixed scale
\begin{equation}
 Q_\perp^2\equiv |\mathbf p_3|^2
 =\frac{|\mathbf p_4|^2}{\zeta\bar\zeta}.
 \label{eq:five-central-soft-hard-transverse-scale}
\end{equation}
With \(\kappa,\bar\kappa,\widehat X_{34},\widehat X_{45}\) fixed and
nonzero, the transverse part of the exact path is
\begin{equation}
 \begin{aligned}
  \zeta&=\kappa\tau,&
  \bar\zeta&=\bar\kappa\tau,\\
  |\mathbf p_4|^2&=Q_\perp^2\kappa\bar\kappa\,\tau^2,&
  |\mathbf p_5|^2&=Q_\perp^2
   (1+\kappa\tau)(1+\bar\kappa\tau).
 \end{aligned}
 \label{eq:five-central-soft-transverse-path}
\end{equation}
The two transverse-magnitude relations follow respectively from
Eq.~\eqref{eq:five-central-soft-hard-transverse-scale} and transverse
momentum conservation, so they are not independent specifications.  In the
physical region \(\bar\kappa=\kappa^*\).  The longitudinal part of the same
parametrisation gives directly
\begin{equation}
 \begin{aligned}
  X_{34}&\equiv\frac{p_3^+}{p_4^+}
  =\frac{\widehat X_{34}}{x\tau},&
  X_{45}&\equiv\frac{p_4^+}{p_5^+}
  =\frac{\widehat X_{45}\tau}
  {x(1+\kappa\tau)(1+\bar\kappa\tau)}.
 \end{aligned}
 \label{eq:five-central-soft-X-path}
\end{equation}
In the representative frame \(p_4^+\propto\tau\), these invariant relations
imply \(p_3^+\propto x^{-1}\) and
\(p_5^+\propto x(1+\kappa\tau)(1+\bar\kappa\tau)\).  Using
Eq.~\eqref{eq:five-central-soft-transverse-path} and on-shellness then gives
\(p_5^-=|\mathbf p_5|^2/p_5^+\propto x^{-1}\).  Thus \(p_3^+\) and
\(p_5^-\) both become large.

The corresponding exponentials of rapidity differences, written directly in
terms of lightcone components, are
\begin{equation}
 \begin{aligned}
 Y_{34}\equiv e^{y_3-y_4}
 &=\sqrt{\frac{p_3^+p_4^-}{p_3^-p_4^+}}
 =X_{34}\sqrt{\zeta\bar\zeta}
 =\frac{\widehat X_{34}\sqrt{\kappa\bar\kappa}}{x},\\[1mm]
 Y_{45}\equiv e^{y_4-y_5}
 &=\sqrt{\frac{p_4^+p_5^-}{p_4^-p_5^+}}
 =X_{45}\sqrt{\frac{(1+\zeta)(1+\bar\zeta)}
                         {\zeta\bar\zeta}}
 =\frac{\widehat X_{45}}
 {x\sqrt{\kappa\bar\kappa
 (1+\kappa\tau)(1+\bar\kappa\tau)}}.
 \end{aligned}
 \label{eq:five-central-soft-rapidity-ratios}
\end{equation}
The final equalities follow by substituting
Eqs.~\eqref{eq:five-central-soft-transverse-path}
and~\eqref{eq:five-central-soft-X-path} into the component-ratio forms.
Thus the exponentials of the two rapidity gaps both scale as \(x^{-1}\), or
equivalently both gaps grow as \(\log x^{-1}\), whereas the transverse ratio
\(|\mathbf p_4|^2/|\mathbf p_3|^2
=\kappa\bar\kappa\tau^2\) vanishes.  At fixed nonzero
\(\tau\), the limit \(x\to0\) has the ordinary MRK hierarchy.  The joint
limit \(x,\tau\to0\), however, departs from the standard MRK scaling in which all
transverse ratios remain fixed.  We therefore call it the central-soft
rapidity-ordered limit, not a central-soft MRK sublimit.

For the one-parameter asymptotic expansion used by HRF we correlate the two
parameters as
\begin{equation}
 \tau=\delta^a,\qquad x=\delta^b,\qquad 0<a<b.
 \label{eq:five-central-soft-correlated-rates}
\end{equation}
The inequality ensures that both \(X_{34}\) and \(X_{45}\) diverge, but at
different rates:
\begin{equation}
 X_{34}=O(\delta^{-(a+b)}),\qquad
 X_{45}=O(\delta^{-(b-a)}).
 \label{eq:five-central-soft-unequal-X-rates}
\end{equation}
The unequal powers in Eq.~\eqref{eq:five-central-soft-unequal-X-rates} arise
because the transversely soft momentum \(p_4\) lies between the two rapidity
gaps: although \(Y_{34}\) and \(Y_{45}\) both scale as \(x^{-1}\), their
conversion to plus-momentum ratios introduces opposite powers of \(\tau\).

The remaining hard momenta approach a \(2\to2\) Regge configuration with
an additional soft emission.  Indeed, identify the hard subprocess as
\(p_1+p_2\to p_3+p_5\), and set \(s=s_{12}\) and \(t=s_{23}<0\).
In the global \((+,-,\perp)_{(p_2,p_1)}\) lightcone frame,
\(s=p_1^-p_2^+\) and \(-t=p_3^-p_2^+\).  Momentum conservation and
on-shellness therefore give
\begin{align}
 -\frac{t}{s}
 &=\frac{p_3^-}{p_1^-}
 =\frac{1}{1+X_{34}\zeta\bar\zeta
              +X_{34}X_{45}(1+\zeta)(1+\bar\zeta)}\nonumber\\
 &=\frac{x^2}{\widehat X_{34}\widehat X_{45}}
 \left[1+\frac{\kappa\bar\kappa}{\widehat X_{45}}\tau x
          +\frac{x^2}{\widehat X_{34}\widehat X_{45}}\right]^{-1}
 =\frac{\delta^{2b}}{\widehat X_{34}\widehat X_{45}}
  \left[1+O(\delta^{a+b})\right].
 \label{eq:five-central-soft-regge-ratio}
\end{align}
Thus \(-t/s\to0\): \(p_3\) becomes collinear to \(p_2\), while \(p_5\)
becomes collinear to \(p_1\).  The leading Regge ratio is independent of
the rate at which the additional momentum \(p_4\) becomes soft.

A sharp interpretation of the present limit is obtained by considering the
parity-odd invariant \(\epsilon_5\).  Combining its exact expression in
Eq.~\eqref{eq:five-exact-parity-odd-chart} with
Eqs.~\eqref{eq:five-central-soft-hard-transverse-scale}
and~\eqref{eq:five-central-soft-transverse-path} gives
\begin{equation}
 \frac{\epsilon_5}{s_{12}Q_\perp^2}
 =\zeta-\bar\zeta
 =\tau(\kappa-\bar\kappa).
 \label{eq:five-central-soft-gram}
\end{equation}
Near-planar wide-angle scattering sends \(\zeta-\bar\zeta\to0\) at fixed
\(|\mathbf p_4|^2\), whereas the present limit sends
\(\zeta,\bar\zeta\to0\) with their phase ratio generic.  Both approach the
normalised planar limit, but by distinct angular and soft degenerations;
this fact alone does not identify their HRs.

\paragraph{The Fish in the soft limit: order alignment and the HRF.}

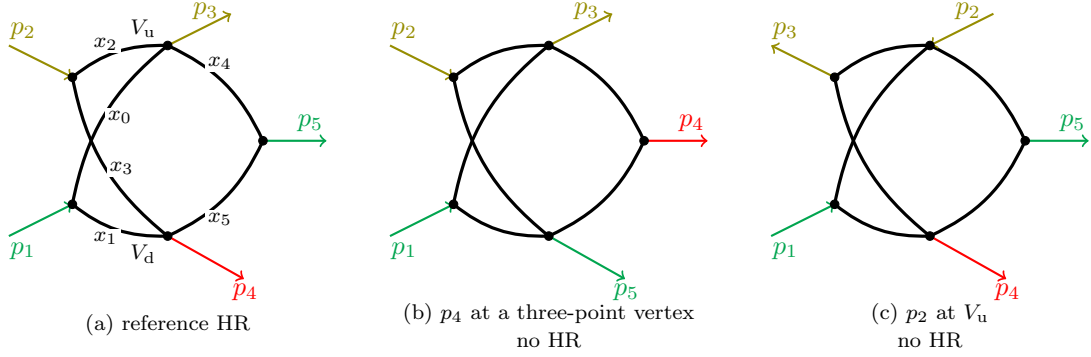
\begin{figure}[H]
\centering
\begin{tikzpicture}[scale=.42]
\begin{scope}[xshift=0cm]
 \coordinate (v1) at (2,3);
 \coordinate (v2) at (2,7);
 \coordinate (v3) at (5,8);
 \coordinate (v4) at (8,5);
 \coordinate (v5) at (5,2);

 \draw[very thick] (v3) to[bend right=20] (v1);
 \draw[very thick] (v5) to[bend left=20] (v1);
 \draw[very thick] (v3) to[bend right=20] (v2);
 \draw[very thick] (v5) to[bend left=20] (v2);
 \draw[very thick] (v3) to[bend left=20] (v4);
 \draw[very thick] (v5) to[bend right=20] (v4);

 \draw[->,thick,Green] (0,2)--(v1);
 \draw[->,thick,olive] (0,8)--(v2);
 \draw[->,thick,olive] (v3)--(7,9);
 \draw[->,thick,red] (v5)--(7.4,.65);
 \draw[->,thick,Green] (v4)--(10,5);
 \foreach \v in {1,2,3,4,5} \node[hrf vertex] at (v\v) {};

 \node[hrf leg label,text=Green] at (.45,1.55) {$p_1$};
 \node[hrf leg label,text=olive] at (.45,8.45) {$p_2$};
 \node[hrf leg label,text=olive] at (6.2,9.1) {$p_3$};
 \node[hrf leg label,text=red] at (7.45,.25) {$p_4$};
 \node[hrf leg label,text=Green] at (9.5,5.55) {$p_5$};

 \node[hrf edge label] at (3.50,5.75) {$x_0$};
 \node[hrf edge label] at (3.05,1.95) {$x_1$};
 \node[hrf edge label] at (3.05,8.05) {$x_2$};
 \node[hrf edge label] at (3.55,4.15) {$x_3$};
 \node[hrf edge label] at (6.65,7.45) {$x_4$};
 \node[hrf edge label] at (6.65,2.55) {$x_5$};
 \node[font=\scriptsize,anchor=east] at (4.95,8.55) {$V_{\rm u}$};
 \node[font=\scriptsize,anchor=east] at (4.95,1.45) {$V_{\rm d}$};
 \node[font=\scriptsize] at (5,-.80) {(a) reference HR};
\end{scope}

\begin{scope}[xshift=12cm]
 \coordinate (v1) at (2,3);
 \coordinate (v2) at (2,7);
 \coordinate (v3) at (5,8);
 \coordinate (v4) at (8,5);
 \coordinate (v5) at (5,2);

 \draw[very thick] (v3) to[bend right=20] (v1);
 \draw[very thick] (v5) to[bend left=20] (v1);
 \draw[very thick] (v3) to[bend right=20] (v2);
 \draw[very thick] (v5) to[bend left=20] (v2);
 \draw[very thick] (v3) to[bend left=20] (v4);
 \draw[very thick] (v5) to[bend right=20] (v4);

 \draw[->,thick,Green] (0,2)--(v1);
 \draw[->,thick,olive] (0,8)--(v2);
 \draw[->,thick,olive] (v3)--(7,9);
 \draw[->,thick,Green] (v5)--(7.4,.65);
 \draw[->,thick,red] (v4)--(10,5);
 \foreach \v in {1,2,3,4,5} \node[hrf vertex] at (v\v) {};

 \node[hrf leg label,text=Green] at (.45,1.55) {$p_1$};
 \node[hrf leg label,text=olive] at (.45,8.45) {$p_2$};
 \node[hrf leg label,text=olive] at (6.2,9.1) {$p_3$};
 \node[hrf leg label,text=Green] at (7.45,.25) {$p_5$};
 \node[hrf leg label,text=red] at (9.5,5.55) {$p_4$};
 \node[font=\scriptsize,align=center] at (5,-.80)
   {(b) \(p_4\) at a three-point vertex\\no HR};
\end{scope}

\begin{scope}[xshift=24cm]
 \coordinate (v1) at (2,3);
 \coordinate (v2) at (2,7);
 \coordinate (v3) at (5,8);
 \coordinate (v4) at (8,5);
 \coordinate (v5) at (5,2);

 \draw[very thick] (v3) to[bend right=20] (v1);
 \draw[very thick] (v5) to[bend left=20] (v1);
 \draw[very thick] (v3) to[bend right=20] (v2);
 \draw[very thick] (v5) to[bend left=20] (v2);
 \draw[very thick] (v3) to[bend left=20] (v4);
 \draw[very thick] (v5) to[bend right=20] (v4);

 \draw[->,thick,Green] (0,2)--(v1);
 \draw[->,thick,olive] (v2)--(0,8);
 \draw[->,thick,olive] (7,9)--(v3);
 \draw[->,thick,red] (v5)--(7.4,.65);
 \draw[->,thick,Green] (v4)--(10,5);
 \foreach \v in {1,2,3,4,5} \node[hrf vertex] at (v\v) {};

 \node[hrf leg label,text=Green] at (.45,1.55) {$p_1$};
 \node[hrf leg label,text=olive] at (.45,8.45) {$p_3$};
 \node[hrf leg label,text=olive] at (6.2,9.1) {$p_2$};
 \node[hrf leg label,text=red] at (7.45,.25) {$p_4$};
 \node[hrf leg label,text=Green] at (9.5,5.55) {$p_5$};
 \node[font=\scriptsize,align=center] at (5,-.80)
   {(c) \(p_2\) at \(V_{\rm u}\)\\no HR};
\end{scope}
\end{tikzpicture}
\caption{External-leg assignments used in the central-soft analysis.  The
internal Fish topology and its orientation are identical in all three
panels; the LP labels shown in panel~(a) are held fixed in panels~(b) and~(c).
The external colours encode only the kinematic roles: Green for the
\(p_1,p_5\) directions, olive for the \(p_2,p_3\) directions and red for the
soft momentum \(p_4\).  Panel~(a) has an HR.  Panel~(b), used in
Eq.~\eqref{eq:five-central-soft-middle-attachment}, moves \(p_4\) to the
remaining three-point vertex.  Panel~(c), used in
Eq.~\eqref{eq:five-central-soft-incoming-endpoint}, keeps \(p_4\) at
\(V_{\rm d}\) but moves the incoming particle \(p_2\) to \(V_{\rm u}\).}
\label{fig:five-central-soft-attachments}
\end{figure}

Let us now study this limit in the context of the Fish topology, which is the
simplest topology where an HR arises.  The three external-leg assignments
needed below are collected in
Fig.~\ref{fig:five-central-soft-attachments}.  Their internal orientation and
LP-parameter labels are fixed; only the external legs are permuted.  We denote
the upper and lower four-point vertices by \(V_{\rm u}\) and \(V_{\rm d}\),
respectively.  Panel~(a) defines the representative HR assignment used in the
construction below, with the soft leg attached to \(V_{\rm d}\).  Panels~(b)
and~(c) are the two negative controls analysed subsequently.

For the representative central-soft assignment in
Fig.~\ref{fig:five-central-soft-attachments}(a) and \((a,b)=(1,2)\),
substitution of the kinematic expansion into the complete \(\mathcal F\)
gives, through native order~\(\delta^{-1}\),
\begin{align*}
 \mathcal F={}&
 -\delta^{-4}\widehat X_{34}\widehat X_{45}Q_\perp^2x_2
   (x_1x_4-x_0x_5)\\
 &+\delta^{-1}Q_\perp^2\Bigl[
   \SLcolour{\widehat X_{45}x_3(x_1x_4-x_0x_5)}
   +\widehat X_{45}x_2(x_1x_4-x_0x_5)\\[-1mm]
 &\hspace{36mm}
   -\widehat X_{34}\kappa\bar\kappa\,x_2x_5(x_0+x_1)
   \Bigr]
 +O(\delta^0).
\end{align*}
At fixed LP parameters the first line is therefore the entire leading
polynomial.  Its possible positive zero
\(x_1x_4-x_0x_5=0\) is not stationary, since
\(
 \partial_{x_1}\mathcal F_{-4}
 =-\widehat X_{34}\widehat X_{45}Q_\perp^2x_2x_4\ne0
\)
in the positive orthant.

This is the role of the asymptotic-order alignment wrapper of
Sec.~\ref{sec:alignment}: it brings terms from different native powers of
\(\delta\) to a common effective leading order before the core HRF search.
Under \(x_e\mapsto\delta^{\phi_e}x_e\), the choice below promotes the red
\(\delta^{-1}\) contribution to weight \(-10\), equal to that of the
\(\delta^{-4}\) term.  The two uncoloured structures at order
\(\delta^{-1}\) instead have weight \(-7\), and every term in
\(O(\delta^0)\) has higher weight.  The two weight-\(-10\) contributions
therefore give the selected face
\begin{align}
 \boldsymbol\phi&=(-3,-3,0,-3,-3,-3),\nonumber\\
 \mathcal F_{-10}^{\rm align}
 &=-\widehat X_{45}Q_\perp^2
 (\widehat X_{34}x_2-x_3)(x_1x_4-x_0x_5).
 \label{eq:five-central-soft-face}
\end{align}
The factorised structure in Eq.~\eqref{eq:five-central-soft-face} is the
central-soft specialisation of the cancellation sector \(s f_1f_2\) in the
spacelike-collinear polynomial~\eqref{eq:five-F0-obstruction}: it describes
the same two-factor Fish pinch locus, but here that locus becomes visible
only after order alignment.  The simultaneous zero of the two displayed
factors is stationary, so this aligned face supplies the starting polynomial
for the core construction.

To express the locus in the common ratio notation, recall from
Eq.~\eqref{eq:five-path-ratios} that
\(r_A=x_0/x_1\), \(r_B=x_2/x_3\) and \(r_C=x_4/x_5\), and let
\(\rho_I\) denote the value of \(r_I\) at the pinch.  Setting the two factors
in Eq.~\eqref{eq:five-central-soft-face} to zero gives
\begin{equation}
 \rho_B=\frac{1}{\widehat X_{34}}>0,
 \qquad \rho_A=\rho_C>0.
 \label{eq:five-central-soft-positive-chamber}
\end{equation}
To relate this result to the generic wide-angle locus, use the normal
polynomials \(f_A,f_B,f_C\) defined in
Eq.~\eqref{eq:five-path-normals}, whose pairwise products appear in the exact
decomposition~\eqref{eq:five-full-stationary-decomposition}.  On
Eq.~\eqref{eq:five-central-soft-positive-chamber}, the two factors of the
aligned face become
\begin{equation}
 x_1x_4-x_0x_5=x_1f_C-x_5f_A,\qquad
 \widehat X_{34}x_2-x_3=\widehat X_{34}f_B.
 \label{eq:five-region-ideal-relation-clean}
\end{equation}
As the kinematics enter the central-soft regime, one of the three independent
normal conditions of the generic planar Fish locus is lost on the aligned
leading face.  Consequently, the generic locus \(f_A=f_B=f_C=0\) is
contained in the larger locus defined by \(f_B=0\) and
\(x_1f_C-x_5f_A=0\).  Equivalently, \(r_B\) is fixed as in
Eq.~\eqref{eq:five-central-soft-positive-chamber}, while the common positive
value \(r_A=r_C\) remains free at this order.  Thus it is the kinematics that
approaches a boundary of the wide-angle domain; the corresponding leading
LP-parameter locus enlarges.  This singular-locus relation does not
determine the region vector.

Returning now to the HRF construction, the order-aligned face in
Eq.~\eqref{eq:five-central-soft-face} is passed to the core algorithm.  In the
notation of the composition law~\eqref{eq:alignment-composition}, it gives the
uniform second-stage scaling and the composed edge vector.  We label
quantities associated with this soft limit by the superscript \({\rm soft}\):
\begin{equation}
 \boldsymbol v_{\rm core}^{\rm soft}=(-1,-1,-1,-1,-1,-1),
 \qquad
 \widetilde{\boldsymbol v}^{\rm soft}
 =\boldsymbol\phi+\boldsymbol v_{\rm core}^{\rm soft}
 =(-4,-4,-1,-4,-4,-4).
 \label{eq:five-central-soft-vector}
\end{equation}

Equation~\eqref{eq:five-central-soft-vector} completes the determination of
the LP-parameter scaling.  To compare the layers of \(\mathcal F\) directly
with those of \(\mathcal U\), we make the explicit scale choice
\(\mu_{\rm LP}^2=s_{12}\) in Eq.~\eqref{eq:LP-representation}.  Thus the
dimensionless LP polynomial is
\(\mathcal P_{s_{12}}=\mathcal U+\mathcal F/s_{12}\).  Unlike the default
fixed-scale convention, this reference scale varies with \(\delta\), and its
scaling is included in the layers below.  In the central-soft limit,
\[
 s_{12}=\widehat X_{34}\widehat X_{45}Q_\perp^2\delta^{-4}
 +O(\delta^{-1}),
 \qquad
 \frac{Q_\perp^2}{s_{12}}
 =\frac{\delta^4}{\widehat X_{34}\widehat X_{45}}
  \bigl[1+O(\delta^3)\bigr].
\]
This kinematic scaling has already been inserted in
Eq.~\eqref{eq:five-central-soft-face}, which is why \(s_{12}\) does not
appear there explicitly.  Applying the composed vector in
Eq.~\eqref{eq:five-central-soft-vector}, the first two nonempty layers of
\(\mathcal F\), expressed in units of \(s_{12}\), are
\begin{align}
 \left.\frac{\mathcal F}{s_{12}}\right|_{
 x_e\mapsto\delta^{\widetilde v_e^{\rm soft}}x_e}
 ={}&
 -\frac{\delta^{-9}}{\widehat X_{34}}
  (\widehat X_{34}x_2-x_3)(x_1x_4-x_0x_5)\nonumber\\
 &-\frac{\delta^{-8}}{\widehat X_{34}\widehat X_{45}}
  x_0x_3x_4+O(\delta^{-7}).
 \label{eq:five-central-soft-physical-F}
\end{align}
The first term is the hard-scale-normalised form of the aligned cancellation
sector in Eq.~\eqref{eq:five-central-soft-face}.  In the
\(\mu_{\rm LP}^2=s_{12}\) normalisation the augmented vector and resolved
weights are
\begin{equation}
 \vec v_{\rm HR,s_{12}}^{\rm soft}
 =(-4,-4,-1,-4,-4,-4;1),
 \qquad (\WSL,\WHR)=(-9,-8).
 \label{eq:five-central-soft-physical-vector}
\end{equation}
The second term in Eq.~\eqref{eq:five-central-soft-physical-F} is the entire
next nonvanishing \(\mathcal F/s_{12}\) layer.  It remains nonzero on the
two-factor locus and has weight \(-8\), exactly the leading
\(\mathcal U\) weight.  In the representative labelling,
\begin{equation}
 \mathcal U_{-8}=x_0x_3+x_1x_3+x_0x_4+x_1x_4+x_3x_4
                 +x_0x_5+x_1x_5+x_3x_5.
 \label{eq:five-central-soft-U}
\end{equation}
The physical layer of \(\mathcal U+\mathcal F/s_{12}\) therefore contains
both this \(\mathcal F/s_{12}\) monomial and \(\mathcal U_{-8}\); because the
former survives on the cancellation locus, this layer is scaleful.  Together
with the promoted factorised terms, its resolved support has affine rank six
and the unique inward normal is precisely
Eq.~\eqref{eq:five-central-soft-physical-vector}.  The hierarchy gap is one.

This layered structure is the soft-limit counterpart of the
spacelike-collinear HR in
Eqs.~\eqref{eq:five-F0-obstruction}--\eqref{eq:five-HR-polynomial}.  After
translating the edge labels and specialising the kinematic coefficients, the
factorised superleading sector is the same.  In the spacelike-collinear
expansion the core HRF acts directly on \(\mathcal F_0\) and isolates the
single-monomial obstruction \(\FObs\); here alignment first exposes the same
cancellation sector with zero obstruction, while restoration of the next
native layer supplies the different monomial \(x_0x_3x_4\).  These different
monomials are tied to the different region vectors selected by the two
kinematic limits.  Thus, while found through different HRF discovery routes, the two
regions share the same cancellation locus and layered structure: in each case
an uncancelled monomial lies one weight above the common cancellation sector
and joins \(\mathcal U\) in the physical layer.

\paragraph{External-leg conditions.}

The attachment scan yields a simple characterisation of the Fish seed in
this limit.  An HR occurs precisely when the soft particle \(p_4\) is
attached to one of the two four-point vertices and the two incoming particles
\(p_1,p_2\) are attached to two distinct three-point vertices.  The remaining
outgoing particles occupy the unused three- and four-point vertices.  This
criterion accounts for all six inequivalent attachments with an HR and is
unchanged for the tested choices \((a,b)=(1,2),(1,3),(2,3)\) of the
exponents defined in Eq.~\eqref{eq:five-central-soft-correlated-rates}.

The reason for requiring the soft particle to be emitted from a four-point
vertex is already visible on the leading face exposed by the natural alignment
vector \(\boldsymbol\phi=(-3,-3,0,-3,-3,-3)\).  For the complete external-leg
assignment in Fig.~\ref{fig:five-central-soft-attachments}(b), in which
\(p_4\) is emitted from the three-point vertex, the aligned face is
\begin{equation}
 \mathcal F_{\rm mid}^{\rm align}
 =-\widehat X_{45}Q_\perp^2x_1
 \big[(\widehat X_{34}x_2-x_3)x_4
       +\widehat X_{34}x_2x_5\big].
 \label{eq:five-central-soft-middle-attachment}
\end{equation}
The bracket can vanish for positive parameters, but the zero is not
stationary:
\begin{equation}
 \frac{\partial}{\partial x_5}
 \big[(\widehat X_{34}x_2-x_3)x_4
       +\widehat X_{34}x_2x_5\big]
 =\widehat X_{34}x_2>0.
 \label{eq:five-central-soft-middle-derivative}
\end{equation}
Attaching \(p_4\) to a four-point vertex is not by itself sufficient.  In the
complete assignment of Fig.~\ref{fig:five-central-soft-attachments}(c), the
incoming particle \(p_2\) is attached to the other four-point vertex.  The
aligned face has, up to a nonzero overall coefficient and graph reflection,
the form
\begin{equation}
 \mathcal F_{\rm in,end}^{\rm align}
 \propto
 (\widehat X_{34}x_2+x_3)(x_1x_4-x_0x_5).
 \label{eq:five-central-soft-incoming-endpoint}
\end{equation}
The first factor is strictly positive in the positive orthant.  Setting only
the second factor to zero does not make the polynomial stationary, so this
attachment has no positive Landau pinch.

By contrast, the reference assignment in
Fig.~\ref{fig:five-central-soft-attachments}(a), in which both incoming
particles are attached to three-point vertices, gives the factorised aligned
face already displayed in Eq.~\eqref{eq:five-central-soft-face}, and hence the
HR certified above.  Here every three-point vertex carries an energetic
external particle, so all three two-edge paths can support the collinear
splittings required by the pinch.  Moving either incoming particle to the
second four-point vertex replaces the difference
\(\widehat X_{34}x_2-x_3\) by the strictly positive sum
\(\widehat X_{34}x_2+x_3\) in
Eq.~\eqref{eq:five-central-soft-incoming-endpoint}, eliminating the pinch.

For the hard-vertex-reflected configuration, the fixed-edge presentation is,
up to a nonzero overall coefficient,
\begin{align}
 \mathcal F_{\star,\mathrm{opp}}^{\rm soft}
 &\propto -(x_2-\widehat X_{34}x_3)(x_1x_4-x_0x_5),
 \nonumber\\
 \boldsymbol\phi_{\mathrm{opp}}
 &=(-3,-3,-3,0,-3,-3),
 &
 \boldsymbol v_{\rm core,opp}^{\rm soft}
 &=(-1,-1,-1,-1,-1,-1),
 \nonumber\\
 \vec v_{\mathrm{HR,opp},s_{12}}^{\rm soft}
 &=(-4,-4,-4,-1,-4,-4;1).
 \label{eq:five-central-soft-opposite-vector}
\end{align}
The reflected external-leg assignment is analysed independently with the same
fixed edge labels.  The graph automorphism
\(V_{\rm u}\leftrightarrow V_{\rm d}\) relates its certificate to the
representative one.

\paragraph{Power counting in parameter space.}

We continue in the \(\mu_{\rm LP}^2=s_{12}\) normalisation used in
Eqs.~\eqref{eq:five-central-soft-physical-F}
and~\eqref{eq:five-central-soft-physical-vector}.  Normalising the first
factor by the nonzero kinematic ratio \(X_{34}\), define the local
cancellation factors
\begin{equation}
 f_1^{\rm soft}=x_2-X_{34}^{-1}x_3
 =x_2-\widehat X_{34}^{-1}\delta^3x_3,
 \qquad f_2^{\rm soft}=x_1x_4-x_0x_5.
 \label{eq:five-central-soft-local-cancellation-factors}
\end{equation}
The second form of \(f_1^{\rm soft}\) follows from
Eqs.~\eqref{eq:five-central-soft-X-path}
and~\eqref{eq:five-central-soft-correlated-rates} with \((a,b)=(1,2)\), for
which \(X_{34}=\widehat X_{34}\delta^{-3}\).
The two factors have generic weights \(-1\) and \(-8\), respectively.  The
HRF gap fixes one additional power of suppression for their product.  We use
the endpoint chart that assigns this power to \(f_1^{\rm soft}\), so that its
resolved weight is \(0\), while \(f_2^{\rm soft}\) retains weight \(-8\).
The resolved polynomial \(\mathcal P_{s_{12}}\) therefore has weight
\(-8\), and its integrand factor scales as
\begin{equation}
 \mathcal P_{s_{12}}^{-D/2}
 \sim\bigl(\delta^{-8}\bigr)^{-D/2}=\delta^{4D}.
 \label{eq:five-central-soft-integrand-power}
\end{equation}

This asymmetric assignment of the cancellation depth is a choice of
dissection chart, not a statement that the physical pinch has only
codimension one.  The invariant component is still
\(f_1^{\rm soft}=f_2^{\rm soft}=0\).  The radial--ratio dissection of
Sec.~\ref{sec:five-point-sc-charts} can likewise represent the unit gap by a
radial variable controlling the product and a ratio variable distributing the
suppression between \(f_1^{\rm soft}\) and \(f_2^{\rm soft}\).  Its
pullback gives the vector in
Eq.~\eqref{eq:five-central-soft-physical-vector}; the endpoint chart above is
the dissection used for the certified power count.

For unit propagator powers, the parameter-space measure is counted separately
in the same chart.  Use
\((f_1^{\rm soft},x_0,x_1,x_3,x_4,x_5)\) as local coordinates.  Since
\(x_2=f_1^{\rm soft}+\widehat X_{34}^{-1}\delta^3x_3\), replacing
\(dx_2\) by \(df_1^{\rm soft}\) at fixed \(x_3\) has unit Jacobian.  The
resolved weight of \(df_1^{\rm soft}\) is zero, while the other five
differentials each have weight \(-4\).  Thus the full parameter-space measure
scales as \(\delta^{-20}\).  Combining it with the cancellation-aware
integrand in Eq.~\eqref{eq:five-central-soft-integrand-power} gives
\begin{equation}
 \frac{I_{\rm soft}^{\rm scalar}}{(s_{12})^{D-6}}
 \sim\delta^{4D-20}=\delta^{-4-8\epsilon},
 \qquad D=4-2\epsilon.
\label{eq:five-central-soft-parameter-power}
\end{equation}

\paragraph{Momentum-space comparison.}

The momentum-space reconstruction uses the same \(s_{12}\) normalisation.
For the representative external assignment choose from the outset the
mode-adapted independent variables
\(r=q_0\) and \(\ell_G=q_0-q_4=p_3-q_2\).  The exact routing is then
\begin{equation}
 (q_0,q_1,q_2,q_3,q_4,q_5)
 =(r,p_1-r,p_3-\ell_G,\ell_G+p_2-p_3,
   r-\ell_G,p_5-r+\ell_G).
 \label{eq:five-central-soft-routing}
\end{equation}
The region vector in Eq.~\eqref{eq:five-central-soft-physical-vector} gives
\[
 \frac{q_e^2}{s_{12}}
 \sim(\delta^4,\delta^4,\delta,\delta^4,\delta^4,\delta^4).
\]
The component-level reconstruction contains more information than these
virtualities.  The consequences of the parameter-space cancellations enter
through the certified region vector, which fixes the six virtuality powers
above.  To obtain a consistent momentum-space realisation, we substitute the
external scaling in
Eqs.~\eqref{eq:five-central-soft-transverse-path}--
\eqref{eq:five-central-soft-correlated-rates}, with \((a,b)=(1,2)\), into
the routing~\eqref{eq:five-central-soft-routing} and impose momentum
conservation on the branch realising those virtualities.  The resulting
compatible edge-momentum scalings are given in
Table~\ref{tab:five-central-soft-edge-scalings}.  In the global
\((+,-,\perp)_{(p_2,p_1)}\) lightcone frame, where \(p_2\) carries only plus
momentum and \(p_1\) only minus momentum, an
entry \((a_e,b_e,c_e)_{(p_2,p_1)}\) means
\[
 \left(\frac{q_e^+}{\sqrt{s_{12}}},
       \frac{q_e^-}{\sqrt{s_{12}}},
       \frac{|q_{e\perp}|}{\sqrt{s_{12}}}\right)_{(p_2,p_1)}
 \sim(\delta^{a_e},\delta^{b_e},\delta^{c_e})_{(p_2,p_1)}.
\]
These are valuations of a compatible edge-momentum routing at the pinch, not
the local widths of independent loop-integration variables.
\begin{table}[H]
\centering
\small
\begin{tabular}{@{}c c c l@{}}
\toprule
edge & \((a_e,b_e,c_e)_{(p_2,p_1)}\) & \(q_e^2/s_{12}\) & momentum scaling\\
\midrule
\(q_0\) & \((4,0,2)\) & \(\delta^4\) & \((p_1\text{-collinear})^2\)\\
\(q_1\) & \((4,0,2)\) & \(\delta^4\) & \((p_1\text{-collinear})^2\)\\
\(q_2\) & \((0,1,2)\) & \(\delta\)   & longitudinally dominated, \(p_2\)-directed\\
\(q_3\) & \((3,1,2)\) & \(\delta^4\) &
\(\text{soft}\times(p_1\text{-collinear})\)\\
\(q_4\) & \((4,0,2)\) & \(\delta^4\) & \((p_1\text{-collinear})^2\)\\
\(q_5\) & \((4,0,2)\) & \(\delta^4\) & \((p_1\text{-collinear})^2\)\\
\bottomrule
\end{tabular}
\caption{Complete edge-momentum scalings of a compatible routing for the representative
hidden-region assignment.  The component exponents and virtualities are
normalised to the hard scale \(s_{12}\).  They complement, and should not be
confused with, the local loop-integration widths in
Eq.~\eqref{eq:five-central-soft-hard-scale-widths}.}
\label{tab:five-central-soft-edge-scalings}
\end{table}
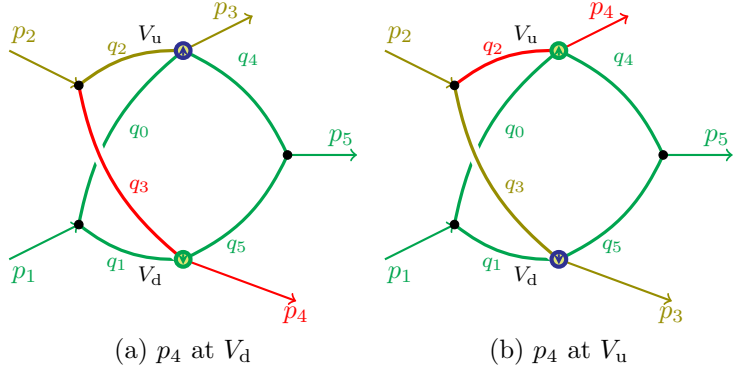
\begin{figure}[H]
\centering
\begin{tikzpicture}[scale=.46]
\begin{scope}[xshift=0cm]
 \coordinate (v1) at (2,3);
 \coordinate (v2) at (2,7);
 \coordinate (v3) at (5,8);
 \coordinate (v4) at (8,5);
 \coordinate (v5) at (5,2);

 \draw[very thick,Green] (v3) to[bend right=20] (v1);
 \draw[very thick,Green] (v5) to[bend left=20] (v1);
 \draw[very thick,olive] (v3) to[bend right=20] (v2);
 \draw[white,line width=5pt] (v5) to[bend left=20] (v2);
 \draw[very thick,red] (v5) to[bend left=20] (v2);
 \draw[very thick,Green] (v3) to[bend left=20] (v4);
 \draw[very thick,Green] (v5) to[bend right=20] (v4);

 \draw[->,thick,Green] (0,2)--(v1);
 \draw[->,thick,olive] (0,8)--(v2);
 \draw[->,thick,olive] (v3)--(7,9);
 \draw[->,thick,red] (v5)--(8.25,.8);
 \draw[->,thick,Green] (v4)--(10,5);

 \foreach \v in {1,2,4} \node[hrf vertex] at (v\v) {};

 \node[hrf leg label,text=Green] at (.45,1.55) {$p_1$};
 \node[hrf leg label,text=olive] at (.45,8.45) {$p_2$};
 \node[hrf leg label,text=olive] at (6.25,9.15) {$p_3$};
 \node[hrf leg label,text=red] at (8.25,.4) {$p_4$};
 \node[hrf leg label,text=Green] at (9.55,5.5) {$p_5$};

 \node[font=\scriptsize,text=Green,anchor=west] at (3.15,5.68) {$q_0$};
 \node[font=\scriptsize,text=Green,anchor=north] at (3.10,2.35) {$q_1$};
 \node[font=\scriptsize,text=olive,anchor=south] at (3.12,7.58) {$q_2$};
 \node[font=\scriptsize,text=red,anchor=west] at (3.15,4.10) {$q_3$};
 \node[font=\scriptsize,text=Green,anchor=south] at (6.88,7.25) {$q_4$};
 \node[font=\scriptsize,text=Green,anchor=north] at (6.55,2.86) {$q_5$};
 \GlauberVertex[Blue]{(v3)}{90}
 \GlauberVertex[Green]{(v5)}{-90}
 \node[font=\scriptsize,anchor=east] at (4.72,8.48) {$V_{\rm u}$};
 \node[font=\scriptsize,anchor=east] at (4.72,1.52) {$V_{\rm d}$};
 \node[font=\small] at (5,-.65) {(a) \(p_4\) at \(V_{\rm d}\)};
\end{scope}

\begin{scope}[xshift=10.8cm]
 \coordinate (v1) at (2,3);
 \coordinate (v2) at (2,7);
 \coordinate (v3) at (5,8);
 \coordinate (v4) at (8,5);
 \coordinate (v5) at (5,2);

 \draw[very thick,Green] (v3) to[bend right=20] (v1);
 \draw[very thick,Green] (v5) to[bend left=20] (v1);
 \draw[very thick,red] (v3) to[bend right=20] (v2);
 \draw[white,line width=5pt] (v5) to[bend left=20] (v2);
 \draw[very thick,olive] (v5) to[bend left=20] (v2);
 \draw[very thick,Green] (v3) to[bend left=20] (v4);
 \draw[very thick,Green] (v5) to[bend right=20] (v4);

 \draw[->,thick,Green] (0,2)--(v1);
 \draw[->,thick,olive] (0,8)--(v2);
 \draw[->,thick,red] (v3)--(7,9);
 \draw[->,thick,olive] (v5)--(8.25,.8);
 \draw[->,thick,Green] (v4)--(10,5);

 \foreach \v in {1,2,4} \node[hrf vertex] at (v\v) {};

 \node[hrf leg label,text=Green] at (.45,1.55) {$p_1$};
 \node[hrf leg label,text=olive] at (.45,8.45) {$p_2$};
 \node[hrf leg label,text=red] at (6.25,9.15) {$p_4$};
 \node[hrf leg label,text=olive] at (8.25,.4) {$p_3$};
 \node[hrf leg label,text=Green] at (9.55,5.5) {$p_5$};

 \node[font=\scriptsize,text=Green,anchor=west] at (3.15,5.68) {$q_0$};
 \node[font=\scriptsize,text=Green,anchor=north] at (3.10,2.35) {$q_1$};
 \node[font=\scriptsize,text=red,anchor=south] at (3.12,7.58) {$q_2$};
 \node[font=\scriptsize,text=olive,anchor=west] at (3.15,4.10) {$q_3$};
 \node[font=\scriptsize,text=Green,anchor=south] at (6.88,7.25) {$q_4$};
 \node[font=\scriptsize,text=Green,anchor=north] at (6.55,2.86) {$q_5$};
 \GlauberVertex[Blue]{(v5)}{-90}
 \GlauberVertex[Green]{(v3)}{90}
 \node[font=\scriptsize,anchor=east] at (4.72,8.48) {$V_{\rm u}$};
 \node[font=\scriptsize,anchor=east] at (4.72,1.52) {$V_{\rm d}$};
 \node[font=\small] at (5,-.65) {(b) \(p_4\) at \(V_{\rm u}\)};
\end{scope}

\end{tikzpicture}
\caption{Momentum-space realisation of the two hard-vertex assignments in the
central-soft rapidity-ordered limit.  Olive marks the edge carrying the
leading \(p_2,p_3\) lightcone direction, green the
\((p_1\text{-collinear})^2\) edges, and red the internal
\(\text{soft}\times(p_1\text{-collinear})\) edge adjacent to the soft
external momentum \(p_4\).  The complete component scalings for panel~(a)
are given in
Table~\ref{tab:five-central-soft-edge-scalings}; panel~(b) follows by the
reflection described in the text.
The GreenYellow discs follow the oriented Glauber-flow convention defined
in Fig.~\ref{fig:crown-regge-glauber} and used in
Ref.~\cite{Chen:2026dnj}.  Thus the upper and lower arrows point in opposite
directions.  In panel~(a)
the upper blue rim records the hard complementary channel and the lower green
rim the \(p_1\)-collinear one; these rim colours exchange in panel~(b).  In
panel~(a), the symbols at \(V_{\rm u}\) and \(V_{\rm d}\) correspond to
\(q_0-q_4=\ell_G\) and \(q_1-q_5=p_1-p_5-\ell_G\), respectively.
Panel~(b) is the hard-vertex-reflected assignment.}
\label{fig:five-central-soft-mode-reflection}
\end{figure}
Panel~(a) of Fig.~\ref{fig:five-central-soft-mode-reflection} corresponds
directly to the compatible edge modes in
Table~\ref{tab:five-central-soft-edge-scalings}.  The colours encode their
leading kinematic association, while the table records the complete
component scaling.  Panel~(b) is obtained by the hard-vertex-reflected
permutation \(q_2\leftrightarrow q_3\), together with
\(q_0\leftrightarrow q_1\) and \(q_4\leftrightarrow q_5\), and therefore
requires no second table.  The olive and red scalings exchange positions,
while all green edges retain \((p_1\text{-collinear})^2\).

Here \((p_1\text{-collinear})^2\) denotes the standard
\(p_1\)-collinear scaling with small parameter \(\delta^2\), so its exponent
vector is \(2(2,0,1)=(4,0,2)\).  Similarly, the
\(\text{soft}\times(p_1\text{-collinear})\) entry obeys
\((3,1,2)=(1,1,1)+(2,0,1)\).
The origin of the remaining entries is transparent in
Eq.~\eqref{eq:five-central-soft-routing}.  The momenta \(q_0=r\) and
\(q_1=p_1-r\), as well as \(q_4=r-\ell_G\) and
\(q_5=p_5-r+\ell_G\), share the
\((p_1\text{-collinear})^2\) scaling.  The momentum
\(q_2=p_3-\ell_G\) retains the leading \(p_2,p_3\) lightcone direction but
has a larger minus component and is therefore longitudinally dominated.
In \(q_3=\ell_G+p_2-p_3\), cancellation of the leading \(p_2,p_3\) plus
components gives the \(\text{soft}\times(p_1\text{-collinear})\) scaling.
All six propagators are infrared relative to the hard scale, with \(q_2\)
the least suppressed.

The Glauber momentum flow shown in Fig.~\ref{fig:five-central-soft-mode-reflection}
is \(\ell_G=q_0-q_4\) at \(V_{\rm u}\).  The exact routing in
Eq.~\eqref{eq:five-central-soft-routing} carries the same loop through
\(V_{\rm d}\), where the transfer is
\(q_1-q_5=p_1-p_5-\ell_G\).  Thus the two four-point vertices lie on one
Glauber exchange, rather than defining two independent loops.

Having identified a compatible momentum-mode assignment and its routing, we now turn
to the widths of the loop integrations around the pinch.  These are not
given by Table~\ref{tab:five-central-soft-edge-scalings}.  To determine a
width, hold the other loop components fixed
and regard each active propagator as a pole in the chosen integration
component \(u\).  Locally,
\[
 \mathcal D_e=A_eu+B_e+i0,
 \qquad
 u_e=-\frac{B_e}{A_e}-\frac{i0}{A_e}.
\]
Two propagators form a pinching pole pair when the signs of their coefficients
\(A_e\) are opposite: their poles then approach the real contour from
opposite sides.  The scaling of the separation between their real parts fixes
the local integration width \(\Delta u\).  This is a momentum-space pole
analysis performed after the parameter-space cancellation analysis has fixed
the propagator virtualities, not an additional cancellation condition.  The
general longitudinal pole rule and the transverse homogeneity rule used below
are collected in Sec.~\ref{sec:pole-pinches-local-widths}.

The following estimates use both kinds of information collected in
Table~\ref{tab:five-central-soft-edge-scalings}.  The component scalings fix
the pole coefficients \(A_e\), while the virtuality scalings fix the size of
the corresponding denominators; their ratio determines each pole
displacement and hence the local width.

In the routing~\eqref{eq:five-central-soft-routing}, the denominators
associated with \((q_0,q_1)\) are both of order \(\delta^4\), while their
coefficients of \(r^+\) are of order one.  Their opposite-side poles
therefore fix \(\Delta r^+\sim\delta^4\).  The unpinched minus component
has \(\Delta r^-\sim\delta^0\), set by the large minus component of
\(p_1\), while the transverse width is \(\delta^2\).  The pair
\((q_4,q_5)\) forms the \(p_5\)-collinear path, since \(q_4+q_5=p_5\).
Because \(p_5\) becomes collinear to \(p_1\) in this limit, the same global
lightcone frame applies: the two denominators are of order \(\delta^4\),
while their coefficients of \(\ell_G^+\) are of order one and have opposite
signs.  Hence \(\Delta\ell_G^+\sim\delta^4\).

To determine the remaining longitudinal width of \(\ell_G\), use the
\(q_2\) and \(q_3\) relations in Eq.~\eqref{eq:five-central-soft-routing},
\[
 q_2=p_3-\ell_G,
 \qquad
 q_3=\ell_G+p_2-p_3.
\]
Regarded as functions of \(\ell_G^-\), these two denominators supply the
pole pair in the global \(\ell_G^-\) contour.  Their coefficients are
\(-q_2^+\sim-\delta^0\) and
\(+q_3^+\sim\delta^3\), which have opposite signs on the physical branch.
Combining these coefficients with the corresponding virtualities, the two
pole displacements have the same scaling,
\[
 \frac{q_2^2}{q_2^+}\sim\delta,
 \qquad
 \frac{q_3^2}{q_3^+}\sim\frac{\delta^4}{\delta^3}=\delta.
\]
They therefore pinch the contour over the common width
\(\Delta\ell_G^-\sim\delta\).

The transverse width is not determined by another pole pair.  A transverse
displacement changes the three relevant
denominators according to
\begin{equation}
 \left.\Delta\mathcal D_e\right|_\perp
 \sim q_{e\perp}\mathbin\cdot\Delta\ell_{G\perp}
      +|\Delta\ell_{G\perp}|^2,
 \qquad e=3,4,5.
 \label{eq:five-central-soft-transverse-width-balance}
\end{equation}
The propagator \(q_2\) is excluded here because it is longitudinally
dominated: \(q_2^+q_2^-\sim\delta\), whereas
\(q_{2\perp}^2\sim\delta^4\).  Transverse changes of order \(\delta^4\)
are therefore subleading to its virtuality and do not determine the width.
For the three propagators obeying the balanced on-shell condition in
Eq.~\eqref{eq:five-central-soft-transverse-width-balance},
Table~\ref{tab:five-central-soft-edge-scalings} gives
\(|q_{e\perp}|\sim\delta^2\) and \(q_e^2\sim\delta^4\).  Requiring the
transverse displacement to preserve these virtuality orders, and taking the
widest allowed support, gives
\(|\Delta\ell_{G\perp}|\sim\delta^2\).  This is the explicit example of the
general transverse homogeneity condition in
Eq.~\eqref{eq:transverse-width-homogeneity}.  In units of
\(\sqrt{s_{12}}\), the complete result is
\begin{equation}
 \Delta r\sim(\delta^4,1,\delta^2)_{(p_2,p_1)},\qquad
 \Delta\ell_G\sim(\delta^4,\delta,\delta^2)_{(p_2,p_1)}.
 \label{eq:five-central-soft-hard-scale-widths}
\end{equation}
Here the triples again denote plus, minus and transverse components in the
global \((p_2,p_1)\) frame.  At the
pinch, \(r=q_0\sim(\delta^4,1,\delta^2)_{(p_2,p_1)}\), as shown in
Table~\ref{tab:five-central-soft-edge-scalings}; its fluctuation therefore has
the same scaling as its value in this routing.  This is the balanced on-shell
\((p_1\text{-collinear})^2\) mode.  The exact routing similarly gives
\(\ell_G\vert_{\rm pinch}\sim
(\delta^4,\delta,\delta^2)_{(p_2,p_1)}\), again matching its
fluctuation width.  In this case, however,
\(\ell_G^+\ell_G^-\sim\delta^5\), whereas
\(\ell_{G\perp}^2\sim\delta^4\), so the mode is transverse dominated and hence
Glauber.  Thus, in this mode-adapted independent-loop basis, the compatible
routing and fluctuation scalings agree for both loops even though their mode
types differ.

It is the integration widths in
Eq.~\eqref{eq:five-central-soft-hard-scale-widths}, rather than the central
edge modes themselves, that determine the loop-measure factors in the power
count.  The two loop measures scale as
\(\delta^{2D}\) and \(\delta^{2D+1}\), respectively.  The six inverse
propagators contribute \(\delta^{-21}\).  Therefore
\begin{equation}
 \frac{I_{\rm soft}^{\rm scalar}}{(s_{12})^{D-6}}
 \sim
 \underbrace{\delta^{2D}}_{r\text{ measure}}\,
 \underbrace{\delta^{2D+1}}_{\ell_G\text{ measure}}\,
 \underbrace{\delta^{-21}}_{\text{six propagators}}
 =\delta^{4D-20},
 \label{eq:five-central-soft-power}
\end{equation}
in agreement with the parameter-space result
Eq.~\eqref{eq:five-central-soft-parameter-power}.  The restricted support
identified by HRF is already encoded in the narrower \(\ell_G\) integration
width; it must not be included as an additional factor.

Restoring the dimensionful normalisation and using
\(s_{12}/Q_\perp^2\sim\delta^{-4}\), this common parameter- and
momentum-space power count becomes
\begin{equation}
 I_{\rm soft}^{\rm scalar}
 \sim (s_{12})^{D-6}\delta^{4D-20}
 \sim (Q_\perp^2)^{D-6}\delta^4.
 \label{eq:five-central-soft-dimensionful-power}
\end{equation}

In summary, this soft-limit example requires order alignment before the core
HRF construction exposes the two-factor Fish pinch.  Within the Fish
topology, the external-leg analysis requires both incoming particles to be
attached to three-point vertices and the soft particle to a four-point
vertex.  The composed vector in
Eq.~\eqref{eq:five-central-soft-physical-vector} gives the parameter-space
power count in Eq.~\eqref{eq:five-central-soft-parameter-power}; its
momentum-space realisation contains one balanced on-shell
\((p_1\text{-collinear})^2\) loop and one transverse-dominated Glauber loop,
and leads to the matching power count in
Eqs.~\eqref{eq:five-central-soft-power} and
\eqref{eq:five-central-soft-dimensionful-power}.  The comparison with the
spacelike-collinear Fish is complementary: after translating the edge
labels, the factorised superleading sector and cancellation locus coincide,
but the spacelike-collinear HR is exposed by obstruction removal whereas the
present soft-limit HR first requires order alignment.  The two kinematic
limits select different resolved monomials and hence different region
vectors.  The relation to the generic wide-angle near-planar locus was
established above in
Eqs.~\eqref{eq:five-central-soft-positive-chamber} and
\eqref{eq:five-region-ideal-relation-clean}: the selected rapidity-ordered
face retains two combinations of the three normal conditions and therefore
has an enlarged leading locus.  Thus the same Fish seed and related
cancellation geometry underlie both expansions, while their region vectors,
momentum modes and scalar powers remain distinct.

\subsubsection{Comparison of the certified five-point regions}

Before turning to the six-point examples, we compare the four certified HRs
found for the Fish seed.  Their organising relation is the singular locus:
each belongs to the positive planar Fish family of
Eq.~\eqref{eq:five-positive-planar-chamber}, or to a controlled boundary
stratum of its closure.  Table~\ref{tab:five-point-region-comparison}
separates this inherited geometry from the expansion-dependent total vectors,
scalar powers, momentum-space integration widths and correlated support
selected by the asymptotic trajectory and the permutation of external legs.

\begin{table}[H]
\centering
\footnotesize
\setlength{\tabcolsep}{2.5pt}
\begin{tabular}{@{}
 >{\raggedright\arraybackslash}p{0.17\textwidth}
 >{\centering\arraybackslash}p{0.25\textwidth}
 >{\centering\arraybackslash}p{0.13\textwidth}
 >{\centering\arraybackslash}p{0.33\textwidth}@{}}
\toprule
Expansion & total edge vector & scalar power & momentum-space widths and support\\
\midrule
Spacelike collinear &
\(\begin{gathered}
 (-2,-1,-2,-2,-2,-1;1)\\[-1mm]
 \text{Eq.~}\eqref{eq:five-HR-vector}
\end{gathered}\) &
\(\delta^{-1-4\epsilon}\) &
\(\begin{gathered}
 \Delta\ell_s\sim(\delta,\delta,\delta)_{(p_2,p_1)},\\[-1mm]
 \Delta\ell_G\sim(\delta,\delta^2,\delta)_{(p_2,p_1)}
\end{gathered}\)\\
Near-planar wide angle &
\(\begin{gathered}
 (-2,-2,-2,-2,-2,-2;1)\\[-1mm]
 \text{Eq.~}\eqref{eq:five-near-planar-vector}
\end{gathered}\) &
\(\lambda^{-1-4\epsilon}\) &
\(\begin{gathered}
 \Delta q_I\sim(1,\lambda^2,\lambda)_{(P_I,\bar P_I)},\ I=A,B,\\[-1mm]
 \mathcal V_{\parallel}^{\rm corr}\sim\lambda^3
\end{gathered}\)\\
MRK plus near-planarity, \(a>0\), \(b>0\) &
\(\begin{gathered}(-2a,-2a,b-2a,\\[-1mm]
                   -2a,-2a,-2a;1)\\[-1mm]
                   \text{Eq.~}\eqref{eq:five-mrk-planar-vector}
   \end{gathered}\) &
\(\delta^{-a+2b-4a\epsilon}\) &
\(\begin{gathered}
 \Delta q_A\sim(\delta^{2a+b},\delta^{-b},\delta^a)_{(\bar p_1,p_1)},\\[-1mm]
 \Delta q_C\sim(\delta^{2a+b},\delta^{-b},\delta^a)_{(\bar p_5,p_5)},\\[-1mm]
 \mathcal V_{\parallel}^{\rm corr}\sim\delta^{3a+b}
\end{gathered}\)\\
Central-soft rapidity ordered &
\(\begin{gathered}
 (-4,-4,-1,-4,-4,-4;1)\\[-1mm]
 \text{Eq.~}\eqref{eq:five-central-soft-physical-vector}
\end{gathered}\) &
\(\begin{gathered}
 \delta^{4D-20}\\[-1mm]
 =\delta^{-4-8\epsilon}
\end{gathered}\) &
\(\begin{gathered}
 \Delta r\sim(\delta^4,1,\delta^2)_{(p_2,p_1)},\\[-1mm]
 \Delta\ell_G\sim(\delta^4,\delta,\delta^2)_{(p_2,p_1)}
\end{gathered}\)\\
\bottomrule
\end{tabular}
\caption{Comparison of the certified five-point hidden regions.  Each row
uses the LP normalisation of its defining subsection, and the equation below
each vector fixes the edge convention and representative permutation of
external legs.  The last column gives loop-integration widths in the
mode-adapted lightcone frame identified by its subscript.  The factor
\(\mathcal V_{\parallel}^{\rm corr}\) denotes the additional longitudinal
restriction generated by cross-loop correlations; where it is absent, there
is no such cross-loop correlation.}
\label{tab:five-point-region-comparison}
\end{table}

Two distinctions are important when reading the last column.  First, the
preceding momentum-space tables record compatible edge-momentum scalings,
whereas Table~\ref{tab:five-point-region-comparison} records independent-loop
integration widths; the frame subscript in
\((+,-,\perp)_{(n_+,n_-)}\) names the plus and minus reference directions.
Second, the
component widths determine the \(D\)-dependent part of the measure power,
whereas \(\mathcal V_{\parallel}^{\rm corr}\) contributes only to its integer
part.  For the central-soft row, \(\mu_{\rm LP}^2=s_{12}\), the widths are in
units of \(\sqrt{s_{12}}\), and the listed scalar power multiplies
\((s_{12})^{D-6}\); in fixed \(Q_\perp\) units it becomes \(\delta^4\), as in
Eq.~\eqref{eq:five-central-soft-dimensionful-power}.

The four cases in Table~\ref{tab:five-point-region-comparison} are distinct
regions, with different kinematic expansions, active cancellation ideals,
vectors and momentum-space realisations.  Their stronger common feature is that their
singular loci belong to the positive planar Fish family defined by
Eq.~\eqref{eq:five-positive-planar-chamber}, or to a controlled boundary
stratum of its closure.  The near-planar wide-angle example is the
nondegenerate interior case.  The spacelike-collinear limit reaches the
\(s_{23}=0\) boundary with the surviving ratio positive, whereas the
timelike-collinear range fails precisely this positivity test.  The
simultaneous MRK--planar family approaches \(\rho_B=0\) from positive
values as in Eq.~\eqref{eq:five-mrk-planar-ratios}, whereas the central-soft
limit retains Eq.~\eqref{eq:five-central-soft-positive-chamber} on an aligned
face.
Within the simultaneous MRK--planar family, only the ratio \(b/a\)
distinguishes inequivalent asymptotic trajectories.  The equal-rate
representative has \(a=b\), while the boundary \(b=0\) removes the MRK
hierarchy and recovers the near-planar wide-angle row with
\(\lambda=\delta^a\).  For \(b>0\), only the representative of
Eq.~\eqref{eq:five-mrk-planar-vector} and its hard-vertex-reflected partner pass
the positivity test.  The interchange \(p_4\leftrightarrow p_5\) is
immaterial in the spacelike-collinear and near-planar wide-angle limits.  In
the central-soft limit, by contrast, the HR requires the soft leg \(p_4\) at
a four-point Fish vertex and the two incoming legs at distinct three-point
vertices; the directly reflected vector is given in
Eq.~\eqref{eq:five-central-soft-opposite-vector}.

Two complementary relations organise the four rows.  The first pair comprises
the spacelike-collinear and central-soft limits, both factorisation limits in
which an underlying \(2\to2\) hard process emerges from the \(2\to3\)
kinematics.  After translating the edge labels, their factorised
superleading sectors have the same two-factor form and define the same
single-generator pinch.  Their HRF discovery routes nevertheless differ:
the spacelike-collinear construction isolates a nontrivial obstruction,
whereas the central-soft construction first requires order alignment.  Their
momentum-space realisations accordingly combine a Glauber loop with,
respectively, a soft or a balanced on-shell collinear loop.

The near-planar wide-angle and simultaneous MRK--planar limits form the second
pair.  Both retain the three-normal, three-generator planar
locus.  The MRK hierarchy drives \(\rho_B\to0\) and narrows the correlated
longitudinal support, while a path-adapted independent-loop basis retains
matching average and marginal scalings and generates no independent Glauber
loop.  Together, these two pairs of limits distinguish inherited cancellation
geometry from the expansion-dependent region vector and momentum modes.

This positive-chamber organisation explains why the same Fish seed supports
HRs in all four limits without identifying the resulting regions: the
singular locus is inherited from the planar family, whereas the certified
region vector belongs to the trajectory approaching that family or one of
its boundary strata.  In the central-soft example, order alignment selects
the active part of the inherited ideal before the core HRF construction.
The scalar powers in Table~\ref{tab:five-point-region-comparison} are
reproduced in both parameter and momentum space.

\subsection{Six-point integrals: the twisted-box seed and its hexagon realisation}
\label{sec:example-alignment}

\begin{figure}[htbp]
\centering
\begin{tikzpicture}[scale=.77]
\begin{scope}[xshift=-4.35cm]
 \coordinate (a2) at (-1.80,1.15);
 \coordinate (a1) at (-1.80,-1.15);
 \coordinate (a4) at (1.15,2.05);
 \coordinate (a5) at (1.15,.62);
 \coordinate (a3) at (1.15,-.62);
 \coordinate (a6) at (1.15,-2.05);

 \draw[hrf internal] (a2)--node[hrf edge label,pos=.48,above] {$x_4$}(a4);
 \draw[hrf internal,dashed,Blue] (a4)--node[hrf edge label,pos=.50,right] {$x_3$}(a5);
 \draw[white,line width=4pt] (a1)--(a5);
 \draw[hrf internal] (a1)--node[hrf edge label,pos=.76,above] {$x_2$}(a5);
 \draw[hrf internal,dashed,Blue] (a3)--node[hrf edge label,pos=.50,right] {$x_0$}(a6);
 \draw[hrf internal] (a6)--node[hrf edge label,pos=.48,below] {$x_1$}(a1);
 \draw[hrf internal] (a3)--node[hrf edge label,pos=.72,below] {$x_5$}(a2);

 \foreach \v in {1,2,3,4,5,6} \node[hrf vertex] at (a\v) {};
 \draw[->,thick] (-3.00,1.15)--(a2);
 \draw[->,thick] (-3.00,-1.15)--(a1);
 \draw[->,thick] (a4)--(2.35,2.45);
 \draw[->,thick] (a5)--(2.35,.92);
 \draw[->,thick] (a3)--(2.35,-.92);
 \draw[->,thick] (a6)--(2.35,-2.45);
 \node[hrf leg label] at (-3.18,1.39) {$p_2$};
 \node[hrf leg label] at (-3.18,-1.39) {$p_1$};
 \node[hrf leg label] at (2.75,2.65) {$p_4$};
 \node[hrf leg label] at (2.75,1.08) {$p_5$};
 \node[hrf leg label] at (2.75,-1.08) {$p_3$};
 \node[hrf leg label] at (2.75,-2.65) {$p_6$};
 \node[font=\small] at (0,-3.05) {(a) twisted hexagon};
\end{scope}

\begin{scope}[xshift=4.35cm]
 \coordinate (b2) at (-1.65,1.10);
 \coordinate (b1) at (-1.65,-1.10);
 \coordinate (bu) at (1.35,1.82);
 \coordinate (bd) at (1.35,-1.82);

 \draw[hrf internal] (b2)--node[hrf edge label,pos=.48,above] {$x_4$}(bu);
 \draw[hrf internal] (b2)--node[hrf edge label,pos=.60,below] {$x_5$}(bd);
 \draw[white,line width=4pt] (b1)--(bu);
 \draw[hrf internal] (b1)--node[hrf edge label,pos=.58,above] {$x_2$}(bu);
 \draw[hrf internal] (b1)--node[hrf edge label,pos=.48,below] {$x_1$}(bd);

 \node[hrf vertex] at (b2) {};
 \node[hrf vertex] at (b1) {};
 \node[hrf vertex,fill=Blue,draw=Blue] at (bu) {};
 \node[hrf vertex,fill=Blue,draw=Blue] at (bd) {};
 \draw[->,thick] (-2.85,1.10)--(b2);
 \draw[->,thick] (-2.85,-1.10)--(b1);
 \draw[->,thick] (bu)--(2.52,2.25);
 \draw[->,thick] (bu)--(2.52,1.42);
 \draw[->,thick] (bd)--(2.52,-1.42);
 \draw[->,thick] (bd)--(2.52,-2.25);
 \node[hrf leg label] at (-3.03,1.34) {$p_2$};
 \node[hrf leg label] at (-3.03,-1.34) {$p_1$};
 \node[hrf leg label] at (2.92,2.43) {$p_4$};
 \node[hrf leg label] at (2.92,1.56) {$p_5$};
 \node[hrf leg label] at (2.92,-1.56) {$p_3$};
 \node[hrf leg label] at (2.92,-2.43) {$p_6$};
 \node[font=\small] at (0,-3.05) {(b) twisted-box boundary};
\end{scope}
\end{tikzpicture}
\caption{The six-point topology and its self-crossing boundary.  Panel (a)
defines the LP parameters \(x_0,\ldots,x_5\) on the twisted hexagon, whose
vertices are all three-point.  Contracting the dashed-blue edges \(x_0\) and
\(x_3\) gives panel (b), with active parameters
\(x_1,x_2,x_4,x_5\) and two four-point hard vertices shown in blue.  The
crossings of drawn lines are not graph vertices.}
\label{fig:twisted-hexagon-topologies}
\end{figure}
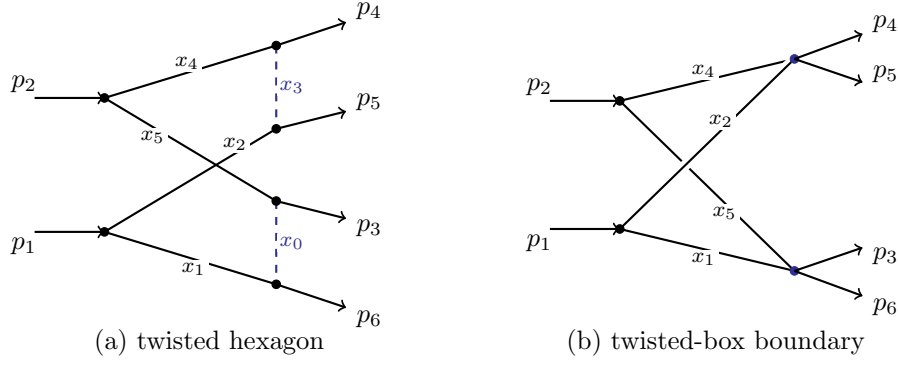

At six points, the first hidden regions in our survey appear already at one
loop.  The relevant graph is a massless hexagon with the non-planar
permutation of external legs displayed in
Fig.~\ref{fig:twisted-hexagon-topologies}(a),
rather than the usual cyclic planar ordering.  We refer to this graph as the
\emph{twisted hexagon}.  Contracting \(x_0\) and \(x_3\) gives the twisted-box
boundary in Fig.~\ref{fig:twisted-hexagon-topologies}(b).  The contraction
merges the \((p_3,p_6)\) and \((p_4,p_5)\) pairs into two four-point vertices
without reducing the loop order.  The twisted box in panel (b) is the minimal
six-point HR seed identified here.  The twisted hexagon in panel (a) is its
simplest six-propagator realisation, and the same externally labelled seed can
occur as a contraction minor of more complicated graphs.

The figure fixes both the permutation of external legs and the LP-parameter
convention.  Throughout this six-point section, \(p_1,p_2\) are
future-directed incoming momenta and \(p_3,\ldots,p_6\) are future-directed
outgoing momenta, so
\(p_1+p_2=p_3+p_4+p_5+p_6\).  For any index set \(I\), define
\begin{equation}
 s_I\equiv
 \left(
 -\sum_{i\in I\cap\{1,2\}}p_i
 +\sum_{j\in I\cap\{3,4,5,6\}}p_j
 \right)^2 .
 \label{eq:six-point-mandelstam-convention}
\end{equation}
Thus, for example,
\(s_{145}=(-p_1+p_4+p_5)^2\), whereas
\(s_{36}=(p_3+p_6)^2\).  In these variables the graph polynomials are
\begin{align}
 \mathcal F={}&
 \NegInv{s_{16}}x_0x_2+\NegInv{s_{156}}x_0x_3
 +\NegInv{s_{23}}x_0x_4
 +\NegInv{s_{15}}x_1x_3+\NegInv{s_{145}}x_1x_4
 +\PosInv{s_{36}}x_1x_5\nonumber\\
 &+\PosInv{s_{45}}x_2x_4+\NegInv{s_{245}}x_2x_5
 +\NegInv{s_{24}}x_3x_5,
 \label{eq:twisted-hexagon-F}\\[-1mm]
 \mathcal U={}&x_0+x_1+x_2+x_3+x_4+x_5.
 \label{eq:twisted-hexagon-U}
\end{align}
Here and below, invariant labels are colour-coded by their physical-sheet
sign: \PosInv{green} denotes positive and \NegInv{burgundy} negative.  All
coefficients in Eq.~\eqref{eq:twisted-hexagon-F} have a fixed sign throughout
the physical \(2\to4\) sheet.  In particular, the three-particle invariants
shown there each contain one incoming and two outgoing momenta and are
negative.  Indeed, in the lightcone frame defined by
\((+,-,\perp)_{(p_2,p_1)}\), any sum \(K\) of outgoing momenta obeys
\(0\leq K^+\leq p_2^+\) and \(0\leq K^-\leq p_1^-\).  It follows that
\((-p_i+K)^2\leq0\) for \(i=1,2\), with strict inequality at wide angle.

\subsubsection{The wide-angle self-crossing hidden region}
\label{sec:six-point-wide-angle-self-crossing}
The twisted hexagon has no positive pinch at generic massless wide-angle
kinematics.  A special wide-angle surface is nevertheless sufficient.  In the
physical self-crossing configuration of Ref.~\cite{Dixon:2016selfcross}, each
incoming particle splits into two collinear lines which undergo two separate
hard scatterings.  Let the two pairs of collinear daughters of \(p_1\) and
\(p_2\) carry the longitudinal fractions \(x,1-x\) and \(y,1-y\), with
\(0<x,y<1\).  For the external assignment in
Fig.~\ref{fig:twisted-hexagon-topologies}, the exact double-scattering
configuration obeys
\begin{equation}
 \begin{aligned}
  xp_1+yp_2&=p_4+p_5,&
  (1-x)p_1+(1-y)p_2&=p_3+p_6 .
 \end{aligned}
 \label{eq:hexagon-self-crossing-subprocesses}
\end{equation}
The collinear assignments follow from the same Coleman--Norton argument used
for the previous wide-angle examples, the four-point Crown and the five-point
Fish.  At a positive physical Landau pinch, the
four active internal propagators represent on-shell classical
trajectories~\cite{Coleman:1965cn,Collins:2020euz}.  Momentum conservation at
either incoming three-point vertex decomposes the future-directed null
momentum \(p_i\) into two future-directed null momenta; both daughters must
therefore lie on the \(p_i\) ray.  The two hard vertices then join one daughter
from each incoming direction, realising the two subprocesses in
Eq.~\eqref{eq:hexagon-self-crossing-subprocesses}.  Consequently, each
outgoing pair has zero net transverse momentum in the incoming
centre-of-mass frame.  The same geometry has a dual Wilson-loop description:
the two non-adjacent null edges associated with the incoming momenta intersect,
and their intersection divides the first into fractions \(x\) and \(1-x\),
and the second into \(y\) and \(1-y\).

Momentum conservation in the two \(2\to2\) subprocesses fixes the invariant
values at the pinch:
\begin{align}
 \PosInv{s_{36}^{(0)}}&=(1-x)(1-y)\PosInv{s_{12}},&
 \PosInv{s_{45}^{(0)}}&=xy\,\PosInv{s_{12}},\nonumber\\
 \NegInv{s_{145}^{(0)}}&=-y(1-x)\PosInv{s_{12}},&
 \NegInv{s_{245}^{(0)}}&=-x(1-y)\PosInv{s_{12}},\nonumber\\
 x\,\NegInv{s_{15}^{(0)}}&=y\,\NegInv{s_{24}^{(0)}},&
 (1-x)\NegInv{s_{16}^{(0)}}&=(1-y)\NegInv{s_{23}^{(0)}}.
 \label{eq:hexagon-self-crossing-pinched-invariants}
\end{align}
For the present cyclic assignment, define the three six-point cross ratios
\begin{equation}
 U_\times\equiv
 \frac{\PosInv{s_{36}}\PosInv{s_{45}}}
      {\NegInv{s_{145}}\NegInv{s_{245}}},
 \qquad
 V_\times\equiv
 \frac{\NegInv{s_{16}}\NegInv{s_{24}}}
      {\NegInv{s_{156}}\NegInv{s_{245}}},
 \qquad
 W_\times\equiv
 \frac{\NegInv{s_{15}}\NegInv{s_{23}}}
      {\NegInv{s_{145}}\NegInv{s_{156}}}.
 \label{eq:hexagon-self-crossing-ratios}
\end{equation}
These are the cross ratios conventionally denoted by \(u,v,w\) in
Ref.~\cite{Dixon:2016selfcross}; capital letters and the subscript \(\times\)
distinguish them from the lightcone transverse coordinate \(w\) introduced
below.  For generic six-point kinematics these three cross ratios are
independent.  Using the invariant relations in \(x\) and \(y\) above, one may
verify explicitly that \(U_\times=1\) and \(V_\times/W_\times=1\); in the
latter ratio, \(s_{156}\) cancels.  The physical self-crossing surface is
therefore
\begin{equation}
 U_\times=1,\qquad V_\times=W_\times.
 \label{eq:hexagon-self-crossing-surface}
\end{equation}
These conditions are the invariant image of the double-scattering
configuration rather than an additional kinematic assumption.  The first is
the singular condition that makes the twisted-box
polynomial factorise (see
Eq.~\eqref{eq:hexagon-wide-angle-boundary-factorisation} below); the second
selects its physical six-point
self-crossing branch.  Correspondingly, the standard six-point discriminant\footnote{Writing
the six dual cusps as null rays \(X_i\) in the six-dimensional conformal
embedding space, \(\Delta_6\) is proportional, after projective normalisation,
to \(-\det(X_i\cdot X_j)\). Thus \(\sqrt{\Delta_6}\) measures the
normalised embedding-space pseudo-volume of the six rays, and
\(\Delta_6=0\) means that they are linearly dependent; see
Ref.~\cite{Bourjaily:2018aeq}.}
\begin{equation}
 \Delta_6=(1-U_\times-V_\times-W_\times)^2
           -4U_\times V_\times W_\times
 \label{eq:hexagon-cross-ratio-discriminant}
\end{equation}
reduces to \((V_\times-W_\times)^2\) at \(U_\times=1\).  A convenient
physical regulator separates the two scattering points by a small spacelike
transverse recoil \(\boldsymbol\Delta_\perp\).  Following the general
self-crossing parametrisation of Ref.~\cite{Dixon:2016selfcross}, set
\begin{equation}
 \Lambda_\times\equiv \PosInv{s_{12}}xy(1-x)(1-y),\qquad
 \sigma_\times\equiv
 \frac{|\boldsymbol\Delta_\perp|^2}{\Lambda_\times}\to0^+ .
 \label{eq:hexagon-self-crossing-expansion}
\end{equation}
The part of the departure from the pinch that survives on the boundary
\(x_0=x_3=0\) relevant below is
\begin{align}
 \PosInv{s_{36}(\sigma_\times)}
 &=\PosInv{s_{36}^{(0)}}-\sigma_\times\Lambda_\times
   +\mathcal O(\sigma_\times^2),&
 \PosInv{s_{45}(\sigma_\times)}
 &=\PosInv{s_{45}^{(0)}}-\sigma_\times\Lambda_\times
   +\mathcal O(\sigma_\times^2),\nonumber\\
 \NegInv{s_{145}(\sigma_\times)}
 &=\NegInv{s_{145}^{(0)}}-\sigma_\times\Lambda_\times
   +\mathcal O(\sigma_\times^2),&
 \NegInv{s_{245}(\sigma_\times)}
 &=\NegInv{s_{245}^{(0)}}-\sigma_\times\Lambda_\times
   +\mathcal O(\sigma_\times^2).
 \label{eq:hexagon-self-crossing-recoil-invariants}
\end{align}
The two relations in the final line of
Eq.~\eqref{eq:hexagon-self-crossing-pinched-invariants} are exact at the
pinch.  Under a generic transverse recoil their deviations begin at
relative order
\(|\boldsymbol\Delta_\perp|/\sqrt{\Lambda_\times}
=\mathcal O(\sqrt{\sigma_\times})\).  The affected invariants multiply
\(x_0\) or \(x_3\) in Eq.~\eqref{eq:twisted-hexagon-F}, so these earlier
corrections vanish identically on the boundary and do not enter its HRF
analysis.

Equation~\eqref{eq:hexagon-self-crossing-recoil-invariants} gives
\(U_\times=1-\sigma_\times+\mathcal O(\sigma_\times^2)\).  Along the
one-parameter path used here we hold the nonsingular cross ratio fixed,
\(V_\times=W_\times=V_\times^{(0)}\).  Expansion of
Eq.~\eqref{eq:hexagon-cross-ratio-discriminant} then gives
\[
 \Delta_6=-4\sigma_\times V_\times^{(0)}
 \bigl(1-V_\times^{(0)}\bigr)+\mathcal O(\sigma_\times^2),
 \qquad 0<V_\times^{(0)}<1.
\]
Thus \(\Lambda_\times\) sets the normalisation of \(\sigma_\times\), while
the finite coefficient is fixed by the nonsingular kinematics on the
self-crossing surface.  All cyclic two-particle invariants \(s_{i,i+1}\)
remain finite and nonzero in this wide-angle limit.

The input to HRF is the complete unfactorised LP polynomial of
Eq.~\eqref{eq:twisted-hexagon-F}; the self-crossing parametrisation enters
through its Mandelstam coefficients.  The boundary scan identifies
\(x_0=x_3=0\), and on this stratum Part~I extracts the restricted leading
polynomial shown in the first line below.  The derivative harvest then finds
the two factors displayed in the second line:
\begin{align}
 \mathcal F_{\mathrm{SL,WA}}
 &=\NegInv{s_{145}^{(0)}}x_1x_4+\PosInv{s_{36}^{(0)}}x_1x_5
 +\PosInv{s_{45}^{(0)}}x_2x_4+\NegInv{s_{245}^{(0)}}x_2x_5
 \nonumber\\
 &=\frac{1}{\PosInv{s_{36}^{(0)}}}
 \underbrace{(\NegInv{s_{145}^{(0)}}x_4+\PosInv{s_{36}^{(0)}}x_5)}_{f_{1,\mathrm{WA}}}
 \underbrace{(\PosInv{s_{36}^{(0)}}x_1+\NegInv{s_{245}^{(0)}}x_2)}_{f_{2,\mathrm{WA}}}.
 \label{eq:hexagon-wide-angle-boundary-factorisation}
\end{align}
Here the second equality follows from the first condition in
Eq.~\eqref{eq:hexagon-self-crossing-surface}.  In particular, expanding the
factorised form replaces its only apparent rational coefficient according to
\(s_{145}^{(0)}s_{245}^{(0)}/s_{36}^{(0)}=s_{45}^{(0)}\), so the result is precisely the polynomial
in the first line.  Each displayed factor combines a positive and a negative
term and therefore has a positive zero.

Pairing these harvested factors retains the generator
\(g_{1,\mathrm{WA}}=f_{1,\mathrm{WA}}f_{2,\mathrm{WA}}\).  Part~II determines
the active scaling \((-1,-1,-1,-1)\), and restoring the contracted
coordinates gives
\begin{equation}
 \vec v_{\rm HR,WA}
 =(\boldsymbol v_{\rm HR,WA};1)
 =(0,-1,-1,0,-1,-1;1),
 \qquad
 (\WSL,\WHR)=(-2,-1).
 \label{eq:hexagon-wide-angle-region-vector}
\end{equation}
The zero entries for \(x_0\) and \(x_3\) have the usual hard-subgraph
interpretation: the corresponding momenta have
\(q_0^2,q_3^2\sim\sigma_\times^0\) and are absorbed into the two four-point
hard vertices of Fig.~\ref{fig:twisted-hexagon-topologies}(b).  Thus neither
edge is a Glauber propagator in the wide-angle region.  The status of these
two resolved edges changes in the NMRK and DSC specialisations below.
The two contributions to the resolved-leading layer on this stratum are
\begin{align}
 \left.\mathcal U\right|_{x_0=x_3=0}
 &=x_1+x_2+x_4+x_5,\nonumber\\
 \left.\mathcal F_1\right|_{x_0=x_3=0}
 &=-\Lambda_\times
   \bigl(x_1x_4+x_1x_5+x_2x_4+x_2x_5\bigr)\nonumber\\
 &=-\Lambda_\times(x_1+x_2)(x_4+x_5).
 \label{eq:hexagon-wide-angle-resolved-pieces}
\end{align}
Here \(\mathcal F_1\) is the coefficient of \(\sigma_\times\) in the
physical expansion of \(\mathcal F\).
Applying the vector in Eq.~\eqref{eq:hexagon-wide-angle-region-vector} after
restricting the complete LP polynomial to \(x_0=x_3=0\) gives
\begin{equation}
\begin{aligned}
 \mathcal P_{\mathrm{WA}}^{(\HR)}(\x;\sigma_\times,\s)
 &\equiv
 \left.\mathcal P(\sigma_\times^{\boldsymbol v_{\rm HR,WA}}\x;
            \sigma_\times,\s)\right|_{x_0=x_3=0}\\
 &=\sigma_\times^{-2}\SLcolour{\mathcal F_{\mathrm{SL,WA}}}
  +\sigma_\times^{-1}\HRcolour{\left.
       \bigl(\mathcal U+\mathcal F_1\bigr)\right|_{x_0=x_3=0}}
  +\Othercolour{\mathcal O(\sigma_\times^0)}.
\end{aligned}
\label{eq:hexagon-wide-angle-complete-layers}
\end{equation}
Thus the restricted \(\mathcal F_1\) contribution is part of the
resolved-leading layer and supplies the expansion-parameter direction needed
by the lower-facet certificate.  Dissecting in the two local normals defined
by \(f_{1,\mathrm{WA}}\) and \(f_{2,\mathrm{WA}}\) produces two endpoint
charts; the complete restricted LP support
has the required codimension-one lower facet in each, providing the final
certificate.  For unit propagator powers the
resulting scalar contribution scales as
\begin{equation}
 I_\times^{\rm scalar}\sim
 \sigma_\times^{D/2-3}=\sigma_\times^{-1-\epsilon},
 \qquad D=4-2\epsilon.
 \label{eq:hexagon-self-crossing-scalar-power}
\end{equation}

The momentum-space interpretation is especially transparent in a routing
adapted to the double scattering.  Momentum conservation at the vertices gives
\begin{equation}
 \ell_\times
 \equiv q_1-(1-x)p_1
 =xp_1-q_2
 =q_4-yp_2
 =(1-y)p_2-q_5 .
 \label{eq:hexagon-wide-angle-mode-routing}
\end{equation}
On the self-crossing pinch, the double-scattering assignments in
Eq.~\eqref{eq:hexagon-self-crossing-subprocesses} imply separately that
\(\ell_\times^\star=0\).  The two contracted propagators have hard
virtualities there:
\begin{equation}
 \bigl(q_0^\star\bigr)^2
 =(1-x)\NegInv{s_{16}^{(0)}}
 =(1-y)\NegInv{s_{23}^{(0)}}
 =\mathcal O(\Lambda_\times),
 \qquad
 \bigl(q_3^\star\bigr)^2
 =x\NegInv{s_{15}^{(0)}}
 =y\NegInv{s_{24}^{(0)}}
 =\mathcal O(\Lambda_\times).
 \label{eq:hexagon-wide-angle-hard-contracted-virtualities}
\end{equation}
This hard scaling shows that, in this region, the twisted-hexagon realisation
can be contracted to the twisted box considered here.
The four active entries \(v_e=-1\) in
Eq.~\eqref{eq:hexagon-wide-angle-region-vector} fix their virtualities to
\begin{equation}
 \frac{q_e^2}{\Lambda_\times}\sim\sigma_\times,
 \qquad e\in\{1,2,4,5\}.
 \label{eq:hexagon-wide-angle-active-virtualities}
\end{equation}
Expanding these denominators in the shifted loop momentum of
Eq.~\eqref{eq:hexagon-wide-angle-mode-routing} gives
\begin{align}
 q_1^2&= 2(1-x)p_1^-\ell_\times^+
          +\ell_\times^2,&
 q_2^2&=-2x p_1^-\ell_\times^+
          +\ell_\times^2,\nonumber\\
 q_4^2&= 2y p_2^+\ell_\times^-
          +\ell_\times^2,&
 q_5^2&=-2(1-y)p_2^+\ell_\times^-
          +\ell_\times^2,
 \label{eq:hexagon-wide-angle-linearised-denominators}
\end{align}
The opposite longitudinal coefficients exhibit the \(q_1,q_2\) pinch in
\(\ell_\times^+\) and the \(q_4,q_5\) pinch in \(\ell_\times^-\), while
\(\ell_\times^2=2\ell_\times^+\ell_\times^-
-|\ell_{\times\perp}|^2\).  Since
\(p_1^-,p_2^+=\mathcal O(\sqrt{\Lambda_\times})\), the virtualities in
Eq.~\eqref{eq:hexagon-wide-angle-active-virtualities} require
\(\ell_\times^\pm=\mathcal O(\sigma_\times\sqrt{\Lambda_\times})\) and
\(|\ell_{\times\perp}|=\mathcal O(\sqrt{\sigma_\times\Lambda_\times})\).
The latter is precisely the physical recoil scale in
Eq.~\eqref{eq:hexagon-self-crossing-expansion}.  This is the same
local-denominator analysis used for the fluctuation widths in the discussion
leading to Eq.~\eqref{eq:five-central-soft-hard-scale-widths}.  In the present
one-loop problem it is the complete momentum-space reconstruction: since
\(\ell_\times^\star=0\), the loop-momentum scaling and its local integration
width coincide,
\begin{equation}
 \frac{\bigl(\Delta\ell_\times^+,\Delta\ell_\times^-,
       |\Delta\ell_{\times\perp}|\bigr)}{\sqrt{\Lambda_\times}}
 \sim\bigl(\sigma_\times,\sigma_\times,
           \sqrt{\sigma_\times}\bigr)_{(p_2,p_1)} .
 \label{eq:hexagon-wide-angle-loop-width}
\end{equation}
Thus the single integration loop is transverse dominated and hence Glauber,
although none of the individual propagators is Glauber: the four active
propagators are collinear and \(q_0,q_3\) are hard.  Finally,
\begin{equation}
 \dd^D\ell_\times\sim\sigma_\times^{D/2+1},
 \qquad
 \prod_{e\in\{1,2,4,5\}}\frac{1}{q_e^2}
 \sim\sigma_\times^{-4},
 \qquad
 I_\times^{\rm scalar}\sim\sigma_\times^{D/2-3},
 \label{eq:hexagon-wide-angle-momentum-power}
\end{equation}
which confirms Eq.~\eqref{eq:hexagon-self-crossing-scalar-power} directly in
momentum space.

We shall now approach this surface along two further kinematic limits:
zero-recoil NMRK and the double
spacelike-collinear (DSC) limit.  In each limit, order alignment selects a
leading face whose two cancellation factors inherit the zero sets of
\(f_{1,\mathrm{WA}}\) and \(f_{2,\mathrm{WA}}\); exact normalised identities
on the common pinch are displayed below.  Thus both
composite-limit HRs inherit the wide-angle cancellation locus, although their
total region vectors and occupied layers depend on the expansion.

We first define the six-point variables and the two limiting paths.  We then
analyse the inherited one-loop cancellation mechanism in each limit, displaying the
alignment vector, the factorised aligned face, the relative and combined
scaling vectors, and the final validation against the complete graph
polynomial.  We then turn to two-loop graphs which contain the twisted-box
seed as an externally labelled contraction minor.

\subsubsection{Six-point lightcone variables and the zero-recoil NMRK surface}

We use the minimal lightcone variables of Ref.~\cite{Byrne:2025phh}.  Take
legs 1 and 2 incoming, set \(P=p_4^+\), and define the longitudinal ratios by
\begin{equation}
 p_3^+=PX_{34},\qquad p_4^+=P,\qquad
 p_5^+=\frac{P}{X_{45}},\qquad
 p_6^+=\frac{P}{X_{45}X_{56}}.
 \label{eq:six-mslcv-plus}
\end{equation}
Writing \(Q=q_1^\perp\bar q_1^\perp=|q_1^\perp|^2\), the transverse
coordinates are
\begin{align}
 p_4^\perp&=-\frac{q_1^\perp}{z-1},&
 p_5^\perp&=\frac{zq_1^\perp}{w(z-1)},\nonumber\\
 p_6^\perp&=\frac{z(w-1)q_1^\perp}{w(z-1)},&
 p_3^\perp&=-p_4^\perp-p_5^\perp-p_6^\perp,
 \label{eq:six-mslcv-transverse}
\end{align}
with barred equations for the conjugate transverse components.  Thus the
four real coordinates \(Q,X_{34},X_{45},X_{56}\), together with the complex
transverse coordinates \(z,w\) and their barred partners, parametrise generic
massless six-point kinematics up to an irrelevant longitudinal boost.  No
asymptotic limit has yet been taken.

Central NMRK is the hierarchy
\begin{equation}
 X_{34}=\frac{\widehat X_{34}}{\eta},\qquad
 X_{56}=\frac{\widehat X_{56}}{\eta},\qquad
 X_{45},Q,z,\bar z,w,\bar w=O(1),\qquad \eta\to0^+.
 \label{eq:nmrk-scaling-example}
\end{equation}
On its physical sheet, \(\bar z=z^*\) and \(\bar w=w^*\).  The limit studied
in Sec.~\ref{sec:six-point-nmrk-seed} combines the NMRK hierarchy in
Eq.~\eqref{eq:nmrk-scaling-example} with the zero-recoil condition
\begin{equation}
 w=z,\qquad \bar w=\bar z
 \label{eq:nmrk-zero-recoil-surface}
\end{equation}
which is not part of generic NMRK.  It imposes
\(p_4^\perp+p_5^\perp=0\).  The self-crossing cross ratio obeys\footnote{This
cross-ratio parametrisation arises in the analysis of the two-gluon
central-emission vertex; see App.~B.1, in particular Eq.~(B.9), of
Ref.~\cite{Byrne:2022cev}.}
\begin{equation}
 1-U_\times=
 \frac{X_{45}(w-z)(\bar w-\bar z)}
 {(1+X_{45})(w\bar w+X_{45}z\bar z)}.
 \label{eq:nmrk-cross-ratio-boundary}
\end{equation}
Substituting the Mandelstam invariants obtained from the parametrisation in
Eqs.~\eqref{eq:six-mslcv-plus} and~\eqref{eq:six-mslcv-transverse} into
Eq.~\eqref{eq:hexagon-self-crossing-ratios} also gives
\(V_\times=W_\times\) when \(w=z\) and
\(\bar w=\bar z\).  Thus Eq.~\eqref{eq:nmrk-zero-recoil-surface} places the
kinematics on the complete self-crossing surface
Eq.~\eqref{eq:hexagon-self-crossing-surface}.  The zero-recoil NMRK trajectory
therefore lies on the same self-crossing singular surface as the wide-angle
HR, while \(\eta\to0\) generates a rapidity hierarchy along that surface
rather than measuring the distance from it.
For the positive HRF chart below we use
\begin{equation}
 Q,\widehat X_{34},X_{45},\widehat X_{56},
 (1-z)(1-\bar z)>0.
 \label{eq:nmrk-domain-example}
\end{equation}

\subsubsection{Twisted-hexagon HR in zero-recoil NMRK}
\label{sec:six-point-nmrk-seed}

\begin{figure}[htbp]
\centering
\begin{tikzpicture}[scale=.98]
 \coordinate (v4) at (-2.20,2.00);
 \coordinate (v5) at ( .70,2.00);
 \coordinate (v3) at (1.50,.35);
 \coordinate (v2) at (1.50,-.35);
 \coordinate (v6) at ( .70,-2.00);
 \coordinate (v1) at (-2.20,-2.00);

 \draw[hrf internal,draw=Purple,densely dotted,line width=1.15pt]
   (v5)--node[hrf edge label,pos=.50,left,text=Purple] {$x_0$}(v6);
 \draw[hrf internal] (v6)--node[hrf edge label,pos=.50,below] {$x_1$}(v1);
 \draw[white,line width=4pt] (v1)--(v2);
 \draw[hrf internal] (v1)--node[hrf edge label,pos=.62,below] {$x_2$}(v2);
 \draw[hrf internal] (v2)--node[hrf edge label,pos=.50,right] {$x_3$}(v3);
 \draw[white,line width=4pt] (v3)--(v4);
 \draw[hrf internal] (v3)--node[hrf edge label,pos=.62,above] {$x_4$}(v4);
 \draw[hrf internal] (v4)--node[hrf edge label,pos=.50,above] {$x_5$}(v5);
 \foreach \v in {1,...,6} \node[hrf vertex] at (v\v) {};

 \draw[->,thick,OliveGreen] (-3.35,2.00)--(v4);
 \draw[->,thick,Green] (-3.35,-2.00)--(v1);
 \draw[->,thick,OliveGreen] (v5)--(3.00,2.00);
 \draw[->,thick,LimeGreen] (v3)--(3.00,.35);
 \draw[->,thick,TealBlue] (v2)--(3.00,-.35);
 \draw[->,thick,Green] (v6)--(3.00,-2.00);
 \node[hrf leg label,text=OliveGreen] at (-3.55,2.25) {$p_2$};
 \node[hrf leg label,text=Green] at (-3.55,-2.25) {$p_1$};
 \node[hrf leg label,text=OliveGreen] at (3.20,2.25) {$p_3$};
 \node[hrf leg label,text=LimeGreen] at (3.20,.60) {$p_4$};
 \node[hrf leg label,text=TealBlue] at (3.20,-.10) {$p_5$};
 \node[hrf leg label,text=Green] at (3.20,-2.25) {$p_6$};

 \GlauberVertex[OliveGreen]{(v5)}{90}
 \GlauberVertex[Green]{(v6)}{90}
\end{tikzpicture}
\caption{Momentum-space interpretation of the twisted-hexagon HR in NMRK.
The embedding displays the rapidity ordering
\(p_3\gg(p_4,p_5)\gg p_6\).  The upper pair \((p_2,p_3)\) and lower
pair \((p_1,p_6)\) have matching colours; the two line crossings are not
additional vertices.  The purple dotted edge labelled \(x_0\) carries
\(q_0\) and is the explicit Glauber propagator in both zero-recoil NMRK
and generic DSC.  The oriented discs at the \(p_3\) and \(p_6\) vertices
mark its two incident vertices and follow the convention defined in
Fig.~\ref{fig:crown-regge-glauber}.  Their olive and green rims record the two
complementary collinear channels.  The incoming three-point vertices carrying \(p_1\) and \(p_2\)
are the factor-localisation vertices: they localise the two factor equations
on the adjacent pairs \((x_1,x_2)\) and \((x_4,x_5)\), respectively.}
\label{fig:one-loop-hexagon}
\end{figure}
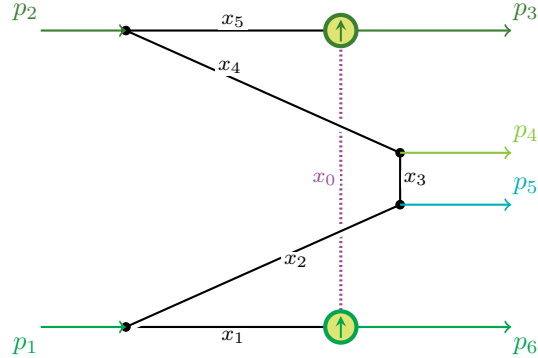

Substitution of Eq.~\eqref{eq:nmrk-scaling-example} at fixed LP
parameters gives a strict leading term
\begin{equation}
 \mathcal F_{0,\mathrm{NMRK}}=
 \eta^{-2}Q\widehat X_{34}X_{45}\widehat X_{56}\,x_1x_5
 \qquad (\text{zero recoil}),
 \label{eq:nmrk-strict-leading-monomial}
\end{equation}
which is a single monomial and cannot reveal the pinch.  The relevant
monomials occupy different native \(\eta\)-orders.  Following
Sec.~\ref{sec:alignment}, HRF therefore searches the exposed lower faces of
the augmented \(\eta\)-dependent support.  This search selects the mixed-order
face whose normal may be represented by
\begin{equation}
 \boldsymbol\phi_{\mathrm{NMRK}}=(-1,-1,-2,-1,-2,-1).
 \label{eq:nmrk-alignment-vector}
\end{equation}
To display the structure of the selected face, strip off its leading
\(\eta\)-power and a nonzero overall kinematic factor.  Denoting equality up
to this factor by \(\doteq\), collect the LP parameters adjacent to the two
incoming vertices into \(\boldsymbol x_{p_1}\) and
\(\boldsymbol x_{p_2}\), and write
\begin{equation}
 \mathcal F^{[\boldsymbol\phi_{\mathrm{NMRK}}]}
 \doteq \boldsymbol x_{p_1}^{T}M\boldsymbol x_{p_2},
 \qquad
 \boldsymbol x_{p_1}=\binom{x_2}{x_1},\qquad
 \boldsymbol x_{p_2}=\binom{x_4}{x_5},
 \label{eq:nmrk-bilinear-face}
\end{equation}
where, abbreviating \(X=X_{45}\),
\begin{equation}
 M=
 \begin{pmatrix}
 \dfrac{(w+Xz)(\bar w+X\bar z)}{z\bar z}
 &-\widehat X_{34}X\dfrac{w\bar w+Xz\bar z}{z\bar z}
 \\[2mm]
 -(w-1)(\bar w-1)X(1+X)\widehat X_{56}\qquad
 &(w-1)(\bar w-1)\widehat X_{34}X^2\widehat X_{56}
 \end{pmatrix}.
 \label{eq:nmrk-bilinear-matrix}
\end{equation}
The nontrivial stationary conditions are therefore
\begin{equation}
 M\boldsymbol x_{p_2}=0,\qquad
 M^{T}\boldsymbol x_{p_1}=0,
 \label{eq:nmrk-bilinear-stationarity}
\end{equation}
while
\begin{equation}
 \det M=
 -\frac{(w-1)(\bar w-1)}{z\bar z}\,
 \widehat X_{34}X^3\widehat X_{56}
 (w-z)(\bar w-\bar z).
 \label{eq:nmrk-bilinear-determinant}
\end{equation}
For generic NMRK this determinant is nonzero, so
Eq.~\eqref{eq:nmrk-bilinear-stationarity} admits only
\(\boldsymbol x_{p_1}=\boldsymbol x_{p_2}=0\), not a
solution in the positive
orthant.  On the additional surface
Eq.~\eqref{eq:nmrk-zero-recoil-surface}, the matrix has rank one and the
bilinear polynomial factorises.  This matrix form merely makes the
coefficient condition explicit.  In the actual run the selected face is
passed to the core HRF algorithm, whose derivative harvest recovers the two
cancellation factors in the usual way.  Define
\begin{equation}
 f_{1,\mathrm{NMRK}}=(1+X_{45})x_4-\widehat X_{34}X_{45}x_5,\qquad
 f_{2,\mathrm{NMRK}}=(1+X_{45})x_2
 -(1-z)(1-\bar z)X_{45}\widehat X_{56}x_1.
 \label{eq:nmrk-factors-example}
\end{equation}
The aligned face is then
\begin{equation}
 \mathcal F_{\star,\mathrm{NMRK}}
 \equiv\mathcal F^{[\boldsymbol\phi_{\mathrm{NMRK}}]}
 =\mathcal F_{\mathrm{SL,NMRK}}
 =Q\widehat X_{34}\widehat X_{56}
 f_{1,\mathrm{NMRK}}f_{2,\mathrm{NMRK}},
 \qquad \mathcal F_{\mathrm{Obs,NMRK}}=0.
 \label{eq:nmrk-decomposition-example}
\end{equation}
Both linear factors have positive zeros in the domain
Eq.~\eqref{eq:nmrk-domain-example}.  Homogeneity of the aligned face alone
leaves a uniform freedom and therefore does not yet fix the core scaling.
The Part~II hierarchy inequalities, applied to the complete aligned LP
polynomial in the original LP variables, fix the remaining core scaling
quoted below.  We then pass to local cancellation-resolving coordinates to
give an independent lower-facet certificate and to make the composition of
the core and alignment scalings transparent.

Here the local certificate is compact enough to display explicitly.  In one
physical sign sector take
\begin{equation}
 t_1=f_{1,\mathrm{NMRK}},\qquad
 t_2=f_{2,\mathrm{NMRK}},\qquad t_1,t_2\geq0,
\end{equation}
and solve
\begin{equation}
 x_4=\frac{\widehat X_{34}X_{45}x_5+t_1}{1+X_{45}},
 \qquad
 x_1=\frac{(1+X_{45})x_2-t_2}
 {(1-z)(1-\bar z)X_{45}\widehat X_{56}}.
 \label{eq:nmrk-local-coordinates}
\end{equation}
The tangential coordinates are \((x_0,x_2,x_3,x_5)\).  Here the \(x_e\)
denote the aligned LP parameters, after applying
Eq.~\eqref{eq:nmrk-alignment-vector}.  Define the local scaling exponents by
\begin{equation}
 x_e\mapsto\eta^{v^{\rm loc}_{x_e}}x_e
 \quad(e\in\{0,2,3,5\}),\qquad
 t_i\mapsto\eta^{v^{\rm loc}_{t_i}}t_i
 \quad(i\in\{1,2\}).
 \label{eq:nmrk-local-rescaling}
\end{equation}
In the edge-compatible ordered local coordinate system
\((x_0,t_2,x_2,x_3,t_1,x_5)\), where \(t_2\) occupies the slot of \(x_1\)
and \(t_1\) the slot of \(x_4\), the transformed complete LP polynomial has
the exact lower-facet normal
\begin{equation}
 \vec v_{\rm loc,NMRK}
 \equiv
 (v^{\rm loc}_{x_0},v^{\rm loc}_{t_2},v^{\rm loc}_{x_2},v^{\rm loc}_{x_3},
  v^{\rm loc}_{t_1},v^{\rm loc}_{x_5};1)
 =(-1,-2,-4,-1,-2,-4;1).
 \label{eq:nmrk-local-vector}
\end{equation}
The last entry is the normalised \(\eta\)-component of the augmented normal.
The Jacobian is nonzero and, up to sign, proportional to
\([(1-z)(1-\bar z)X_{45}(1+X_{45})\widehat X_{56}]^{-1}\).  Pulling this normal back to
the aligned LP parameters reproduces the core scaling fixed by Part~II.
Composing it with the alignment vector in
Eq.~\eqref{eq:nmrk-alignment-vector} then gives the certified vector in the
original coordinates,
\begin{align}
 \boldsymbol v_{\rm core}&=(-1,-4,-4,-1,-4,-4),\nonumber\\
 \boldsymbol v_{\HR}&=\boldsymbol\phi_{\mathrm{NMRK}}
 +\boldsymbol v_{\rm core}
 =(-2,-5,-6,-2,-6,-5),\nonumber\\
 \vec v_{\HR}&=(\boldsymbol v_{\HR};1),
 \qquad (\WSL,\WHR)=(-10,-6).
 \label{eq:nmrk-vector-example}
\end{align}
The edge-compatible ordering in Eq.~\eqref{eq:nmrk-local-vector} makes the
cancellation depth directly visible:
\[
 \boldsymbol v_{\rm loc,NMRK}
 =\boldsymbol v_{\rm core}+(0,2,0,0,2,0).
\]
Only the two normal entries differ.  In those slots \(t_2\) replaces \(x_1\)
and \(t_1\) replaces \(x_4\); each local normal width is suppressed by two
powers of \(\eta\) relative to the corresponding edge-parameter scaling.
Each transverse factor is suppressed by two powers, so their product
\(f_{1,\mathrm{NMRK}}f_{2,\mathrm{NMRK}}\) first contributes four powers
above the individual
superleading monomials, precisely at \(\WHR\).  The resolved lower facet
contains this first nonzero contribution from
\(\mathcal F_{\mathrm{SL,NMRK}}\), together with two
\(\mathcal U\) monomials and seven monomials from the remaining
\(\mathcal F\) polynomial.  Its ten exponent points have affine rank six,
and the inverse-map and pullback checks are exact.  Thus the equality of the
resolved \(\mathcal F_{\mathrm{SL,NMRK}}\) weight and \(\WHR\) is observed
in the dissected
polynomial rather than supplied as an assumption.  Generic NMRK, for which
the determinant above is nonzero, has no corresponding certificate for this
seed.

Dissection enters once more, now for power counting rather than for
determining the vector: the local normal widths determine the measure near
the cancellation locus.  For unit propagator powers the
unrestricted edge measure has
weight \(\sum_e(v_{\HR})_e=-26\).  The actual local measure instead uses the
normal widths displayed in the comparison above.  The
alignment Jacobian has weight
\(\sum_e(\phi_{\mathrm{NMRK}})_e=-8\), while the six local variables in
Eq.~\eqref{eq:nmrk-local-vector} have total weight \(-14\).  The exact local
measure therefore has weight \(-8-14=-22\).  Equivalently, the two changes
\(-4\to-2\) in that comparison restore four powers
relative to the unrestricted edge measure.  Since
\(\mathcal P\) has resolved weight \(-6\),
\begin{equation}
 I_{\mathrm{NMRK}}^{\rm scalar}\sim
 \eta^{-22}\bigl(\eta^{-6}\bigr)^{-D/2}
 =\eta^{3D-22}.
 \label{eq:nmrk-one-loop-parameter-power}
\end{equation}

Turning to momentum space, the inverse edge scalings give propagator virtuality powers
\((2,5,6,2,6,5)\), whose sum is 26.  A momentum-space realisation of the
complete power in Eq.~\eqref{eq:nmrk-one-loop-parameter-power} must therefore
have local one-loop measure \(\dd^D\ell\sim\eta^{3D+4}\).  In the incoming-momentum
lightcone frame \((+,-,\perp)_{(p_2,p_1)}\), where \(p_2\) defines the plus
direction and \(p_1\) the minus direction, write
\[
 (\Delta\ell^+,\Delta\ell^-,|\Delta\ell_\perp|)_{(p_2,p_1)}
 \sim(\eta^a,\eta^b,\eta^c)_{(p_2,p_1)}.
\]
The corresponding measure scales as
\(\eta^{a+b+(D-2)c}\).
Matching this with \(\eta^{3D+4}\) gives
\begin{equation}
 c=3,\qquad a+b=10,\qquad a+b-2c=4>0.
 \label{eq:nmrk-one-loop-glauber-width}
\end{equation}
Here the \(D\)-dependent part fixes \(c=3\), while the remaining power fixes
\(a+b=10\).

We now reconstruct the component widths directly from the pinching propagator
pairs, thereby confirming Eq.~\eqref{eq:nmrk-one-loop-glauber-width} and fixing
the individual longitudinal exponents \(a\) and \(b\).  Normalise momenta to
the growing beam scale, so that
\(p_2^+\) and \(p_1^-\) are order one.  The two central momenta and the
corresponding self-crossing splitting fractions are then of order \(\eta\).
Let \(q_e\) denote the momentum through the edge labelled by \(x_e\).
Orienting \((q_1,q_2)\) away from the \(p_1\) vertex and
\((q_4,q_5)\) away from the \(p_2\) vertex gives the exact routing
\[
 p_1=q_1+q_2,\qquad p_2=q_4+q_5,\qquad
 q_0=q_1-p_6=p_3-q_5,\qquad
 q_3=q_2-p_5=p_4-q_4.
\]
The edge values at the pinch obey
\begin{equation}
 q_1^-,q_5^+\sim\eta^0,
 \qquad q_2^-,q_4^+\sim\eta.
 \label{eq:nmrk-pinched-longitudinal-components}
\end{equation}
In the same hard-normalised frame, the two resolved \(t\)-channel
propagators consequently have the average momentum scalings
\begin{equation}
 q_0\sim(\eta^2,\eta^2,\eta)_{(p_2,p_1)},\qquad
 q_3\sim(\eta,\eta,\eta)_{(p_2,p_1)}.
 \label{eq:nmrk-resolved-transfer-values}
\end{equation}
Thus \(q_0\) is Glauber whereas \(q_3\) is soft, even though both have
virtuality of order \(\eta^2\).  Component scaling, rather than virtuality
alone, distinguishes the two modes.
Combining these coefficients with the virtuality powers
\((2,5,6,2,6,5)\), the two members of each pole pair give the same width,
\begin{equation}
 \Delta\ell^+\sim
 \frac{|q_1^2|}{|q_1^-|}\sim
 \frac{|q_2^2|}{|q_2^-|}\sim\eta^5,
 \qquad
 \Delta\ell^-\sim
 \frac{|q_5^2|}{|q_5^+|}\sim
 \frac{|q_4^2|}{|q_4^+|}\sim\eta^5.
 \label{eq:nmrk-longitudinal-pole-widths}
\end{equation}
Positivity of the LP parameters at the pinch places the two poles in each pair on
opposite sides of the contour.  The active transverse momenta vanish at the
pinch, so their first fluctuation is quadratic; the order-\(\eta^6\)
virtualities of \(q_2\) and \(q_4\) require
\(|\Delta\ell_\perp|\sim\eta^3\).  Thus the symmetric hard-normalised frame
gives
\begin{equation}
 \Delta\ell\sim(\eta^5,\eta^5,\eta^3)_{(p_2,p_1)}.
 \label{eq:nmrk-direct-pole-widths}
\end{equation}
A longitudinal boost shifts the first two entries oppositely but preserves
their sum.

The pole analysis consequently gives
\(\dd^D\ell\sim\eta^{5+5+3(D-2)}=\eta^{3D+4}\).  Including the six
propagators completes the momentum-space power count,
\[
 I_{\mathrm{NMRK}}^{\rm scalar}
 \sim\frac{\dd^D\ell}{\prod_{e=0}^{5}q_e^2}
 \sim\eta^{3D+4-26}=\eta^{3D-22},
\]
in direct agreement with Eq.~\eqref{eq:nmrk-one-loop-parameter-power}.
Independently of the frame choice, the inequality
\(a+b-2c=4>0\) in Eq.~\eqref{eq:nmrk-one-loop-glauber-width} identifies the
unique loop width as Glauber.  Its average \(q_0\) momentum is also Glauber,
so \(x_0\) is an explicit Glauber propagator.  The two oriented markers in
Fig.~\ref{fig:one-loop-hexagon} are at its incident \(p_3\) and \(p_6\)
vertices.  They are distinct from the two incoming factor-localisation
vertices.

\subsubsection{Twisted-hexagon HR in generic DSC}

The DSC limit uses the momentum-twistor chart of
Ref.~\cite{Duhr:2025lyg}, not the lightcone transverse variables \(z,w\)
above.  Here \(a\) is the hard seed modulus,
\(\tau_i=(1-\xi_i)/\xi_i\) encode the two longitudinal momentum fractions,
and \(z_D,\bar z_D\) are the two twistor roots that remain after the
double-collinear deformation.  The expansion parameter is
\(\varepsilon\to0^+\).  On the physical DSC sheet the symbols
\(z_D,\bar z_D\) are independent real variables rather than a conjugate
pair.  A convenient connected real slice is
\begin{equation}
 a<0,\qquad -1<\tau_1<0,\qquad \tau_2<-1,\qquad
 z_D<0,\qquad \bar z_D<0.
 \label{eq:dsc-domain-example}
\end{equation}
With \(p_1,p_2\) incoming, \(p_3,\ldots,p_6\) outgoing and the invariants
defined as in Eq.~\eqref{eq:six-point-mandelstam-convention}, the adjacent
invariants have the following small-\(\varepsilon\) expansions:
\begin{align}
 s_{12}&=-\frac{\tau_2}{a(1+\tau_1)(1+\tau_2)}+O(\varepsilon),&
 s_{23}&=\frac{\varepsilon^2\tau_1}{a\bar z_D}+O(\varepsilon^3),\nonumber\\
 s_{34}&=\frac{1}{(a-1)(1+\tau_2)}+O(\varepsilon),&
 s_{45}&=-\frac1a,\nonumber\\
 s_{56}&=\frac{\tau_1}{(a-1)(1+\tau_1)}+O(\varepsilon),&
 s_{16}&=\frac{\varepsilon^2\tau_2z_D}{a}+O(\varepsilon^3).
 \label{eq:dsc-invariants-example}
\end{align}
Thus \(s_{23},s_{16}<0\), while
\(s_{12},s_{34},s_{45},s_{56}>0\), as required for \(2\to4\) scattering
with legs 1 and 2 incoming.  The two limits therefore use the same graph and
physical scattering channel, but they are distinct paths in six-point
kinematic space.  In terms of the invariant cross ratios
Eq.~\eqref{eq:hexagon-self-crossing-ratios}, the DSC chart obeys
\begin{equation}
 1-U_\times=O(\varepsilon^2),\qquad
 V_\times=O(\varepsilon^2),\qquad
 W_\times=O(\varepsilon^2),\qquad
 V_\times-W_\times=O(\varepsilon^2).
 \label{eq:dsc-self-crossing-endpoint}
\end{equation}
It therefore approaches the endpoint
\((U_\times,V_\times,W_\times)=(1,0,0)\) of the self-crossing surface,
rather than remaining on its interior at finite \(\varepsilon\).  The DSC
roots \(z_D,\bar z_D\) consequently have no notational or algebraic
identification with the NMRK transverse coordinates \(z,w\).

As in the NMRK example, the cancelling monomials of
Eq.~\eqref{eq:twisted-hexagon-F} lie at different fixed-parameter orders.
HRF therefore selects the alignment vector
\begin{equation}
 \boldsymbol\phi_{\mathrm{DSC}}=(-2,-2,-2,-1,-2,-2).
 \label{eq:dsc-alignment-vector}
\end{equation}
Part~I of the core HRF algorithm, applied to this selected face, identifies
the cancellation factors
\begin{equation}
 f_{1,\mathrm{DSC}}=(1+\tau_2)x_4+x_5,\qquad
 f_{2,\mathrm{DSC}}=\tau_1x_1+(1+\tau_1)x_2.
 \label{eq:dsc-factors-example}
\end{equation}
In terms of these factors, the aligned face is
\begin{equation}
 \mathcal F_{\star,\mathrm{DSC}}
 \equiv\mathcal F^{[\boldsymbol\phi_{\mathrm{DSC}}]}
 =\mathcal F_{\mathrm{SL,DSC}}
 =-\frac{f_{1,\mathrm{DSC}}f_{2,\mathrm{DSC}}}
 {a(1+\tau_1)(1+\tau_2)},
 \qquad \mathcal F_{\mathrm{Obs,DSC}}=0.
 \label{eq:dsc-decomposition-example}
\end{equation}
The physical domain in Eq.~\eqref{eq:dsc-domain-example} makes the two
positive zeros explicit:
\begin{equation}
 x_5=-(1+\tau_2)x_4>0,\qquad
 x_2=-\frac{\tau_1}{1+\tau_1}x_1>0.
 \label{eq:dsc-positive-orthant-pinch-example}
\end{equation}
For crossed double-timelike collinear splittings \(\tau_i>0\), both factors
are strictly positive for positive \(x_i\), so they have no positive zeros
and the pinch disappears.

Part~II of the core HRF algorithm, applied to the aligned face alone, gives a
unique candidate core vector, which is uniform:
\begin{equation}
 \boldsymbol v_{\rm core,DSC}^{(0)}=(-1,-1,-1,-1,-1,-1)
 \label{eq:dsc-provisional-vector}
\end{equation}
This face-only result is provisional.  Returning to the full native
polynomial, Part~II rescales both the LP parameters and the kinematics and
tests every occupied layer.  With
\(\ideal_D=\langle f_{1,\mathrm{DSC}},f_{2,\mathrm{DSC}}\rangle\), the first
three occupied layers of the accepted rescaled polynomial are
\begin{equation}
 \mathcal F^{(\HR)}=
 \underbrace{\varepsilon^{-4}\SLcolour{\bigl(\mathcal F_{-4}\in\ideal_D^2\bigr)}}_{
  \operatorname{ord}_{\ideal_D}=2}
 +\underbrace{\varepsilon^{-3}\Othercolour{\bigl(\mathcal F_{-3}\in
  \ideal_D\setminus\ideal_D^2\bigr)}}_{\operatorname{ord}_{\ideal_D}=1}
 +\underbrace{\varepsilon^{-2}\HRcolour{\bigl(\mathcal F_{-2}\notin\ideal_D\bigr)}}_{
  \operatorname{ord}_{\ideal_D}=0}.
 \label{eq:dsc-two-layer-colours}
\end{equation}
The intervening \(\varepsilon^{-3}\) layer contains 52 nonzero terms.  Since
\(\mathcal F_{-3}\in\ideal_D\setminus\ideal_D^2\), it vanishes to first order
on the common pinch.  At the same scaling,
\begin{equation}
 \mathcal U^{(\HR)}
 =\varepsilon^{-2}\HRcolour{\mathcal U_{-2}}+\Othercolour{x_3},
 \qquad
 \mathcal U_{-2}=x_0+x_1+x_2+x_4+x_5,
 \label{eq:dsc-U-leading}
\end{equation}
Together with Eq.~\eqref{eq:dsc-U-leading}, the common resolved weight
\[
 -4+2=-3+1=-2+0=-2
\]
of the three occupied ideal layers fixes the corrected relative vector and
the final original-coordinate answer:
\begin{align}
 \boldsymbol v_{\rm core,DSC}&=(0,0,0,1,0,0),&
 \boldsymbol v_{\HR,\mathrm{DSC}}=\boldsymbol\phi_{\mathrm{DSC}}
 +\boldsymbol v_{\rm core,DSC}
   &=(-2,-2,-2,0,-2,-2),\nonumber\\
 \vec v_{\HR,\mathrm{DSC}}&=(\boldsymbol v_{\HR,\mathrm{DSC}};1),&
 (\WSL,\WHR)&=(-4,-2).
 \label{eq:dsc-vector-example}
\end{align}
The gap is two, and a local dissection pulls back to the same vector.  This
is why asymptotic-order alignment cannot be treated as a cosmetic
preprocessing step: adding
\(\boldsymbol\phi_{\mathrm{DSC}}+
\boldsymbol v_{\rm core,DSC}^{(0)}\) would give both the wrong
vector and the wrong cancellation depth.

The corresponding momentum audit is even more direct.  The unrestricted
edge measure has weight \(-10\); the two first-order normal restrictions
restore two powers, and the resolved LP weight is \(-2\):
\begin{equation}
 I_{\mathrm{DSC}}^{\rm scalar}\sim
 \varepsilon^{-10+2}\bigl(\varepsilon^{-2}\bigr)^{-D/2}
 =\varepsilon^{D-8}.
 \label{eq:dsc-one-loop-parameter-power}
\end{equation}
The propagator virtuality powers are
\((2,2,2,0,2,2)\), again with sum 10.  Thus the loop measure is
\(\varepsilon^{D+2}\).  In the incoming-momentum lightcone frame
\((+,-,\perp)_{(p_2,p_1)}\), where \(p_2\) defines the plus direction and
\(p_1\) the minus direction, write
\[
 (\Delta\ell^+,\Delta\ell^-,|\Delta\ell_\perp|)_{(p_2,p_1)}\sim
 (\varepsilon^{a_{\mathrm{DSC}}},\varepsilon^{b_{\mathrm{DSC}}},
  \varepsilon^{c_{\mathrm{DSC}}})_{(p_2,p_1)}
\]
The exponents then satisfy
\begin{equation}
 c_{\mathrm{DSC}}=1,\qquad
 a_{\mathrm{DSC}}+b_{\mathrm{DSC}}=4,\qquad
 a_{\mathrm{DSC}}+b_{\mathrm{DSC}}-2c_{\mathrm{DSC}}=2>0.
 \label{eq:dsc-one-loop-glauber-width}
\end{equation}
The individual component widths follow directly from the pinching propagator
pairs.  Normalise momenta to the hard beam scale, as in the NMRK analysis.
Because the DSC splitting fractions \(\tau_1\) and \(\tau_2\) are fixed, the
two lines in each pinching pair carry finite nonzero longitudinal fractions:
\(q_1^-,q_2^-,q_4^+,q_5^+\sim\varepsilon^0\).  Combining these coefficients
with the order-\(\varepsilon^2\) virtualities gives
\begin{equation}
 \Delta\ell^+\sim
 \frac{|q_1^2|}{|q_1^-|}\sim
 \frac{|q_2^2|}{|q_2^-|}\sim\varepsilon^2,
 \qquad
 \Delta\ell^-\sim
 \frac{|q_4^2|}{|q_4^+|}\sim
 \frac{|q_5^2|}{|q_5^+|}\sim\varepsilon^2.
 \label{eq:dsc-longitudinal-pole-widths}
\end{equation}
Positivity of the LP parameters at the pinch places the two poles in each
pair on opposite sides of the contour.  The transverse virtuality is also of
order \(\varepsilon^2\), so
\(|\Delta\ell_\perp|\sim\varepsilon\).  The symmetric hard-normalised frame
therefore gives
\begin{equation}
 (\Delta\ell^+,\Delta\ell^-,|\Delta\ell_\perp|)_{(p_2,p_1)}
 \sim(\varepsilon^2,\varepsilon^2,\varepsilon)_{(p_2,p_1)}.
 \label{eq:dsc-direct-pole-widths}
\end{equation}
For the average propagator momenta, the routing introduced in the NMRK
analysis instead gives
\begin{equation}
 q_0\sim(\varepsilon^2,\varepsilon^2,\varepsilon)_{(p_2,p_1)},\qquad
 q_3\sim(1,1,1)_{(p_2,p_1)}.
 \label{eq:dsc-resolved-transfer-values}
\end{equation}
Accordingly, \(x_0\) is again an explicit Glauber propagator, while \(x_3\)
is hard and is absorbed into the four-point hard vertex associated with the
\((p_4,p_5)\) subprocess.  In DSC the Glauber scaling of the average
\(q_0\) momentum coincides with its local integration width.
The resulting measure
\(\varepsilon^{2+2+(D-2)}=\varepsilon^{D+2}\) confirms
Eq.~\eqref{eq:dsc-one-loop-glauber-width} and fixes
\(a_{\mathrm{DSC}}=b_{\mathrm{DSC}}=2\).  A longitudinal boost shifts these
two exponents oppositely while preserving their sum.  Thus the DSC loop is
Glauber, as in the NMRK example, and the marked Glauber vertices
in Fig.~\ref{fig:one-loop-hexagon} are incident to \(x_0\).  The incoming
three-point vertices carrying \(p_1\) and \(p_2\) are the separate
factor-localisation vertices.  The transverse width
and longitudinal-to-transverse imbalance nevertheless differ from NMRK.

\subsubsection{Inheritance from the wide-angle self-crossing region}

We label the wide-angle factors in
Eq.~\eqref{eq:hexagon-wide-angle-boundary-factorisation} so that
\(f_{1,\mathrm{WA}}\) uses \((x_4,x_5)\) and \(f_{2,\mathrm{WA}}\) uses
\((x_1,x_2)\), matching the NMRK and DSC labels.  Only on the positive
double-scattering pinch of
Eqs.~\eqref{eq:hexagon-self-crossing-subprocesses}
and~\eqref{eq:hexagon-self-crossing-pinched-invariants}, after taking the
leading aligned restriction and identifying the same fractions \(x,y\), do
their monic forms obey
\begin{align}
 \frac{f_{1,\mathrm{WA}}}{s_{145}^{(0)}}
 &=\frac{f_{1,\mathrm{NMRK}}}{1+X_{45}}
  =\frac{f_{1,\mathrm{DSC}}}{1+\tau_2}
  =x_4-\frac{1-y}{y}x_5,
 \nonumber\\
 \frac{f_{2,\mathrm{WA}}}{s_{245}^{(0)}}
 &=\frac{f_{2,\mathrm{NMRK}}}{1+X_{45}}
  =\frac{f_{2,\mathrm{DSC}}}{1+\tau_1}
  =x_2-\frac{1-x}{x}x_1,
 \qquad\text{on the common positive pinch.}
 \label{eq:hexagon-wide-angle-factor-inheritance}
\end{align}
The coefficient identifications responsible for these equalities are
\begin{equation}
 \frac{\widehat X_{34}X_{45}}{1+X_{45}}
 =-\frac{1}{1+\tau_2}=\frac{1-y}{y},
 \qquad
 \frac{(1-z)(1-\bar z)X_{45}\widehat X_{56}}{1+X_{45}}
 =-\frac{\tau_1}{1+\tau_1}=\frac{1-x}{x}.
 \label{eq:composite-factor-map}
\end{equation}
Equation~\eqref{eq:hexagon-self-crossing-subprocesses} identifies \((1-y)/y\)
as the ratio of the \(p_2\)-collinear fractions feeding \((p_3,p_6)\) and
\((p_4,p_5)\), respectively; \((1-x)/x\) is the analogous \(p_1\) ratio.
Thus Eq.~\eqref{eq:composite-factor-map} expresses the same two pinch
momentum-sharing ratios in NMRK and DSC variables.  All these equalities are
pinch-restricted: there Eq.~\eqref{eq:hexagon-wide-angle-factor-inheritance}
is an identity in the LP parameters, not merely an equality of zero sets, but
neither equation maps the full NMRK and DSC parametrisations.
Their occupied layers, and hence their physical vectors, are different:
\begin{table}[H]
\centering
\begin{tabular}{@{}lccc@{}}
\toprule
Expansion & certified edge vector & \((\WSL,\WHR)\) & cancellation depth\\
\midrule
wide-angle self-crossing & \((0,-1,-1,0,-1,-1)\) & \((-2,-1)\) & 1\\
zero-recoil NMRK & \((-2,-5,-6,-2,-6,-5)\) & \((-10,-6)\) & 4\\
generic DSC & \((-2,-2,-2,0,-2,-2)\) & \((-4,-2)\) & 2\\
\bottomrule
\end{tabular}
\caption{Comparison of the three certified one-loop twisted-hexagon regions.}
\label{tab:twisted-hexagon-one-loop-comparison}
\end{table}
In every row the displayed edge vector becomes the homogeneous LP vector by
appending the final entry \(1\).  Equation~\eqref{eq:hexagon-wide-angle-factor-inheritance}
shows that the two composite limits inherit the same seed cancellation
geometry; the distinct resolved layers show why the three HRs must not be
identified.

Finally, the positive Landau equations make the loop diagnosis a genuine
pinch statement.  For a one-loop graph they include
\(\sum_e x_eq_e^+=\sum_e x_eq_e^-=0\).  Since every active \(x_e>0\), the
nonzero \(q_e^+\) and \(q_e^-\) cannot each have a common sign.  The
corresponding \(\ell^-\) and \(\ell^+\) poles therefore approach from
opposite half-planes.  The purple dotted \(x_0\) edge in
Fig.~\ref{fig:one-loop-hexagon} displays the corresponding explicit Glauber
propagator.  Its two incident vertices carry the oriented markers.  The incoming
\(p_1\) and \(p_2\) three-point vertices are the two factor-localisation
vertices.

\subsubsection{Two-loop topology test}

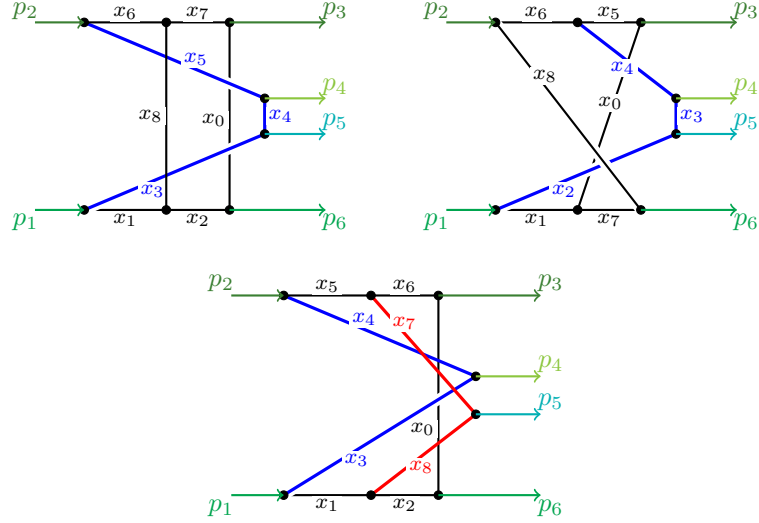
\begin{figure}[H]
\centering
\begin{tikzpicture}[scale=.62]
 \coordinate (v4) at (-2.20,2.00);
 \coordinate (v7) at (-.45,2.00);
 \coordinate (v5) at ( .90,2.00);
 \coordinate (v3) at (1.65,.38);
 \coordinate (v2) at (1.65,-.38);
 \coordinate (v6) at ( .90,-2.00);
 \coordinate (v8) at (-.45,-2.00);
 \coordinate (v1) at (-2.20,-2.00);

 \draw[hrf internal] (v5)--node[hrf edge label,pos=.52,left] {$x_0$}(v6);
 \draw[hrf internal] (v1)--node[hrf edge label,pos=.50,below] {$x_1$}(v8);
 \draw[hrf internal] (v8)--node[hrf edge label,pos=.50,below] {$x_2$}(v6);
 \draw[white,line width=4pt] (v1)--(v2);
 \draw[hrf internal,very thick,blue] (v1)--
   node[hrf edge label,pos=.38,below,text=blue] {$x_3$}(v2);
 \draw[hrf internal,very thick,blue] (v2)--
   node[hrf edge label,pos=.50,right,text=blue] {$x_4$}(v3);
 \draw[white,line width=4pt] (v3)--(v4);
 \draw[hrf internal,very thick,blue] (v3)--
   node[hrf edge label,pos=.38,above,text=blue] {$x_5$}(v4);
 \draw[hrf internal] (v4)--node[hrf edge label,pos=.50,above] {$x_6$}(v7);
 \draw[hrf internal] (v7)--node[hrf edge label,pos=.50,above] {$x_7$}(v5);
 \draw[hrf internal] (v7)--node[hrf edge label,pos=.50,left] {$x_8$}(v8);
 \foreach \v in {1,...,8} \node[hrf vertex] at (v\v) {};
 \draw[->,thick,OliveGreen] (-3.25,2)--(v4);
 \draw[->,thick,Green] (-3.25,-2)--(v1);
 \draw[->,thick,OliveGreen] (v5)--(2.95,2);
 \draw[->,thick,LimeGreen] (v3)--(2.95,.38);
 \draw[->,thick,TealBlue] (v2)--(2.95,-.38);
 \draw[->,thick,Green] (v6)--(2.95,-2);
 \node[hrf leg label,text=OliveGreen] at (-3.45,2.25) {$p_2$};
 \node[hrf leg label,text=Green] at (-3.45,-2.25) {$p_1$};
 \node[hrf leg label,text=OliveGreen] at (3.15,2.25) {$p_3$};
 \node[hrf leg label,text=LimeGreen] at (3.15,.63) {$p_4$};
 \node[hrf leg label,text=TealBlue] at (3.15,-.13) {$p_5$};
 \node[hrf leg label,text=Green] at (3.15,-2.25) {$p_6$};
\end{tikzpicture}
\hspace{5mm}
\begin{tikzpicture}[scale=.62]
 \coordinate (v4) at (-2.20,2.00);
 \coordinate (v7) at (-.45,2.00);
 \coordinate (v5) at ( .90,2.00);
 \coordinate (v3) at (1.65,.38);
 \coordinate (v2) at (1.65,-.38);
 \coordinate (v6) at ( .90,-2.00);
 \coordinate (v8) at (-.45,-2.00);
 \coordinate (v1) at (-2.20,-2.00);

 \draw[hrf internal] (v5)--node[hrf edge label,pos=.48,above] {$x_0$}(v8);
 \draw[hrf internal] (v1)--node[hrf edge label,pos=.50,below] {$x_1$}(v8);
 \draw[white,line width=4pt] (v1)--(v2);
 \draw[hrf internal,very thick,blue] (v1)--
   node[hrf edge label,pos=.38,below,text=blue] {$x_2$}(v2);
 \draw[hrf internal,very thick,blue] (v2)--
   node[hrf edge label,pos=.50,right,text=blue] {$x_3$}(v3);
 \draw[hrf internal,very thick,blue] (v3)--
   node[hrf edge label,pos=.54,below,text=blue] {$x_4$}(v7);
 \draw[hrf internal] (v7)--node[hrf edge label,pos=.50,above] {$x_5$}(v5);
 \draw[hrf internal] (v4)--node[hrf edge label,pos=.50,above] {$x_6$}(v7);
 \draw[hrf internal] (v8)--node[hrf edge label,pos=.50,below] {$x_7$}(v6);
 \draw[white,line width=4pt] (v4)--(v6);
 \draw[hrf internal] (v4)--node[hrf edge label,pos=.34,above] {$x_8$}(v6);
 \foreach \v in {1,...,8} \node[hrf vertex] at (v\v) {};
 \draw[->,thick,OliveGreen] (-3.25,2)--(v4);
 \draw[->,thick,Green] (-3.25,-2)--(v1);
 \draw[->,thick,OliveGreen] (v5)--(2.95,2);
 \draw[->,thick,LimeGreen] (v3)--(2.95,.38);
 \draw[->,thick,TealBlue] (v2)--(2.95,-.38);
 \draw[->,thick,Green] (v6)--(2.95,-2);
 \node[hrf leg label,text=OliveGreen] at (-3.45,2.25) {$p_2$};
 \node[hrf leg label,text=Green] at (-3.45,-2.25) {$p_1$};
 \node[hrf leg label,text=OliveGreen] at (3.15,2.25) {$p_3$};
 \node[hrf leg label,text=LimeGreen] at (3.15,.63) {$p_4$};
 \node[hrf leg label,text=TealBlue] at (3.15,-.13) {$p_5$};
 \node[hrf leg label,text=Green] at (3.15,-2.25) {$p_6$};
\end{tikzpicture}

\vspace{3mm}
\begin{tikzpicture}[scale=.66]
 \coordinate (v4) at (-2.20,2.00);
 \coordinate (v7) at (-.45,2.00);
 \coordinate (v5) at ( .90,2.00);
 \coordinate (v3) at (1.65,.38);
 \coordinate (v2) at (1.65,-.38);
 \coordinate (v6) at ( .90,-2.00);
 \coordinate (v8) at (-.45,-2.00);
 \coordinate (v1) at (-2.20,-2.00);

 \draw[hrf internal] (v5)--node[hrf edge label,pos=.67,left] {$x_0$}(v6);
 \draw[hrf internal] (v1)--node[hrf edge label,pos=.50,below] {$x_1$}(v8);
 \draw[hrf internal] (v8)--node[hrf edge label,pos=.50,below] {$x_2$}(v6);
 \draw[white,line width=4pt] (v1)--(v3);
 \draw[hrf internal,very thick,blue] (v1)--
   node[hrf edge label,pos=.38,below,text=blue] {$x_3$}(v3);
 \draw[hrf internal,very thick,blue] (v3)--
   node[hrf edge label,pos=.58,above,text=blue] {$x_4$}(v4);
 \draw[hrf internal] (v4)--node[hrf edge label,pos=.50,above] {$x_5$}(v7);
 \draw[hrf internal] (v7)--node[hrf edge label,pos=.50,above] {$x_6$}(v5);
 \draw[hrf internal,very thick,red] (v7)--
   node[hrf edge label,pos=.32,above,text=red] {$x_7$}(v2);
 \draw[hrf internal,very thick,red] (v2)--
   node[hrf edge label,pos=.52,below,text=red] {$x_8$}(v8);
 \foreach \v in {1,...,8} \node[hrf vertex] at (v\v) {};
 \draw[->,thick,OliveGreen] (-3.25,2)--(v4);
 \draw[->,thick,Green] (-3.25,-2)--(v1);
 \draw[->,thick,OliveGreen] (v5)--(2.95,2);
 \draw[->,thick,LimeGreen] (v3)--(2.95,.38);
 \draw[->,thick,TealBlue] (v2)--(2.95,-.38);
 \draw[->,thick,Green] (v6)--(2.95,-2);
 \node[hrf leg label,text=OliveGreen] at (-3.45,2.25) {$p_2$};
 \node[hrf leg label,text=Green] at (-3.45,-2.25) {$p_1$};
 \node[hrf leg label,text=OliveGreen] at (3.15,2.25) {$p_3$};
 \node[hrf leg label,text=LimeGreen] at (3.15,.63) {$p_4$};
 \node[hrf leg label,text=TealBlue] at (3.15,-.13) {$p_5$};
 \node[hrf leg label,text=Green] at (3.15,-2.25) {$p_6$};
\end{tikzpicture}
\caption{Two-loop graphs in the same rapidity-ordered embedding as
Fig.~\ref{fig:one-loop-hexagon}.  The upper-left graph is the planar
hexagon--box.  The upper-right and lower graphs have the same
hexagon--pentagon internal topology but different external-leg attachments.
In the upper two configurations both central emissions lie on the same
\(t\)-channel path, shown in blue, and contraction of the additional edges
recovers the externally labelled twisted-box seed.  In the lower configuration
the emissions lie on separate paths, shown in blue and red, and the externally
labelled twisted-box seed is not a contraction minor.  Black dots denote vertices.}
\label{fig:two-loop-six-point-graphs}
\end{figure}

Having established the one-loop twisted-hexagon realisation of the seed, we
now ask which two-loop graphs retain its cancellation mechanism.  The nearly horizontal chains in
Fig.~\ref{fig:one-loop-hexagon} are the \((23)\) and \((16)\) collinear
sectors.  A \(t\)-channel exchange is an internal path connecting these two
chains.  The central emissions \(p_4,p_5\) may lie on the same such path or
on two separate paths.  Figure~\ref{fig:two-loop-six-point-graphs} defines
the three tested graph configurations and all their edge parameters.

Table~\ref{tab:six-point-two-loop-topology-audit} summarises the completed
audits.
\begin{table}[H]
\centering
\small
\setlength{\tabcolsep}{5pt}
\begin{tabular}{@{}lcccc@{}}
\toprule
Graph & emission paths & \shortstack{labelled seed\\minor?} & zero-recoil NMRK & generic DSC\\
\midrule
twisted hexagon & same & yes & 1 interior \([1]\) & 1 interior\\
planar hexagon--box & same & yes & 13 interior \(+\) 12 boundary \([7]\) & 1 interior\\
hexagon--pentagon & same & yes & 3 interior \(+\) 6 boundary \([4]\) & none\\
hexagon--pentagon & separate & no & none \([0]\) & none\\
\bottomrule
\end{tabular}
\caption{Certified HR counts in the one- and two-loop six-point topology
audit.  The square-bracketed NMRK entry counts distinct Part-I generators
that supply at least one certified HR.  ``Labelled seed minor'' uses the
externally labelled sense defined in
Sec.~\ref{sec:boundary}.}
\label{tab:six-point-two-loop-topology-audit}
\end{table}
The entries count inequivalent certified HRs for the displayed permutation
of external legs.  Two certificates are identified when they have the same
boundary stratum, exact cancellation ideal, normalised pullback vector and
\(W_{\rm HR}\).  The value of \(W_{\rm SL}\), and hence the cancellation depth
assigned to a particular Part-I presentation, does not split an HR.
``Interior'' means that every edge parameter is active; ``boundary'' refers
in this case to a certified codimension-one (one-edge contraction) stratum.
Only codimension-one boundary strata were included in this two-loop audit;
deeper boundary strata were not tested.  The generator count is a coarser
classification: it measures how many distinct Part-I cancellation
structures supply HRs, but one generator can resolve into several HRs.

A staged candidate is a structural Part-I class specified by its restricted
face polynomial and generator set.  It is not counted as an HR unless the
source-aware Part-II construction is applied to the full source
decomposition, with every native kinematic layer of the original
\(\mathcal F\), together with \(\mathcal U\), retained, to determine the
total vector, and the final dissection produces a scaleful lower-facet
certificate.  The thirteen planar hexagon--box
candidates produce 41 such certificates.  After the above identification
they give 25 HRs, 13 interior and 12 boundary, supplied by seven distinct
generators.  Two of these HRs each admit two different \(W_{\rm SL}\) and
cancellation-depth presentations, giving 27 strict certificate-presentation
classes but no additional HRs.  For the same-path hexagon--pentagon, three
of the eight staged candidates have no common pinch in the positive domain,
one has no certified facet after the complete dissection, and the remaining
four produce nine HRs, three interior and six boundary, supplied by four
distinct generators.  The separate-path graph has no staged candidate.

Within this sample, placing both emissions on one exchange is necessary:
the separate-exchange graph has no staged candidate in either limit and no
externally labelled twisted-box contraction minor.  For the same-path graphs,
contracting the additional edges preserves the external momentum labels and
their attachment to the seed vertices, so the seed HR is inherited on the
corresponding contraction stratum.  Since this is a deeper boundary, it lies
outside the interior and codimension-one strata tested in the two-loop audit;
the absence of a generic-DSC entry for the same-path hexagon--pentagon therefore
does not exclude this inherited boundary HR.  Its certified NMRK regions show
in addition that the seed mechanism extends away from the contraction stratum
in that limit.

\subsection{Algorithmic capabilities demonstrated by the examples}
\label{sec:capabilities-demonstrated}

The examples above demonstrate the successive capabilities collected in the
algorithm summary of Sec.~\ref{sec:workflow}.  The first is to recover
cancellation factors from derivatives, as described in
Sec.~\ref{sec:decomposition}.  In the Crown, individual derivatives mix terms
weighted by the independent invariants \(s_{12}\) and \(s_{23}\).  Separating
the corresponding kinematic components exposes the four factors in
Eq.~\eqref{eq:crown-F0} and permits their pairing into two generators.  The
Crown descendants in Sec.~\ref{sec:example-crown-boundaries} show that the
same derivative harvest also identifies general multi-term cancellation
factors and the generators constructed from them.

A second capability is demonstrated by the nonzero obstruction in the
spacelike-collinear five-point Fish.  At the common zero of \(f_1\) and
\(f_2\), the term \(\FObs=sz(1+\chi)x_2x_3x_6\) in
Eq.~\eqref{eq:five-F0-obstruction} remains nonzero throughout the physical
domain~\eqref{eq:five-domain}.  Thus the strict leading polynomial
\(\mathcal F_0\) does not itself admit the stationary Landau solution; the
stationary conditions~\eqref{eq:SLpinch} apply only to the selected
\(\FSL=s f_1f_2\).  Part~I must therefore determine the decomposition
\(\mathcal F_\star=\FSL+\FObs\) of
Eqs.~\eqref{eq:candidate-obstruction}--\eqref{eq:obstruction} before Part~II
can determine the non-uniform vector in Eq.~\eqref{eq:five-HR-vector}, under
which the monomials of \(\FSL\) become individually superleading.  The
five-point seed and its vertex correction also introduce kinematic variables
inside the cancellation factors; compare
Eqs.~\eqref{eq:five-F0-obstruction} and~\eqref{eq:five-vertex-factors}.  Thus a
factorisation is physically admissible only if the factors have a common zero
at positive LP parameters in the stated kinematic domain, as required by
Eq.~\eqref{eq:commonzero}.

The near-planar wide-angle example demonstrates how factors can be recovered
when no individual derivative displays them.  Here the saturation procedure
of Eq.~\eqref{eq:selected-gradient-saturation}, which permits division only by
quantities known to be nonzero in the chosen sector, becomes the explicit
calculation in Eqs.~\eqref{eq:five-near-planar-saturation-data}
and~\eqref{eq:five-near-planar-saturated-ideal}.  It recovers the three factors
in Eq.~\eqref{eq:five-near-planar-factors}, and hence the three pair-product
generators appearing in the decomposition
Eq.~\eqref{eq:five-coplanar-factorisation}.

Correlating two kinematic limits can make the stationary ratios rate
dependent.  For the reference assignment in the simultaneous MRK and
near-planar limit of Sec.~\ref{sec:five-point-mrk-planar}, all three ratios in
Eq.~\eqref{eq:five-mrk-planar-ratios} remain positive at every finite
\(x>0\), while \(\rho_B\to0\) as the MRK limit is approached.  Thus this
particular positive solution approaches the boundary of the positive LP
orthant but does not leave it.  This behaviour does not determine the result
for another external-leg assignment: each inequivalent permutation must be
tested separately unless it is related to the first by a symmetry.  In the
example, the two assignments in
Fig.~\ref{fig:five-mrk-planar-attachments}(a,b) retain positive stationary
ratios, whereas the \(p_3\leftrightarrow p_5\) assignment in panel~(c) gives
the negative ratios in
Eq.~\eqref{eq:five-mrk-planar-p3-p5-negative-ratios}; its stationary solution
therefore lies outside the positive LP orthant and it has no HR.

The same correlated limit presents a separate hierarchy issue.  In
Eq.~\eqref{eq:five-mrk-planar-F-layers}, generator terms occur both at
\(W_{\rm SL}\) and at an intervening nominal weight \(W_k\), and both
contributions vanish on the same pinch.  The Part-II hierarchy must therefore
include both occupied cancellation layers.

Asymptotic-order alignment is needed when the relevant cancelling monomials
start at different powers of the expansion parameter.  Its general
construction and the composition of the alignment and core vectors are given
in Sec.~\ref{sec:alignment}, especially
Eq.~\eqref{eq:alignment-composition}.  The central-soft example in
Sec.~\ref{sec:five-point-central-soft-status} and the six-point NMRK example in
Sec.~\ref{sec:six-point-nmrk-seed} illustrate this operation.  The DSC example
adds a further lesson: after alignment, the omitted layers of the full LP
polynomial must be restored.  The intervening layer in
Eq.~\eqref{eq:dsc-two-layer-colours} vanishes on the same pinch and changes the
provisional face-only scaling to the final vector in
Eq.~\eqref{eq:dsc-vector-example}, following the general test in
Eq.~\eqref{eq:finalhierarchy}.

Across these examples, the leading nonzero dependence normal to the pinch and
every restored contribution that cancels on the same locus enter at the same
final scaling, even when they originate at different orders of the kinematic
expansion.  This saturation is an empirical regularity of the asymptotic
hierarchy, not an assumption of the construction:
Eq.~\eqref{eq:finalhierarchy} requires a resolved contribution only to enter at
\(\WHR\) or above.  The distinct polynomial sources must nevertheless be
retained separately until their combined scaling has been checked.

Finally, the boundary wrapper acts around rather than inside the core
construction.  As described in Sec.~\ref{sec:boundary} and illustrated by the
Crown descendants and the two-loop audit in
Table~\ref{tab:six-point-two-loop-topology-audit}, it scans contraction strata
and compares their restricted ideals with the parent ideal.

Independently, every candidate that passes the algebraic stages, whether
interior or boundary, undergoes the final dissection certificate described in
Sec.~\ref{sec:scaleful}.  This maps each relevant local lower-facet normal back
to the original LP variables and thereby certifies the scaling; it is not part
of the construction of the cancellation generators.  The capabilities are
therefore cumulative: the algebraic steps reconstruct information absent from
the original Newton polytope, while the local lower-facet test supplies the
final geometric certificate.

Allowing massive internal propagators would relax the standing massless
multiaffinity assumption.  The resulting quadratic edge-parameter dependence
changes the possible orders of cancellation and the local facet problem.  It is a
genuine extension of this framework, not another example in the present
tour.

\section{From parameter regions to momentum modes}
\label{sec:momentum-reconstruction}

The output of HRF is a parameter-space region.  Its translation into momentum
space is essential for going beyond individual integrals to amplitudes and
observables: physical intuition about scale separation, and the formulation
and proof of factorisation, are naturally expressed in terms of momentum
modes.  Momentum-space reconstruction is therefore the bridge from
parameter-space region finding to amplitude-level physics.  It is a logically
subsequent problem, rather than an additional stage of the region-finding
algorithm.  We address that problem here, after the examples, by distinguishing
three questions that are easily conflated: the virtuality of each propagator,
the average loop-momentum configuration at the pinch and its associated edge
momenta, and the local integration widths of the independent loop fluctuations.
The second identifies the momentum modes in the usual EFT sense; the third
determines the integration measure
and hence enters power counting.

In the Schwinger representation, before performing the loop integrations, the
loop-momentum integral is Gaussian at fixed proper times.  For a region
\(\vec v\), we implement the
scaling in Eq.~\eqref{eq:schwinger-inverse-virtuality} by writing
\(Q^2\widetilde x_e=\delta^{v_e}\widehat x_e\), with
\(\widehat x_e=O(1)\).  At fixed values of the projective ratios of the
\(\widehat x_e\), the centre of the resulting Gaussian defines the conditional
average \(\langle\boldsymbol\ell\rangle_{\vec v}\).  Here and below
\(\langle\,\cdot\,\rangle_{\vec v}\) denotes the normalised Gaussian average
at fixed rescaled proper-time ratios and fixed external kinematics, after the
region scaling has been imposed.  Its
componentwise powers define the \emph{average-momentum scaling}.  The standard
deviations obtained from the same Gaussian covariance define the marginal
integration widths.  Agreement between the componentwise average-momentum
scaling and the marginal-width scaling is not assumed; it is tested separately
in the examples.\footnote{A systematic mixed-representation definition of
these averages, standard deviations and cross-loop covariances, together with
a general momentum-reconstruction algorithm for facet and hidden regions,
will be presented in forthcoming work~\cite{GardiZhu:2026regions}.  That work will
also place the faithful-routing and average--width correspondence observed
below on a general footing and distinguish the role of the balanced on-shell condition
for facet and hidden regions.}
Throughout this section, \(q_e\) and
\(\mathcal D_e\) denote the oriented edge momenta and propagator denominators introduced in
Eqs.~\eqref{eq:edge-momentum-routing} and \eqref{eq:momentum-integral}.

\subsection{Inherited virtualities and component constraints}

Let the certified original-coordinate HR vector be
\(\vHR=(\boldsymbol v_{\HR};1)=(v_1,\ldots,v_N;1)\).  Inverse
LP-parameter scaling fixes the
relative virtuality powers of the denominators in
Eq.~\eqref{eq:momentum-integral}, as derived in
Ref.~\cite[Sec.~2.3, especially Eqs.~(46)--(48), and App.~B]{Gardi:2022khw},
\begin{equation}
 \frac{\mathcal D_e}{\mu_{\rm LP}^2}\sim\delta^{-v_e}
 \equiv\delta^{\kappa_e},
 \qquad \kappa_e=-v_e,
 \label{eq:inverse-schwinger-rule}
\end{equation}
in the LP normalisation of Eq.~\eqref{eq:LP-representation}.  By default
\(\mu_{\rm LP}\) is fixed.  If instead the chosen reference scale varies in
the limit --- as in the central-soft rapidity-ordered limit of
Sec.~\ref{sec:five-point-central-soft-status}, where
\(\mu_{\rm LP}^2=s_{12}\) grows as \(\delta^{-4}\) --- virtuality powers relative to a
fixed scale are obtained by adding the common power carried by
\(\mu_{\rm LP}^2\) to every \(\kappa_e\).  HRF also supplies the cancellation ideal and the gap
\(\WHR-\WSL\).  In momentum space this gap reappears as restricted support
normal to the pinch.  Thus HRF determines the virtualities and the amount of
local restriction, but not a unique set of loop momenta or their separate
lightcone components.  For each mode-complete example presented in
Sec.~3, we obtain a faithful loop routing by combining the virtuality constraints with
componentwise momentum conservation in an exact kinematic chart.  The
pinching poles and local denominator homogeneity then determine the widths;
any residual cross-loop restriction is determined from the joint
momentum-space support and independently checked against the LP power count.

Choose null reference directions \(n_+\) and \(n_-\).  We record this choice
on every potentially ambiguous component triple by the subscript
\((+,-,\perp)_{(n_+,n_-)}\), where the entries name the reference directions
defining the plus and minus components, respectively.  An edge value is then
written as
\begin{equation}
 q_e\sim(\delta^{a_e},\delta^{b_e},\delta^{c_e})_{(n_+,n_-)}
 \equiv(q_e^+,q_e^-,|q_{e\perp}|)_{(n_+,n_-)}.
 \label{eq:edge-component-powers}
\end{equation}
In the absence of a cancellation between the leading longitudinal and
transverse terms,
\begin{equation}
 \kappa_e=\min(a_e+b_e,2c_e).
 \label{eq:direct-virtuality-rule}
\end{equation}
When
\begin{equation}
 a_e+b_e=2c_e,
 \label{eq:on-shell-mode-condition}
\end{equation}
the longitudinal and transverse contributions have homogeneous scaling.  We
refer to Eq.~\eqref{eq:on-shell-mode-condition} as the \emph{balanced on-shell
condition} for a momentum mode.  The qualifier ``balanced'' emphasises this
homogeneity and does not imply a small virtuality; the condition applies
equally to UV modes.  The complementary Glauber hierarchy,
\(q_e^+q_e^-\ll q_{e\perp}^2\), has long been recognised in the classic
factorisation analyses~\cite{Collins:1984kg,Collins:1985ue}; see in particular
Ref.~\cite[Sec.~8.4, especially Eq.~(137)]{Collins:1989gx}.  The standard hard,
jet and soft scalings used here are reviewed in
Ref.~\cite[Sec.~2.3, especially Eqs.~(2.42)--(2.45)]{Gardi:2022khw}, while
Ref.~\cite[Sec.~4, especially Eqs.~(4.15)--(4.17)]{Gardi:2025ule} gives explicit
facet-region reconstructions based on the balanced on-shell condition.
If the common power in Eq.~\eqref{eq:on-shell-mode-condition} is smaller than
\(\kappa_e\), the longitudinal and transverse terms in
\(q_e^2=q_e^+q_e^- -q_{e\perp}^2\) must cancel at leading order.  Such an
enhancement can occur for individual edges even in facet regions, and does so
in our examples.  It cannot be inferred from the local triple
\((a_e,b_e,c_e)\) alone; whether and where it occurs is determined by the
global solution of the scaling constraints and componentwise momentum
conservation.

At each vertex, momentum conservation is imposed component by component.
At valuation level this requires the smallest exponent in every signed sum
to occur at least twice.  We then verify the signs and leading coefficients
in an exact kinematic chart.  Equations~\eqref{eq:direct-virtuality-rule} and
\eqref{eq:on-shell-mode-condition}, together with componentwise momentum
conservation, constrain the component powers but do not by themselves provide
a general existence or uniqueness theorem.  In the examples we use the
balanced on-shell condition for independent loop variables whenever it is compatible
with the certified HR data; we do not impose it on every loop, since a
transverse-dominated Glauber mode necessarily violates it.  Leading cancellations
between longitudinal and transverse terms can occur in either facet or hidden
regions and can be obscured by a lightcone frame not aligned with the relevant
collinear direction.  For HRs, moreover, the cancellation equations can
correlate leading coefficients between momentum components or between loops.
Such information is lost when only the powers of \(\delta\) are retained.

\subsection{Loop bases, routing freedom and correlated support}

The edge virtualities inherited from HRF do not by themselves define a
momentum region.  Reconstructing the corresponding momentum region is a global
problem and requires a consistent set of independent loop
variables.\footnote{This is a general issue, not one specific to hidden
regions.  The need for such a faithful choice of independent loop momenta was
observed and analysed explicitly for ordinary facet regions in the two-loop
soft anomalous dimension in Ref.~\cite[Sec.~4]{Gardi:2025ule}: an arbitrary
independent routing need not make the region scaling manifest.}  Choose
these variables as chords of a spanning tree, or use any affine rerouting by
external momenta, and express every edge momentum in that basis.  Two such
bases describe the same region when their local fluctuations are related by
a nonsingular linear transformation.  In particular, a combination such as
\(q_i-q_j\) may be the natural loop variable even when both individual edge
momenta look soft or collinear.

It is useful to write the distinction between the centre and the local
support explicitly.  Collecting the independent loop components into
\(\boldsymbol\ell\), let \(\Sigma_\ell\) denote their Gaussian covariance at
fixed projective ratios of the \(\widehat x_e\).  In a fixed loop-momentum
basis the covariance depends on the proper times, whereas the external
kinematics enter the centre
\(\langle\boldsymbol\ell\rangle_{\vec v}\).  A local region may then be
written as
\begin{equation}
 \begin{aligned}
 \Sigma_\ell
 &=T_{\widehat{\x}}(\delta)
   T_{\widehat{\x}}^{\rm T}(\delta),
 &\qquad
 \boldsymbol\ell
 &=\langle\boldsymbol\ell\rangle_{\vec v}
   +T_{\widehat{\x}}(\delta)\boldsymbol u,
 \\[1mm]
 \langle\boldsymbol u\boldsymbol u^{\rm T}\rangle_{\vec v}
 &=\boldsymbol 1,
 &
 \dd^{LD}\boldsymbol\ell
 &=\bigl|\det T_{\widehat{\x}}(\delta)\bigr|
   \,\dd^{LD}\boldsymbol u.
 \end{aligned}
 \label{eq:loop-centre-width-matrix}
\end{equation}
Here \(T_{\widehat{\x}}(\delta)\) is a matrix square root of the
Gaussian covariance.  This is the standard Gaussian whitening
transformation~\cite{Kessy:2018whitening}; such a factorisation exists for the
nondegenerate Gaussian at finite \(\delta\).  It defines
\(\boldsymbol u=O(1)\) with unit
covariance, whose components are therefore independent.  The matrix
\(T_{\widehat{\x}}(\delta)\) carries the complete local
covariance structure, including both the component standard deviations and
the correlations between loop variables.  A product of separately
listed component widths gives the measure only after a mode-adapted change of
coordinates has made those correlations explicit, for example by putting
\(T_{\widehat{\x}}(\delta)\) in triangular form.  The component powers of
\(\langle\boldsymbol\ell\rangle_{\vec v}\) do not determine
\(T_{\widehat{\x}}(\delta)\), and the marginal widths
alone do not determine its determinant.  The examples exhibit three distinct
possibilities:
average and marginal scalings can agree in a suitable loop basis; an
individual dependent edge need not represent either member of that basis;
and, even when they agree loop by loop, the joint support can acquire an
additional suppression through correlations between several loops.  The last
mechanism occurs in the wide-angle Crown and in the near-planar five-point
examples.

There is also a harmless longitudinal-frame freedom.  A boost changes
\begin{equation}
 a\longrightarrow a+\chi,\qquad
 b\longrightarrow b-\chi,\qquad
 c\longrightarrow c,
 \label{eq:longitudinal-boost-freedom}
\end{equation}
and therefore preserves \(a+b\), all virtualities and the distinction
between longitudinal and transverse dominance, whether the triple describes
an average momentum or an integration width.  A result which determines only
\(a+b\) and \(c\) can consequently establish this invariant class without
fixing the separate plus and minus powers.  We distinguish this diagnosis
from a \emph{mode-complete} solution, for which an exact chart and pole
analysis determine a faithful component-level momentum routing and its local
integration widths.

No constraint beyond Eqs.~\eqref{eq:inverse-schwinger-rule}--
\eqref{eq:direct-virtuality-rule}, vertex conservation, the already certified
cancellation equations and the physical kinematic chart is imposed.  The
choice of a convenient loop basis and the use of the balanced on-shell condition where
it is compatible with the certified HR organise the solutions but do not
change their invariant content.

\subsection{Pole pinches and local integration widths}
\label{sec:pole-pinches-local-widths}

The scaling of a momentum combination at the pinch and that of its local
integration width are a priori distinct and must be determined independently.
In the faithful routings of our examples they agree loop by loop, although the
joint support may still contain additional cross-loop correlations.  To
determine the widths, express every denominator in a mode-adapted loop basis.
If a denominator is linear in the selected component,
\begin{equation}
 \mathcal D_e=A_e\ell^+ +B_e+i0,
 \qquad
 \ell_e^+=-\frac{B_e}{A_e}-\frac{i0}{A_e},
 \label{eq:light-cone-pole-rule}
\end{equation}
then active denominators with opposite signs of \(A_e\) place the poles
\(\ell_e^+\) on opposite sides of the real \(\ell^+\) axis.  If
\(A_e\sim\delta^{\alpha_e}\) and
\(\mathcal D_e\sim\delta^{\kappa_e}\), their approach restricts the
fluctuation to
\begin{equation}
 \Delta\ell^+\sim\delta^{\kappa_e-\alpha_e}.
 \label{eq:light-cone-pinch-width}
\end{equation}
A denominator may depend algebraically on \(\ell^+\) and nevertheless be
subleading in this local fluctuation.  Such a denominator does not generate
the leading pinch.

The central-soft Fish analysis leading to
Eq.~\eqref{eq:five-central-soft-hard-scale-widths} illustrates this rule
explicitly: the \((q_0,q_1)\) and \((q_4,q_5)\) pairs determine the two plus
widths, while \((q_2,q_3)\) determines the remaining minus width.

Transverse widths are determined instead by homogeneity of the local
denominator expansion.  Suppose
\(q_{e\perp}\sim\delta^{c_e}\),
\(\mathcal D_e\sim\delta^{\kappa_e}\), and a transverse loop fluctuation
obeys \(\Delta\ell_\perp\sim\delta^\gamma\).  If
\(\Delta q_{e\perp}=C_e\Delta\ell_\perp\), with an order-one routing
coefficient \(C_e\), then generically
\begin{equation}
 \left.\Delta\mathcal D_e\right|_\perp
 \sim q_{e\perp}\mathbin\cdot C_e\Delta\ell_\perp
      +|C_e\Delta\ell_\perp|^2
 \sim\delta^{\min(c_e+\gamma,2\gamma)},
 \qquad
 \min(c_e+\gamma,2\gamma)\geq\kappa_e.
 \label{eq:transverse-width-homogeneity}
\end{equation}
The inequality must hold for every active denominator.  The transverse width
is the widest support, equivalently the smallest \(\gamma\), satisfying all
these conditions, with equality for at least one denominator.  If the linear
term vanishes, its first nonzero replacement in the Taylor expansion is used.
For several transverse loop variables the conditions are solved jointly,
including mixed terms and correlated directions.  In
Eq.~\eqref{eq:five-central-soft-transverse-width-balance}, for example,
\(c_e=2\) and \(\kappa_e=4\) give \(\gamma=2\).

For an independent loop fluctuation in a specified lightcone frame,
\begin{equation}
 (\Delta\ell^+,\Delta\ell^-,|\Delta\ell_\perp|)_{(n_+,n_-)}
 \sim(\delta^a,\delta^b,\delta^c)_{(n_+,n_-)},
 \label{eq:loop-fluctuation-widths}
\end{equation}
we use the invariant classification of its integration width
\begin{equation}
 \begin{array}{lll}
 a+b=2c &:& \text{balanced on-shell mode},\\
 a+b<2c &:& \text{longitudinal dominated},\\
 a+b>2c &:& \text{transverse dominated (Glauber)}.
 \end{array}
 \label{eq:mode-classification}
\end{equation}
These labels apply here to the width.  When the average momentum has the
same component powers, they also classify its mode; otherwise the average mode
and the width must be stated separately.  The last line concerns an
independent integration variable, not an individual propagator: a Glauber loop
may or may not be carried by an explicit Glauber propagator.  The drawing
convention is defined at its first use in
Fig.~\ref{fig:crown-regge-glauber}.  A marker denotes a vertex through which
the loop flows, and its arrow follows the local \(t\)-channel orientation.

\subsection{What the examples establish}

The examples of Sec.~3 now form a useful hierarchy.  Their momentum-space
content is summarised in Table~\ref{tab:comparative-glauber-audit}.
\begin{table}[!t]
\centering
\small
\begin{tabular}{@{}
 >{\raggedright\arraybackslash}p{.25\textwidth}
 >{\raggedright\arraybackslash}p{.29\textwidth}
 >{\raggedright\arraybackslash}p{.36\textwidth}@{}}
\toprule
region & independent loop modes & Glauber structure and correlations\\
\midrule
wide-angle Crown~\cite{Gardi:2024axt} and its audited boundary descendants
& Crown: three collinear loops; descendants add soft or collinear loops
& no Glauber loop; correlated longitudinal path sharing\\
\addlinespace[.45em]
Regge Crown~\cite{Gardi:2024axt}
& two collinear loops; one Glauber loop
& two marked Glauber vertices; opposite local orientations
when opened in the \(t\) channel\\
\addlinespace[.45em]
two-loop spacelike-collinear Fish family
& one soft loop; one Glauber loop
& one marked Glauber vertex; all six-, seven- and eight-propagator UT integrals
of Ref.~\cite{Chen:2026dnj} inherit the Fish seed modes\\
\addlinespace[.45em]
vertex-corrected spacelike-collinear Fish
& one soft loop; one collinear loop; one Glauber loop
& one marked Glauber vertex; the correction adds the collinear loop\\
\addlinespace[.45em]
five-point near-planar Fish
& two balanced on-shell collinear loops
& no Glauber loop; the dependent third collinear path restricts their joint support\\
\addlinespace[.45em]
five-point simultaneous MRK and near-planar Fish
& two balanced on-shell, path-adapted collinear loops
& no Glauber loop; correlated, MRK-narrowed longitudinal support\\
\addlinespace[.45em]
five-point central-soft rapidity-ordered Fish
& one balanced on-shell \((p_1\text{-collinear})^2\) loop; one Glauber loop
& two marked Glauber vertices; restricted support is encoded in the Glauber width\\
\addlinespace[.45em]
six-point wide-angle self-crossing twisted box
& one Glauber loop
& no Glauber propagator; the four active propagators are collinear and the
two four-point vertices are hard\\
\addlinespace[.45em]
six-point twisted hexagon in generic DSC or in zero-recoil NMRK
& one Glauber loop
& one Glauber propagator \(x_0\) with its two incident vertices marked\\
\bottomrule
\end{tabular}
\caption{Independent loop modes and Glauber structure of the certified HRs.}
\label{tab:comparative-glauber-audit}
\end{table}
Table~\ref{tab:comparative-glauber-audit} records the independent loop modes of
the examples and the diagrammatic realisation of each Glauber loop.  In every
listed HR that exhibits Glauber scaling, the
Glauber mode forms an independent integration loop rather than merely a flow
through the diagram.  The marked vertices identify where this loop flows;
their arrows record its local \(t\)-channel orientation.  The presence of an
explicit Glauber propagator is a separate property.  The wide-angle twisted
box illustrates the alternative: its Glauber loop is realised by the common
fluctuation of four collinear propagators around two hard vertices, with no
explicit Glauber propagator.

The five-point spacelike-collinear example demonstrates mode inheritance.
Ref.~\cite{Chen:2026dnj} computed the HR for every member of a complete basis
of two-loop UT integrals, containing six, seven or eight propagators.  Each
inherits the mode-adapted Fish-seed loop momenta, reproduced in
Eq.~\eqref{eq:five-sc-modes},
\begin{equation}
 \ell_s=q_5\sim(\delta,\delta,\delta),
 \qquad
 \ell_G=q_1-q_5\sim(\delta,\delta^2,\delta).
 \label{eq:five-sc-glauber-rerouting-summary}
\end{equation}
Thus \(\ell_s\) is soft and \(\ell_G\) is Glauber.  Inspecting the propagators
individually would miss the latter mode.
The Glauber loop has one marked collinear-to-soft vertex, as shown in
Fig.~\ref{fig:five-sc-mode}.  The vertex-corrected descendant preserves this
soft--Glauber pair and adds one collinear loop, so the two cases are listed
separately in Table~\ref{tab:comparative-glauber-audit}.

In the central-soft rapidity-ordered example, using \(s_{12}\) as the hard
scale, the realisation has one balanced on-shell loop and one Glauber loop,
\begin{equation}
 \Delta r\sim(\delta^4,1,\delta^2),
 \qquad
 \Delta\ell_G\sim(\delta^4,\delta,\delta^2),
 \qquad \ell_G=q_0-q_4.
 \label{eq:five-central-soft-mode-summary}
\end{equation}
The first satisfies the balanced on-shell condition in
Eq.~\eqref{eq:on-shell-mode-condition}, while the second is transverse dominated.
The momentum-transfer fluctuations at the two four-point vertices differ only
by a sign, and the corresponding markers in
Fig.~\ref{fig:five-central-soft-mode-reflection} have opposite local
orientations.  The longitudinal components of the Glauber loop's
integration width are unequal: \(\Delta\ell_G^+\sim\delta^4\) is narrower than
\(\Delta\ell_G^-\sim\delta\), realising in momentum space the restricted
support found by HRF.

In the displayed mode-adapted frames,
Eqs.~\eqref{eq:five-sc-glauber-rerouting-summary} and
\eqref{eq:five-central-soft-mode-summary} therefore provide two forms of
unequal longitudinal Glauber width, respectively
\((\delta,\delta^2)\) and \((\delta^4,\delta)\); the vertex-corrected Fish
inherits the former.  By contrast, the Regge Crown in
Fig.~\ref{fig:crown-regge-glauber}, the wide-angle twisted box in
Eq.~\eqref{eq:hexagon-wide-angle-loop-width}, and the zero-recoil NMRK and DSC
regions in Eqs.~\eqref{eq:nmrk-direct-pole-widths} and
\eqref{eq:dsc-direct-pole-widths} have equal longitudinal widths in their
symmetric frames.

\paragraph{Faithful routing, average--width correspondence and cross-mode
correlations.}
\label{par:faithful-routing-average-width}
The mode-complete examples support a two-part conclusion.  First, in every
such example we find a faithful basis of independent loop momenta, with
appropriate local lightcone directions, in which the conditional
average-momentum scaling and the marginal integration width agree
componentwise.  This correspondence is found for balanced on-shell modes and also for
the transverse-dominated Glauber modes.  It is a statement about an adapted
loop basis, not about an arbitrary edge momentum; the deeper saddle value of
the dependent edge \(q_2\) in the simultaneous MRK--planar Fish is an
explicit illustration of this distinction.

Second, several HR examples show that the marginal modes can be mutually
correlated.  Momentum
conservation among several collinear sectors may restrict their joint
longitudinal support without changing any one marginal width.  If the width
of loop \(\ell\) is
\((\delta^{a_\ell},\delta^{b_\ell},\delta^{c_\ell})\), the integration-volume
power separates as
\begin{equation}
 \Omega_{\rm meas}(D)=
 \sum_{\ell=1}^{L}\bigl[a_\ell+b_\ell+(D-2)c_\ell\bigr]
 +\gamma_{\rm corr},
 \qquad
 \mathcal V_{\parallel}^{\rm corr}\sim\delta^{\gamma_{\rm corr}}.
 \label{eq:width-correlation-measure-split}
\end{equation}
Thus the marginal widths determine the \(D\)-dependent part of the measure,
whereas the cross-mode correlation changes only its integer power.  The
wide-angle Crown and near-planar Fish exhibit this second effect; for the
near-planar Fish \(\gamma_{\rm corr}=3\) in powers of \(\lambda\), while the
simultaneous MRK--near-planar family has
\(\gamma_{\rm corr}=3a+b\).  By contrast, in the local soft--Glauber examples
the restriction is already absorbed into the narrower Glauber width and no
additional joint-support factor is required.

For the NMRK and DSC one-loop six-point examples, we reconstructed the
component widths explicitly from the pinching poles.  The NMRK reconstruction
is given in Eqs.~\eqref{eq:nmrk-pinched-longitudinal-components}--
\eqref{eq:nmrk-direct-pole-widths}, and the DSC reconstruction in
Eqs.~\eqref{eq:dsc-longitudinal-pole-widths}--
\eqref{eq:dsc-direct-pole-widths}.  In the symmetric hard-normalised frame
their widths give
\begin{align}
 \text{zero-recoil NMRK}:\quad
 &(a_{\mathrm{NMRK}},b_{\mathrm{NMRK}},c_{\mathrm{NMRK}})=(5,5,3),
 \nonumber\\
 &a_{\mathrm{NMRK}}+b_{\mathrm{NMRK}}
   -2c_{\mathrm{NMRK}}=4>0,\nonumber\\[1mm]
 \text{DSC}:\quad
 &(a_{\mathrm{DSC}},b_{\mathrm{DSC}},c_{\mathrm{DSC}})=(2,2,1),
 \nonumber\\
 &a_{\mathrm{DSC}}+b_{\mathrm{DSC}}
   -2c_{\mathrm{DSC}}=2>0.
 \label{eq:six-point-glauber-summary}
\end{align}
For a single loop with no additional support correlation, its measure exponent
may be written as
\[
 a+b+(D-2)c=Dc+(a+b-2c).
\]
Thus \(a+b-2c\) is the extra suppression relative to a balanced on-shell
width with the same transverse exponent.  In NMRK and DSC it equals the
corresponding cancellation depth \(\WHR-\WSL\), namely \(4\) and \(2\) in
Table~\ref{tab:twisted-hexagon-one-loop-comparison}.  The wide-angle member
obeys the same relation: Eq.~\eqref{eq:hexagon-wide-angle-loop-width} gives
\((a,b,c)=(1,1,\tfrac12)\) in powers of \(\sigma_\times\), so both the
transverse-dominance excess and the cancellation depth are \(1\).  The
momentum-space width therefore already incorporates the restriction measured
by the cancellation depth; the latter must not be included as a separate
factor.
A longitudinal boost acts as in
Eq.~\eqref{eq:longitudinal-boost-freedom}, shifting the first two exponents
oppositely but preserving their sum.  Each region contains one Glauber loop
and one explicit Glauber propagator, \(x_0\), whose incident \(p_3\) and
\(p_6\) vertices carry the two oriented markers in
Fig.~\ref{fig:one-loop-hexagon}.  The incoming three-point vertices carrying
\(p_1\) and \(p_2\), selected by the cancellation factors, are instead the
factor-localisation vertices.  The other
resolved transfer, \(q_3\), is soft in NMRK and hard in DSC.

\subsection{Power counting as a consistency check}
\label{sec:power-counting-certificate}

Once a loop basis and all local widths are known, momentum-space power
counting follows directly from Eq.~\eqref{eq:momentum-integral}:
\begin{equation}
 I_{\rm mom}\sim
 \frac{\prod_{r=1}^{L}\dd^D\ell_r}
      {\prod_e \mathcal D_e^{\nu_e}}.
 \label{eq:momentum-power-general}
\end{equation}
The measure uses the fluctuation widths of the independent loop variables,
not the values at the pinch of a convenient set of edge momenta.  The
propagator virtualities are inherited from the parameter-space vector, while
momentum conservation and the pole analysis determine the component widths
and their correlations.  The resulting momentum-space count must agree with
the LP count, including the reduced support associated with the cancellation
depth.  Detailed calculations are given alongside the corresponding examples
in Sec.~\ref{sec:applications}.

This comparison does more than label a mode.  It checks that the loop basis
is complete, that no cancellation-normal width has been counted twice, and
that a putative Glauber value is genuinely an integration direction.  It is
therefore a strong optional consistency check for a newly discovered HR, while the
existence and scaling of the HR itself remain parameter-space conclusions of
the core algorithm.

\section{Topological organisation of Landshoff hidden regions}
\label{sec:topological-organization}

The examples above reveal a common origin: every HR studied here can be
associated with a wide-angle multiple-hard-scattering, or Landshoff, seed.  In
the graph-theoretic direction, the seed is exposed by a label-preserving
contraction of a larger topology; in kinematic space, a specialised region descends from the
relevant wide-angle limit through further kinematic specialisation.
This motivates a conjectural organising principle: the presence of a topology
supporting such a Landshoff ancestor is necessary for an HR.  It is not
sufficient.  Whether the region is realised in a particular diagram and limit
depends on additional data, including the external-leg assignment, positivity
in the physical domain and the kinematic path by which the singular surface is
approached.  Together these determine whether the pinch is reached and which
region vector results.

The six-point twisted box illustrates this dependence.  Its positive
wide-angle pinch occurs on the physical self-crossing surface rather than at
generic wide-angle kinematics, and its NMRK and DSC regions are
specialisations of that locus.  More generally, the same topology may realise
different HRs on collinear or high-energy faces of kinematic space.  Topology
therefore organises candidate mechanisms; it does not by itself determine the
physical sheet, the required kinematic surface or the region vector.

\begin{figure}[!t]
\centering
\begin{tikzpicture}[scale=.72]
 \foreach \i/\y/\p in {1/1.7/1,2/0/2,3/-1.7/5}{
   \coordinate (a32\i) at (0,\y);
   \draw[hrf internal] (-.8,\y)--(a32\i);
   \node[font=\scriptsize,anchor=east] at (-.95,\y) {$p_{\p}$};
   \node[font=\scriptsize,anchor=south east]
     at ($(a32\i)+(-.08,.08)$) {$A_{\i}$};
 }
 \foreach \a/\y/\p in {1/.85/3,2/-.85/4}{
   \coordinate (b32\a) at (3.0,\y);
   \draw[hrf internal] (b32\a)--(3.8,\y);
   \node[font=\scriptsize,anchor=west] at (3.95,\y) {$p_{\p}$};
   \node[font=\scriptsize,anchor=south west]
     at ($(b32\a)+(.08,.08)$) {$B_{\a}$};
 }
 \foreach \i in {1,2,3}{
   \foreach \a in {1,2}{\draw[hrf internal] (a32\i)--(b32\a);}
 }
 \foreach \v in {a321,a322,a323,b321,b322}{\node[hrf vertex] at (\v) {};}
 \node[font=\small] at (1.5,-2.25) {(a) \(K_{3,2}\)};
\end{tikzpicture}
\hspace{18mm}
\begin{tikzpicture}[scale=.65]
 \foreach \i/\y in {1/2.4,2/.8,3/-.8,4/-2.4}{
   \coordinate (a42\i) at (0,\y);
   \draw[hrf internal] (-.8,\y)--(a42\i);
   \node[font=\scriptsize,anchor=east] at (-.95,\y) {$p_{\i}$};
   \node[font=\scriptsize,anchor=south east]
     at ($(a42\i)+(-.08,.08)$) {$A_{\i}$};
 }
 \foreach \a/\y in {1/.9,2/-.9}{
   \coordinate (b42\a) at (3.0,\y);
   \node[font=\scriptsize,anchor=south west]
     at ($(b42\a)+(.08,.08)$) {$B_{\a}$};
 }
 \foreach \i in {1,2,3,4}{
   \foreach \a in {1,2}{\draw[hrf internal] (a42\i)--(b42\a);}
 }
 \foreach \v in {a421,a422,a423,a424,b421,b422}{\node[hrf vertex] at (\v) {};}
 \node[font=\small] at (1.5,-3.0) {(b) \(K_{4,2}\)};
\end{tikzpicture}

\vspace{3mm}

\begin{tikzpicture}[scale=.72]
 \foreach \i/\y in {1/1.7,2/0,3/-1.7}{
   \coordinate (a33\i) at (0,\y);
   \draw[hrf internal] (-.8,\y)--(a33\i);
   \node[font=\scriptsize,anchor=east] at (-.95,\y) {$p_{\i}$};
   \node[font=\scriptsize,anchor=south east]
     at ($(a33\i)+(-.08,.08)$) {$A_{\i}$};
 }
 \foreach \a/\y/\p in {1/1.7/4,2/0/5,3/-1.7/6}{
   \coordinate (b33\a) at (3.0,\y);
   \draw[hrf internal] (b33\a)--(3.8,\y);
   \node[font=\scriptsize,anchor=west] at (3.95,\y) {$p_{\p}$};
   \node[font=\scriptsize,anchor=south west]
     at ($(b33\a)+(.08,.08)$) {$B_{\a}$};
 }
 \foreach \i in {1,2,3}{
   \foreach \a in {1,2,3}{\draw[hrf internal] (a33\i)--(b33\a);}
 }
 \foreach \v in {a331,a332,a333,b331,b332,b333}{\node[hrf vertex] at (\v) {};}
 \node[font=\small] at (1.5,-2.25) {(c) \(K_{3,3}\)};
\end{tikzpicture}
\hspace{18mm}
\begin{tikzpicture}[scale=.65]
 \foreach \i/\y in {1/2.4,2/.8,3/-.8,4/-2.4}{
   \coordinate (a43\i) at (0,\y);
   \draw[hrf internal] (-.8,\y)--(a43\i);
   \node[font=\scriptsize,anchor=east] at (-.95,\y) {$p_{\i}$};
   \node[font=\scriptsize,anchor=south east]
     at ($(a43\i)+(-.08,.08)$) {$A_{\i}$};
 }
 \foreach \a/\y in {1/1.8,2/0,3/-1.8}{
   \coordinate (b43\a) at (3.0,\y);
   \node[font=\scriptsize,anchor=south west]
     at ($(b43\a)+(.08,.08)$) {$B_{\a}$};
 }
 \foreach \i in {1,2,3,4}{
   \foreach \a in {1,2,3}{\draw[hrf internal] (a43\i)--(b43\a);}
 }
 \foreach \v in {a431,a432,a433,a434,b431,b432,b433}{\node[hrf vertex] at (\v) {};}
 \node[font=\small] at (1.5,-3.0) {(d) \(K_{4,3}\)};
\end{tikzpicture}
\caption{Four higher complete-bipartite incidence cores used in this section;
the minimal \(K_{2,2}\) member is the twisted-box seed in
Fig.~\ref{fig:twisted-hexagon-topologies}(b).  The edge joining the
explicitly named vertices \(A_i\) and \(B_a\) carries
the LP parameter \(x_{ia}\).  The external half-edges make the
kinematic distinction explicit: panels (a) and (c) are five- and six-point
graphs, whereas panels (b) and (d) are both four-point graphs.  Panel (a)
uses the near-planar Fish permutation of external legs in
Fig.~\ref{fig:five-point-near-planar-attachment}; panel (b) is the Crown of
Fig.~\ref{fig:crown-graph}.  Panel (d) has the incidence topology of the
baryon--baryon independent-scattering diagram in Fig.~14 of
Ref.~\cite{Botts:1989kf}.}
\label{fig:complete-bipartite-cores}
\end{figure}
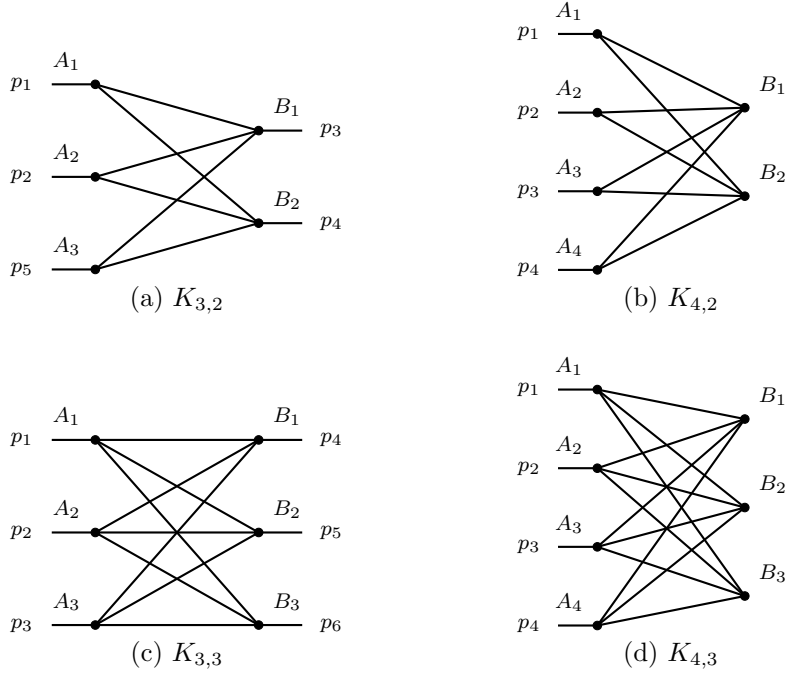

The minimal member of the complete-bipartite family relevant here is the
one-loop \(K_{2,2}\) twisted-box seed in
Fig.~\ref{fig:twisted-hexagon-topologies}(b).
Figure~\ref{fig:complete-bipartite-cores} displays four higher
complete-bipartite incidence cores.  We denote
the vertices in the two parts by
\begin{equation}
 \{A_1,\ldots,A_m\}\qquad\hbox{and}\qquad
 \{B_1,\ldots,B_n\},
 \label{eq:Kmn-vertex-partitions}
\end{equation}
and assign the LP parameter \(x_{ia}\) to the edge joining \(A_i\)
to \(B_a\).  Thus the pair of vertex labels, rather than a separate edge
number, names every edge, and we use this \(x_{ia}\) notation throughout the
present section.  The external half-edges shown in the figure do not carry
an \(x_{ia}\) label.

Although the two parts of a complete-bipartite graph are combinatorially
interchangeable, its physical Landshoff realisation selects a preferred
orientation.  In all certified configurations considered here, the \(A_i\)
are external collinear or multi-collinear factor-localisation vertices: the
row variables \(x_{ia}\) encode how the attached momentum is shared among the
incident on-shell lines.  The cancellation ideal constrains these sharing
fractions and their compatibility between different \(A_i\); an individual
normal may therefore compare several \(A_i\) rather than belong to a single
vertex.  The \(B_a\), by contrast, represent the hard-scattering components
and may be internal or carry additional external legs.  In specialised limits
they may carry oriented Glauber flow, or resolve into vertices joined by an
explicit Glauber propagator.  This physical interpretation explains why all
\(A_i\) carry external legs in the configurations studied here, whereas
external attachments to the \(B_a\) are optional.

The external half-edges in Fig.~\ref{fig:complete-bipartite-cores} display
several realisations of this division.  Every vertex of \(K_{3,2}\) and
\(K_{3,3}\) carries an external leg, so these are respectively five- and
six-point problems.  By contrast, in the displayed \(K_{4,2}\) and
\(K_{4,3}\) configurations only the four \(A_i\) vertices carry external legs,
while the \(B_a\) vertices are internal; both are therefore four-point
problems.  The \(K_{4,2}\) core also admits the five- and six-point
configurations analysed in Sec.~\ref{sec:K42-external-attachments}.
Increasing the second partition in the right-hand column raises the loop
order without enlarging the external kinematic space.  In the left-hand
column it also introduces additional external directions and the associated
momentum-conservation and Gram constraints.  The complete-bipartite language
thus organises the internal incidence structure, but it does not make the
four-, five- and six-point kinematics equivalent.

\subsection{Complete-bipartite incidence cores}

The simplest established wide-angle seeds have a common incidence
description.  After the external half-edges are suppressed, the six-point
twisted box is \(K_{2,2}\), the five-point Fish is \(K_{3,2}\), and the Crown
is \(K_{4,2}\).  Here ``wide angle'' includes special physical kinematic
surfaces: the twisted-box pinch requires
Eq.~\eqref{eq:hexagon-self-crossing-surface}, and the Fish requires the
near-planar condition \(\Gamma_5=0\), whereas the Crown pinch exists in an
open wide-angle domain.  With the vertex convention introduced above, the
three established seeds all have \(n=2\), so their path ratios provide the
most direct introduction.  The two edges from \(A_i\) to \(B_1\) and \(B_2\)
carry parameters \(x_{i1}\) and \(x_{i2}\).  Define
\begin{equation}
 r_i=\frac{x_{i1}}{x_{i2}},\qquad
 R_i=\frac{r_i}{r_m}
     =\frac{x_{i1}x_{m2}}{x_{i2}x_{m1}},
 \qquad i=1,\ldots,m-1.
 \label{eq:Km2-path-cycle-ratios}
\end{equation}
At a physical Landau pinch, the normalised pair
\(z_i=x_{i2}/(x_{i1}+x_{i2})\) and \(1-z_i\) encodes the momentum fractions
carried by the two collinear partons emerging from \(A_i\) towards \(B_1\)
and \(B_2\).

We call the subset of LP-parameter space on which these path splittings are
mutually compatible the \emph{incidence locus}.  It is the locus associated
with the bipartite incidence pattern before the complete stationary and
vanishing Landau equations are imposed.  In its undeformed form, mutual
compatibility means
\begin{equation}
 R_i=1,\quad i<m
 \qquad\Longleftrightarrow\qquad
 (r_1,\ldots,r_m)=c(1,\ldots,1).
 \label{eq:Km2-undeformed-incidence-locus}
\end{equation}
The \(m-1\) quantities \(R_i-1\) are normal coordinates associated with a
basis of independent graph cycles.  The collective ratio \(c\), by contrast,
is tangent to the incidence locus: varying it changes the common momentum
sharing while preserving every relative compatibility condition.

The Crown is the simplest case.  Its four path ratios in
Eq.~\eqref{eq:crown-path-ratios} obey three independent relative conditions.
The four binomial cancellation factors in Eq.~\eqref{eq:crown-F0} form a
redundant cyclic set of equations for the same codimension-three locus.  Their
common value \(c=\rho\) remains free, so the incidence locus is already the
LP-parameter part of the physical pinch.

The other two seeds separate the incidence count from the codimension of the
full physical pinch.  On the near-planar surface \(\Gamma_5=0\), the Fish
factors in Eq.~\eqref{eq:five-near-planar-factors} impose
\(r_I=\rho_I\), \(I=A,B,C\).  Two independent relative equations,
\begin{equation}
 \frac{r_A}{r_C}=\frac{\rho_A}{\rho_C},
 \qquad
 \frac{r_B}{r_C}=\frac{\rho_B}{\rho_C},
 \label{eq:fish-deformed-incidence-ratios}
\end{equation}
define its kinematically deformed incidence locus
\(\boldsymbol r=c\boldsymbol\rho\).  The remaining stationary equation fixes
the tangent coordinate to \(c=1\).  Thus the Fish has two incidence normals
but three independent LP-parameter normals \(f_A,f_B,f_C\) at the physical
pinch.  The separate kinematic condition \(\Gamma_5=0\) is required because
the graph polynomial evaluated at the stationary point is proportional to
\(\Gamma_5\), as shown in Eq.~\eqref{eq:five-stationary-gram}.

For the twisted box, the single relative condition between its two path
ratios defines a codimension-one incidence locus.  The two hard-vertex
momentum-balance factors \(f_{1,\mathrm{WA}}\) and \(f_{2,\mathrm{WA}}\) in
Eq.~\eqref{eq:hexagon-wide-angle-boundary-factorisation} also fix the
collective ratio, giving a codimension-two pinch within the active
\(K_{2,2}\) parameter space on the self-crossing
surface~\eqref{eq:hexagon-self-crossing-surface}.  The three counts within
the corresponding active incidence cores may therefore be summarised as
\begin{equation}
 \begin{array}{c|c|c}
  & \operatorname{codim} C_{\rm inc}
  & \operatorname{codim}_{\rm core} C_{\rm pinch}\\ \hline
  K_{4,2}\ \text{(Crown)} &3&3\\
  K_{3,2}\ \text{(Fish)} &2&3\\
  K_{2,2}\ \text{(twisted box)} &1&2
 \end{array}
 \label{eq:seed-cycle-normal-counts}
\end{equation}

We now generalise this construction to \(K_{m,n}\).  Its edge parameters form
the \(m\times n\) matrix
\begin{equation}
 X=(x_{ia}),\qquad i=1,\ldots,m,\qquad a=1,\ldots,n.
 \label{eq:Kmn-edge-matrix}
\end{equation}
The undeformed incidence locus is parametrised by
\begin{equation}
 C_{\rm inc}:\qquad x_{ia}=u_iw_a .
 \label{eq:Kmn-rank-one-parametrisation}
\end{equation}
The variables \(u_i\) and \(w_a\), modulo the redundancy
\(u_i\mapsto c u_i,\ w_a\mapsto c^{-1}w_a\), parametrise motion tangent to
\(C_{\rm inc}\).  A set of local normal coordinates is provided by
\begin{equation}
 R_{ia}=\frac{x_{ia}x_{mn}}{x_{in}x_{ma}},
 \qquad i<m,\quad a<n.
 \label{eq:Kmn-cycle-ratios}
\end{equation}
The index ranges give \((m-1)(n-1)\) such ratios.  They are independent on
the positive chart, so \(R_{ia}-1\) provide the same number of local normal
coordinates to \(C_{\rm inc}\).
Each \(R_{ia}\) is associated with the four-cycle
\(A_i-B_a-A_m-B_n-A_i\), is invariant under row and column rescalings
\(x_{ia}\mapsto\alpha_i\beta_a x_{ia}\).  The incidence locus is the common
level set \(R_{ia}=1\); equivalently, on the positive chart,
\begin{equation}
 R_{ia}=1
 \quad\Longleftrightarrow\quad
 x_{ia}x_{mn}-x_{in}x_{ma}=0.
 \label{eq:Kmn-ratio-minor-equivalence}
\end{equation}

More invariantly, let
\begin{equation}
 I_2(X)\equiv
 \left\langle
 x_{ia}x_{jb}-x_{ib}x_{ja}
 \ \middle|\ i<j,\ a<b
 \right\rangle
 \label{eq:Kmn-determinantal-ideal}
\end{equation}
denote the ideal generated by all \(2\times2\) minors of \(X\).
Equation~\eqref{eq:Kmn-rank-one-parametrisation} guarantees that every such
minor vanishes.  Conversely, because \(x_{mn}>0\), the vanishing of the
reference minors in Eq.~\eqref{eq:Kmn-ratio-minor-equivalence} gives
\(x_{ia}=x_{in}x_{ma}/x_{mn}\).  Defining
\(u_i=x_{in}/x_{mn}\) and \(w_a=x_{ma}\) then reconstructs
Eq.~\eqref{eq:Kmn-rank-one-parametrisation}.  Thus the factorised
parametrisation and the simultaneous cancellation equations describe the
same incidence locus in the positive domain.

Although \(I_2(X)\) generally has more generators, its zero locus in the
positive domain is \(C_{\rm inc}\).  The parametrisation in
Eq.~\eqref{eq:Kmn-rank-one-parametrisation} uses \(m+n\) variables with one
rescaling redundancy, so
\(\dim C_{\rm inc}=m+n-1\).  Its codimension in the \(mn\)-dimensional
space of matrices \(X\) is therefore \((m-1)(n-1)\), in agreement with the
count following Eq.~\eqref{eq:Kmn-cycle-ratios}.  Since \(K_{m,n}\) is
connected and has \(E=mn\) edges and \(V=m+n\) vertices, the usual graph
relation \(L=E-V+1\) gives
\begin{equation}
 L(K_{m,n})=E-V+1
 =mn-(m+n)+1=(m-1)(n-1)
 =\operatorname{codim} C_{\rm inc}.
 \label{eq:Kmn-determinantal-codimension}
\end{equation}
The equality says that the number of normal coordinates of the incidence
locus agrees with the loop order.  It does not count all independent
cancellation factors of the physical pinch, as the Fish and twisted-box
examples demonstrate.  The complete physical pinch must additionally satisfy
the stationary and vanishing equations and positivity in the physical
domain.  In particular, kinematics can deform the incidence normals to
\(R_{ia}=\rho_{ia}(\boldsymbol s)\), with polynomial representatives
\begin{equation}
 f_{ia}=x_{ia}x_{mn}-\rho_{ia}x_{in}x_{ma}.
 \label{eq:Kmn-deformed-minors}
\end{equation}

\subsection{External-leg configurations of
\texorpdfstring{\(K_{4,2}\)}{K4,2}}
\label{sec:K42-external-attachments}

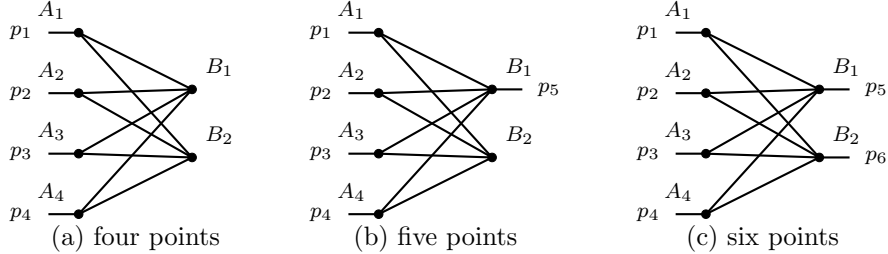
\begin{figure}[htbp]
\centering
\begin{tikzpicture}[scale=.50]
 \foreach \i/\y in {1/2.4,2/.8,3/-.8,4/-2.4}{
   \coordinate (a4\i) at (0,\y);
   \draw[hrf internal] (-.8,\y)--(a4\i);
   \node[font=\scriptsize,anchor=east] at (-.95,\y) {$p_{\i}$};
   \node[font=\scriptsize,anchor=south east]
     at ($(a4\i)+(-.08,.08)$) {$A_{\i}$};
 }
 \foreach \a/\y in {1/.9,2/-.9}{
   \coordinate (b4\a) at (3.0,\y);
   \node[font=\scriptsize,anchor=south west]
     at ($(b4\a)+(.08,.08)$) {$B_{\a}$};
 }
 \foreach \i in {1,2,3,4}{
   \foreach \a in {1,2}{\draw[hrf internal] (a4\i)--(b4\a);}
 }
 \foreach \v in {a41,a42,a43,a44,b41,b42}{\node[hrf vertex] at (\v) {};}
 \node[font=\small] at (1.5,-3.05) {(a) four points};
\end{tikzpicture}
\hspace{5mm}
\begin{tikzpicture}[scale=.50]
 \foreach \i/\y in {1/2.4,2/.8,3/-.8,4/-2.4}{
   \coordinate (a5\i) at (0,\y);
   \draw[hrf internal] (-.8,\y)--(a5\i);
   \node[font=\scriptsize,anchor=east] at (-.95,\y) {$p_{\i}$};
   \node[font=\scriptsize,anchor=south east]
     at ($(a5\i)+(-.08,.08)$) {$A_{\i}$};
 }
 \foreach \a/\y in {1/.9,2/-.9}{
   \coordinate (b5\a) at (3.0,\y);
   \node[font=\scriptsize,anchor=south west]
     at ($(b5\a)+(.08,.08)$) {$B_{\a}$};
 }
 \draw[hrf internal] (b51)--(3.8,.9);
 \node[font=\scriptsize,anchor=west] at (3.95,.9) {$p_5$};
 \foreach \i in {1,2,3,4}{
   \foreach \a in {1,2}{\draw[hrf internal] (a5\i)--(b5\a);}
 }
 \foreach \v in {a51,a52,a53,a54,b51,b52}{\node[hrf vertex] at (\v) {};}
 \node[font=\small] at (1.5,-3.05) {(b) five points};
\end{tikzpicture}
\hspace{5mm}
\begin{tikzpicture}[scale=.50]
 \foreach \i/\y in {1/2.4,2/.8,3/-.8,4/-2.4}{
   \coordinate (a6\i) at (0,\y);
   \draw[hrf internal] (-.8,\y)--(a6\i);
   \node[font=\scriptsize,anchor=east] at (-.95,\y) {$p_{\i}$};
   \node[font=\scriptsize,anchor=south east]
     at ($(a6\i)+(-.08,.08)$) {$A_{\i}$};
 }
 \foreach \a/\y/\p in {1/.9/5,2/-.9/6}{
   \coordinate (b6\a) at (3.0,\y);
   \draw[hrf internal] (b6\a)--(3.8,\y);
   \node[font=\scriptsize,anchor=west] at (3.95,\y) {$p_{\p}$};
   \node[font=\scriptsize,anchor=south west]
     at ($(b6\a)+(.08,.08)$) {$B_{\a}$};
 }
 \foreach \i in {1,2,3,4}{
   \foreach \a in {1,2}{\draw[hrf internal] (a6\i)--(b6\a);}
 }
 \foreach \v in {a61,a62,a63,a64,b61,b62}{\node[hrf vertex] at (\v) {};}
 \node[font=\small] at (1.5,-3.05) {(c) six points};
\end{tikzpicture}
\caption{Three external-leg configurations of the same three-loop \(K_{4,2}\)
incidence core.  The four-point configuration (a) is the Crown.  In (b), only
\(B_1\) carries an additional external momentum; in (c), both \(B_1\) and
\(B_2\) do.  The change in external kinematics changes the existence and
location of the positive Landau pinch even though the internal graph is
unchanged.}
\label{fig:K42-attachment-chain}
\end{figure}

The external-leg configuration is part of the definition of a Landshoff seed.
The same \(K_{4,2}\) incidence core gives three sharply different Landau
problems, shown in Fig.~\ref{fig:K42-attachment-chain}.  The familiar Crown
has external legs only at the four \(A_i\) vertices.  Attaching a leg also
at one or both \(B_a\) vertices instead gives five- and six-point
kinematics, without changing the loop order.

For all three configurations, introduce the path coordinates
\begin{equation}
 h_i=x_{i1}+x_{i2},\qquad
 z_i=\frac{x_{i2}}{h_i},\qquad 0<z_i<1,
 \qquad i=1,\ldots,4.
 \label{eq:K42-path-coordinates}
\end{equation}
These coordinates are equivalent to the Crown path-ratio representation in
Eq.~\eqref{eq:crown-path-ratios}, since
\(r_i=x_{i1}/x_{i2}=(1-z_i)/z_i\).
For the six-point configuration we retain the paper-wide convention that
\(p_1,p_2\) are incoming and \(p_3,\ldots,p_6\) are outgoing.  To keep the
following algebra symmetric, define the corresponding all-outgoing
representatives \(\widehat p_r=\eta_r p_r\), where
\(\eta_1=\eta_2=-1\) and \(\eta_3=\cdots=\eta_6=1\), so that
\(\sum_{r=1}^6\widehat p_r=0\).
For massless internal propagators, the exact Symanzik polynomials are
\begin{align}
 \mathcal U&=e_3(\boldsymbol h)
 =\sum_{i<j<k}h_i h_j h_k,\nonumber\\
 \mathcal F&=\left(\prod_{i=1}^4h_i\right)
 \left[
   Q(\boldsymbol z)^2
   +\sum_{i=1}^4z_i(1-z_i)\widehat p_i^{\,2}
 \right],
 \qquad
 Q(\boldsymbol z)=\widehat p_5+\sum_{i=1}^4z_i\widehat p_i .
 \label{eq:K42-six-point-UF}
\end{align}
Here \(Q(\boldsymbol z)\) measures the residual four-momentum imbalance at
\(B_1\) for the prospective pinch assignment in which the propagators joining
\(A_i\) to \(B_1\) and \(B_2\) carry the fractions \(z_i\widehat p_i\) and
\((1-z_i)\widehat p_i\), respectively.  Overall external momentum conservation
implies that the corresponding imbalance at \(B_2\) is
\(\widehat p_6+\sum_{i=1}^4(1-z_i)\widehat p_i=-Q(\boldsymbol z)\).
For general \(\boldsymbol z\), neither imbalance need vanish; the ratio Landau
equations below select the values for which they do.  In particular, at
massless on-shell kinematics,
\(\mathcal F_0=(\prod_i h_i)Q(\boldsymbol z)^2\).  Let
\begin{equation}
 G_{ij}=\widehat p_i\mathbin{\cdot}\widehat p_j,\qquad
 b_i=\widehat p_i\mathbin{\cdot}\widehat p_5 .
\label{eq:K42-six-point-Gram-data}
\end{equation}
The ratio Landau equations are
\(\widehat p_i\mathbin{\cdot}Q(\boldsymbol z)=0\), \(i=1,\ldots,4\).
When \(\det G\ne0\), the four momenta \(\widehat p_1,\ldots,
\widehat p_4\) form a basis of four-dimensional momentum space.  Writing
\(\widehat p_5=\sum_jc_j\widehat p_j\) and contracting with this basis gives
\(G\boldsymbol c=\boldsymbol b\).  Substitution into Eq.~\eqref{eq:K42-six-point-UF}
gives
\(Q(\boldsymbol z)=\sum_j(c_j+z_j)\widehat p_j\), so the stationary equations
are \(G(\boldsymbol c+\boldsymbol z)=0\).  Inverting \(G\) therefore gives the
unique solution
\begin{equation}
 \boldsymbol z_\star=-G^{-1}\boldsymbol b,\qquad
 Q(\boldsymbol z_\star)=0,
 \label{eq:K42-six-point-stationary-point}
\end{equation}
At this point the residual imbalances vanish at both \(B_1\) and \(B_2\), so
momentum is conserved separately at the two vertices.  The quantities \(z_{\star i}\) and
\(1-z_{\star i}\) are therefore the momentum fractions carried by the two
propagators at the pinch.  Equation~\eqref{eq:K42-six-point-stationary-point}
is also simply the expansion of \(-\widehat p_5\) in that basis.  The
corresponding invariant compatibility condition is the five-vector Gram
relation
\[
 0=\det\!\begin{pmatrix}
 G & \boldsymbol b\\
 \boldsymbol b^{T} & \widehat p_5^{\,2}
 \end{pmatrix}
 =\det G\left(\widehat p_5^{\,2}
   -\boldsymbol b^{T}G^{-1}\boldsymbol b\right).
\]
Thus, in addition to \(G\boldsymbol c=\boldsymbol b\), it states
\(\widehat p_5^{\,2}=\boldsymbol b^{T}G^{-1}\boldsymbol b
=\boldsymbol c^{T}G\boldsymbol c\).  This Gram relation is automatic for
physical six-point kinematics in four dimensions (with
\(\widehat p_6\) fixed by momentum conservation); it is not an additional
special kinematic hypersurface analogous to \(\Gamma_5=0\) in the Fish
example.  The condition \(\det G\ne0\) merely selects the open chart in
which the first four momenta provide a basis.  The first-sheet condition is
the transparent set of inequalities
\begin{equation}
 0<-(G^{-1}\boldsymbol b)_i<1,\qquad i=1,\ldots,4.
 \label{eq:K42-six-point-positivity}
\end{equation}
These inequalities define a nonempty open domain of physical \(2\to4\)
kinematics.

The Hessian determines the local normal form rather than the existence of
the stationary solution.  Setting
\(\boldsymbol z=\boldsymbol z_\star+\Delta\boldsymbol z\), Eq.~\eqref{eq:K42-six-point-stationary-point}
gives
\begin{align}
 Q(\boldsymbol z_\star+\Delta\boldsymbol z)
 &=\sum_{i=1}^4\Delta z_i\widehat p_i,\nonumber\\
 \mathcal F_0(\boldsymbol z_\star+\Delta\boldsymbol z)
 &=\left(\prod_{k=1}^4h_k\right)
   \sum_{i,j=1}^4G_{ij}\Delta z_i\Delta z_j,\nonumber\\
 \left.
 \frac{\partial^2\mathcal F_0}{\partial z_i\partial z_j}
 \right|_{\boldsymbol z_\star}
 &=2\left(\prod_{k=1}^4h_k\right)G_{ij}.
 \label{eq:K42-six-point-Hessian}
\end{align}
Since \(G\) is non-singular, all four ratio deviations are genuine normal
coordinates.  To translate this statement into the cancellation-factor
language of HRF, define the local factors
\[
 f_i^{(6)}\equiv h_i\Delta z_i
 =x_{i2}
  +\frac{\bigl(\operatorname{adj}(G)\boldsymbol b\bigr)_i}{\det G}\,h_i
 =x_{i2}-z_{\star i}(x_{i1}+x_{i2}).
\]
The denominator is harmless on the chart \(\det G\ne0\); it may be cleared to
obtain an equivalent polynomial representative without changing the local
cancellation ideal.  Since, in the on-shell limit considered,
\(G_{ii}=\widehat p_i^{\,2}=0\), the polynomial can be written exactly as
\[
 \mathcal F_0
 =2\sum_{1\leq i<j\leq4}G_{ij}
   \left(\prod_{k\ne i,j}h_k\right)f_i^{(6)}f_j^{(6)}.
\]
The natural graph-dependent HRF presentation therefore contains the six
generators \(g_{ij}^{(6)}=f_i^{(6)}f_j^{(6)}\), with \(i<j\).  These are six
generator polynomials, not six independent normal constraints: their common
pinch component is defined by the four independent factors \(f_i^{(6)}\).
For the Crown, by contrast, the four factors in Eq.~\eqref{eq:crown-F0}
describe only three independent ratio normals.  A common shift of all four
path ratios is tangent to its pinch locus, so the fourth factor vanishes once
three independent ones do, and only the two disjoint-support products shown
in Eq.~\eqref{eq:crown-F0} occur.  Here the stationary point is isolated in
the four ratio variables, so no ratio-tangent direction remains; the four
path scales \(h_i\) are the tangent coordinates.

The scaling is fixed before introducing a local chart.  The local LP analysis
uses the same external-virtuality expansion as for the Crown.  At finite
regulator, take the four legs attached to \(A_i\) slightly off shell,
\(\widehat p_i^{\,2}=\delta\mu_i\), with fixed nonzero \(\mu_i\); the
massless on-shell point is recovered as \(\delta\to0\).  The legs attached to
\(B_1\) and \(B_2\) are kept on shell in the explicit certificate below.  An
order-\(\delta\) continuation of their virtualities would add further terms at
resolved-leading weight without changing the scaling.  The Part-II homogeneity
equations force a common edge weight \(v\).  Since
\(\mathcal F_0\) and \(\mathcal F_1\) have degree four whereas \(\mathcal U\)
has degree three, equality at resolved-leading weight gives
\(1+4v=3v\), and hence \(v=-1\).  Together with the Part-I decomposition
above, Part~II therefore determines the HR vector and both weights
simultaneously:
\begin{align}
 \mathcal F_\star&=\mathcal F_0=\mathcal F_{\rm SL},
 &\mathcal F_{\rm Obs}&=0,\nonumber\\
 \vHR
 &=(-1,-1,-1,-1,-1,-1,-1,-1;1),
 &(\WSL,\WHR)&=(-4,-3).
 \label{eq:K42-six-point-HR-data}
\end{align}
In particular, every edge parameter scales as \(x_{ia}\sim\delta^{-1}\).

The normal weights now follow directly from the second line of
Eq.~\eqref{eq:K42-six-point-Hessian}.  Write
\(\Delta z_i\sim\delta^{\alpha_i}\).  The core scaling just found gives
\(h_k\sim\delta^{-1}\), so the prefactor \(\prod_k h_k\) has weight \(-4\),
and its term proportional to \(G_{ij}\Delta z_i\Delta z_j\) has weight
\(-4+\alpha_i+\alpha_j\).  Requiring this term at the resolved-leading weight
\(\WHR=-3\) gives \(\alpha_i+\alpha_j=1\).  In the generic domain all six
off-diagonal terms occur; equivalently, all six generator products
\(g_{ij}^{(6)}\) must be resolved.  The resulting equations for every \(i<j\)
have the unique solution \(\alpha_i=1/2\) for all four factors.  Writing the path sums of
Eq.~\eqref{eq:K42-path-coordinates} as \(h_i=\delta^{-1}H_i\), with \(H_i\)
of order one, therefore gives the chart
\begin{align}
 x_{i1}&=\delta^{-1}H_i
   \bigl(1-z_{\star i}-\delta^{1/2}y_i\bigr),\nonumber\\
 x_{i2}&=\delta^{-1}H_i
   \bigl(z_{\star i}+\delta^{1/2}y_i\bigr)
 \label{eq:K42-six-point-local-chart}
\end{align}
with \(H_i,y_i=\mathcal O(1)\).  The same powers can be reconstructed from the
collinear momentum scaling, but they have been derived here from the
parameter-space HRF conditions.  Substituting the chart into the exact
polynomials in Eq.~\eqref{eq:K42-six-point-UF} gives the resolved-leading layer
\begin{equation}
 \mathcal P_{\rm loc}
 =\delta^{-3}\left[
 e_3(\boldsymbol H)
 +\left(\prod_{k=1}^4H_k\right)
  \left(
   \sum_{i,j=1}^4G_{ij}y_i y_j
   +\sum_{i=1}^4\mu_i z_{\star i}(1-z_{\star i})
  \right)
 +\mathcal O(\delta^{1/2})\right].
 \label{eq:K42-six-point-local-leading-layer}
\end{equation}
Here the quadratic term in the \(y_i\) comes from \(\mathcal F_0\).  The final
sum in parentheses is the contribution of the explicit \(p_i^2\) term in
Eq.~\eqref{eq:K42-six-point-UF}, with
\(\widehat p_i^{\,2}=\delta\mu_i\), evaluated at the pinch.  If this final,
\(y_i\)-independent sum were absent, the remaining two terms would be
homogeneous under
\(H_i\mapsto\lambda H_i\), \(y_i\mapsto\lambda^{-1/2}y_i\): both
\(e_3(\boldsymbol H)\) and
\((\prod_iH_i)\sum_{i,j}G_{ij}y_i y_j\) scale as \(\lambda^3\).  The
off-shell contribution instead scales as \(\lambda^4\), and therefore breaks
this residual homogeneity and makes the local region scaleful.  The leading
exponent support in
Eq.~\eqref{eq:K42-six-point-local-leading-layer} has full affine rank in the
eight local coordinates, thereby certifying the Part-II solution in
Eq.~\eqref{eq:K42-six-point-HR-data}.
The four cancellation-normal widths supply the factor \(\delta^{4/2}\) in the
measure.  For unit propagator powers and \(D=4-2\epsilon\), the scalar power
count is
\begin{equation}
 I_{K_{4,2}}^{\rm scalar}\sim
 \delta^{-8}\,\delta^{4/2}
 \bigl(\delta^{-3}\bigr)^{-D/2}
 =\delta^{-3\epsilon}.
 \label{eq:K42-six-point-power-count}
\end{equation}
The six-point \(K_{4,2}\) configuration is therefore a certified
three-loop six-point wide-angle Landshoff seed.

The five-point configuration behaves differently.  Removing the \(p_6\) leg
sets \(\widehat p_6=0\), and momentum conservation gives
\(\widehat p_5=-\sum_i\widehat p_i\).
Equation~\eqref{eq:K42-six-point-stationary-point}
then reduces to
\begin{equation}
 \sum_{i=1}^4(z_i-1)\widehat p_i=0.
 \label{eq:K42-five-point-balance}
\end{equation}
At generic five-point kinematics the four \(\widehat p_i\) are independent, so the
only solution is \(z_i=1\).  Equivalently, all \(x_{i1}\) vanish.  The pinch
has reached an LP-parameter boundary and the corresponding contraction is a
massless vacuum banana, hence scaleless.  There is therefore no generic
wide-angle five-point HR for this configuration.  Removing instead the leg
attached to the other \(B\) vertex gives the equivalent boundary \(z_i=0\).

On the coplanar surface, however, \(\det G=0\).  If
\(\sum_i c_i\widehat p_i=0\),
Eq.~\eqref{eq:K42-five-point-balance} admits
\begin{equation}
 z_i=1+t\,c_i .
 \label{eq:K42-five-point-coplanar-family}
\end{equation}
Physical coplanar \(2\to3\) configurations exist for which an interval of
\(t\) places every \(z_i\) strictly between zero and one.  The rank-three
normal quadratic form then gives a near-coplanar five-point HR.  Thus the
same topology has no interior HR at generic five-point kinematics, but does
have one after expanding about the coplanar Gram surface.

This mechanism has a close parallel in the wide-angle Fish of
Sec.~\ref{sec:five-point-near-planar-seed}.  Both problems require the
Gram-degenerate surface \(\Gamma_5=0\), but for different algebraic reasons.
For the Fish, the ratio-stationary point exists at generic kinematics, while
the polynomial evaluated there is proportional to \(\Gamma_5\), as in
Eq.~\eqref{eq:five-stationary-gram}; planarity removes this nonzero
obstruction.  Here generic stationarity instead confines the pinch to the
\(z_i=1\) boundary, while the rank loss at \(\Gamma_5=0\) supplies the null
direction that carries it into the positive interior.  In both cases the
physical five-point pinch has three independent ratio-normal directions,
although in the present \(K_{4,2}\) problem the fourth ratio direction is
tangent to the coplanar family in Eq.~\eqref{eq:K42-five-point-coplanar-family}.

This comparison suggests that massless five-point wide-angle HRs may require
Gram-degenerate kinematics, but the examples do not establish such a general
theorem.  All certified five-point Fish regions in Sec.~\ref{sec:example-five-point}
approach its positive planar singular family or a boundary stratum of its
closure.  Near-planarity need not, however, be imposed as a separate
expansion: the spacelike-collinear and central-soft limits reach that locus
or its boundary through their collinear or soft degeneration.

The near-planar five-point \(K_{4,2}\) region may likewise serve as the
ancestor of regions in further composite kinematic limits.  Establishing
such descendants would require analysing each specialised expansion and its
positive cancellation locus separately; we do not pursue that question here.

Taken together, these configurations show that the external-leg attachment
is part of the Landau classification: the same \(K_{4,2}\) incidence core
leads to distinct stationary loci and positivity properties.  Each attachment
must therefore be tested separately against the positivity, scaling, facet
and scalefulness conditions of Sec.~2.

\subsection{Further complete-bipartite cases}

The \(K_{4,3}\) graph in Fig.~\ref{fig:complete-bipartite-cores}(d) is the
natural next complete-bipartite continuation of the Crown.  This incidence
topology already appears in Fig.~14 of
Botts and Sterman~\cite{Botts:1989kf} as the independent-scattering diagram
for baryon--baryon elastic scattering: its four external baryonic vertices
are connected to three hard quark--quark scattering subgraphs.  The
corresponding scalar topology does support a generic wide-angle Landshoff
HR.  To see this directly, attach the four external legs to the four-vertex
partition and take the on-shell expansion
\(p_i^2=\delta P_i^2\), with fixed \(P_i^2>0\).  The leading two-forest
polynomial can be described without displaying its 72 monomials.  First make
the relation between the twelve edge variables and the ratio coordinates
explicit.  Retaining a symmetric row--column normalisation, an exact chart
of the positive orthant is
\begin{equation}
 X=(x_{ia})=
 \begin{pmatrix}
  u_1w_1R_{11}&u_1w_2R_{12}&u_1w_3\\
  u_2w_1R_{21}&u_2w_2R_{22}&u_2w_3\\
  u_3w_1R_{31}&u_3w_2R_{32}&u_3w_3\\
  u_4w_1&u_4w_2&u_4w_3
 \end{pmatrix}.
 \label{eq:K43-explicit-matrix-chart}
\end{equation}
Here the \(u_i\) and \(w_a\) are the tangent variables introduced in
Eq.~\eqref{eq:Kmn-rank-one-parametrisation}; setting all \(R_{ia}=1\) reduces
Eq.~\eqref{eq:K43-explicit-matrix-chart} to that rank-one parametrisation.
The chart has the single redundancy
\(u_i\mapsto\alpha u_i\), \(w_a\mapsto w_a/\alpha\).  If it is fixed by
\(w_3=1\), then \(u_i=x_{i3}\), \(w_a=x_{4a}/x_{43}\) for \(a=1,2\), and the
six \(R_{ia}\) are precisely the ratios in
Eq.~\eqref{eq:Kmn-cycle-ratios}.  Thus
Eq.~\eqref{eq:K43-explicit-matrix-chart} is an exact change of coordinates on
the positive orthant: the six tangent coordinates
\((u_1,\ldots,u_4,w_1,w_2)\) and the six normal ratios \(R_{ia}\) account for
the twelve original variables \(x_{ia}\).  On \(C_{\rm inc}\), the six
conditions \(R_{ia}=1\) leave the tangent coordinates free, and hence
\[
 \operatorname{codim}C_{\rm inc}=6=(4-1)(3-1)
 =12-7+1=L,
\]
in agreement with the general count following
Eq.~\eqref{eq:Kmn-cycle-ratios}.  The last equality is the connected-graph
loop formula \(L=E-V+1\), with twelve edges and seven vertices.

For a spanning two-forest \(\mathcal T_{ij|kl}\) which separates external
vertices \(i,j\) from \(k,l\), define
\begin{equation}
 \Phi_{ij|kl}
 =\sum_{\mathcal T_{ij|kl}}
   \prod_{(r,a)\notin\mathcal T_{ij|kl}}x_{ra} .
 \label{eq:K43-channel-forest-polynomials}
\end{equation}
Each of the three channel polynomials contains 24 monomials.  Using
\(s_{13}=-(s_{12}+s_{23})\), the superleading polynomial is
\begin{align}
 \FSL=\mathcal F_0
 &=
 s_{12}\Phi_{12|34}+s_{23}\Phi_{14|23}
 +s_{13}\Phi_{13|24}
 \nonumber\\
 &=
 s_{12}\bigl(\Phi_{12|34}-\Phi_{13|24}\bigr)
 +s_{23}\bigl(\Phi_{14|23}-\Phi_{13|24}\bigr),
 \qquad \FObs=0 .
 \label{eq:K43-FSL-channel-form}
\end{align}
This is the compact expression for \(\FSL\) used below.  On the rank-one
incidence locus \(C\) parametrised in
Eq.~\eqref{eq:Kmn-rank-one-parametrisation}, substituting
\(x_{ia}=u_iw_a\) into the 24 monomials of each channel polynomial gives the
same positive polynomial,
\begin{equation}
 \left.\Phi_{12|34}\right|_C
 =\left.\Phi_{14|23}\right|_C
 =\left.\Phi_{13|24}\right|_C
 =2\!\left(\prod_{i=1}^4u_i\right)
 e_3(\boldsymbol u)(w_1w_2w_3)^2(w_1+w_2+w_3),
 \label{eq:K43-common-channel-forest}
\end{equation}
where \(e_3(\boldsymbol u)=\sum_{i<j<k}u_i u_j u_k\).  The common result is
symmetric in the four \(u_i\) and independent of the \(2|2\) channel
partition.  Through Eq.~\eqref{eq:K43-FSL-channel-form},
Eq.~\eqref{eq:K43-common-channel-forest} gives
\(\mathcal F_0|_C=0\).  Direct differentiation of the full channel
polynomials, followed by restriction to \(C\), shows that their first normal
derivatives agree there as well.  Equivalently, after localising to the
positive chart,
\begin{equation}
 \Phi_{12|34}-\Phi_{13|24},\quad
 \Phi_{14|23}-\Phi_{13|24}
 \ \in\ I_2(X)^2 .
 \label{eq:K43-channel-differences-square-ideal}
\end{equation}
Membership in \(I_2(X)^2\) means that both channel differences and all their
first derivatives vanish on \(C\).  Equation~\eqref{eq:K43-FSL-channel-form}
therefore gives \(\mathcal F_0|_C=d\mathcal F_0|_C=0\), so \(C\) is the
positive Landau stationary locus
\begin{equation}
 x_{ia}=u_i w_a>0,
 \qquad
 R_{ia}=\frac{x_{ia}x_{43}}{x_{i3}x_{4a}}=1,
 \qquad i=1,2,3,\qquad a=1,2.
 \label{eq:K43-positive-rank-one-locus}
\end{equation}
The six displayed ratios are independent normal coordinates.  Their
quadratic layer is nondegenerate throughout the physical domain.  With the
tangent coordinates \(u_i,w_a\) held fixed, define the normal Hessian by
\[
 (H_R)_{ia,jb}
 :=\left.
 \frac{\partial^2\mathcal F_0}{\partial R_{ia}\,\partial R_{jb}}
 \right|_{R_{kc}=1},
 \qquad i,j=1,2,3,\qquad a,b=1,2.
\]
Its determinant is
\begin{align}
 \det H_R={}&4s_{12}^2s_{23}^2(s_{12}+s_{23})^2(w_1w_2w_3)^{12}
 \bigl(w_1w_2+w_1w_3+w_2w_3\bigr)^3
 \prod_{i=1}^4u_i^6
 \nonumber\\[-1mm]
 &\times
 \bigl(u_1u_2u_3+u_1u_2u_4+u_1u_3u_4+u_2u_3u_4\bigr)^6,
 \label{eq:K43-normal-Hessian}
\end{align}
which cannot vanish for \(u_i,w_a>0\) and \(s_{12}>-s_{23}>0\).  The six normal
stationarity equations are consequently independent, and \(C\) is locally
the full positive Landau component rather than a sublocus of a larger
stationary component.

The Part-II homogeneity equations give a uniform edge weight \(v\).  Since
\(\mathcal U\) has weight \(6v\), whereas
\(\delta\mathcal F_1\) has weight \(1+7v\), their balance fixes \(v=-1\)
and the resolved-leading weight to \(-6\).  Individual monomials of
\(\mathcal F_0=\FSL\) then have weight \(-7\).  Its first nonzero normal layer is
quadratic, so a normal width \(R_{ia}-1\sim\delta^\alpha\) enters at weight
\(-7+2\alpha\).  Equating this to \(-6\) fixes \(\alpha=1/2\).  The resulting
local scaling is
\begin{equation}
 x_{ia}=\delta^{-1}u_i w_a R_{ia},
 \qquad R_{ia}=1+\delta^{1/2}y_{ia},
 \qquad
 \vHR=(\underbrace{-1,\ldots,-1}_{12\text{ edge components}};1),
 \label{eq:K43-HR-scaling}
\end{equation}
where ratios not listed in Eq.~\eqref{eq:K43-positive-rank-one-locus} are
fixed to one.  The three contributions at weight \(-6\) are therefore
\(\mathcal U\), \(\delta\mathcal F_1\), and the quadratic normal layer of
\(\mathcal F_0\):
\begin{equation}
 \mathcal P
 =\delta^{-6}\left[
   \mathcal U\big|_C+\mathcal F_1\big|_C
   +\frac12\,\boldsymbol y^T H_R\boldsymbol y
   +\mathcal O(\delta^{1/2})\right],
 \qquad (\WSL,\WHR)=(-7,-6).
 \label{eq:K43-local-leading-polynomial}
\end{equation}
Here \(C\) denotes the rank-one locus.  The leading polynomial has affine
exponent rank twelve in the thirteen local variables
\((u_i,w_a,y_{ia})\); the sole null direction is the coordinate redundancy
\(u_i\mapsto\alpha u_i\), \(w_a\mapsto w_a/\alpha\).  There is therefore no
additional rescaling which could make the region scaleless.  Finally, for
unit propagator powers and \(D=4-2\epsilon\), the twelve edge measures, the
six cancellation widths and \(\mathcal P^{-D/2}\) give
\begin{equation}
 I_{K_{4,3}}^{\rm scalar}\sim
 \delta^{-12}\,\delta^{6/2}\,
 \bigl(\delta^{-6}\bigr)^{-D/2}
 =\delta^{3-6\epsilon}.
 \label{eq:K43-scalar-power-count}
\end{equation}
Thus the positive Landau locus, the local lower facet and the absence of an
extra scaling null direction all agree: \(K_{4,3}\) is a certified
six-loop wide-angle Landshoff seed.

The \(K_{3,3}\) calculation gives a complementary warning.  To test whether
its two-forest polynomial can exhibit cancellation beyond quadratic order,
it is enough to find one consistent kinematic slice.  We choose the slice
invariant under simultaneous permutations of the two three-vertex
partitions,
\begin{equation}
 s_{i,3+a}=\begin{cases}p,&i=a,\\ q,&i\ne a,\end{cases}
 \qquad
 \operatorname{Gram}_5=-\frac{3}{16}p^2(p-4q)^2(p+2q),
 \label{eq:K33-symmetric-kinematics}
\end{equation}
which is realised at the symmetric point of a planar cyclic \(3\to3\)
kinematic family.  This restriction is used only to establish the existence
of an algebraic higher-order stationary locus, not to classify general
\(K_{3,3}\) kinematics.

On the nondegenerate Gram branch \(p=4q\), write the edge-parameter matrix
\(X=(x_{ia})\) in the same symmetry class as
\(x_{ii}=A\) and \(x_{ia}=B\) for \(i\ne a\).  The exact polynomial and a
representative of each of the two derivative orbits reduce to
\begin{align*}
 \left.\mathcal F\right|_{\rm sym}
 &=3qB^2(A+2B)^3,\\
 \left.\frac{\partial\mathcal F}{\partial x_{ii}}\right|_{\rm sym}
 &=3qB^2(A+2B)^2,&
 \left.\frac{\partial\mathcal F}{\partial x_{ia}}\right|_{\rm sym}
 &=qB(A+2B)^2(A+5B),\qquad i\ne a.
\end{align*}
For \(qB\ne0\), the stationary branch is \(A=-2B\).  Removing the projective
scale gives
\begin{equation}
 X_\star\propto
 \begin{pmatrix}-2&1&1\\1&-2&1\\1&1&-2\end{pmatrix}.
 \label{eq:K33-cubic-stationary-ray}
\end{equation}
There \(\mathcal F\), its gradient and its Hessian vanish, while the third
derivative in a transverse direction does not.  Hence the polynomial vanishes
to cubic order.
The rows and columns of \(X_\star\) sum to zero, however, so no nonzero point
on this ray lies in the positive orthant.  The two-forest structure therefore
permits a higher-order cancellation algebraically, but positivity still
excludes it from the first sheet.

Table~\ref{tab:complete-bipartite-landshoff-status} summarises the certified
cases and separates them from algebraic stationary loci which fail the
positive-orthant requirement.
\begin{table}[H]
\centering
\small
\begin{tabular}{@{}
 >{\centering\arraybackslash}p{.12\textwidth}
 >{\raggedright\arraybackslash}p{.35\textwidth}
 >{\raggedright\arraybackslash}p{.39\textwidth}@{}}
\toprule
core and loop order & locus and domain & present status\\
\midrule
\(K_{2,2}\)\par{\scriptsize \(L=1\)}
& hard-vertex momentum-balance locus; two linear defining factors; positive
  on the twisted-box boundary at the physical self-crossing surface
  \eqref{eq:hexagon-self-crossing-surface}
& certified six-point wide-angle HR; no positive pinch at generic wide-angle
  kinematics.  Its NMRK and DSC specialisations inherit the same two
  factors\\\addlinespace[3pt]
\(K_{3,2}\)\par{\scriptsize \(L=2\)}
& hard-vertex momentum-balance locus; kinematically deformed path ratios;
  positive on the near-planar surface \(\Gamma_5=0\)
& certified five-point Fish Landshoff HR
  (Figs.~\ref{fig:complete-bipartite-cores}(a) and
  \ref{fig:five-point-near-planar-attachment}); no generic wide-angle
  HR\\\addlinespace[3pt]
\(K_{4,2}\), 4 pt\par{\scriptsize \(L=3\)}
& hard-vertex momentum-balance locus; common path ratio; positive at generic
  wide-angle kinematics
& certified four-point Crown HR
  (Figs.~\ref{fig:complete-bipartite-cores}(b) and
  \ref{fig:crown-graph}); the same core organises its
  audited boundary descendants and Regge HRs\\\addlinespace[3pt]
\(K_{4,2}\), 5 pt\par{\scriptsize \(L=3\)}
& hard-vertex momentum-balance locus \(Q(\boldsymbol z)=0\); no positive
  interior solution at generic wide-angle kinematics; positive on the
  coplanar Gram surface
& the generic stationary point lies on a scaleless LP-parameter boundary;
  the coplanar solution is a certified HR,
  Eq.~\eqref{eq:K42-five-point-coplanar-family}\\\addlinespace[3pt]
\(K_{4,2}\), 6 pt\par{\scriptsize \(L=3\)}
& hard-vertex momentum-balance locus \(Q(\boldsymbol z)=0\); positive at
  generic wide-angle kinematics
& certified six-point Landshoff HR in the open physical domain
  \eqref{eq:K42-six-point-positivity}\\\addlinespace[3pt]
\(K_{3,3}\)\par{\scriptsize \(L=4\)}
& higher-order algebraic stationary locus; special planar slice; mixed-sign
  stationary parameter matrix
& not a first-sheet Landau singularity and cannot generate an HR; see
  Fig.~\ref{fig:complete-bipartite-cores}(c)\\\addlinespace[3pt]
\(K_{4,3}\)\par{\scriptsize \(L=6\)}
& rank-one incidence locus \(x_{ia}=u_i w_a\); positive at generic
  wide-angle kinematics
& certified four-point Landshoff HR with six independent quadratic
  cancellation normals; this is the Botts--Sterman independent-scattering
  topology; see
  Fig.~\ref{fig:complete-bipartite-cores}(d)\\
\bottomrule
\end{tabular}
\caption{Status of the complete-bipartite Landshoff organisation.}
\label{tab:complete-bipartite-landshoff-status}
\end{table}

\subsection{Incidence cores and physical hidden regions}
\label{sec:incidence-cores-physical-HRs}

The topological analysis separates three levels intertwined in the examples.
The incidence locus \(C_{\rm inc}\) records the
path-sharing compatibility intrinsic to a complete-bipartite core.  Once the
collinear lines are removed, the \(B_a\) components are mutually disconnected
hard subdiagrams --- the defining Landshoff structure.  The paths from each
factor-localisation vertex \(A_i\) distribute its collinear momentum among
them, and the incidence conditions enforce compatible momentum fractions
across the hard components.  The
rank-one form in Eq.~\eqref{eq:Kmn-rank-one-parametrisation} has one normal
coordinate for each independent graph cycle, and hence codimension equal to
the loop order, as shown in Eq.~\eqref{eq:Kmn-determinantal-codimension}.
The full physical pinch additionally obeys the remaining stationary and
vanishing Landau equations and must lie in the positive physical domain.
Only after a kinematic expansion is specified do the aligned LP polynomial
and the HRF hierarchy determine the active cancellation ideal, region vector
and cancellation depth.  Topology therefore organises the possible
mechanism, but does not by itself determine the HR.

The three established \(K_{m,2}\) seeds exhibit this distinction
quantitatively in Eq.~\eqref{eq:seed-cycle-normal-counts}.  For the Crown,
the incidence locus is already the LP-parameter part of the physical pinch,
and both have codimension three.  The Fish has two incidence normals, whereas
a third stationary equation fixes the collective path-sharing coordinate;
the separate condition \(\Gamma_5=0\) is also needed to remove the kinematic
obstruction.  The twisted box has one incidence normal, while its two
hard-vertex momentum-balance factors also fix the collective ratio and hence
define a codimension-two pinch on the self-crossing surface.  The two factors
illustrate that topology carries the incidence path-compatibility, not
necessarily the complete factor set or its scaling.

The external-leg analysis of \(K_{4,2}\) gives a direct control on the scope
of this organisation.  The same incidence core supports the generic
four-point Crown HR and a generic six-point HR, but its five-point
configuration has no positive interior solution at generic wide-angle
kinematics and acquires one only on the coplanar Gram surface.  The
\(K_{3,3}\) example provides a complementary control: its two-forest
polynomial admits a cubic stationary locus algebraically, but the stationary
matrix has mixed signs and therefore lies outside the first sheet.  Thus the
incidence topology, the external-leg assignment, the kinematic surface and
positivity all enter the Landau classification.

New wide-angle \(K_{4,2}\) and \(K_{4,3}\) seeds are certified; their
specialised descendants remain to be analysed.  Contraction minors instead
relate different graphs.  When the contracted graph reproduces the seed with
compatible external attachments, the seed loop-momentum routing lifts directly
and its HR survives on the corresponding, possibly deeper, contraction
stratum.  This label compatibility is automatic for the symmetric Crown but is
a genuine restriction for the Fish and twisted-box descendants.

The Crown gives the strongest quantitative evidence: a Crown minor occurs in
all 1081 four-loop topologies selected by the mixed-sign screen
of Ref.~\cite{Gardi:2024axt}.  The present audit certifies representative
descendants and excludes all 16 topologies without it; the remaining Crown
graphs inherit the boundary HR from the Crown minor, even though those
strata were not enumerated topology by topology.  The Regge audit supports
extending this Crown-minor criterion beyond the audited loop orders.

These results motivate two complementary conjectural principles.  The first is
topological: the minimal massless wide-angle Landshoff seeds that support HRs
possess a complete-bipartite incidence core, with path-sharing or hard-vertex
momentum-balance coordinates providing the natural normals.  This includes
seeds whose HRs are confined to special physical kinematic surfaces, such as
the Fish on a near-planar surface and the twisted box on a self-crossing
surface; it does not assert that their pinches persist at generic kinematics.
Within the audited four-point class, the more specific Crown-minor conjecture
is an inheritance criterion of this topological type: a Crown contraction
minor controls both wide-angle and Regge HRs.

The second principle concerns the local analytic type of the pinch.  Every
genuine interior HR encountered so far arises from an ordinary transverse
first-sheet pinch: after the stationary conditions remove the constant and
linear terms, the graph polynomial retains a nonzero quadratic dependence on
displacements normal to the pinch surface.  If
\(I=\langle f_1,\ldots,f_r\rangle\) is the saturated ideal generated by local
normal coordinates, the invariant conjecture is
\begin{equation}
 \mathcal F_{\rm SL}\in I^2\setminus I^3.
 \label{eq:ordinary-positive-HR-conjecture}
\end{equation}
In words, \(\mathcal F_{\rm SL}\) vanishes exactly to quadratic order normal to
the pinch; cubic and higher-order terms may be present as well.
The two parts have different status.  For a smooth pinch component, the Landau
equations prove membership in \(I^2\), as shown in
Appendix~\ref{app:squared-ideal}.  The conjectural statement is that the class
in \(I^2/I^3\) is nonzero, or equivalently that the transverse quadratic form
does not vanish completely.  At the level of the presentations used by HRF,
every generator needed in the examples contains precisely two distinct
cancellation factors with non-overlapping LP-parameter support.  This is
compatible with quadratic transverse vanishing, but is not its invariant
content: generator presentations are not unique, and terms in \(I^3\) may
coexist with the nonzero quadratic class in \(I^2/I^3\).

The \(K_{3,3}\) stationary locus shows why the physical qualification matters:
its two-forest polynomial reaches cubic transverse order only outside the
positive orthant.  Thus any proof of the nonvanishing quadratic class must use
more than multiaffinity alone; it must also incorporate graph-polynomial
support, connectivity and the positive physical domain.

\section{Conclusions and outlook}
\label{sec:conclusions}

We have formulated the Hidden Region Finder as a procedure for addressing a
central problem in applying the Method of Regions to non-Euclidean Feynman
integrals: how to find and certify the complete set of regions when
scaling and cancellations jointly determine which terms control the
asymptotic expansion.  The geometric Method of Regions
retains the exponent support of the graph polynomial but is insensitive to
the signs and kinematic dependence of its coefficients.  HRF
supplies this missing information by extracting candidate cancellation
factors together with the scaling that enhances their constituent monomials
relative to the remaining terms.  On the pinch, these enhanced monomials
cancel, so that their sum contributes at the same order as any non-cancelling
obstruction and the other dominant terms of the complete Lee--Pomeransky (LP)
polynomial.  A candidate is retained
only when the cancellation factors have a common zero in the positive LP domain
and the resulting integral is scaleful.  A local dissection around the
cancellation locus identifies the relevant lower-facet normals.  Their
consistent pullback to the original LP parameters provides the final
geometric certificate.

The examples show why each part of this construction is needed.  Hidden
regions may have a nonzero obstruction, a non-uniform scaling, multi-term
polynomial cancellation factors or several generators.  The five-point
sector already exhibits almost this entire range.  The Fish (the non-planar
double box) was identified in the recent Letter~\cite{Chen:2026dnj} as the
unique two-loop source of kinematics-dependent factorisation-violating
contributions in the spacelike-collinear limit.  In that limit it features a
nonzero obstruction and a non-uniform vector, and both the seed and its
vertex-corrected descendant involve kinematics-dependent cancellation factors
that must be tested in the physical domain.  For the same seed in the
near-planar wide-angle limit, suitable combinations of polynomial derivatives
are required to reveal all cancellation factors, yielding several generators.
In its simultaneous MRK and near-planar limit, contributions from several
orders of the kinematic expansion cancel on the same pinch and all affect the
final scaling.  Its central-soft limit separately requires asymptotic orders
to be aligned before the cancellation structure can be determined.  Alignment
makes the cancelling terms comparable, while restoring the remaining
contributing orders ensures that the full kinematic expansion determines the
final scaling.  In all such examples the active cancelling contributions
saturate the hierarchy bound and enter at the same resolved weight.  This is
not a requirement of HRF: Eq.~\eqref{eq:finalhierarchy} allows a resolved
contribution to enter above \(\WHR\).  The observed saturation is instead an
empirical finding and supplies the dominant balance through which the combined
leading support becomes scaleful.

Independent of this hierarchy problem, a distinct completeness issue is
whether an HR appears only on a graph boundary, after one or more propagators
have been contracted.  The absence of an interior candidate therefore does
not imply the absence of an HR: a complete analysis must also scan these
contraction boundaries.  An HR of a larger graph can originate in a seed
subgraph exposed by contracting additional propagators.  When inequivalent
external attachments are not related by a symmetry, they must also be
preserved; this defines an externally labelled contraction minor.  For the
Crown this qualification is automatic, and
Ref.~\cite{Gardi:2024axt} demonstrated this mechanism for enlarged three-loop
graphs with additional propagators, at the same loop order as the seed; here
we extend it to representative four- and five-loop descendants.  The same
mechanism exposes the twisted-box seed inside the six-point twisted hexagon
and compatible descendants.  Its wide-angle HR lies on the self-crossing
surface; in the six-point NMRK and DSC specialisations, order alignment and
restoration of all contributing orders determine its realisation.

\medskip
\noindent\textbf{Structural conclusions and conjectures.}\par\nopagebreak\smallskip\nopagebreak

Taken together, the examples suggest two complementary organising principles.
First, massless wide-angle HRs originate from Landshoff seeds with
complete-bipartite incidence cores.  The evidence is not confined to the
Crown: the mechanism recurs in the five-point Fish and six-point twisted-box
families, in the additional complete-bipartite seeds, and in examples extending
through six loops.  Its appearance across external multiplicities, loop orders
and kinematic limits suggests that it is a widespread source of HRs, with
potential relevance to many scattering processes.  Within the audited
four-point class, the principle is realised concretely through inheritance from
a Crown contraction minor.  Second, every genuine interior HR found
here is associated with an ordinary transverse first-sheet pinch: after the
stationary conditions remove the constant and linear terms, a nonzero
quadratic dependence normal to the pinch remains.  In every admissible HRF
decomposition found here, each required generator has the form \(f_i f_j\),
with two distinct cancellation factors of non-overlapping LP-parameter
support.  The persistence of this pair-product structure across all examples
is a valuable empirical regularity.  These principles constrain, but do not
alone determine, the
existence of an HR, which also depends on the external-leg assignment, the
kinematic limit and positivity.  Both principles are formulated and supported
in Sec.~\ref{sec:incidence-cores-physical-HRs}.  The local ideal-theoretic
statement underlying the second is explained in
Appendix~\ref{app:squared-ideal}.

The Fish topology in the spacelike-collinear limit also shows that one HR can
admit several local coordinate descriptions.  A fixed cancellation depth may
be distributed continuously among several directions normal to the pinch.  A
centred chart treats their simultaneous approach, whereas endpoint charts
isolate limiting directions; all describe the same HR after translation back to the original LP
parameters.  Endpoint charts can be particularly efficient for evaluating
integrals, but may overlap ordinary facet regions and display rapidity
divergences absent from a centred presentation.  Thus the separation into
regional contributions, including its overlap prescription, must be made
after the local coordinates are chosen.  The individual regional terms and
their regulators may therefore depend on the dissection chart, while their
properly combined expansion does not.

\medskip
\noindent\textbf{Seed inheritance and kinematic specialisation.}\par\nopagebreak\smallskip\nopagebreak

The HRs of the Crown, Fish and twisted-box seeds in wide-angle kinematics have
a common Coleman--Norton backbone.  At the positive physical pinch, the active
massless propagators describe on-shell classical
trajectories~\cite{Coleman:1965cn,Collins:2020euz}.  At each external
splitting vertex the two null propagator momenta are related by momentum
conservation to the attached null external momentum; on the physical branch
this places the internal lines on its collinear ray.  The momentum fractions
carried by the collinear daughters must then satisfy momentum conservation
simultaneously at both hard vertices.  In parameter space, this requirement is
encoded by the path-sharing or hard-vertex momentum-balance cancellation
factors.  Thus the physical Landau pinches
determine the collinear-mode assignments of all three seed families.

The six-point family gives the clearest chain of kinematic inheritance.  The
full one-loop twisted hexagon has no positive pinch at generic wide-angle
kinematics, but its contraction stratum \(x_0=x_3=0\) exposes the
\(K_{2,2}\) twisted-box seed, which has a certified HR on the physical
self-crossing surface.  The HRs of this seed in zero-recoil NMRK and DSC are
aligned degenerations of its wide-angle cancellation locus.  Their construction
requires asymptotic-order alignment before the cancellation structure and
scaling can be determined.  In the DSC presentation, contributions from more
than one order of the kinematic expansion must also be retained.  Thus the
cancellation geometry is inherited after specialisation, but the final region
vector and cancellation depth must be determined anew in the specialised
expansion.

The momentum-space reconstruction sharpens the comparison among three limits
of the twisted-box seed.  Its wide-angle, zero-recoil NMRK and DSC HRs all
contain one independent Glauber loop.
The two hard-vertex momentum-balance factors constrain its two longitudinal
directions.  At wide angle the Glauber loop
is the common fluctuation of four collinear propagators and no individual
propagator is Glauber, whereas in NMRK and DSC \(x_0\) is an explicit Glauber
propagator.  The degeneration therefore changes the support and diagrammatic
realisation of the loop rather than creating it.  Across the three regions
the transverse-dominance excess \(a+b-2c\) equals the cancellation depth,
with respective values \(1\), \(4\) and \(2\), as shown in
Sec.~\ref{sec:momentum-reconstruction}.

The Crown topology supplies a complementary pattern.  Its HR in generic
wide-angle kinematics has no Glauber loop, whereas the HR in its Regge
specialisation develops one.  The Crown contraction minor organises both the
wide-angle and three-channel
Regge audits, but the Regge singular locus is defined by fewer equations and its
region vector is fixed anew by the specialised kinematics.  This is a second
example in which the wide-angle seed organises the limiting region without
making the two HRs equivalent.

For the Fish topology, the HR in near-planar wide-angle kinematics provides
the origin of HRs in a variety of specialised limits that expose near-planar
configurations.  These include the spacelike-collinear, central-soft and
simultaneous MRK--near-planar limits.  The corresponding HRs approach the
positive planar singular family or strata of its closure, but their
cancellation factors, region vectors and momentum modes differ.
A controlled derivation of these rapidity-ordered regions from the wide-angle
limit of the Fish remains open, because matching the singular loci is not enough: one must
also match the scalings and every contributing order of the specialised
expansion.

These examples separate two inheritance mechanisms.  Graph-theoretic
inheritance can expose a seed through a compatible contraction minor, both in
graphs with additional propagators at the same loop order and in higher-loop
topologies.  Kinematic specialisation and order alignment can instead carry a
physical pinch into a Regge, collinear or correlated limit.  The mechanisms can occur together, as in the twisted
hexagon, but they are logically distinct operations.  In either
case common ancestry does not identify the resulting HRs; positivity, the
external-leg assignment and the remaining edges decide whether the inherited
mechanism is realised.  Both mechanisms are encoded in the two-forest
polynomial, but the relevant cancellations and scaling are selected by the
particular kinematic expansion.

\medskip
\noindent\textbf{Momentum-space conclusions.}\par\nopagebreak\smallskip\nopagebreak

The translation from a certified parameter region to momentum space is a
separate step.\footnote{A systematic reconstruction for facet and hidden
regions is being developed in forthcoming work~\cite{GardiZhu:2026regions}.}
The parameter-space vector directly supplies the propagator
virtuality scalings, but it does not determine a basis of loop momenta or their
separate lightcone components.  Momentum-space reconstruction uses these
virtualities, together with momentum conservation and the pole structure, to
determine the conditional average loop momenta and, separately, the local
widths of their independent fluctuations.  The first conclusion is an
existence statement: in every reconstructed example we find a faithful routing
--- an adapted basis of independent loop momenta with appropriate local
lightcone directions --- in which the componentwise scaling of the average
momentum agrees with the corresponding integration width.  This agreement is
an outcome of the reconstruction, not an a priori assumption.
Linear independence alone is not sufficient, and an arbitrary loop routing
need not display this correspondence componentwise.
This includes Glauber loops,
which may be carried by a difference of edge momenta even when no individual
propagator is Glauber, or by an explicit Glauber propagator, as in the NMRK and
DSC examples.  The examples therefore provide no counterexample to
this correspondence, although they do not establish it in general.  These
widths determine the \(D\)-dependent part of the momentum-space
measure, as made explicit in Eq.~\eqref{eq:width-correlation-measure-split}.

The second conclusion concerns longitudinal correlations imposed by Landshoff
multiple scattering.  Momentum conservation correlates the fractions carried
along distinct collinear paths.  In the explicit momentum-space
reconstructions, this restricts the joint longitudinal support of the loop
modes for the Crown in wide-angle kinematics and for the Fish both in
near-planar kinematics and in the simultaneous MRK--near-planar limit.  The
incidence analysis of Sec.~\ref{sec:topological-organization} exhibits the same
path-compatibility structure in every positive complete-bipartite Landshoff
example: its defining conditions enforce compatible momentum fractions across
the disconnected hard subdiagrams.  With several independent loop modes this
becomes a cross-loop restriction.  In the one-loop twisted-box family, the
same path compatibility is encoded within the component widths of its single
loop.

The correlated support enters through a Jacobian among longitudinal
components.  It can therefore change only the integer part of the power count,
not its \(D\)-dependent coefficient, which is fixed by the transverse
integration dimensions.  When a Glauber loop is present, this same
longitudinal restriction is already incorporated in the integration width
discussed above.  Both equal and unequal longitudinal-width scalings occur.
The spacelike-collinear Fish family, its vertex-corrected descendant and the
central-soft Fish have unequal plus and minus widths.  The Regge Crown and the
wide-angle, zero-recoil NMRK and DSC twisted-box regions instead have equal
longitudinal widths in their symmetric frames.  In either pattern the
restriction is already contained in the Glauber integration width, so no
additional cross-mode factor is required.

Returning to parameter space, the correlated longitudinal support reconstructed
in momentum space is already visible before dissection as the cancellation
encoded by the incidence conditions.  The parameter-space cancellation and
the momentum-space support restriction have a common origin in the global
compatibility among distinct collinear paths.
This interpretation is an empirical conclusion
from the examples, not a general theorem, and both the average--width
correspondence and the separation of any residual correlation require an
appropriate independent-loop basis.  Whenever a complete momentum-space
reconstruction was obtained, direct power counting agreed with the
parameter-space count, including the restricted support associated with the
cancellation depth.
Because the virtuality scaling is supplied by the parameter-space region and
the component reconstruction also uses physical mode and pinch information,
this agreement is a strong consistency test rather than a wholly independent
derivation or part of the existence test for an HR.

\noindent\begin{minipage}{\textwidth}
\medskip
\noindent\textbf{Scope and outlook.}\par\nopagebreak\smallskip\nopagebreak

The present construction assumes massless propagators and scalar unit
numerators.  Numerators can alter powers but not denominator-defined regions.
Once those regions have been identified, incorporating numerator power
counting is relatively straightforward, but is essential for determining
their consequences in gauge-theory amplitudes.  Internal masses instead
introduce quadratic dependence on individual parameters and permit new
patterns of cancellation and scaling.  Treating masses is therefore a
principal extension of HRF, with direct applications to heavy-quark pair
production near threshold, where
potential modes play a central role.
\end{minipage}\par\medskip

Two complementary momentum-space programmes, both drawing on the
parameter-space method in different ways, are under way.  The first will
systematically convert the parameter-space region vectors obtained from the
geometric Method of Regions and HRF into faithful loop-momentum routings,
widths and correlations for facet and hidden regions
\cite{GardiZhu:2026regions}.  In parallel, building upon the wide-angle
graphical algorithms of Refs.~\cite{Gardi:2022khw,Ma:2023hrt}, a direct
all-order graphical description of both classes in the massless \(2\to2\)
Regge limit is being developed \cite{GardiHerzogJonesMa:2026regge}.

The construction of asymptotic expansions is one of several applications of
Landau analysis; another is the study of the discontinuity and cut structure
of Feynman integrals.  These applications are related, and we envisage using
the first-sheet analysis developed here to explore their connections in future
work.

On the theoretical side, priorities are to understand the two inheritance
mechanisms more systematically: the transmission of seed HRs through
compatible contraction minors, which is most transparent in momentum space,
and the relation between HRs obtained from different kinematic specialisations
of the same wide-angle seed.  A separate structural question is whether the
observed \(f_i f_j\) generator form and the absence of a purely cubic leading
transverse cancellation on the positive first sheet are universal, and how
the order of transverse vanishing controls the region scaling.  Broader
interior and boundary surveys provide a practical means of testing these
inheritance patterns and cancellation structures across further topologies
and kinematic limits.

On the applications side, this work is relevant to kinematic expansions of
massless scattering processes more generally, opening a broad range of
applications.  The Regge limit is one important example, relating to the
general question of how QCD amplitudes behave at high energy.  Its classical
foundations are gluon Reggeisation and BFKL evolution
\cite{Lipatov1976,Fadin:1975cb,Kuraev:1976ge,Kuraev:1977fs,Balitsky:1978ic}.
Modern analyses have approached this problem both through Glauber-operator
formulations in SCET
\cite{Rothstein:2016bsq,Moult:2022lfy,Gao:2024qsg} and through Reggeisation
and fixed-order amplitudes
\cite{DelDuca:2001gu,Bret:2011xm,Caron-Huot:2017fxr,Caola:2021izf,
Falcioni:2021dgr,Falcioni:2021buo}.  Within this broader
programme, one theoretical question is the appearance of Regge cuts, to which
HRs are expected to be relevant.  In the planar theory Regge-cut contributions
first arise at six points and have been studied extensively in the multi-Regge limit
\cite{Bartels:2008ce,Dixon:2012yy,Caron-Huot:2019vjl}; beyond the planar limit,
Regge-cut contributions occur already at four and five points
\cite{Caron-Huot:2017fxr,Falcioni:2021dgr,Caron-Huot:2020vlo,Abreu:2024xoh}.

A second direction concerns multi-collinear and soft limits and
spacelike-collinear splitting amplitudes at higher perturbative orders.
Spacelike-collinear limits are particularly interesting since, in contrast to
their timelike counterparts, they do not admit strict collinear factorisation
\cite{Catani:2011st,Dixon:2019lnw,Henn:2024qjq,Buccioni:2026mfg}.  The recent
Letter~\cite{Chen:2026dnj} showed that the entire kinematics-dependent
factorisation-violating contribution at two loops originates in a single HR,
and proposed a broader physical role for HRs: they provide the mechanism by
which kinematic limits related by crossing can cease to be analytically
connected.  For the spacelike- and timelike-collinear limits this mechanism is
direct: the HR occurs in the former and is absent in the latter.  A natural
direction is therefore to test this principle at higher perturbative orders
and in other crossing-related limits, including multi-collinear and soft
limits, by determining when crossing changes the HR content and thereby the
analytic relation between the corresponding expansions.

\newpage
HRF is directly applicable to the asymptotic analysis of jet-observable cross
sections.  For limits involving soft or collinear final-state particles, the
expansion acts directly on the external-particle degrees of freedom describing
real radiation.  At cross-section level, it must be combined with phase-space
integration and measurement restrictions, which determine which
configurations contribute.  A complementary real--virtual connection is
provided by the spacelike-collinear HR identified using HRF in the
Letter~\cite{Chen:2026dnj}: it is the virtual counterpart of the region that
restores PDF factorisation in gap-between-jets observables
\cite{Becher:2024kmk,Becher:2025igg,Becher:2026kbr}.  This connection may help
to expose the region structure underlying Glauber-sensitive logarithms in jet
observables more generally~\cite{Dasgupta:2025cgl,Banfi:2025mra}.

\section*{Code availability}

The Hidden Region Finder implementation, its user guide, the application
notebooks and the regression tests used in this work are publicly available
in Ref.~\cite{Gardi:HRFSoftware}.  The results reported here correspond to
release \texttt{v1.0.1}, distributed under the BSD 3-Clause License.

\section*{Acknowledgements}

I thank Franz Herzog, Stephen Jones, Yao Ma,  Wen Chen, Rourou Ma, Yang Zhang and
Zehao Zhu for many useful discussions and  collaboration on related projects. This research has been supported by the STFC Consolidated Grant
``Particle Physics at the Higgs Centre''. I would like to thank the CERN theoretical physics department for hospitality as a Scientific Associate during the final stages of writing this paper. I also acknowledge the substantial
assistance of OpenAI Codex with constructing, debugging and running the HRF
code and with analysing its results, as well as with manuscript editing and
consistency checks. I take full responsibility for the scientific content and final text.

\appendix
\begingroup
\raggedbottom
\section{Landau conditions and the squared cancellation ideal}
\label{app:squared-ideal}

This appendix isolates the local statement behind the product-generator
construction in Sec.~2.  Fix the kinematics and, for a boundary region, first
restrict to the corresponding contraction stratum.  Let \(A\) denote the
space of active LP parameters and let \(C\subset A\) be a smooth
component of a pinch locus contained in the positive orthant.  Suppose that
the selected cancellation factors
\begin{equation}
 \ideal_C=\langle f_1,\ldots,f_r\rangle
 \label{eq:appendix-normal-ideal}
\end{equation}
generate the radical ideal of \(C\) locally and have independent gradients
there.  We show that the Landau equations on \(C\) are equivalent to
quadratic vanishing in the square of its local defining ideal.

\paragraph{The transverse-coordinate argument.}
The implicit-function theorem allows the \(f_i\) to be completed to local
coordinates \((f_1,\ldots,f_r,y_1,\ldots,y_{N-r})\), where the \(y_a\) are
tangent to \(C\).  In these coordinates a function regular near \(C\) has
the expansion
\begin{equation}
 \FSL(f,y)
 =A(y)+\sum_{i=1}^r f_i B_i(y)
  +\sum_{1\leq i\leq j\leq r}f_if_j C_{ij}(f,y).
 \label{eq:normal-Taylor-expansion}
\end{equation}
The condition \(\FSL|_C=0\) sets \(A(y)=0\).  The stationary Landau
conditions set every derivative in a transverse direction to zero on \(C\), and hence
\(B_i(y)=0\).  The tangential derivatives already vanish because the
restriction of \(\FSL\) to \(C\) is identically zero.  Consequently
\begin{equation}
 \FSL
 =\sum_{i\leq j}f_if_j C_{ij}(f,y)
 \in\ideal_C^2.
 \label{eq:normal-Taylor-square}
\end{equation}
This gives an intuitive interpretation of the result: a stationary zero has
neither a constant displacement from the zero locus nor a linear displacement
transverse to it.  Its first possible nonzero term is quadratic in the
cancellation factors.

The individual transverse coordinates are not canonical.  They may be
changed by an invertible local transformation, and a factorised quadratic
term can be approached anisotropically.  Consequently, the weights assigned
to separate \(f_i\), and the facets exposed in a particular dissection chart,
are not invariants.  What is invariant is the common locus \(C\), the lowest
nonzero degree in its transverse Taylor expansion, and the scaling obtained after
pulling the resolved normal data back to the original LP parameters.  Ratio
coordinates parameterise the corresponding projectivised normal directions;
their endpoint charts may expose the bounding facets of a centred common
face, as in Sec.~\ref{sec:five-point-sc-charts}.

\paragraph{The ideal-theoretic statement.}
For completeness, let \(\mathcal O_{A,p}\) be the local ring of the ambient
active-parameter space at a point \(p\in C\), and use the same symbol
\(\ideal_C\) for the corresponding local ideal.  Working in this local ring
means restricting attention to a neighbourhood of \(p\) and allowing division
by functions that are nonzero there.  Because \(C\) is smoothly
embedded in the smooth space \(A\), the conormal sequence is short exact\footnote{A
sequence \(0\to A_1\to A_2\to A_3\to0\) is short exact when the first map is
injective, the second is surjective, and the image of the first equals the
kernel of the second.}
\cite{StacksConormalSequence}:
\begin{equation}
 0\longrightarrow \ideal_C/\ideal_C^2
 \xrightarrow{\ d\ }
 \Omega_A\big|_C
 \longrightarrow \Omega_C\longrightarrow0.
 \label{eq:conormal-exact-sequence}
\end{equation}
This sequence separates ambient differentials normal to \(C\), represented by
\(\ideal_C/\ideal_C^2\), from the intrinsic differentials along \(C\),
represented by \(\Omega_C\).
If \(\FSL|_C=0\), then \(\FSL\in\ideal_C\).  Its class in
\(\ideal_C/\ideal_C^2\) is mapped by the first arrow to
\(d\FSL|_C\).  The stationary Landau equations make this image zero, and
injectivity therefore gives
\begin{equation}
 [\FSL]=0\quad\hbox{in }\ideal_C/\ideal_C^2,
 \qquad\Longleftrightarrow\qquad
 \FSL\in\ideal_C^2.
 \label{eq:conormal-Landau-square}
\end{equation}
Conversely, the product rule shows immediately that every element of
\(\ideal_C^2\) and all of its first derivatives vanish on \(C\).  Thus, at a
smooth component, the Landau conditions and membership in the square of the
local defining ideal are locally equivalent.  Since all active parameters are nonzero, the
same statement follows if the Landau equations are written with logarithmic
derivatives \(x_e\partial_{x_e}\).

\paragraph{Localisation and limitations.}
Equation~\eqref{eq:conormal-Landau-square} is a local result, which is exactly
what is required for a specified physical component of the pinch.  In a
global polynomial ring it may take the form
\begin{equation}
 h\FSL\in\ideal_C^2,
 \qquad h|_C\ne0,
 \label{eq:localised-square-membership}
\end{equation}
or, equivalently, membership in the saturation of \(\ideal_C^2\) by factors
known not to vanish on that component.  The regular multipliers in
Eq.~\eqref{eq:idealpresentation} are the local coefficients produced in this
way.

Smoothness and radicality are essential to this formulation.  On a singular
or nonreduced component the conormal map in
Eq.~\eqref{eq:conormal-exact-sequence} need not be injective, so vanishing of
a function and its first derivatives does not by itself prove membership in
the ordinary square of an arbitrarily chosen defining ideal.  In such a case
one must first resolve the relevant smooth positive component, for example in
local dissection coordinates, or use the appropriate differential or
symbolic-power formulation.  For HRF this makes the Jacobian-rank test on the
candidate cancellation factors part of the certificate for applying the
two-factor construction.

Finally, the Landau equations establish only
\(\FSL\in\ideal_C^2\).  They do not imply that the quadratic class is
nonzero.  The stronger condition
\begin{equation}
 [\FSL]\ne0\quad\hbox{in }\ideal_C^2/\ideal_C^3
 \label{eq:nonzero-quadratic-class}
\end{equation}
is the statement that the transverse Hessian is nonvanishing.  It
distinguishes an ordinary quadratic pinch from a higher-order degeneracy and
is the genuinely conjectural part of Eq.~\eqref{eq:ordinary-positive-HR-conjecture}.

\endgroup
\clearpage
\section{Notation summary}

\begingroup
\normalsize
\renewcommand{\arraystretch}{1.10}
\setlength{\LTleft}{\fill}
\setlength{\LTright}{\fill}
\setlength{\LTcapwidth}{\textwidth}
\begin{longtable}{@{}p{0.38\textwidth}p{0.57\textwidth}@{}}
\caption{Principal notation used throughout the paper.}
\label{tab:notation-summary}\\
\toprule
Symbol & Meaning\\
\midrule
\endfirsthead
\multicolumn{2}{@{}l}{\tablename~\thetable\ (continued)}\\[2pt]
\toprule
Symbol & Meaning\\
\midrule
\endhead
\midrule
\multicolumn{2}{r}{Continued on next page}\\
\endfoot
\bottomrule
\endlastfoot
\(\delta\) & kinematic expansion parameter\\
\(\s\) & real kinematic coordinates; their physical domain \(\K\) is specified
  by explicit inequalities\\
\(x_e,\quad \x\equiv(x_e)_{e\in E}\) & independent LP edge parameter and the vector of all such parameters; each \(x_e\) is integrated over \((0,\infty)\)\\
\(q_e,\ \mathcal D_e=q_e^2+i0\) & oriented edge momentum and massless
  propagator denominator\\
\(\mu_{\rm LP}\) & auxiliary reference mass in the dimensionless LP
  polynomial; fixed as \(\delta\to0\) unless stated otherwise\\
\(\mathcal P_{\mu_{\rm LP}}=\mathcal U+\mathcal F/\mu_{\rm LP}^2\) &
  dimensionless Lee--Pomeransky polynomial; the scale label is suppressed
  for the default fixed choice\\
\(X[\mathcal G;\alpha]\) & graph-dependent quantity; the optional
  second entry distinguishes inequivalent regions or branches of the same graph\\
\(\mathcal P^{(R)}[\mathcal G]\) & LP polynomial of \(\mathcal G\) after the
  rescaling associated with region \(R\); parenthesised superscripts denote
  rescaled expansions, not graph identity\\
\(\deg_x\mathcal U=L,\ \deg_x\mathcal F=L+1\) & massless graph-polynomial
  degrees; both polynomials are linear in each individual \(x_e\)\\
\(m_i=c_i(\s)\delta^{a_i}\x^{\boldsymbol r_i}\) & monomial of
  \(\mathcal P\)\\
\(\boldsymbol r_i\) & length-\(N\) LP-parameter exponent vector of \(m_i\)\\
\(\vec r_i=(\boldsymbol r_i;a_i)\) & augmented exponent point\\
\(\vec v_R=(\boldsymbol v_R;1)\) & normalised augmented region vector;
  \(\boldsymbol v_R\) gives the LP-parameter exponents\\
\(\mathcal F_0\) & strict fixed-\(\x\) leading polynomial\\
\(\mathcal F^{[\boldsymbol\phi]}\) & face selected by the preliminary alignment vector
  \(\boldsymbol\phi\)\\
\(\vec\phi=(\boldsymbol\phi;1),\quad x_e=\delta^{\phi_e}y_e\) & augmented
  alignment normal and the associated first transformation\\
\(\mathcal F_\star\) & HRF starting polynomial, either
  \(\mathcal F_0\) or \(\mathcal F^{[\boldsymbol\phi]}\)\\
\(\mathcal M_\star\) & monomial support of the HRF starting polynomial\\
\(\MObs\) & obstruction monomial support selected for removal\\
\(\MSL=\mathcal M_\star\setminus\MObs\) & candidate superleading monomial
  support\\
\(\FSL\) & superleading cancellation sector\\*
\(\FObs\) & obstruction polynomial\\
\(f_j\) & non-monomial cancellation factor\\
\(C\) & selected smooth component of the common positive cancellation locus\\
\(\ideal_C\) & radical local ideal of the selected smooth pinch component\\
\(\ideal_C^2\) & square of the local defining ideal, generated locally by all
  pairwise products \(f_if_j\)\\
\(\operatorname{supp}_x(f_j)\) & LP parameters on which \(f_j\) depends\\
\(g_k=f_{i_k}f_{j_k}\) & massless pair generator with distinct factors and
  disjoint LP-parameter support\\
\(G=\{g_1,\ldots,g_{n_{\rm gen}}\}\) & candidate generator set, retained when
  it yields an admissible decomposition\\
\(n_{\rm gen}=|G|\) & number of generators in the selected presentation\\
\(\mathcal I_{\rm gen}(G)\) & candidate ideal generated by \(G\), retained
  only if the obstruction search finds an admissible decomposition\\
\(\boldsymbol v_{\rm core}^{(0)}\) & provisional face-only HRF scaling in the aligned
  variables\\
\(y_e=\delta^{(v_{\rm core})_e}z_e\) & corrected second-stage HRF scaling after all LP
  layers are restored\\
\(\boldsymbol v_{\HR}=\boldsymbol\phi+\boldsymbol v_{\rm core}\) & composition of the
  two edge-parameter transformations\\
\(\vHR=(\boldsymbol v_{\HR};1)\) & normalised total hidden-region vector; the
  last component is appended only after \(\boldsymbol v_{\HR}\) has been
  formed as the sum above, and the vector becomes certified after the final
  dissection\\
\(\mathcal P_{W_k}\) & sum of monomials in the occupied LP layer of nominal
  weight \(W_k\)\\
\(\WSL\) & common individual weight of SL monomials\\
\(\WHR\) & first resolved nonvanishing LP weight after cancellation\\
\(\WHR-\WSL\) & cancellation depth\\
\end{longtable}
\endgroup

\bibliographystyle{JHEP}
\bibliography{biblio}

\end{document}